\documentclass[11pt]{book}
\usepackage[utf8]{inputenc}
\usepackage{enumerate}
\usepackage{xcolor}
\usepackage{amsmath}
\usepackage{amsfonts}
\usepackage{amssymb}
\usepackage{capt-of}
\usepackage[a4paper]{geometry}
\usepackage{float}
\usepackage{graphicx}
\usepackage{subfig}
\usepackage{authblk}
\usepackage{graphics, setspace}
\usepackage[title]{appendix}
\usepackage{hyperref}
\usepackage{braket}
\usepackage{comment}
\usepackage[main=english,spanish]{babel}
\newcommand{\mathsym}[1]{{}}
\newcommand{\unicode}[1]{{}}
\usepackage[backend=biber,
style=numeric,sorting=none
]{biblatex}
\usepackage[automake,toc]{glossaries}
\bibliography{references}
\hypersetup{
	colorlinks   = true, 
	urlcolor     = blue, 
	linkcolor    = blue, 
	citecolor   = blue 
}

\makeglossaries
\newacronym{pbc}{PBC}{periodic boundary conditions}
\newacronym{dbc}{DBC}{Dirichlet boundary conditions}
\newacronym{abc}{APBC}{anti-periodic boundary conditions}
\newacronym{nbc}{NBC}{Neumann boundary conditions}
\newacronym{zbc}{ZBC}{Zaremba boundary conditions}
\newacronym{pccbc}{PCCBC}{perfect colour conductor boundary conditions}
\newacronym{mc}{MC}{Monte Carlo}
\newacronym{wzw}{WZW}{Wess-Zumino-Witten}
\newacronym{gpu}{GPU}{graphics processing unit}
\newacronym{cpu}{CPU}{central processing unit}
\newacronym{nkp}{KNP}{Karabali-Nair parametrization}
\newacronym{uv}{UV}{ultraviolet}

\begin{document}
	\begin{titlepage}
		\newgeometry{,bottom=0.1cm}
		\begin{center}
			\includegraphics[width=0.6\textwidth]{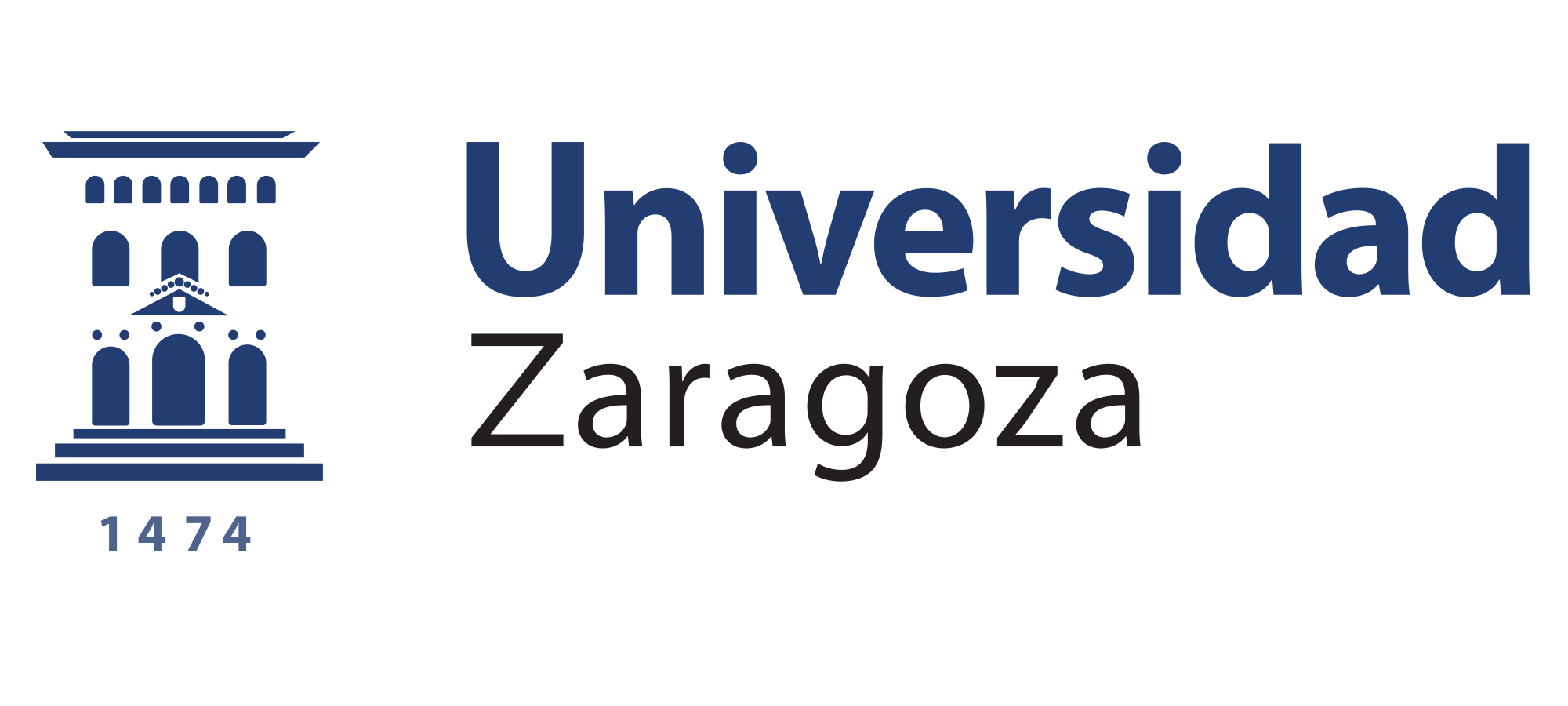}
		\end{center}
		\centering
		\vspace{0.8cm}
		\huge	\framebox[3\width]{PhD thesis} \\
		\vspace{3cm}
		\LARGE Properties of the quantum vacuum in non-abelian gauge theories\\
		\vspace{3cm}
		\normalsize	Author:\\
		\vspace{0.3cm}
		\LARGE Fernando Ezquerro Sastre\\
		\vspace{3cm}
		\normalsize	Advisors:\\
		\vspace{0.4cm}
		\large Dr. Manuel Asorey Carballeira\\
		\large Dr. Eduardo Follana Adín\\
		\vspace{2cm}
		\normalsize Programa de Doctorado en Física\\
		Escuela de Doctorado\\
		\vspace{1cm}
		2026\\
		\vspace{0.4cm}
		{\small
			\centering
			Repositorio de la Universidad de Zaragoza – Zaguan\quad
			\url{http://zaguan.unizar.es}\par
		}
	\end{titlepage}
    \pagenumbering{Roman}
{   \hypersetup{linkcolor=black}
    \tableofcontents
}
	\chapter*{Agradecimientos/Acknowledgements}
	\addcontentsline{toc}{chapter}{Agradecimientos/Acknowledgements}

En primer lugar agradecer a mis directores Manolo y Eduardo por su dedicación, disponibilidad y por transmitirme parte de su incasable pasión por la física. No podía haber elegido unos mejores directores.

Special thanks to Paolo Facchi, Giuliano Angelone and the rest of the Bari quantum theory group for welcoming me and making me feel at home during my stay. It was a very productive and pleasant time in Bari.

A mis padres y mi hermano por su constante apoyo durante estos años de doctorando, sin vosotros esta tesis no habría sido posible.

The author acknowledges financial support by Gobierno de Aragón through the grant
ORDEN CUS/581/2020, and partially through: grant 2020-E21\textunderscore17R and 2022-E21\textunderscore23R of the Aragon Government, Grupo Teórico de Física de altas energías; by Spanish MINECO/FEDER Grants No.
PGC2022-126078NB- C21 funded by MCIN/AEI/10.13039/ 501100011033 and
“ERDF A way of making Europe”; Quantum Spain project of the QUANTUM ENIA of Ministerio de Economía y Transformación Digita; and the European Union, NextGenerationEU Recovery and Resilience Program on ’Astrofısica y Fısica de Altas Energıas , CEFCA-CAPA-ITAINNOVA.
	\chapter*{Abstract}
	\addcontentsline{toc}{chapter}{Abstract}
	In this work we analyze the quantum vacuum properties of non-abelian gauge theories. We calculate the energy of the quantum vacuum by non-perturbative methods using Monte Carlo simulations, focusing on the contribution of boundary effects to the Casimir energy. In particular, we analyze the dependence of the vacuum energy on the types of boundary conditions. The main goal is to clarify the behaviour of this energy for large separation $L$ between the boundaries of the domain  where the  fields are confined. Usually this Casimir energy decreases polynomially with $L$ for massless theories and exponentially for massive theories. Since gauge theories interpolate between these two regimes, being massless in the ultraviolet regime and massive in the infrared regime, one expects a very special change of behaviour from the perturbative to the non-perturbative approaches. In pure gauge theories there is evidence of the existence of glueball states in the low energy spectrum with a non-vanishing mass, the second goal will be testing if the mass of the lightest glueball is responsible for the exponential decay of the Casimir energy of gauge theories. The answer to these two questions is the main objective of the present work.

In \autoref{chp:scalar_cont} we compute the effective action of a massive scalar field when it is confined between two homogeneous infinite walls in the low temperature regime for general  boundary conditions. We also obtain analytically the Casimir energy and show how the rate of the exponential decay with the distance between the boundary walls in the Casimir energy allows the classification of the boundary conditions in two different families. If the boundary conditions involve independent constrains in each wall (e.g. Dirichlet boundary conditions)
this exponential decay in the Casimir energy is twice as fast than when the boundary conditions interconnect the two boundary walls (e.g periodic boundary conditions).

In \autoref{chp:scalar_lattice} we calculate the same observable quantities on  
the lattice formalism to be able to use later the same strategy for non-abelian gauge theories.  We find the same vacuum structure and Casimir energy than in the continuum case. In particular, the same classification of boundary conditions according to the rate of the exponential decay of the Casimir energy as in the analytic continuum approach is found.

In \autoref{chp:su2_21}   we address the calculation of the Casimir energy for $SU(2)$ Yang-Mills theory in 2+1 dimensions. We analyze two different boundary conditions: periodic and perfect colour conductor boundary conditions. We find a similar behaviour  of the quantum vacuum  as in the  massive scalar case for each of them. This Casimir energy has an exponential decay with the distance between the boundary walls characteristic of massive theories. By fitting this Casimir energy  we find that for both boundary conditions the mass that drives this exponential decay is smaller than the lightest glueball of the theory. Moreover, the behaviour of this mass found for the different boundary conditions does not follow the pattern of the massive scalar fields, thus excluding the description of the low energy regime of Yang-Mills theory by a free massive scalar field mode as advocated by D. Karabali and V. P. Nair in 2+1 dimensions.

Finally, in \autoref{chp:su2_31} we perform the same analysis for $SU(2)$ Yang-Mills theory in 3+1 dimensions. The renormalization is easier due to the form of the energy density in these 3+1 dimensions that makes the bulk contributions vanish. The Casimir energy also exhibits an exponential decay with the distance between the boundary walls for both boundary conditions. But, the rate of decay found for the different boundary conditions does not follow the pattern of the massive scalar fields, hence, excluding the description of the low energy regime of Yang-Mills theory in 3+1 dimensions by means of an effective free massive scalar field. Furthermore, the mass that drives this exponential decay is again much smaller than the lightest glueball in the theory. 
{   \hypersetup{linkcolor=black}
		\printglossary[type=\acronymtype,title={Acronyms and abbreviations},toctitle={Acronyms and abbreviations}]
}

  \numberwithin{equation}{chapter}
  	
	\chapter{Introduction}
\pagenumbering{arabic}
\setcounter{page}{1}

Gauge theories constitute the basis of the interaction theory on the Standard Model of particle physics providing a unified framework for describing three of the four fundamental interactions: electromagnetic, weak and strong interactions. The key point of the gauge theories is that the Lagrangian and physical observables must be invariant under the gauge transformations $\Omega$ on the gauge fields $A_\mu$
\begin{equation}
	A_\mu\rightarrow \Omega\ A_\mu \Omega^{-1}+i\ \Omega\ \partial_\mu \Omega^{-1}.
\end{equation}
In the non-abelian case, the self interacting nature of gauge fields makes the analytic treatment of these theories very challenging. Moreover, in the low energy regime  the perturbative methods fail to explain phenomena like quark confinement and the existence of a mass gap, in spite of the fact that the gauge fields are massless {\it ab-initio}. Thus,
 alternative techniques are required. Quark confinement means that quarks are never observed in isolation due to the appearance of a linear potential $V(r)\simeq\sigma r$ that grows with separation $r$  between the pairs of quarks $\bar q$-$q$. This potential  is generated by the non-perturbative dynamics of the theory. An analytic derivation of these phenomena is still missing. Closely related to this issue is the so called \textit{mass gap problem}. Pure non-abelian gauge theories must exhibit a mass gap, i.e. the lightest state above the vacuum should have a positive lower bound $\Delta$ on its energy $E_{g}>\Delta> E_0=0$. Probing this property in 3+1 dimensions was included as one of the seven Clay Millennium Prize Problems.

 Despite the lack of improvements in the analytic approach, since the introduction of the non-abelian gauge theories on the lattice by K. Wilson in 1974 \cite{wilson1974confinement} a large number of positive results have been obtained by non-perturbative simulations. Evidences of the existence of a confinement mechanism and a mass gap in the glueball spectra are among the most relevant ones.

Apart from these numerical results very little is known of the vacuum structure of gauge theories. A  way to get more information about its structure is to confine the field in a domain where the boundary can have different types of boundary conditions. Observing the reaction of the quantum vacuum to this type of external perturbation might give us some insights about its deep structure.

In this work we will focus our effort on a particular low energy phenomenon in quantum theories, the Casimir effect. This was one of the first and clearest examples of boundary effects in quantum field theory. When a quantum field is confined between two parallel plates, the renormalized vacuum energy becomes dependent on the boundary conditions  imposed by the plates. The dependence of this  energy on the distance between the boundary plates generates a force between them. This effect was first predicted by H. Casimir in 1948 \cite{Casimir:1948dh} for the electromagnetic case. Although the Casimir effect is a tiny quantity, it has been measured in different set ups for  electromagnetic fields \cite{M1sparnaay1957attractive,M2sparnaay1958measurements,M3lamoreaux1997demonstration,M4PhysRevLett.81.4549,M5PhysRevLett.87.211801,M6Dalvit}.

A considerable progress has undergone in the understanding and computation of the Casimir effect for different setups and models (see for e.g. \cite{bordag2009advances,milton2001casimir} for some general reviews). Remarkable results were found in Ref. \cite{Boundary_general_2013} where the vacuum energy in arbitrary dimensions and general boundary conditions was obtained for a massless scalar field. Regarding the dependence on the temperature, the massless scalar field was studied in Ref. \cite{Munoz-Castaneda:2020wif} for general boundary conditions in 3+1 dimensions.

Nevertheless, the Casimir effect has been much less studied for interacting theories \cite{symanzik1981schrodinger,barone2004radiative}. In recent years some interesting progress has been achieved in the analytic approach to gauge theories in 2+1 dimensions. In this approach developed by  D. Karabali and V. P. Nair in Refs. \cite{karabali1996gauge,karabali1998planar} the gauge invariant degrees of freedom are parametrized by a massive scalar field. Some interesting non-perturbative results can be derived from this approach. In particular, the prediction that the Casimir energy is given by that of this effective massive scalar field. Some numerical simulations seem to support this conjecture \cite{chernodub2018casimir,karabali2018casimir}.

The main goal of this work is to compute the Casimir vacuum energy of non-abelian gauge theories by means of non-perturbative Monte Carlo simulations, with particular emphasis on its dependence on the distance $L$  between the boundary walls. Usually this Casimir energy decreases polynomially with $L$ for massless theories and exponentially for massive theories. Since gauge theories interpolate between these two regimes, being massless in the 
ultraviolet regime and massive in the infrared regime one expects a very special change of behaviour from the perturbative to the non-perturbative approaches. In the latter we can also compute the mass scale that drives the exponential decay and compare it with the glueball spectra of the theory. Another interesting goal is the analysis of the dependence of the Casimir energy on the boundary conditions which can provide information about the internal structure of the quantum vacuum. Specially, our attention will be focused on the dependence pattern on the boundary conditions to confirm or exclude the Karabali-Nair conjecture that states that the low energy regime of $SU(2)$ Yang-Mills theories is governed by a massive scalar field.

The thesis is organized as follows:

\renewcommand\chapterautorefname{Chapter}
 \autoref{chp:scalar_cont} focuses on the computation of the effective action in the low temperature regime of a massive scalar field confined between two homogeneous infinite walls for general boundary conditions. The Casimir energy has been obtained  by using zeta function renormalization  which permits a detailed analysis of its asymptotic behaviour. 
 
 In \autoref{chp:scalar_lattice} we develop the lattice formalism for the massive scalar field. Different boundary conditions are implemented on the lattice and the vacuum response to the boundary conditions is analyzed. Finally, the Casimir energy is computed and compared with the continuum results.
 
 In \autoref{chp:gauge_theory} we  introduce the lattice formalism for gauge theories and some key  aspects of Monte Carlo simulations. Two different boundary conditions for gauge theories are implemented: periodic and perfect colour conductor boundary conditions.
 
 In \autoref{chp:su2_21} we develop  non-perturbative computations of the Casimir energy by Monte Carlo simulations of Yang-Mills theories in 2+1 dimensions with $SU(2)$ gauge group and different boundary conditions. The results for the Casimir energy are compared with those of  massive scalar fields and the mass that drives the exponential decay of the Casimir energy with the distance is analyzed.
 
In \autoref{chp:su2_31} we perform similar computations for 3+1 dimensional space-times. The Casimir energy and its exponential behaviour is analyzed for the different types of boundary conditions.
 
 In \autoref{chp:Ca} we present the main results and conclusions of this work.
 
 In \autoref{sec:Bessel} we summarize  some existing relations  between the spectral integrals, Bessel functions and polylogarithms that are used in \autoref{chp:scalar_cont}. 
 
 In \autoref{ch:21_values} we collect all the numerical results from the  Monte Carlo simulations of non-abelian gauge theory in 2+1 dimensions that are used in \autoref{chp:su2_21}. 
 
  In \autoref{ch:31_values} we compile all the numerical results of  Monte Carlo simulations of non-abelian gauge theory in 3+1 dimensions that are used in \autoref{chp:su2_31}.
 \renewcommand\chapterautorefname{chapter}

    \chapter{Casimir energy}\label{chp:scalar_cont}

In order to compare the Casimir energy of non-abelian gauge theories, which is the purpose of this work and will be computed by Monte Carlo (\acrshort*{mc}) simulations later on,  with the Casimir energy of scalar fields we first need to calculate this Casimir energy for different boundary conditions. Furthermore, since the thermal fluctuations are unavoidable in \acrshort*{mc} simulations, we need to understand how the temperature can affect the calculation of the Casimir energy. In this chapter we shall calculate the Casimir energy and the thermal contributions in the low temperature limit of the vacuum energy of a massive scalar field in arbitrary spatial dimensions $D$ where the fields are confined between two homogeneous parallel walls satisfying some  boundary conditions.

 In \autoref{sec:Effective_action} we present the setup and renormalization techniques that will be used to obtain the Casimir energy from the divergent free energy. Then, in \autoref{sec:Low_temp} we shall work on the low temperature limit and obtain the finite zero temperature contributions as well as some temperature dependent terms. In \autoref{sec:Casimir_cont} we finally obtain the Casimir energy formulas in the general $D$ dimensional case, and study the asymptotic behaviour for large distances between the walls. In \autoref{sec:Particular_conditions} we analyze some particular boundary conditions that will be of interest later on. Finally, in \autoref{sec:21_continuum} and \autoref{sec:31_continuum} we will focus on the 2+1 and 3+1 dimensional cases, which are the ones that will be deeply analyzed in the case of non-abelian gauge theories.

\section{Effective action}\label{sec:Effective_action}
\begin{figure}[H]
	\centering
	\includegraphics[width=.5\textwidth]{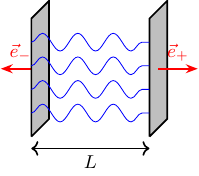}
	\caption{Schematic representation of the Casimir energy framework, where a massive scalar field is confined between two homogeneous infinite walls.}
	\label{fig:Boundary_wall_representation}
\end{figure}

We consider a free massive scalar field $\psi$ with mass $m$ in $D$ dimensions confined between two homogeneous infinite walls of dimension $D-1$ perpendicular to the transverse $D$ dimension in our coordinates $(x_1,x_2,\ldots,x_D)$, i.e. the normal vector to the walls is of the form $\mathbf e_\pm=(0,0,\ldots,\pm 1)$ (see \autoref{fig:Boundary_wall_representation}). The field satisfies  some boundary conditions at the walls which for simplicity will be  located at $\mathbf x^\pm=(x_1,x_2,\ldots,\pm L/2)$.

The homogeneity of the boundary walls imposes that the boundary conditions are invariant under longitudinal translations in the walls, therefore, they have to be homogeneous boundary conditions on each wall. In other words, they just relate the values of the fields at each wall in a way that does not  depend on  their position in the wall. Homogeneous local boundary conditions should define the domain of a self-adjoint extension of the spatial Laplacian operator $-\boldsymbol{\nabla}^2$ (which is the relevant operator for the calculation of the free energy of  free massive scalar fields). These selfadjoint extensions are completely characterized  by 2x2 unitary matrices $U$\cite{asorey2005global} in the following way
\begin{equation}
	\varphi - i\delta\dot\varphi = U(\varphi +i\delta\dot\varphi ),
	\label{eq:bccc}
\end{equation}
where
$\delta$ is an arbitrary scale parameter and
\begin{equation}
	\varphi=\begin{pmatrix}
		\varphi(L/2)  \\ \varphi(-L/2)  \\ 
	\end{pmatrix},
	\quad
	\dot\varphi=\begin{pmatrix}
		\dot\varphi(L/2)  \\ \dot\varphi(-L/2)  \\ 
	\end{pmatrix},
\end{equation}
are the boundary values  $\varphi(\pm L/2)= \psi(t,\mathbf{x}, \pm  L/2)$ of  
the fields  $\psi$   and  their outward normal derivatives 
$\dot\varphi(\pm L/2)=\pm\partial_D\psi(t,\mathbf{x}, \pm L/2)$   
on the walls. We shall assume that $\delta=1$ for simplicity and use the standard parametrization of $U(2)$ matrices
\begin{eqnarray}\label{eq:U2}
	U(\alpha,\gamma,{{\bf n}})&=&{\mathrm e}^{i\alpha}\left(\mathbb{I }\cos\gamma+i{{\bf n}}\cdot\boldsymbol{\sigma}\,\sin\gamma \right);
	\label{eq:parametrization}\quad  {\alpha\in[0,2\pi],\,\, \gamma\in[-\pi/2,\pi/2]},
\end{eqnarray}
where ${{\bf n}}\in S^2$ is a unit vector and $\boldsymbol{\sigma}$ the Pauli matrices. Also, since the scalar field is real only the boundary conditions where $U=U^T$ (which implies in the parametrization that $n_2=0$) have to be considered in order to keep the relationship in \eqref{eq:bccc} between the real fields  \cite{asorey2006casimir}. This space of boundary conditions has to be further restricted in order to preserve the non-negativity in the spectrum of the operator $\boldsymbol{-\nabla}^2$, by the inequality $0\leq \alpha+\gamma \leq \pi$ \cite{asorey2005global}.

In order to describe the finite temperature $T\not=0$ effects, we shall use the Euclidean formalism. In this formalism, the Euclidean time direction is compactified into a circle that has a radius $\beta/(2\pi)=1/(2\pi T)$ with periodic boundary conditions $\psi(t+\beta,\mathbf x)=\psi(t,\mathbf x)$. Taking all of this into account, the partition function is given by the determinant
\begin{equation}\label{eq:det}
	Z(\beta)=\text{det}\left(-\partial_0^2-\boldsymbol{\nabla}^2+m^2\right)^{-1/2},
\end{equation}
with $\boldsymbol{\nabla}^2$ the spatial $D$ dimensional Laplacian and $\partial_0$ the Euclidean time derivative. This determinant in $\eqref{eq:det}$ of the second order differential operator $-\partial_0^2-\boldsymbol{\nabla}^2+m^2$ is clearly ultraviolet (\acrshort*{uv}) divergent. In a similar manner that was done in Refs. \cite{asorey2013thermodynamics,asorey2015topological}, we will use zeta function renormalization methods \cite{zeta_reg,zeta_reg_2,irina_zeta} in order to regularize this determinant and obtain the finite contributions to the free energy. In these renormalization methods, we define the zeta function in terms of the trace of a second order differential operator  involved in the partition function \eqref{eq:det}, i.e.  
\begin{equation}
	\zeta(L,m,\beta;s)= \text{tr}\left[\mu^{2s}\left(-\partial_0^2-\boldsymbol{\nabla}^2+m^2\right)^{-s}\right],
\end{equation}
where $\mu$ is a scalar parameter needed for making the zeta function dimensionless and also encodes the standard ambiguity of zeta function renormalization techniques \footnote{See e.g. \cite{zeta_amb,zeta_amb_2} for a detailed discussion and comparison with other renormalization methods.}. Later, this ambiguity represented by $\mu$ will be fixed by a renormalization prescription. Once we have this zeta function we can define the effective action as the logarithm of the partition function \eqref{eq:det},
\begin{equation}\label{eq:effaction}
	S_{\text{eff}}=-\log Z=-\frac{1}{2}\frac{d}{ds}\zeta \left(s\right)|_{s=0}.
\end{equation}

In the framework  previously described, the eigenvalues of the operator $-\partial_0^2-\boldsymbol{\nabla}^2$ are given by the sum of the square of temporal modes $2\pi l/\beta$, the continuous spatial modes $\mathbf q=(q_1,q_2,\ldots,q_{D-1})$ in the direction parallel to the boundary walls (normal to $\mathbf e_{\pm}$ ) and the discrete spatial modes $k_i$ in the perpendicular direction to the boundary walls (parallel to $\mathbf e_{\pm}$) which depend on the boundary conditions parametrized by $U$, i.e. 
\begin{equation}
	\lambda_{l,i}= \left(\frac{2\pi l}{\beta}\right)^2+\mathbf {q}^2+k_i^2+m^2 \hspace{2cm}l \in \mathbb Z,\mathbf{q} \in \mathbb R^{D-1}, i\in {\mathbb N}.
\end{equation} 

Thus, the functional trace in the zeta function can be expressed with a sum of the discrete temporal and spatial modes, and the integral of the continuous spatial modes
\begin{equation}\label{eq:ini}
	\zeta(L,m,\beta;s)=\mu^{2s}\frac{A}{(2\pi)^{D-1}}\sum_{l=-\infty}^\infty\sum_i \int d^{D-1}q\left(\left(\frac{2\pi l}{\beta}\right)^2+\mathbf q^2+k_i^2+m^2\right)^{-s},
\end{equation}
where $A$ is the infinite $D-1$ volume of the boundary walls. Using the generalized spherical coordinates on $\mathbf q$ and integrating the angular variables we arrive at
\begin{align}\nonumber
	\zeta(L,m,\beta;s)&=\frac{A\mu^{2s}}{2^{D-2}\pi^{\frac{D-1}{2}}\Gamma\left(\frac{D-1}{2}\right)}\\ \label{eq:ini_2}
	&\times\sum_{l=-\infty}^\infty\sum_i \int_{-\infty}^\infty dq\ q^{D-2}\left(\left(\frac{2\pi l}{\beta}\right)^2+q^2+k_i^2+m^2\right)^{-s},
\end{align}
where we have used that the area of a $D-2$ dimensional sphere is
\begin{equation}
	\Sigma_{D-2}=\frac{2\pi^{\frac{D-1}{2}}}{\Gamma\left(\frac{D-1}{2}\right)}.
\end{equation}
Now, the continuous $q$ can be integrated by using the analytic extension of the zeta function \footnote{Notice how for the $D=1$ case we recover the correct expression, that describes the one dimensional case with boundary conditions, by setting $A=1$.}
\begin{equation}\label{eq:zeta_D}
	\zeta(L,m,\beta;s)=\mu^{2s}\frac{A\Gamma\left(s-\frac{D-1}{2}\right)}{2^{D-1}\pi^{\frac{D-1}{2}}\Gamma\left(s\right)}\sum_{l=-\infty}^\infty\sum_i \left(\left(\frac{2\pi l}{\beta}\right)^2+k_i^2+m^2\right)^{-s+\frac{D-1}{2}},
\end{equation}
leaving only the sum in temporal and discrete spatial modes. The spectrum of eigenvalues (including the multiplicities) corresponding to the discrete spatial modes is given by the zeros of the spectral function \cite{Boundary_general_2013,munoz2015qft}
\begin{equation}\label{eq:spectral_function}
	h^L_U(k)=2i\left(\sin(kL)\left((k^2-1)\cos \gamma +(k^2+1)\cos \alpha \right)-2k\sin \alpha \cos (kL)-2kn_1\sin \gamma\right),
\end{equation}
aside from the zero modes of the spectrum which can be incorrectly described by the spectral function. But, since these zero modes are independent of the distance between the walls $L$, their contributions will be canceled by the renormalization prescription that we will use \eqref{eq:zeta_combination} and this will be not be an issue in our calculations. Also, the apparent lack of dimensional consistency in the powers of $k$ is due to the choice of $\delta=1$, since they are actually powers of $k\delta$. By using the Cauchy theorem we can write this sum in the discrete spatial modes as a contour integral of the derivative of the logarithm of the spectral function
\begin{equation}\label{eq:zeta_integral}
	\zeta(L,m,\beta;s)=\mu^{2s}\frac{A\Gamma\left(s-\frac{D-1}{2}\right)}{2^{D}\pi^{\frac{D+1}{2}}\Gamma\left(s\right)i}\sum_{l=-\infty}^\infty \oint dk \left(\left(\frac{2\pi l}{\beta}\right)^2+k^2+m^2\right)^{-s+\frac{D-1}{2}}\frac{d}{dk}\log h^L_U(k),
\end{equation}
with the contour being a thin strip enclosing the positive real axis, in which all the zeros of the spectral function $h_U(k)$ are located. 

Once we have integrated the spatial continuous modes by means of the analytic extension of the zeta function, the \acrshort*{uv} divergences that are left in \eqref{eq:zeta_D} arise from the zero temperature vacuum energy. In this zero temperature limit, these divergences have the following asymptotic behaviour with $L$ \cite{Boundary_general_2013} in the effective action
\begin{equation}\label{eq:divergences_energy}
	S_{\text{eff}}^{l=0}= \beta E_0 =A\beta LC_0(m)+A\beta C_1(m)+A\beta\ C_c(m,L)+\ldots. 
\end{equation}
where $E_0$ is the vacuum energy, $C_0(m)$ the divergent bulk vacuum energy density, $C_1(m)$  the divergent energy density on the boundary walls and $C_c(m,L)$ is the finite coefficient of the Casimir energy. In summary, we have two kinds of divergences. The first one that is the linear  dependent term with $L$ of $E_0$ and the second one that is independent of $L$. We can add a renormalization prescription that cancels all these divergences and has a physical meaning by redefining the effective action as
\begin{equation}\label{eq:zeta_combination}
	S^{\text{ren}}_{\text{eff}}=-
	\frac{1}{2}\frac{d}{ds} \zeta_{\text{ren}}
	(L,m,\beta;s) 
	\Bigr|_{s=0}\ ,
\end{equation}
where
\begin{equation}\label{eq:zeta_re}
	\zeta_{\text{ren}}(L,m,\beta;s)=\lim_{L_0\rightarrow \infty}\left(\zeta(L,m,\beta;s)+\zeta(2L_0+L,m,\beta;s)-2\zeta(L_0+L,m,\beta;s)\right),
\end{equation}
in terms of an auxiliary length $L_0 \gg L$. The finite contributions from these extra terms vanish when applying the $L_0\rightarrow \infty$ limit. The physical condition  fixing our renormalization prescription is that we not only cancel the divergent contributions which do not depend or are linear with $L$, but also the associated finite contributions, i.e. the prescription requires that $E_0$ vanishes in the limit $L\to \infty$. This means, that the only remaining terms are the finite contributions that go to zero as the distance between the boundary walls does and describe the modification of the effective action due to the presence of the walls, which is exactly the Casimir energy we want to  calculate. Notice that this renormalization prescription is not strictly necessary for getting a finite result from \eqref{eq:zeta_D}, this is already given by the zeta function regularization\footnote{This can be seen in the Appendices of \cite{ezquerro2024casimir,ezquerro2025casimir} where the calculation of the Casimir energy using only the zeta function regularization gives a finite result.}.

\section{Low temperature regime}\label{sec:Low_temp}
Like it was explained in the introduction of this chapter, we are interested in the low temperature regime\footnote{See \cite{ezquerro2025temperature} for an analysis on the high temperature limit and the matching between these two regimes.} $\beta/L >1$ that will be the equivalent situation on the lattice of taking the time dimension large ($\beta$ large). Let us deal first with the sum of the temporal modes and later with the spatial ones. We come back to \eqref{eq:zeta_D} and rewrite it as
 \begin{align}\nonumber
	\zeta(L,m,\beta;s)&=\left(\frac{\mu \beta}{2\pi}\right)^{2s}\frac{A\pi^{\frac{D-1}{2}}\Gamma\left(s-\frac{D-1}{2}\right)}{\beta^{D-1}\Gamma\left(s\right)}\\\label{eq:ini_4}
	&\times\sum_{l=-\infty}^\infty\sum_i \left( l^2+\left(\frac{k_i \beta}{2\pi}\right)^2+\left(\frac{m \beta}{2\pi}\right)^2\right)^{-s+\frac{D-1}{2}}.
\end{align}
 We use the Cahen-Mellin integral formula,
 \begin{equation}
 	a^{-p}=\frac{1}{\Gamma(p)}\int_{0}^{\infty}t^{p-1}e^{at} dt,
 \end{equation}
 which transforms our equation into
  \begin{equation}\label{eq:ini_5}
 	\zeta(L,m,\beta;s)=\left(\frac{\mu \beta}{2\pi}\right)^{2s}\frac{A\pi^{\frac{D-1}{2}}}{\beta^{D-1}\Gamma\left(s\right)}\sum_{l=-\infty}^\infty\sum_i\int_{0}^{\infty} dt\ t^{s-\frac{D+1}{2}}\	e^{-\left( l^2+\left(\frac{k_i \beta}{2\pi}\right)^2+\left(\frac{m \beta}{2\pi}\right)^2\right)t},
 \end{equation}
 and then use the Poisson formula for the sum in $l$,
 \begin{equation}\label{eq:Poisson sum_2d}
 	\sum_{n=-\infty}^\infty e^{-2\pi \alpha n^2}=\frac{1}{\sqrt{2\alpha }}\sum_{n=-\infty}^{\infty}e^{-\frac{\pi  n^2}{2\alpha}},
 \end{equation}
 which gives us
 \begin{equation}\label{eq:ini_6}
 	\zeta(L,m,\beta;s)=\left(\frac{\mu \beta}{2\pi}\right)^{2s}\frac{A\pi^{\frac{D}{2}}}{\beta^{D-1}\Gamma\left(s\right)}\sum_{l=-\infty}^\infty\sum_{i}\int_{0}^{\infty} dt\ t^{s-\frac{D}{2}-1}\	e^{-\left(\left(\frac{k_i \beta}{2\pi}\right)^2+\left(\frac{m \beta}{2\pi}\right)^2\right)t-\frac{(\pi l)^2}{t}}.
 \end{equation}
 We can actually compute this integral by separating the $l=0$ mode from the rest
 \begin{align}\nonumber
 	\zeta(L,m,\beta;s)&=\left(\frac{\mu \beta}{2\pi}\right)^{2s}\frac{A\pi^{\frac{D}{2}}}{\beta^{D-1}\Gamma\left(s\right)}\left(\Gamma \left(s-\frac{D}{2}\right)\sum_i\left(\left(\frac{k_i \beta}{2\pi}\right)^2+\left(\frac{m \beta}{2\pi}\right)^2\right)^{\frac{D}{2}-s}+\right.\\
 	&\left. +4\sum_{i}\sum_{l=1}^\infty\left(\pi l\right)^{-\frac{D}{2}+s}\left(\left(\frac{k_i \beta}{2\pi}\right)^2+\left(\frac{m \beta}{2\pi}\right)^2\right)^{\frac{D}{4}-\frac{s}{2}}K_{\frac{D}{2}-s}\left(\beta l\sqrt{k_i^2+m^2}\right)\right),\label{eq:temp_int}
 \end{align}
the first term ($l=0$) of the integration is the corresponding to the zero temperature contribution since it is linear with $\beta$ \footnote{When computing the free energy from the effective action we have to divide by $\beta$, therefore losing the dependence with $\beta$ on this term.}. The second one ($l\not =0$) contains the temperature dependent contributions that have a more complicated dependence in $\beta$ and where $K_q$ is the second type modified Bessel function (defined in \autoref{sec:Bessel}) .
\subsection{Zero temperature term}\label{sec:zero_temp}
Let us now focus on the zero temperature term (the first one in \eqref{eq:temp_int})
\begin{equation}
	\zeta^{l=0}(L,m,\beta;s)=\mu^{2s}\frac{A\beta  \Gamma \left(s-\frac{D}{2}\right)}{2^{D}\pi^{\frac{D}{2}}\Gamma(s)}\sum_{i} \left(k_i^2+m^2\right)^{\frac{D}{2}-s}.
\end{equation}
We recover the spectral function \eqref{eq:spectral_function} and rewrite the sum in the spatial discrete modes as the contour integral \eqref{eq:zeta_integral}
\begin{equation}
	\zeta^{l=0}(L,m,\beta;s)=\mu^{2s}\frac{A\beta  \Gamma \left(s-\frac{D}{2}\right)}{2^{D+1}\pi^{\frac{D}{2}+1} i \Gamma(s)}\oint dk\ \left(k^2+m^2\right)^{\frac{D}{2}-s}\frac{d}{dk}\log h^L_U(k).
\end{equation}
Finally, we can apply our renormalization prescription by using the renormalized zeta function we defined in \eqref{eq:zeta_re}
\begin{align}\nonumber
	\zeta_{\text{ren}}^{l=0}(L,m,\beta;s)&=\mu^{2s}\frac{A\beta  \Gamma \left(s-\frac{D}{2}\right)}{2^{D+1}\pi^{\frac{D}{2}+1} i \Gamma(s)}\\\label{eq:spectral_pre}
	&\times\lim_{L_0\rightarrow \infty}\oint dk\ \left(k^2+m^2\right)^{\frac{D}{2}-s}\frac{d}{dk}\log \frac{h^L_U(k)h^{2L_0+L}_U(k)}{\left(h^{L_0+L}_U(k)\right)^2},
\end{align}
notice that since the different zeta functions that appear in \eqref{eq:zeta_re} only differ by their distance between the walls, the expression can be reduced to just the logarithm of spectral functions with different distances $L$. Once we have this renormalized zeta function the \acrshort*{uv} divergences from the integral in spatial modes cancel. Therefore, the only possible divergent terms when $s=0$ are the Gamma function $\Gamma(s)$ and, depending on the parity of the spatial dimension $D$, $\Gamma(s-\frac{D}{2})$. This means that we have to analyze separately the even and odd cases.

For the even case $D=2n$, we have that $\Gamma(s-n)$ is divergent when $s=0$, which means that we have a quotient of two divergent terms and we can not directly differentiate and evaluate the expression. Instead, we have to use the  small $s$ expansion for the quotient of the Gamma functions
\begin{equation}
	\frac{\Gamma(s-n)}{\Gamma(s)}=(-1)^{n}\left(\frac{1}{n!}+a(n)\ s\right)+ O(s^2),
\end{equation}
where $a(n)$ is a numeric constant that depends on $n$ but will vanish in the following steps. Moreover, only the linear term is needed in the expansion since the higher order terms will be zero when we derive and evaluate on $s=0$. Now, we have to insert this small $s$ expansion in \eqref{eq:spectral_pre} and, derive and evaluate the terms that depend on $s$ obtaining
\begin{align}\nonumber
	&\frac{d}{ds}\left.\left(\left(\frac{1}{n!}+a(n)\ s\right)\left(k^2+m^2\right)^{n-s}\mu^{2s}\right)\right|_{s=0}=\\
	&-\frac{(k^2+m^2)^n}{n!}\left(\log\left(k^2+m^2\right)-2\log \mu-a(n)n!\right).
\end{align}
Once we have the result of the derivation, the zeta function is
\begin{align}\nonumber
	\frac{d}{ds}\zeta_{\text{ren}}^{l=0}&(L,m,\beta;s)\Bigr|_{s=0}=\frac{(-1)^{n+1}A\beta  }{2^{2n+1}\pi^{n+1}n! i }\lim_{L_0\rightarrow \infty}\oint dk \left(k^2+m^2\right)^n\\\label{eq:even_integral}
  &\times\left(\log\left(k^2+m^2\right)-2\log \mu-a(n)n!\right)
	\left(\frac{d}{dk}\log \frac{h^L_U(k)h^{2L_0+L}_U(k)}{\left(h^{L_0+L}_U(k)\right)^2}\right),
\end{align}
 this expression is finite (apart from the volume factor of the infinite walls) and, as it will be seen later, gives the Casimir energy of the system.
 
\begin{figure}[H]
 	\centering
 	\includegraphics[width=.8\textwidth]{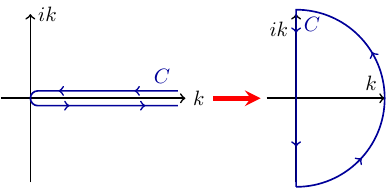}
 	\caption{Deformation of the contour in the integral \eqref{eq:even_integral} 
	using that the integrand is holomorphic in the swept domain of the deformation,
	from the left one to the right one.}
 	\label{fig:Integral_contour}
 \end{figure}

Let us simplify this expression a bit more. First, by using that the integrand is holomorphic, we can extend the integration contour (that was a thin strip around the real line) to an infinite semicircle limited by the imaginary axis on its left (represented in \autoref{fig:Integral_contour}).
 
  In the limit $ik\rightarrow \infty$ the integrand on the semicircle vanishes due to the fact that the ratio of logarithms of spectral functions tends to one. This can be seen in \eqref{eq:expansion_log_spectral} when the asymptotic behaviour of the Casimir energy is analyzed, and therefore the integration over the infinite semicircle becomes zero. Hence, the contour integral can be reduced to just the imaginary axis
\begin{align}\nonumber
	\frac{d}{ds}\zeta_{\text{ren}}^{l=0}&(L,m,\beta;s)\Bigr|_{s=0}=\frac{A\beta  }{2^{2n+1}\pi^{n+1}n! i }\lim_{L_0\rightarrow \infty}\int_{-\infty}^\infty dk \left(k^2-m^2\right)^n\\
	&\times\left(\log\left(m^2-k^2\right)-2\log \mu-a(n)n!\right)
	\left(\frac{d}{dk}\log \frac{h^L_U(ik)h^{2L_0+L}_U(ik)}{\left(h^{L_0+L}_U(ik)\right)^2}\right).
\end{align}

Since the spectral function \eqref{eq:spectral_function} is odd in $k$, the product inside the logarithm is even, as are the rest of the terms of the integral besides the term with derivative of the logarithm. Thus, the integrand is odd in $k$, and the integral from $-\infty$ to $\infty$ would vanish, but we have to take into account the branching point in $k=m$ of the logarithm $\log(m^2-k^2)$. This gives an extra factor $2\pi i$ for the interval $(m,\infty)$, which does not cancel with the contribution from the interval $(-\infty,-m)$ unlike the rest of the term $\log\left(m^2-k^2\right)-2\log \mu-a(n)n!$. Hence, the integral reduces to

\begin{align}\nonumber
	\frac{d}{ds}\zeta_{\text{ren}}^{l=0}(L,m,\beta;s)\Bigr|_{s=0}=\frac{A\beta  }{2^{2n}\pi^{n}n! }\lim_{L_0\rightarrow \infty}\int_{m}^\infty dk &\left(k^2-m^2\right)^n\\
	&\times\left(\frac{d}{dk}\log \frac{h^L_U(ik)h^{2L_0+L}_U(ik)}{\left(h^{L_0+L}_U(ik)\right)^2}\right).
\end{align}

Finally, using the fact that the integral domain begins at $m$, we can take the limit $L_0\rightarrow\infty$ on the spectral function \eqref{eq:spectral_function}
\begin{equation}\label{eq:L0 infinity}
	\lim_{L_\ast\rightarrow \infty} h_U^{L_\ast}(ik)=\lim_{L_\ast\rightarrow \infty}e^{k(L_\ast)}\left((k^2+1)\cos \gamma +(k^2-1)\cos \alpha +2k\sin\alpha \right),
\end{equation}
and consider its value when the distance between the walls goes to infinity as
\begin{equation}\label{eq:spectra_infinity}
	h^\infty_U(ik)\equiv \left((k^2+1)\cos \gamma +(k^2-1)\cos \alpha +2k\sin\alpha \right).
\end{equation}
Using this limit \eqref{eq:L0 infinity} on the quotient of spectral functions with distance $2L_0+L$ and $L_0+L$, simplifies the exponential factors to just $L$ after applying derivation and the infinite distance limit of the spectral function \eqref{eq:spectra_infinity} in the logarithm. This simplifies the expression to the final formula
\begin{equation}\label{eq:zeta_even_Cas}
	\frac{d}{ds}\zeta_{\text{ren}}^{l=0}(L,m,\beta;s)\Bigr|_{s=0}=-\frac{A\beta  }{2^{2n}\pi^{n}n! }\int_{m}^\infty dk \left(k^2-m^2\right)^n\left(L-\frac{d}{dk}\log \frac{h^L_U(ik)}{h^{\infty}_U(ik)}\right).
\end{equation}

Let us now deal with the case when the spatial dimensions are odd $D=2n+1$. The gamma factor  $\Gamma(s-n-\frac{1}{2})$ is finite when $s=0$, therefore the only divergent contribution left in \eqref{eq:spectral_pre} is the $\Gamma(s)$ in the denominator. Thus, it is the only term we have to derive
\begin{equation}
	\frac{d}{ds}\left.\left(\frac{1}{\Gamma(s)}\right)\right|_{s=0}=1.
\end{equation}
Hence, after evaluation in $s=0$ the integral is
\begin{align} \nonumber
	\frac{d}{ds}\zeta_{\text{ren}}^{l=0}(L,m,\beta;s)\Bigr|_{s=0}=\frac{A\beta  \Gamma \left(-n-\frac{1}{2}\right)}{2^{2n+2}\pi^{n+\frac{3}{2}} i }\lim_{L_0\rightarrow \infty}&\oint dk\ \left(k^2+m^2\right)^{n+\frac{1}{2}}\\\label{eq:odd_integral}
	&\times \left(\frac{d}{dk}\log \frac{h^L_U(k)h^{2L_0+L}_U(k)}{\left(h^{L_0+L}_U(k)\right)^2}\right),
\end{align}
which is a finite expression. In the same way as for the even case, we can extend the integration contour to an infinite semicircle limited by the imaginary axis on its left (\autoref{fig:Integral_contour}), and then reduce the integral to just the imaginary axis since it is zero in the semicircle by the same reasoning as in the even case
\begin{align}\nonumber
	\frac{d}{ds}\zeta_{\text{ren}}^{l=0}(L,m,\beta;s)\Bigr|_{s=0}=-\frac{A\beta  \Gamma \left(-n-\frac{1}{2}\right)}{2^{2n+2}\pi^{n+\frac{3}{2}} i }&\lim_{L_0\rightarrow \infty}\int_{-\infty}^\infty dk \left(m^2-k^2\right)^{n+\frac{1}{2}}\\
	&\times \left(\frac{d}{dk}\log \frac{h^L_U(ik)h^{2L_0+L}_U(ik)}{\left(h^{L_0+L}_U(ik)\right)^2}\right).
\end{align}

The integrand has odd parity  as in the even dimensional case and the integral would cancel. But there is the branching point of the square root $\sqrt{m^2-k^2}$ that in the interval $(-\infty,-m)$ gives a change of sign which makes contributions between $(-\infty,-m)$ and $(m,\infty)$ add instead of canceling. Taking  this into account, the integral reduces to
\begin{align}\nonumber
	\frac{d}{ds}\zeta_{\text{ren}}^{l=0}(L,m,\beta;s)\Bigr|_{s=0}=(-1)^{n+1}&\frac{A\beta  \Gamma \left(-n-\frac{1}{2}\right)}{2^{2n+1}\pi^{n+\frac{3}{2}} }\lim_{L_0\rightarrow \infty}\int_{m}^\infty dk \left(k^2-m^2\right)^{n+\frac{1}{2}}\\
	&\times\left(\frac{d}{dk}\log \frac{h^L_U(ik)h^{2L_0+L}_U(ik)}{\left(h^{L_0+L}_U(ik)\right)^2}\right).
\end{align}
Finally, we can use the $L_0 \to \infty$ limit \eqref{eq:L0 infinity} on the quotient of spectral functions which leads to
\begin{equation}\label{eq:zeta_odd_Cas}
	\frac{d}{ds}\zeta_{\text{ren}}^{l=0}(L,m,\beta;s)\Bigr|_{s=0}\!\!=\!(-1)^n\frac{A\beta  \Gamma \left(-n-\frac{1}{2}\right)}{2^{2n+1}\pi^{n+\frac{3}{2}} }\int_{m}^\infty\!  dk \!\left(k^2-m^2\right)^{n+\frac{1}{2}}\!\left(\!L-\!\frac{d}{dk}\log\! \frac{h^L_U(ik)}{h^{\infty}_U(ik)}\right).
\end{equation}

\subsection{Temperature dependent terms}
Let us now compute the temperature dependent part ($l\not =0$) of the zeta function, that was the second term in \eqref{eq:temp_int}
 \begin{align}\nonumber
	\zeta^{l\not =0}(L,m,\beta;s)&=\left(\frac{\mu \beta}{2\pi}\right)^{2s}\frac{4A\pi^{\frac{D}{2}}}{\beta^{D-1}\Gamma\left(s\right)}\\ 
	&\times \sum_{i}\sum_{l=1}^\infty\left(\pi l\right)^{-\frac{D}{2}+s}\left(\left(\frac{k_i \beta}{2\pi}\right)^2+\left(\frac{m \beta}{2\pi}\right)^2\right)^{\frac{D}{4}-\frac{s}{2}}K_{\frac{D}{2}-s}\left(\beta l\sqrt{k_i^2+m^2}\right),
\end{align}
Since the Bessel function $K_n$ goes to zero exponentially as the argument grows (see formula \eqref{eq:Bessel_asymptotic} for the analysis of its asymptotic behaviour), the sum in $l$ and $k_i$ is convergent. Therefore, the only divergent contribution left is the $\Gamma(s)$ of the denominator that, as for the odd dimensional case in \autoref{sec:zero_temp}, is the only term that we have to derive, i.e.
 \begin{equation}
	\frac{d}{ds}\zeta^{l\not =0}(L,m,\beta;s)\Bigr|_{s=0}=\frac{2^{2-\frac{D}{2}}A}{\pi^{\frac{D}{2}}\beta^{\frac{D}{2}-1}}\sum_{i}\sum_{l=1}^\infty \frac{\left(k_i^2+m^2\right)^{\frac{D}{4}}}{l^{\frac{D}{2}}}\ K_{\frac{D}{2}}\left(\beta l\sqrt{k_i^2+m^2}\right).
\end{equation}
Again, we substitute the sum of the spatial modes with a contour integral using the spectral function \eqref{eq:spectral_function}
 \begin{align}\nonumber
	\frac{d}{ds}\zeta^{l\not =0}(L,m,\beta;s)\Bigr|_{s=0}=\frac{2^{1-\frac{D}{2}}A}{\pi^{1+\frac{D}{2}}\beta^{\frac{D}{2}-1}i}&\sum_{l=1}^\infty\oint dk\ \frac{\left(k^2+m^2\right)^{\frac{D}{4}}}{l^{\frac{D}{2}}}\\
	&\times K_{\frac{D}{2}}\left(\beta l\sqrt{k^2+m^2}\right)\frac{d}{dk}\log\left(h^L_U(k)\right),
\end{align}
and the $l\neq  0$ terms of the renormalized function \eqref{eq:zeta_re} are
 \begin{align}\nonumber
	\frac{d}{ds}\zeta^{l\not =0}_{\text{ren}}(L,m,\beta;s)\Bigr|_{s=0}&=\frac{2^{1-\frac{D}{2}}A}{\pi^{1+\frac{D}{2}}\beta^{\frac{D}{2}-1}i}\lim_{L_0\rightarrow \infty}\sum_{l=1}^\infty\oint dk\ \frac{\left(k^2+m^2\right)^{\frac{D}{4}}}{l^{\frac{D}{2}}}\\
	 &\times K_{\frac{D}{2}}\left(\beta l\sqrt{k^2+m^2}\right)
	\left(\frac{d}{dk}\log \frac{h^L_U(k)h^{2L_0+L}_U(k)}{\left(h^{L_0+L}_U(k)\right)^2}\right).
\end{align}
As we did for the $l=0$ term of the zeta function in \autoref{sec:zero_temp}, we can extend the contour to an infinite semicircle limited by the imaginary axis since the integrand is holomorphic (\autoref{fig:Integral_contour}) and reduce it to the imaginary axis \footnote{For the same arguments used for the $l=0$ term, the spectral factor in the integrand vanishes when $ik$ goes to infinity which implies that the integral in the semicircle also vanishes.},
 \begin{align}\nonumber
	\frac{d}{ds}\zeta^{l\not =0}_{\text{ren}}(L,m,\beta;s)\Bigr|_{s=0}&=-\frac{2^{1-\frac{D}{2}}A}{\pi^{1+\frac{D}{2}}\beta^{\frac{D}{2}-1}i}\lim_{L_0\rightarrow \infty}\sum_{l=1}^\infty\int_{-\infty}^\infty dk\ \frac{\left(m^2-k^2\right)^{\frac{D}{4}}}{l^{\frac{D}{2}}}\\\label{eq:temperature_zeta_D} &\times K_{\frac{D}{2}}\left(\beta l\sqrt{m^2-k^2}\right)\left(\frac{d}{dk}\log \frac{h^L_U(ik)h^{2L_0+L}_U(ik)}{\left(h^{L_0+L}_U(ik)\right)^2}\right).
\end{align}
Due to the different behaviour of the branching points of $(m^2-k^2)^{D/4}$ and the Bessel function depending on the parity of $D$, we also have to treat separately the even or odd dimensional cases. For the even spatial dimensions $D=2n$
\begin{align}\nonumber
	\frac{d}{ds}\zeta^{l\not =0}_{\text{ren}}(L,m,\beta;s)\Bigr|_{s=0}&=-\frac{2^{1-n}A}{\pi^{1+n}\beta^{n-1}i}\lim_{L_0\rightarrow \infty}\sum_{l=1}^\infty\int_{-\infty}^\infty dk\ \frac{\left(m^2-k^2\right)^{\frac{n}{2}}}{l^{n}}\\
	 &\times K_{n}\left(\beta l\sqrt{m^2-k^2}\right) \left(\frac{d}{dk}\log \frac{h^L_U(ik)h^{2L_0+L}_U(ik)}{\left(h^{L_0+L}_U(ik)\right)^2}\right),
\end{align}
similarly to the odd dimensional case for the $l=0$ term, the integral is odd and it would vanish, but again we have the existence of the branching point of the square root $\sqrt{k^2-m^2}$ in the argument of the Bessel function which gives a different sign for the $(-\infty,-m)$ and $(m,\infty)$ intervals. Also, the factor $(k^2-m^2)^{\frac{n}{2}}$ depending on the parity of $n$ gives a minus sign when odd and not when is even, i.e. $(-1)^n$. All of this translates into simplifying the integral to just the interval $(m,\infty)$ but with a factor that is the difference (or sum) of the Bessel functions with opposite arguments
\begin{align}\nonumber
	&\frac{d}{ds}\zeta^{l\not =0}_{\text{ren}}(L,m,\beta;s)\Bigr|_{s=0}=\frac{2^{1-n}Ai^{n+1}}{\pi^{1+n}\beta^{n-1}}\lim_{L_0\rightarrow \infty}\sum_{l=1}^\infty\int_{m}^\infty dk\ \frac{\left(k^2-m^2\right)^{\frac{n}{2}}}{l^{n}}\\
	&\times\left(K_{n}\left(i\beta l\sqrt{k^2-m^2}\right)- (-1)^n K_{n}\left(-i\beta l\sqrt{k^2-m^2}\right)\right) \times\left(\frac{d}{dk}\log \frac{h^L_U(ik)h^{2L_0+L}_U(ik)}{\left(h^{L_0+L}_U(ik)\right)^2}\right).
\end{align}
Notice that this expression is always  real. When $n$ is even, the factor $i^{n+1}$ is imaginary but also the difference of Bessel functions $K_n(ix)-K_n(-ix)=2i\ \text{Im} (K_n(ix))$ is purely imaginary ($x=\beta l\sqrt{k^2-m^2}$ in our particular case), making the full expression real. For the case when $n$ is odd, $i^{n+1}$ is real and we have instead $K_n(ix)+K_n(-ix)=2\ \text{Re} (K_n(ix))$, that is also real. Taking this into account, and considering the $L_0\rightarrow \infty$ limit we can write the previous formula in a more compact way by combining together the $i$ factor and the combination of Bessel functions as
 \begin{align}\nonumber
	\frac{d}{ds}\zeta^{l\not =0}_{\text{ren}}(L,m,\beta;s)\Bigr|_{s=0}&=-\frac{2^{2-n}A}{\pi^{1+n}\beta^{n-1}}\sum_{l=1}^\infty\int_{m}^\infty dk\ \frac{\left(k^2-m^2\right)^{\frac{n}{2}}}{l^{n}}\\ \label{eq:zeta_temp_even}
	&\times \text{Re}\left(i^{n+1} K_{n}\left(i\beta l\sqrt{k^2-m^2}\right)\right)
	\left(L-\frac{d}{dk}\log \frac{h^L_U(ik)}{h^{\infty}_U(ik)}\right),
\end{align}

In the odd dimensional case $D=2n+1$ the initial formula \eqref{eq:temperature_zeta_D} transforms into
 \begin{align}\nonumber
	\frac{d}{ds}\zeta^{l\not =0}_{\text{ren}}(L,m,\beta;s)\Bigr|_{s=0}&=-\frac{2^{\frac{1}{2}-n}A}{\pi^{\frac{3}{2}+n}\beta^{n-\frac{1}{2}}i}\lim_{L_0\rightarrow \infty}\sum_{l=1}^\infty\int_{-\infty}^\infty dk\ \frac{\left(m^2-k^2\right)^{\frac{n}{2}+\frac{1}{4}}}{l^{n+\frac{1}{2}}}\\
	 &\times K_{n+\frac{1}{2}}\left(\beta l\sqrt{m^2-k^2}\right)
	\left(\frac{d}{dk}\log \frac{h^L_U(ik)h^{2L_0+L}_U(ik)}{\left(h^{L_0+L}_U(ik)\right)^2}\right).
\end{align}
We can use that the modified Bessel function of the second kind of half-integer order has the following expression \cite{abramowitz1968handbook}
\begin{equation}\label{eq:bessel_semi_integer}
K_{n+\frac{1}{2}}(z)=\frac{e^{-z}}{\sqrt{z}}\sqrt{\frac{\pi}{2}}\sum_{q=0}^n\frac{(n+q)!}{q!(n-q)!}(2z)^{-q},
\end{equation}
which we can use in the integral
 \begin{align}\nonumber
	&\frac{d}{ds}\zeta^{l\not =0}_{\text{ren}}(L,m,\beta;s)\Bigr|_{s=0}=-\frac{2^{-n}A}{\pi^{1+n}\beta^{n}i}\lim_{L_0\rightarrow \infty}\sum_{l=1}^\infty\int_{-\infty}^\infty dk\ \frac{\left(m^2-k^2\right)^{\frac{n}{2}}}{l^{n+1}}\ e^{-\beta l\sqrt{m^2-k^2}}\\ 
	&\times\sum_{q=0}^n\frac{(n+q)!}{q!(n-q)!}\left(2\left(\beta l\sqrt{m^2-k^2}\right)\right)^{-q}
	\left(\frac{d}{dk}\log \frac{h^L_U(ik)h^{2L_0+L}_U(ik)}{\left(h^{L_0+L}_U(ik)\right)^2}\right).
\end{align}
The integrand of the above expression is odd and would cancel the integral but again we have  branching points, in a similar manner as in the even case we have two different contributions. First, in the exponential there is always the branching point of the square root that gives a change of sign between the intervals $(-\infty,-m)$ and $(m,\infty)$. Secondly, we have the factor $(m^2-k^2)^{(n-q)/2}$ that depending on whether $n-q$ is odd or even, it adds or not another change of sign due to the branching point of the square root. Thus we have 
 \begin{align}\nonumber
	&\frac{d}{ds}\zeta^{l\not =0}_{\text{ren}}(L,m,\beta;s)\Bigr|_{s=0}=-\frac{A}{\pi^{1+n}}\sum_{l=1}^\infty\int_{m}^\infty dk\ \left(L-\frac{d}{dk}\log \frac{h^L_U(ik)}{h^{\infty}_U(ik)}\right)  \\ \label{eq:zeta_temp_odd}
	&\times\sum_{q=0}^n\frac{(n+q)!}{q!(n-q)!}\frac{(k^2-m^2)^{\frac{n-q}{2}}i^{n-q+1}}{2^{q+n}\ \beta^{n+q}\ l^{n+q+1}}\left(e^{-i\beta l\sqrt{k^2-m^2}}-(-1)^{n-q}e^{i\beta l\sqrt{k^2-m^2}}\right)
	,
\end{align}
where we have taken the $L_0\rightarrow\infty$ limit. All terms in the sum over $q$ are real. When $n-q$ is even the term $i^{n-q+1}$ is imaginary but we have the difference of exponential with opposite signs that gives a pure imaginary contribution. On the other hand, when $n-q$ is odd the factor  $i^{n-q+1}$ is real and we have a sum of exponentials with opposite signs that gives also a real term.

\section{Casimir energy}\label{sec:Casimir_cont}

Once we have computed all contributions to the renormalized zeta function we can easily obtain the Casimir energy from it. The effective action \eqref{eq:zeta_combination}  gives the free energy  by $F=S_{\text{eff}}/\beta$, and this free energy has two different contributions \cite{bordag2018free},  one that does depend on the temperature ($l\neq 0$) and another that does not ($l=0$) that gives the Casimir energy $F^{l=0}_U(L,m,\beta)=E_U(L,m)$. In the even dimensional case $D=2n$ the Casimir energy is \eqref{eq:zeta_even_Cas}
\begin{equation}\label{eq:Casimir_even}
E_U^{2n}(L,m)	=\frac{A }{2^{2n+1}\pi^{n}\Gamma(n+1) }\int_{m}^\infty dk \left(k^2-m^2\right)^n\left(L-\frac{d}{dk}\log \frac{h^L_U(ik)}{h^{\infty}_U(ik)}\right),
\end{equation}
whereas in the odd case $D=2n+1$ \eqref{eq:zeta_odd_Cas}
\begin{equation} \label{eq:Casimir_odd}
	E^{2n+1}_U(L,m)=(-1)^{n+1}\frac{A  \Gamma \left(-n-\frac{1}{2}\right)}{2^{2n+2}\pi^{n+\frac{3}{2}} }\int_{m}^\infty  dk \left(k^2-m^2\right)^{n+\frac{1}{2}}\left(L-\frac{d}{dk}\log \frac{h^L_U(ik)}{h^{\infty}_U(ik)}\right).
\end{equation}
If we undo the change $D=2n+1$ in the odd case and use Euler's reflection formula 
\begin{equation}
\Gamma(z)\Gamma(1-z)=\frac{\pi}{\sin(\pi z)}
\end{equation}
with $z=-D/2$, we arrive at
\begin{equation}\label{eq:Casimir_energy_continuum}
	E_U(L,m)=\frac{A }{2^{D+1}\pi^{D/2}\Gamma\left(\frac{D}{2}+1\right) }\int_{m}^\infty  dk \left(k^2-m^2\right)^{D/2}\left(L-\frac{d}{dk}\log \frac{h^L_U(ik)}{h^{\infty}_U(ik)}\right),
\end{equation}
which is exactly the same expression obtained by undoing the change $D=2n$ in the even formula \eqref{eq:Casimir_even}. Therefore, the Casimir energy for any spatial dimension $D$ is given by \eqref{eq:Casimir_energy_continuum}.

The rest of the contributions to the free energy $F^{l\not =0}_U(L,m,\beta)$ are the ones that depend on the temperature, in the even dimensional case $D=2n$ we have \eqref{eq:zeta_temp_even}
 \begin{align}\nonumber
	F^{l\not =0}_{2n,U}(L,m,\beta)=\frac{2^{1-n}A}{\pi^{1+n}\beta^{n}}\sum_{l=1}^\infty&\int_{m}^\infty dk\ \frac{\left(k^2-m^2\right)^{\frac{n}{2}}}{l^{n}}\ \text{Re}\left(i^{n+1} K_{n}\left(i\beta l\sqrt{k^2-m^2}\right)\right)\\ \label{eq:Free_energy_even}
	&\times\left(L-\frac{d}{dk}\log \frac{h^L_U(ik)}{h^{\infty}_U(ik)}\right),
\end{align}
and the odd dimensional case $D=2n+1$ \eqref{eq:zeta_temp_odd}
 \begin{align}\nonumber
	&F^{l\not =0}_{2n+1,U}(L,m,\beta)=\frac{A}{2\pi^{1+n}\beta}\sum_{l=1}^\infty\int_{m}^\infty dk\ \left(L-\frac{d}{dk}\log \frac{h^L_U(ik)}{h^{\infty}_U(ik)}\right)  \\ \label{eq:Free_energy_odd}
	&\times\sum_{q=0}^n\frac{(n+q)!}{q!(n-q)!}\frac{(k^2-m^2)^{\frac{n-q}{2}}i^{n-q+1}}{2^{q+n}\ \beta^{n+q}\ l^{n+q+1}}\left(e^{-i\beta l\sqrt{k^2-m^2}}-(-1)^{n-q}e^{i\beta l\sqrt{k^2-m^2}}\right)
	.
\end{align}
We can recover the dependence on the second kind modified Bessel function in the previous formula by using \eqref{eq:bessel_semi_integer}
 \begin{align}\nonumber
	&F^{l\not =0}_{2n+1,U}(L,m,\beta)=\\\nonumber
	&=\frac{A}{2^{n+1/2}\pi^{3/2+n}\beta^{n+1/2}}\sum_{l=1}^\infty\int_{m}^\infty \frac{i^{n+3/2}(k^2-m^2)^{\frac{n}{2}+1/4}}{l^{n+1/2}}dk\ \left(L-\frac{d}{dk}\log \frac{h^L_U(ik)}{h^{\infty}_U(ik)}\right)\\\nonumber
	&\times\left(K_{n+\frac{1}{2}}\left(i\beta l\sqrt{k^2-m^2}\right)+i(-1)^{n}\left(K_{n+\frac{1}{2}}\left(-i\beta l\sqrt{k^2-m^2}\right)\right)\right)\\\nonumber
	&=\frac{2A}{2^{n+1/2}\pi^{3/2+n}\beta^{n+1/2}}\sum_{l=1}^\infty\int_{m}^\infty \frac{(k^2-m^2)^{\frac{n}{2}+1/4}}{l^{n+1/2}}dk\ \left(L-\frac{d}{dk}\log \frac{h^L_U(ik)}{h^{\infty}_U(ik)}\right)\\
	&\times \text{Re}\left(i^{n+3/2} K_{n+\frac{1}{2}}\left(i\beta l\sqrt{k^2-m^2}\right)\right),
\end{align}
and undo the change $D=2n+1$ obtaining
\begin{align}\nonumber
	F^{l\not =0}_U(L,m,\beta)&=\frac{(2/\pi)A}{(2\pi\beta)^{\frac{D}{2}}}\sum_{l=1}^\infty\int_{m}^\infty dk\ \frac{(k^2-m^2)^{\frac{D}{4}}}{l^{\frac{D}{2}}} \left(L-\frac{d}{dk}\log \frac{h^L_U(ik)}{h^{\infty}_U(ik)}\right)\\\label{eq:Free_energy_temperature}
	&\times \text{Re}\left(i^{\frac{D}{2}+1} K_{\frac{D}{2}}\left(i\beta l\sqrt{k^2-m^2}\right)\right).
\end{align}
This is exactly the same formula we would obtain by also undoing the change of the even case $D=2n$ in \eqref{eq:Free_energy_even}, therefore \eqref{eq:Free_energy_temperature} is the general formula for any spatial dimension. 

From these general formulas \eqref{eq:Casimir_energy_continuum} and \eqref{eq:Free_energy_temperature}, it can be seen that all the terms of the free energy go to zero as the distance between the walls $L$ tends to infinity. This is precisely the physical condition we used for fixing the renormalization scheme prescription \eqref{eq:zeta_combination}. Also, we recover the zero temperature energy limit, since when we do the $\beta \rightarrow\infty$ limit the temperature dependent terms of the free energy go to zero $F^{l\not =0}_U\rightarrow 0$.

\subsection{Asymptotic behaviour}
In this section, we  analyze the behaviour of the Casimir energy when the effective distance between the walls is very large  $mL\rightarrow \infty$.

Let us now rewrite the spectral function \eqref{eq:spectral_function} as 
\begin{align}\nonumber
	h_U^L(ik)&=\ e^{kL}\left((k^2+1)\cos\gamma+(k^2-1)\cos\alpha+2k\sin\alpha\right)\\
	&\times\left(1+n_1\sin(\gamma) {\mathcal {A}}\ e^{-kL}+{\mathcal {B}}\ e^{-2kL}\right),
\end{align}
where ${\mathcal {A}}$ and ${\mathcal {B}}$ are
\begin{align}
	&{\mathcal {A}}(k,\alpha,\gamma)=\frac{4k}{(k^2+1)\cos\gamma+(k^2-1)\cos\alpha+2k\sin\alpha}\\
	&{\mathcal {B}}(k,\alpha,\gamma)=\frac{-(k^2+1)\cos\gamma-(k^2-1)\cos\alpha+2k\sin\alpha}{(k^2+1)\cos\gamma+(k^2-1)\cos\alpha+2k\sin\alpha}.
\end{align}
We can use this formula to approximate the logarithm of the quotient of spectral functions 
\begin{equation} \label{eq:expansion_log_spectral}
	\log \frac{h_U^L(ik)}{h_U^\infty(ik)}=kL+n_1\sin\gamma {\mathcal {A}}\ e^{-kL}+\left({\mathcal {B}}-\frac{{\mathcal {A}}'}{2}\right)e^{-2kL}+O(e^{-3kL}),
\end{equation}
where we have expanded in powers of $e^{-kL}$ and defined $\mathcal {A}'=(n_1\sin(\gamma)\mathcal A)^2$.

  We can introduce this expansion into the Casimir energy formula \eqref{eq:Casimir_energy_continuum} giving rise to
\begin{align}\nonumber
	E_U(L,m)&=\frac{A }{2^{D+1}\pi^{D/2}\Gamma\left(\frac{D}{2}+1\right) }\int_{m}^\infty  dk \left(k^2-m^2\right)^{D/2}\\\nonumber
	&\times\frac{d}{dk}\left(n_1\sin\gamma {\mathcal {A}}\ e^{-kL}+\left({\mathcal {B}}-\frac{{\mathcal {A}}'}{2}\right)e^{-2kL}+O(e^{-3kL})\right)\\\nonumber
	&=\frac{AD }{2^{D+1}\pi^{D/2}\Gamma\left(\frac{D}{2}+1\right) }\int_{m}^\infty  dk\ k\left(k^2-m^2\right)^{D/2-1}\\\label{eq:Casimir_expansion}
	&\times\left(n_1\sin\gamma {\mathcal {A}}\ e^{-kL}+\left({\mathcal {B}}-\frac{{\mathcal {A}}'}{2}\right)e^{-2kL}+O(e^{-3kL})\right),
\end{align}
where we have integrated by parts to get rid of the derivative with respect to $k$. We can actually recover all the terms in the expansion of the logarithm \eqref{eq:expansion_log_spectral} in the previous formula, which allows us to write the Casimir energy as a sum of the integrals with powers of the exponential $e^{-kL}$
\begin{equation}
E_U(L,m)=\frac{AD }{2^{D+1}\pi^{D/2}\Gamma\left(\frac{D}{2}+1\right) }\sum_{j=1}^\infty \int_{m}^{\infty}dk\ h(k,n_1,\gamma,\alpha)e^{-jkL}.
\end{equation}
Again there is a difference between the even and odd cases since applying  integration by parts we get
\begin{equation}
	\int_m^\infty dk\ h(k,n_1,\gamma,\alpha)e^{-jkL}=-\left.\frac{h(k,n_1,\gamma,\alpha)}{jL}e^{-jkL}\right |_{m}^\infty+ \int_m^\infty dk\ \frac{h(k,n_1,\gamma,\alpha)'}{jL}e^{-jkL},
\end{equation}
and only in the even case $h$ is regular in $[m,\infty]$ due to the factor $(k^2-m^2)^{D/2-1}$. Therefore, by iterating this process of integrating by parts we can solve the integral in the even case as a power series in $L$ for each exponential order
\begin{equation}
	E_U^{2n}(L,m)=\sum_{j=1}^\infty\sum_{\mu=1}^{\infty}\frac{b_{j,\mu}(m,n_1,\gamma,\alpha)A}{(jL)^\mu}e^{-jmL},
\end{equation}
where
\begin{equation}\label{eq:leading_even}
	b_{1,\mu}=\frac{D\ n_1\sin \gamma  }{2^{D+1}\pi^{D/2}\Gamma\left(\frac{D}{2}+1\right) }\left.\frac{d^\mu}{dk^{\mu}}\left(k(k^2-m^2)^{D/2-1}\mathcal A\right)\right |_{k=m}.
\end{equation}

In the odd dimensional case, we can not use the method of integrating by parts for each exponential order because when deriving the term $\left(k^2-m^2\right)^{D/2-1}$ we end up with a singular contribution when $k=m$. Instead, we can use the saddle point approximation
\begin{equation}
	\int_m^\infty e^{F(x)}dx \simeq e^{F(x_0)}\int_{m}^{\infty}e^{\frac{1}{2} F''(x_0)(x-x_0)^2} dx
\end{equation}
where $x_0$ is the maximum of $F(x)$ in the integration interval. For each exponential order, $F(k)$ is  just the logarithm of the integrand in \eqref{eq:Casimir_expansion}
\begin{equation}\label{eq:saddle_function}
F_j(k)=-jkL+\left(\frac{D}{2}-1\right)\log(k^2-m^2)+w(k,m,n_1,\gamma,\alpha)
\end{equation}
with $w(k,m,n_1,\gamma,\alpha)$ the logarithm of a combination of $\mathcal A$, $\mathcal B$ and their derivatives. Since we are on the large $mL$ limit, this expression is   dominated by the first term $-jkL$, which means that the maximum value is  very close to $m$, i.e.
\begin{equation}\label{eq:maximum}
	x_0=m\left(1+\frac{a}{mL}+O\left(\frac{1}{(mL)^2}\right)\right) \hspace{1cm}\text{and }x_0>m.
\end{equation}
Hence, the integral for each exponential order of $kL$ can be expressed as
\begin{equation}
\int_m^\infty dk\ h(k,n_1,\gamma,\alpha)e^{-jkL}=e^{-jmL}\ G_j(L,m,n_1,\gamma,\alpha),
\end{equation}
where the factor $e^{-jmL}$ is the first term we obtain in $e^{F(x_0)}$ by inserting \eqref{eq:maximum} in \eqref{eq:saddle_function}, and $G_j$ is a function that carries the rest of the contributions. This procedure allows us to rewrite the Casimir energy formula in the same way as in the even case, namely, as a sum of exponentials of $mL$
\begin{equation}
	E_U^{2n+1}(L,m)=\frac{AD }{2^{D+1}\pi^{D/2}\Gamma\left(\frac{D}{2}+1\right) }\sum_{j=1}^\infty G_j(L,m,n_1,\gamma,\alpha)e^{-jmL},
\end{equation}
in which the first term is proportional to $n_1\sin \gamma$
\begin{equation}
	G_1(L,m,n_1,\gamma,\alpha)\propto n_1\sin \gamma.
\end{equation}
This dependence of the first exponential term with $n_1\sin \gamma$ is shared with the even case \eqref{eq:leading_even}, and it just depends on the boundary conditions. Therefore, when the boundary condition implies that $n_1\sin \gamma=0$ the asymptotic behaviour of the Casimir energy is  different than when it is not zero. We can actually rewrite this condition in terms of the trace of the matrix $U$  that parametrizes the boundary condition as
\begin{equation}\label{eq:rate}
	E_U\sim \left\{ \begin{matrix}e^{-mL} &\text{if  }\text{ tr}(U\sigma_1)\not =0\\
		e^{-2mL} &	\text{if  }\text{ tr}(U\sigma_1)=0.  \end{matrix} \right.
\end{equation}
Which means that when $n_1\sin\gamma \not =0$ (which is equivalent to the parametrization of the boundary conditions by the $U(2)$ matrix depending on $\sigma_1$) the Casimir energy decays exponentially as $e^{-mL}$. Whereas, when $n_1\sin\gamma =0$ (the $U(2)$ matrix does not depend on $\sigma_1$) the Casimir energy decays faster with $e^{-2mL}$.

This is one of the main results of this chapter, the boundary conditions are divided in two families depending on the rate of the exponential decay of the Casimir energy with the effective distance between the boundary walls $mL$. The physical condition that separates these two families is how their boundary conditions are imposed, when they relate the values and/or the derivatives at the two boundary walls ($\text{tr}(U\sigma_1)\not =0$) we have the slow exponential decay $e^{-mL}$ , whilst if the boundary conditions imposes independent constraints for each wall ($\text{tr}(U\sigma_1 =0$) the faster decay appears $e^{-2mL}$.

This behaviour was previously observed for some particular boundary conditions like Dirichlet or periodic \cite{cougo1994schwinger,fulling2005mass}, but here is found that is a general rule for these homogeneous boundary conditions along the walls for arbitrary spatial dimensions $D$.

\section{Special  boundary conditions}\label{sec:Particular_conditions}
In this section we shall analyze some particular boundary conditions that will be of interest in the following chapters for the lattice scalar and gauge theories, and also belong to each of the two different families described by their asymptotic behaviour on the Casimir energy.

\subsection{Dirichlet boundary conditions}
\begin{figure}[H]
	\centering
	\includegraphics[width=1\textwidth]{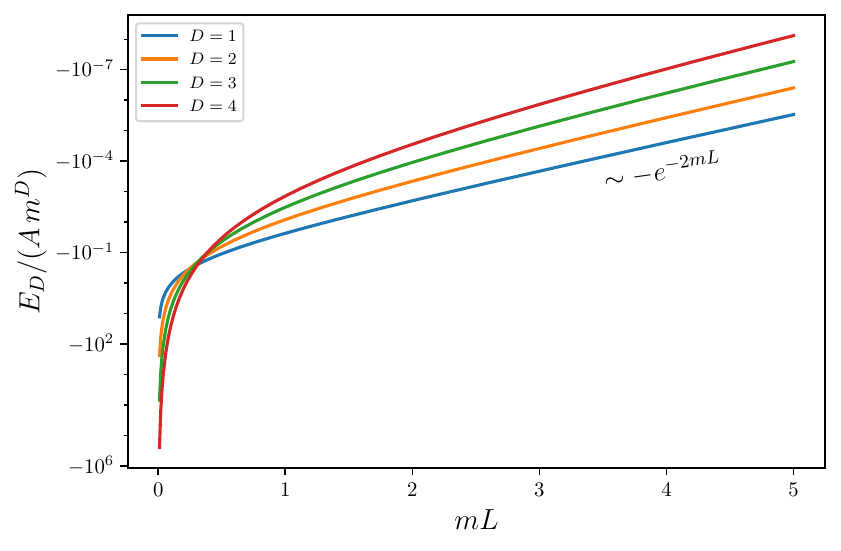}
	\caption{Dimensionless Casimir energy with \acrshort*{dbc}/\acrshort*{nbc} in logarithmic scale as a function of $mL$ for spatial dimensions $D=\{1,2,3,4\}$.}
	\label{fig:plot_Casimir_Dirichlet}
\end{figure}

One of the most well known boundary conditions are the  Dirichlet boundary conditions (\acrshort*{dbc}) which constrain the fields to vanish at the boundary, $\varphi(L/2)=\varphi(-L/2)=0$. This condition corresponds in the general framework of boundary conditions to the parameters $\alpha=\pi$ and $\gamma=0$ which makes the $U(2)$ matrix associated $U_D=-\mathbb{I}$. Once we have the values of the parameters of the boundary condition, we can get the derivative of the logarithm of spectral functions 
\begin{equation}\label{eq:spectral_Dir}
	\frac{d}{dk}\log\left(h^L_{U_D}/h^\infty_{U_D}\right)=L\coth(kL).
\end{equation}
 This quantity   characterizes the boundary condition on the integrals we obtained for the Casimir energy and free energy for every dimension.
Therefore, the Casimir energy is given by \eqref{eq:Casimir_energy_continuum}
\begin{equation}\label{eq:Casimir_Dirichlet_integral}
	E_D(L,m)=\frac{A L }{2^{D+1}\pi^{D/2}\Gamma\left(\frac{D}{2}+1\right) }\int_{m}^\infty  dk \left(k^2-m^2\right)^{D/2}\left(1-\coth(kL)\right).
\end{equation}
In the \autoref{sec:Bessel_integral} of the Appendix it is shown how we can use a sum of modified second kind Bessel functions $K_q(z)$ for simplifying the integral when $m\not =0$ to 
\begin{equation}\label{eq:Casimir_Dirichlet}
	E_D(L,m)=-\frac{Am^{\frac{D+1}{2}}}{2^D\pi^{\frac{D+1}{2}}L^{\frac{D-1}{2}}}\sum_{j=1}^\infty\frac{1}{j^{\frac{D+1}{2}}}K_{\frac{D+1}{2}}(2jmL).
\end{equation}
For the massless case we can get a more explicit expression of the Casimir energy in two different ways. Either solving the integral \eqref{eq:Casimir_Dirichlet_integral} with $m=0$ or doing the $m\rightarrow 0$ limit in \eqref{eq:Casimir_Dirichlet} using the asymptotic behaviour of the Bessel functions $K_q$ when the argument goes to zero \eqref{eq:Bessel_asymptotic_zero}. We obtain the same result in both methods
\begin{equation}\label{eq:Casimir_Dirichlet_massless}
	E_D(L,0)=\frac{A\Gamma\left(\frac{D+1}{2}\right)}{2^{D+1}\pi^{\frac{D+1}{2}}L^D}\text{Li}_{D+1}(1).
\end{equation}
Now we can use the asymptotic behaviour of the Bessel functions \eqref{eq:Bessel_asymptotic} on this expression to study the behaviour of the Casimir energy    in the $mL\rightarrow\infty$ limit. The Casimir energy has the following expansion
\begin{equation}\label{eq:Casimir_asymptotic_Dirichlet}
	E_D(L,m)=-\frac{Am^{D/2}}{2^{D+1}\pi^{D/2}L^{D/2}}e^{-2mL}\left(1+O(1/mL)\right)+O(e^{-4mL}).
\end{equation}
It clearly shows the exponential decay $e^{-2mL}$ which is what was expected since the boundary condition imposes independent conditions at both walls \footnote{Although \acrshort*{dbc} and \acrshort*{nbc} impose the same condition on both walls, they are extreme cases of the so called Robin boundary conditions, that are a family of local boundary conditions that relate the values and derivatives of the fields at each wall independently \cite{angelone2024classical}. \label{fn:DirNeu}}, i.e. tr$(U_D\sigma_1)=0$. This asymptotic behaviour, that is independent of the dimension, can be seen in \autoref{fig:plot_Casimir_Dirichlet}, where in logarithmic scale the curves tend to be parallel as $mL$ grows.

Now, we can insert the spectral term \eqref{eq:spectral_Dir} into the $l\not =0$ free energy terms \eqref{eq:Free_energy_temperature}
\begin{align}\nonumber
	F^{l\not =0}_D(L,m,\beta)&=\frac{(2/\pi)AL}{(2\pi\beta)^{\frac{D}{2}}}\sum_{l=1}^\infty\int_{m}^\infty dk\  \frac{(k^2-m^2)^{\frac{D}{4}}}{l^{\frac{D}{2}}} \left(1-\coth(kL)\right)\\\label{eq:Free_energy_Dirichlet}
	&\times \text{Re}\left(i^{\frac{D}{2}+1} K_{\frac{D}{2}}\left(i\beta l\sqrt{k^2-m^2}\right)\right).
\end{align}
This integral can not be solved or simplified further but one can show that it follows a similar asymptotic behaviour than the Casimir energy by noticing that the spectral term in the $kL\rightarrow \infty$ limit behaves as
\begin{equation}
	1-\coth(kL)\sim e^{-2kL}.
\end{equation}
and check that this  asymptotic behaviour is displayed in \autoref{fig:plot_FreeEnergy_2d} and \autoref{fig:plot_FreeEnergy_3d}.

\subsection{Neumann boundary conditions}
Neumann boundary conditions (\acrshort*{nbc}) are also very well known boundary conditions. They require that the normal derivatives of the fields at the boundaries vanish $\dot{\varphi}(L/2)=\dot{\varphi}(-L/2)=0$. The parameters that characterize these conditions in the general framework of boundary conditions are $\alpha=0$ and $\gamma=0$, which makes the $U(2)$ matrix $U_N=\mathbb I$. The spectral factor is then 
\begin{equation}\label{eq:spectral_Ne}
	\frac{d}{dk}\log\left(h^L_{U_N}/h^\infty_{U_N}\right)=L\coth(kL),
\end{equation}
that is the same as in the \acrshort*{dbc} \eqref{eq:spectral_Dir}. Since this is the factor that characterizes the boundary conditions in the Casimir energy and free energy formulas, this means that it will be exactly the same for \acrshort*{nbc} and \acrshort*{dbc}. Obviously, the asymptotic behaviour is also the same one, which also fulfills our rule \eqref{eq:rate} because tr\,($U_N\sigma_1)=0$ which means that  the boundary conditions imposed are independent at each wall\footref{fn:DirNeu}. 
\subsection{Periodic boundary conditions}
Periodic boundary conditions (\acrshort*{pbc}) correspond to conditions where the values of the fields at both boundaries are equal $\varphi(L/2)=\varphi(-L/2)$ whereas the normal derivatives have opposite signs $\dot \varphi(L/2)=-\dot \varphi(-L/2)$ \footnote{Notice that the opposite sign is due to the normal derivatives having opposite direction, see \autoref{fig:Boundary_wall_representation}.}. These conditions are parametrized by the matrix $U_P=\sigma_1$ which corresponds to $\alpha=\pi/2$, $\gamma=-\pi/2$ and $\bold n=(1,0,0)$, and the logarithm ratio of spectral functions 
satisfies that
\begin{equation}\label{eq:spectral_per}
	\frac{d}{dk}\log\left(h^L_{U_P}/h^\infty_{U_P}\right)=L\coth(kL/2).
\end{equation}

\begin{figure}[H]
 	\centering
 	\includegraphics[width=1\textwidth]{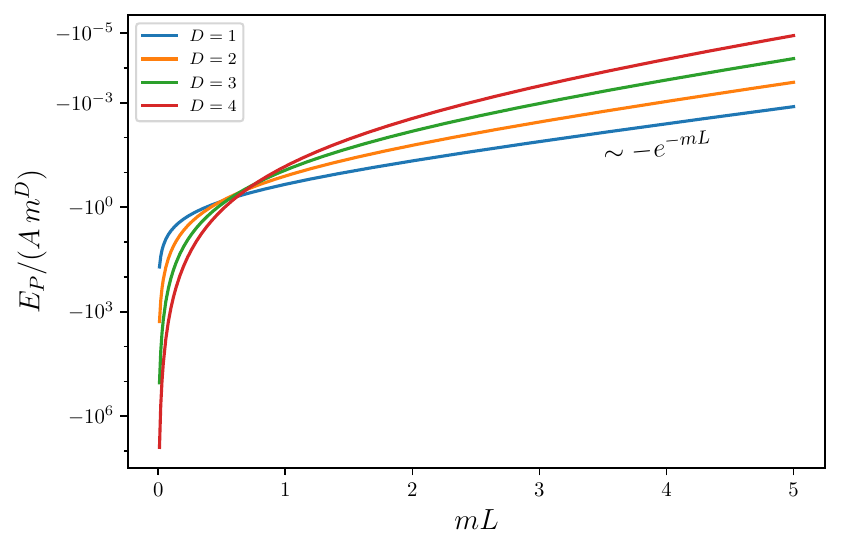}
 		\caption{Dimensionless Casimir energy with \acrshort*{pbc} in logarithmic scale as a function of $mL$ for spatial dimensions $D=\{1,2,3,4\}$.}
 	\label{fig:plot_Casimir_Periodic}
 \end{figure}

As in the \acrshort*{dbc} case, once we know this factor we can calculate the Casimir energy  by using the sum of Bessel functions $K_q$  \eqref{eq:Bessel_periodic_result} getting
\begin{equation}\label{eq:Casimir_periodic}
	E_P(L,m)=-\frac{Am^{\frac{D+1}{2}}}{2^{\frac{D-1}{2}}\pi^{\frac{D+1}{2}}L^{\frac{D-1}{2}}}\sum_{j=1}^\infty\frac{1}{j^{\frac{D+1}{2}}}K_{\frac{D+1}{2}}(jmL).
\end{equation}
In the massless case we can directly integrate the Casimir energy formula or apply the massless limit in the previous formula. In both cases we get
\begin{equation}
	E_P(L,0)=-\frac{A\Gamma\left(\frac{D+1}{2}\right)}{\pi^{\frac{D+1}{2}} L^{D}}\text{Li}_{D+1}(1).
\end{equation}
Once we have the formulas of the Casimir energy in terms of a sum of Bessel functions we can analyze the asymptotic behaviour with $mL$ using the  asymptotic behaviour of Bessel functions \eqref{eq:Bessel_asymptotic}
\begin{equation}\label{eq:Casimir_asymptotic_periodic}
	E_P(L,m)=-\frac{Am^{D/2}}{(2\pi L)^{D/2}}e^{-mL}\left(1+O(1/mL)\right)+O(e^{-2mL}).
\end{equation}	
 We observe that this is the expected asymptotic behaviour since \acrshort*{pbc} belong to the family where the decay is $e^{-mL}$ because tr$(U_P\sigma_1)\not =0$. This comes from the fact that, by definition, \acrshort*{pbc} imposes a relation  between both boundary walls. By plotting the Casimir energy for different spatial dimensions we can see in \autoref{fig:plot_Casimir_Periodic} how this asymptotic behaviour is independent from the dimension.

Finally we can insert the spectral term \eqref{eq:spectral_per} on the rest of the free energy contributions \eqref{eq:Free_energy_temperature} 
\begin{align}\nonumber
	F^{l\not =0}_P(L,m,\beta)&=\frac{(2/\pi)AL}{(2\pi\beta)^{\frac{D}{2}}}\sum_{l=1}^\infty\int_{m}^\infty dk\ \frac{(k^2-m^2)^{\frac{D}{4}}}{l^{\frac{D}{2}}} \left(1-\coth(kL/2)\right)\\\label{eq:Free_energy_periodic}
	&\times \text{Re}\left(i^{\frac{D}{2}+1} K_{\frac{D}{2}}\left(i\beta l\sqrt{k^2-m^2}\right)\right).
\end{align}
We can not analytically integrate this formula or simplify it further, but again, we can use the asymptotic behaviour of the spectral term  with $kL$
\begin{equation}
		1-\coth(kL/2)\sim e^{-kL}
\end{equation}
to hint that it have the same asymptotic behaviour as the Casimir energy. This is shown in \autoref{fig:plot_FreeEnergy_2d} and \autoref{fig:plot_FreeEnergy_3d}, for the 2+1 and 3+1 dimensional cases.

\subsection{Anti-periodic boundary conditions}
\begin{figure}[H]
	\centering
	\includegraphics[width=1\textwidth]{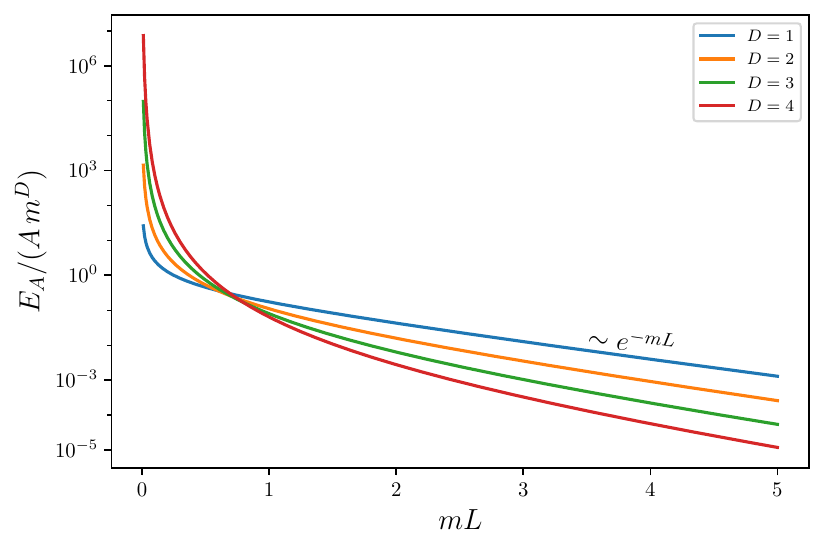}
	\caption{Dimensionless Casimir energy with \acrshort*{abc} in logarithmic scale as a function of $mL$ for spatial dimensions $D=\{1,2,3,4\}$.}
	\label{fig:plot_Casimir_Anti}
\end{figure}

The anti-periodic boundary conditions (\acrshort*{abc}) are given by the conditions $\varphi(L/2)=-\varphi(-L/2)$ and $\dot\varphi(L/2)=\dot\varphi(-L/2)$. The corresponding $U(2)$ matrix is $U_A=-\sigma_1$ and the parameters associated are $\alpha=\pi/2$, $\gamma=\pi/2$ and $\bold n=(1,0,0)$. With these parameters the logarithm of the spectral functions satisfies
\begin{equation}\label{eq:spectral_anti}
	\frac{d}{dk}\log\left(h^L_{U_A}/h^\infty_{U_A}\right)=L\tanh(kL/2).
\end{equation}
Now, by inserting the previous expression into \eqref{eq:Casimir_energy_continuum} and using \eqref{eq:Bessel_anti_result} we obtain the Casimir energy 
\begin{equation}\label{eq:Casimir_anti}
	E_A(L,m)=\frac{Am^{\frac{D+1}{2}}}{2^{\frac{D-1}{2}}\pi^{\frac{D+1}{2}}L^{\frac{D-1}{2}}}\sum_{j=1}^\infty\frac{(-1)^{j+1}}{j^{\frac{D+1}{2}}}K_{\frac{D+1}{2}}(jmL)
\end{equation}
in the massive case, whereas in the massless situation we arrive at
\begin{equation}
	E_A(L,0)=\frac{A(1-2^{-D})\Gamma\left(\frac{D+1}{2}\right)}{\pi^{\frac{D+1}{2}}\ L^{D}}\zeta(1+D).
\end{equation}
We can analyze the asymptotic behaviour as in the previous cases by using \eqref{eq:Bessel_asymptotic}
\begin{equation}
E_A=\frac{Am^{D/2}}{(2\pi L)^{D/2}}e^{-mL}\left(1+O(1/mL)\right)+O(e^{-2mL}).
\end{equation}
The asymptotic behaviour is the expected one ($e^{-mL}$), since these boundary conditions fulfill tr$(U_A\sigma_1)\neq 0$ which implies that the boundary condition connect the values at both boundary walls of the fields and/or its derivatives. This can also be seen in \autoref{fig:plot_Casimir_Anti} where the Casimir energy is displayed for different spatial dimensions. The asymptotic formula is  the same up to a global sign as for \acrshort*{pbc} \eqref{eq:Casimir_asymptotic_periodic}.

 We can introduce as in the previous cases the spectral factor \eqref{eq:spectral_anti} into the formulas for the rest of the free energy terms \eqref{eq:Free_energy_temperature}
\begin{align}\nonumber
	F^{l\not =0}_A(L,m,\beta)&=\frac{(2/\pi)AL}{(2\pi\beta)^{\frac{D}{2}}}\sum_{l=1}^\infty\int_{m}^\infty dk\ \frac{(k^2-m^2)^{\frac{D}{4}}}{l^{\frac{D}{2}}} \left(1-\tanh(kL/2)\right)\\\label{eq:Free_energy_anti}
	&\times \text{Re}\left(i^{\frac{D}{2}+1} K_{\frac{D}{2}}\left(i\beta l\sqrt{k^2-m^2}\right)\right),
\end{align}
which we can not analytically simplify any further. Again, the asymptotic behaviour of the spectral term
\begin{equation}
	1-\tanh(kL/2)\sim e^{-kL}
\end{equation}
suggest the same exponential behaviour as we will observe later in \autoref{fig:plot_FreeEnergy_2d} and \autoref{fig:plot_FreeEnergy_3d} for the 2+1 and 3+1 dimensional cases.

\subsection{Zaremba boundary conditions}
Let us now consider another type of boundary conditions where the behaviour of both boundaries is different, Zaremba boundary conditions (\acrshort*{zbc}). One of the wall satisfies \acrshort*{dbc} whereas the other one does it with \acrshort*{nbc},  i.e., $\varphi(L/2)=0$ and $\dot{\varphi}(-L/2)=0$, or $\varphi(-L/2)=0$ and $\dot{\varphi}(L/2)=0$. They have two possible matrix representations $U_Z=\pm \sigma_3$, and the associated parameters are $\alpha=\pi/2$, $\gamma=\mp\pi/2$ and $\bold n=(0,0,1)$. The sign depends on the choice of the wall (left or right) with \acrshort*{dbc}. The logarithm of spectral functions satisfies
\begin{equation}\label{eq:spectral_zaremba}
	\frac{d}{dk}\log\left(h^L_{U_Z}/h^\infty_{U_Z}\right)=L\tanh(kL).
\end{equation}
for both cases.

Once again, with this factor we can compute the Casimir energy \eqref{eq:Casimir_energy_continuum} in terms of the Bessel functions by using \eqref{eq:Bessel_zarembat_result} in the massive case
\begin{equation}\label{eq:Casimir_Zaremba}
	E_Z(L,m)=\frac{Am^{\frac{D+1}{2}}}{2^{D}\pi^{\frac{D+1}{2}}L^{\frac{D-1}{2}}}\sum_{j=1}^\infty\frac{(-1)^{j+1}}{j^{\frac{D+1}{2}}}K_{\frac{D+1}{2}}(2jmL).
\end{equation}
In the massless case instead we have 
\begin{equation}
	E_Z(L,0)=\frac{A(2^{D}-1)\Gamma\left(\frac{D+1}{2}\right)}{2^{2D+1}\pi^{\frac{D+1}{2}}L^{D}}\zeta(1+D).
\end{equation}

\begin{figure}[H]
	\centering
	\includegraphics[width=1\textwidth]{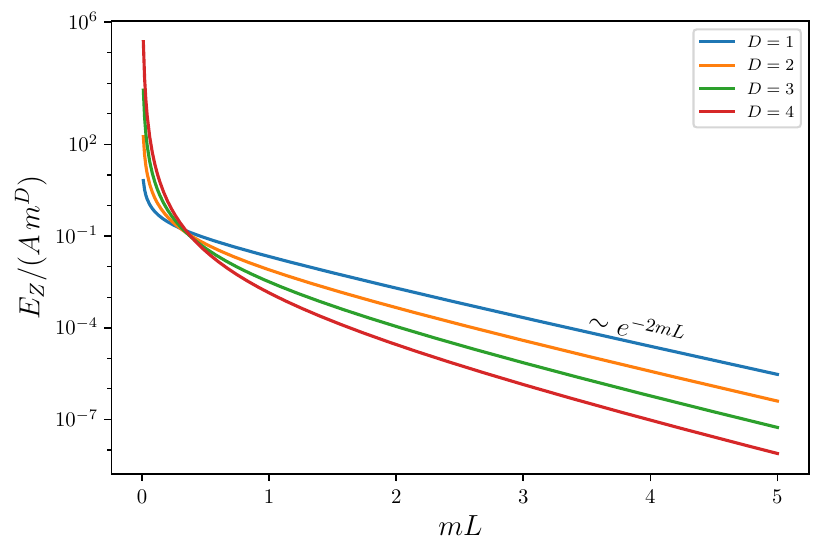}
	\caption{Dimensionless Casimir energy with \acrshort*{zbc} in logarithmic scale as a function of $mL$ for spatial dimensions $D=\{1,2,3,4\}$.}
	\label{fig:plot_Casimir_Zaremba}
\end{figure}

Again we can explore  the asymptotic behaviour in the $mL\rightarrow \infty$ limit by using the asymptotic limit of the Bessel functions \eqref{eq:Bessel_asymptotic}
\begin{equation}
	E_Z(m,L)=\frac{Am^{D/2}}{2^{D+1}(\pi L)^{D/2}}e^{-2mL}\left(1+O(1/mL)\right)+O(e^{-4mL}).
\end{equation}
In fact, we get the same asymptotic behaviour up to a global sign as for \acrshort*{dbc} and \acrshort*{nbc} with $e^{-2mL}$, that is expected  since these boundary conditions are independent at each wall, i.e.  tr$(U_Z\sigma_1)=0$. In \autoref{fig:plot_Casimir_Zaremba} this asymptotic behaviour can be seen for different spatial dimensions. Also, this asymptotic formula is the same as for \acrshort*{dbc}/\acrshort*{nbc}  \eqref{eq:Casimir_asymptotic_Dirichlet} but positive instead of negative coefficient.

Finally, we can introduce the spectral factor \eqref{eq:spectral_zaremba} into the formula of the free energy terms \eqref{eq:Free_energy_temperature}
\begin{align}\nonumber
	F^{l\not =0}_Z(L,m,\beta)&=\frac{(2/\pi)AL}{(2\pi\beta)^{\frac{D}{2}}}\sum_{l=1}^\infty\int_{m}^\infty dk\ \frac{(k^2-m^2)^{\frac{D}{4}}}{l^{\frac{D}{2}}} \left(1-\tanh(kL)\right)\\\label{eq:Free_energy_zaremba}
	&\times \text{Re}\left(i^{\frac{D}{2}+1} K_{\frac{D}{2}}\left(i\beta l\sqrt{k^2-m^2}\right)\right).
\end{align}
Expanding  the integrand in  power series of the spectral term
\begin{equation}
	1-\tanh(kL)\sim e^{-2kL}
\end{equation}
we get the same exponential behaviour than the Casimir energy. This will be explicitly observed later in the 2+1 and 3+1 dimensional cases in \autoref{fig:plot_FreeEnergy_2d} and \autoref{fig:plot_FreeEnergy_3d}.

\section{2+1 dimensional case}\label{sec:21_continuum}
In this section we  study in detail the particular case of $D=2$ spatial dimensions which we will use later to compare with the results on the lattice regularization. This case was explicitly developed in Ref. \cite{ezquerro2024casimir} using the same renormalization techniques. In fact to obtain  the results for the two dimensional theory we have to simply evaluate the formulas we computed in \autoref{sec:Casimir_cont} and set $D=2$.

The Casimir energy corresponds to the formula \eqref{eq:Casimir_energy_continuum}, that gives the result
\begin{equation}
	E_U(L,m)	=\frac{A }{8\pi}\int_{m}^\infty dk \left(k^2-m^2\right)\left(L-\frac{d}{dk}\log \frac{h^L_U(ik)}{h^{\infty}_U(ik)}\right),
\end{equation}
and the temperature dependent terms of the free energy \eqref{eq:Free_energy_temperature} are
\begin{align}\nonumber
	F^{l\not =0}_U(L,m,\beta)&=-\frac{A}{\pi^2\beta}\sum_{l=1}^\infty\int_{m}^\infty dk\ \frac{\sqrt{k^2-m^2}}{l} \left(L-\frac{d}{dk}\log \frac{h^L_U(ik)}{h^{\infty}_U(ik)}\right)\\\label{eq:Free_energy_2d}
	&\times \text{Re}\left(K_{1}\left(i\beta l\sqrt{k^2-m^2}\right)\right).
\end{align}
These results agree with the formulas obtained in Ref. \cite{ezquerro2024casimir}.
\subsection{Special boundary conditions}
Let us now calculate the explicit formulas for the boundary conditions we analyzed in \autoref{sec:Particular_conditions}.

\begin{enumerate}[i)]
	\item {\it Dirichlet and Neumann boundary conditions}. As we mentioned before these two boundary conditions share the same spectral factor \eqref{eq:spectral_Dir} and in consequence have the same free energy. The Casimir energy \eqref{eq:Casimir_Dirichlet} is
	\begin{equation}\label{eq:Dirichlet_2d}
		E_D(L,m)=-\frac{Am^{3/2}}{4\pi^{3/2}L^{1/2}}\sum_{j=1}^\infty\frac{1}{j^{3/2}}K_{3/2}(2jmL),
	\end{equation}
    we can use equation \eqref{eq:Bessel_poly_dir} to simplify this expression by using the polylogarithm functions
    \begin{equation}\label{eq:Cas_dir_2d}
    	E_D(L,m)=-\frac{A}{16\pi L^2}\left(2mL\ \text{Li}_2\left(e^{-2mL}\right)+\text{Li}_3\left(e^{-2mL}\right)\right).
    \end{equation}
In the massless limit this formula reduces to
\begin{equation}\label{eq:Casimir_massless_dir_2d}
	E_D(L,0)=-\frac{A\zeta(3)}{16\pi L^2},
\end{equation}
where we have used Li$_3(1)=\zeta(3)$, and it does also coincides with the expression obtained by setting $D=2$ in the general massless even dimension formula \eqref{eq:Casimir_Dirichlet_massless}. The rest of the free energy is
\begin{align}\nonumber
	F^{l\not =0}_D(L,m,\beta)&=-\frac{AL}{\pi^{2}\beta}\sum_{l=1}^\infty\int_{m}^\infty dk\ \frac{\left(k^2-m^2\right)^{\frac{1}{2}}}{l}\\
	&\times \text{Re}\left(K_1\left(i\beta l\sqrt{k^2-m^2}\right)\right)\left(1-\coth(kL)\right).
\end{align}
\item {\it Periodic boundary conditions.} The Casimir energy \eqref{eq:Casimir_periodic} for this boundary condition is 
\begin{equation}\label{eq:Cas_per_2d}
	E_P(L,m)=-\frac{A}{2\pi L^2}\left(mL\ \text{Li}_2\left(e^{-mL}\right)+\text{Li}_3\left(e^{-mL}\right)\right)
\end{equation}
where we have already used \eqref{eq:Bessel_poly_dir} to express it in terms of polylogarithms functions. When $m=0$ it reduces to
\begin{equation}\label{eq:Casimir_massless_per_2d}
	E_P(L,0)=-\frac{A\zeta(3)}{2\pi L^2},
\end{equation}
and the rest of the free energy terms are
\begin{align}\nonumber
	F^{l\not =0}_P(L,m,\beta)&=-\frac{AL}{\pi^{2}\beta}\sum_{l=1}^\infty\int_{m}^\infty dk\ \frac{\left(k^2-m^2\right)^{\frac{1}{2}}}{l}\\
	&\times \text{Re}\left(K_1\left(i\beta l\sqrt{k^2-m^2}\right)\right)\left(1-\coth(kL/2)\right).
\end{align}
\item {\it Anti-periodic boundary conditions.} The formula of the Casimir energy \eqref{eq:Casimir_anti} simplifies to 
\begin{equation}
	E_A(L,m)=-\frac{A}{2\pi L^2}\left(mL\ \text{Li}_2\left(-e^{-mL}\right)+\text{Li}_3\left(-e^{-mL}\right)\right),
\end{equation}
by using the relationship between the sum of Bessel functions and polylogarithms \eqref{eq:Bessel_poly_anti},
whereas in the massless limit we get
\begin{equation}
	E_A(L,0)=\frac{3A\zeta(3)}{8\pi L^2}
\end{equation}
where we have used that Li$_3(-1)=-3\zeta(3)/4$. The  temperature dependent terms of the free energy are
\begin{align}\nonumber
	F^{l\not =0}_A(L,m,\beta)&=-\frac{AL}{\pi^{2}\beta}\sum_{l=1}^\infty\int_{m}^\infty dk\ \frac{\left(k^2-m^2\right)^{\frac{1}{2}}}{l}\\
	&\times \text{Re}\left(K_1\left(i\beta l\sqrt{k^2-m^2}\right)\right)\left(1-\tanh(kL/2)\right).
\end{align}
\item {\it Zaremba boundary conditions.} Using the same formula than in the \acrshort*{abc} case \eqref{eq:Bessel_poly_anti} the Casimir energy is
\begin{equation}
	E_Z(L,m)=-\frac{A}{16\pi L^2}\left(2mL\ \text{Li}_2\left(-e^{-2mL}\right)+\text{Li}_3\left(-e^{-2mL}\right)\right)
\end{equation}
that when $m=0$ reduces to
\begin{equation}
	E_Z(L,0)=\frac{3A\zeta(3)}{64\pi L^2}.
\end{equation}
The rest of the free energy terms are
\begin{align}\nonumber
	F^{l\not =0}_Z(L,m,\beta)&=-\frac{AL}{\pi^{2}\beta}\sum_{l=1}^\infty\int_{m}^\infty dk\ \frac{\left(k^2-m^2\right)^{\frac{1}{2}}}{l}\\ \label{eq:eq:a}
	&\times \text{Re}\left(K_1\left(i\beta l\sqrt{k^2-m^2}\right)\right)\left(1-\tanh(kL)\right).
\end{align}
\end{enumerate}
\subsection{Asymptotic behaviour}
\begin{figure}[H]
	\centering
	\includegraphics[width=1\textwidth]{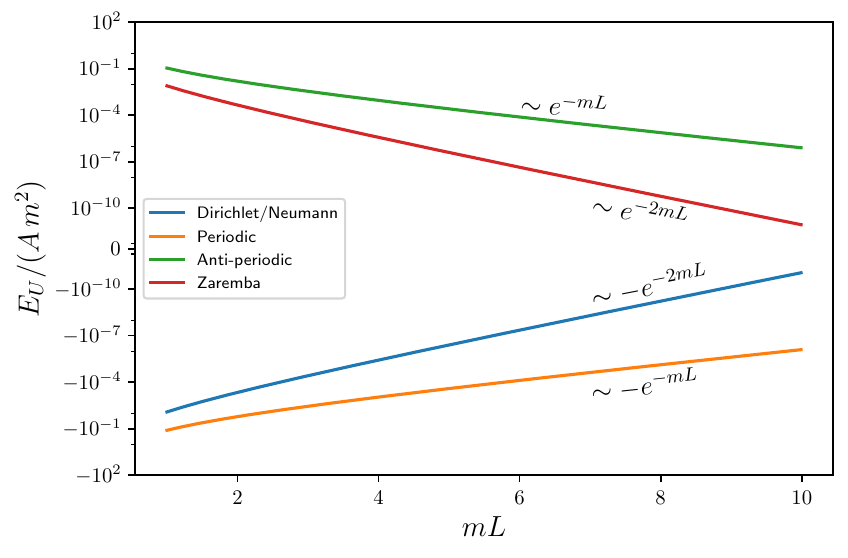}
	\caption{Exponential decay of the dimensionless Casimir energy (in logarithmic scale) as a function of the dimensionless distance $mL$ for different boundary conditions in 2+1 dimensions.}
	\label{fig:plot_Casimir_2d}
\end{figure}

Let us now analyze whether or not we get  the asymptotic behaviour predicted in \autoref{sec:Particular_conditions}. First, in \autoref{fig:plot_Casimir_2d} we display the Casimir energy for the five different boundary conditions we have analyzed. We observe the two different asymptotic behaviours \eqref{eq:rate} where the \acrshort*{dbc}, \acrshort*{nbc} and \acrshort*{zbc} have the expected faster decay whereas \acrshort*{pbc} and \acrshort*{abc} show a  slower decay. Also, we can notice how the boundary conditions have positive or negative energy, this is a well known fact in the literature \cite{Boundary_general_2013} and implies that some boundary conditions produce an attractive force (\acrshort*{dbc}, \acrshort*{nbc} or \acrshort*{pbc}) whereas others are repulsive (\acrshort*{abc} or \acrshort*{zbc}).

\begin{figure}[H]
	\centering
	\includegraphics[width=0.84\textwidth]{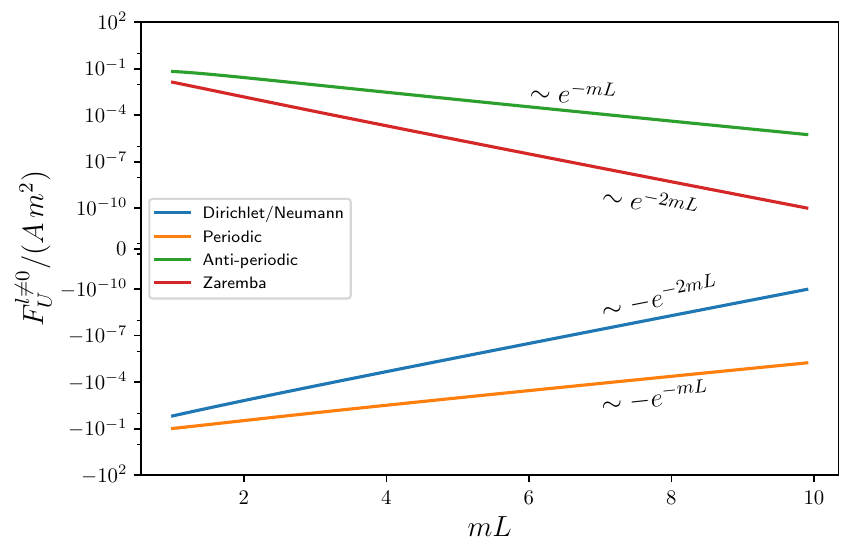}
	\caption{Exponential decay of the dimensionless free energy (in logarithmic scale) as a function of the dimensionless distance $mL$ for different boundary conditions with $m\beta=1$ in 2+1 dimensions.}
	\label{fig:plot_FreeEnergy_2d}
\end{figure}

In \autoref{fig:plot_FreeEnergy_2d} we display the asymptotic behaviour of the temperature dependent part of the free energy with $mL$. We can see that the boundary conditions follow the same asymptotic behaviour than the Casimir energy. \acrshort*{dbc}, \acrshort*{nbc} and \acrshort*{zbc} present the faster decay as $e^{-2mL}$, whereas \acrshort*{pbc} and \acrshort*{abc} show a slower exponential decay with $e^{-mL}$. As was argued before, this is due to the spectral factor on the integrals which is the one that drives this exponential decay behaviour both in the Casimir energy and the temperature dependent terms of the free energy.

\begin{figure}[H]
	\centering
	\includegraphics[width=0.84\textwidth]{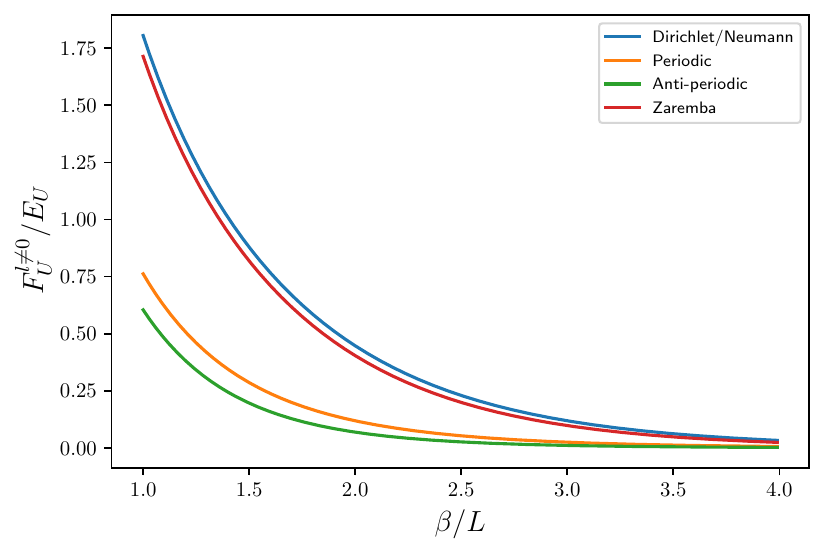}
	\caption{Quotient of the temperature dependent terms of the free energy and the Casimir energy as a function of $\beta/L$ for different boundary conditions with $mL=1$ in 2+1 dimensions.}
	\label{fig:plot_FreeEnergy__temp_2d}
\end{figure}

Finally, we plot in \autoref{fig:plot_FreeEnergy__temp_2d} the quotient of the temperature dependent part of the free energy $F_U^{l\not=0}$ and the Casimir energy $E_U$ with respect to $m\beta$ while fixing $mL$ for the different boundary conditions. \autoref{fig:plot_FreeEnergy__temp_2d} shows how as $\beta$ grows with respect to $L$  the Casimir energy contribution is the dominant term on the free energy for all five of the boundary conditions. This will be very relevant for the lattice simulations, on these we have to deal with the effects of finite temperature but this shows how if there is a low enough temperature (high $\beta$) this effect is much smaller than the Casimir energy we are interested in computing.

\section{3+1 dimensional case}\label{sec:31_continuum}
Let us now analyze the $D=3$ dimensional case that we will also study later on the lattice. In Ref. \cite{ezquerro2025casimir}, the low temperature regime Casimir energy in 3+1 dimensions was studied with a similar methodology than in this thesis, obtaining the same results that we will get in this section by just setting $D=3$ in the formulas we computed in \autoref{sec:Casimir_cont}.
The Casimir energy in the three dimensional case is \eqref{eq:Casimir_energy_continuum}
 \begin{equation}\label{eq:Casimir_energy_3d}
 	E_U(L,m)=\frac{A }{12\pi^{2}}\int_{m}^\infty  dk \left(k^2-m^2\right)^{3/2}\left(L-\frac{d}{dk}\log \frac{h^L_U(ik)}{h^{\infty}_U(ik)}\right).
 \end{equation}
and the rest of the terms of the free energy are \eqref{eq:Free_energy_temperature}
\begin{align}\nonumber
	F^{l\not =0}_U(L,m,\beta)&=-\frac{(2/\pi)A}{(2\pi\beta)^{\frac{3}{2}}}\sum_{l=1}^\infty\int_{m}^\infty dk\ \frac{(k^2-m^2)^{\frac{3}{4}}}{l^{\frac{3}{2}}} \left(L-\frac{d}{dk}\log \frac{h^L_U(ik)}{h^{\infty}_U(ik)}\right)\\ \label{eq:Free_energy_3d_pre}
	&\times \text{Re}\left((1+i) K_{\frac{3}{2}}\left(i\beta l\sqrt{k^2-m^2}\right)\right).
\end{align}
We can then apply the relationship of the Bessel function $K_{3/2}$ with polylogarithms \eqref{eq:Bessel_poly_dir} to simplify the expression
\begin{align}\nonumber
	F^{l\not=0}_U(L,m,\beta)&=-\frac{A}{2\pi^2\beta^{3}}\int_{m}^\infty dk\ \left(L-\frac{d}{dk}\log \frac{h^L_U(ik)}{h^{\infty}_U(ik)}\right)\\\label{eq:Free_energy_3d}
	&\times \left( \left(\beta \sqrt{k^2-m^2}\ \text{Re}\left(\text{Li}_2\left(e^{-i\beta \sqrt{k^2-m^2}}\right)\right)+\text{Im}\left(\text{Li}_3\left(e^{-i\beta \sqrt{k^2-m^2}}\right)\right)\right)\right).
\end{align}
\subsection{Special boundary conditions}
\begin{enumerate}[i)]
	\item {\it Dirichlet and Neumann boundary conditions}. Like in the two dimensional case, these two boundary conditions have the same free energy as a consequence of sharing the same spectral factor. The Casimir energy \eqref{eq:Casimir_energy_continuum} with $D=3$ reduces to
\begin{equation}\label{Casimir_3d_Dirichlet}
	E_D(L,m)=-\frac{Am^{2}}{8\pi^{2}L}\sum_{j=1}^{\infty}\frac{K_{2}(2jmL)}{j^{2}},
\end{equation}
whereas when $m=0$ we have
\begin{equation}\label{eq:massless_3d_dir}
	E_D(L,0)=-\frac{A\pi^2}{1440L^{3}}.
\end{equation}
The remainder part of the free energy  is given by
\begin{align}\nonumber
	F^{l\not=0}_D(L,m,\beta)&=-\frac{AL}{2\pi^2\beta^{3}}\int_{m}^\infty dk\ \left(1-\coth(kL)\right)\\\label{Free_energy_Dirichlet_3d}
	\times& \left( \left(\beta \sqrt{k^2-m^2}\ \text{Re}\left(\text{Li}_2\left(e^{-i\beta \sqrt{k^2-m^2}}\right)\right)+\text{Im}\left(\text{Li}_3\left(e^{-i\beta \sqrt{k^2-m^2}}\right)\right)\right)\right).
\end{align}
\item  {\it  Periodic boundary conditions.} The Casimir energy is 
\begin{equation}\label{eq:Casimir_periodic_31_scalar}
	E_P(L,m)=-\frac{Am^{2}}{2\pi^{2}L}\sum_{j=1}^{\infty}\frac{K_{2}(jmL)}{j^{2}},
\end{equation}
and in the massless case we have instead
\begin{equation}\label{eq:massless_3d_per}
	E_P(L,0)=-\frac{A\pi^2 }{90L^{3}}.
\end{equation}
In the free energy we have the same result as for \acrshort*{dbc} \eqref{Free_energy_Dirichlet_3d} but changing the spectral factor
\begin{align}\nonumber
	&F^{l\not=0}_P(L,m,\beta)=-\frac{AL}{2\pi^2\beta^{3}}\int_{m}^\infty dk\ \left(1-\coth(kL/2)\right)\\\label{Free_energy_periodic_3d}
	&\times \left( \left(\beta \sqrt{k^2-m^2}\ \text{Re}\left(\text{Li}_2\left(e^{-i\beta \sqrt{k^2-m^2}}\right)\right)+\text{Im}\left(\text{Li}_3\left(e^{-i\beta \sqrt{k^2-m^2}}\right)\right)\right)\right).
\end{align}
\item  {\it  Anti-periodic boundary conditions.} The Casimir energy in this case is
\begin{equation}
	E_A(L,m)=\frac{Am^{2}}{2\pi^{2}L}\sum_{j=1}^{\infty}\frac{(-1)^{j+1}}{j^{2}}K_{2}(jmL),
\end{equation}
which when $m=0$ reduces to
\begin{equation}
	E_A(L,0)=\frac{7\pi^2A}{720L^3}.
\end{equation}
The temperature dependent free energy is
\begin{align}\nonumber
	&F^{l\not=0}_A(L,m,\beta)=-\frac{AL}{2\pi^2\beta^{3}}\int_{m}^\infty dk\ \left(1-\tanh(kL/2)\right)\\\label{Free_energy_anti_3d}
	&\times \left( \left(\beta \sqrt{k^2-m^2}\ \text{Re}\left(\text{Li}_2\left(e^{-i\beta \sqrt{k^2-m^2}}\right)\right)+\text{Im}\left(\text{Li}_3\left(e^{-i\beta \sqrt{k^2-m^2}}\right)\right)\right)\right).
\end{align}
\item  {\it  Zaremba boundary conditions.} For this boundary condition the Casimir energy is given by
\begin{equation}
	E_Z(L,m)=\frac{Am^{2}}{8\pi^{2}L}\sum_{j=1}^{\infty}\frac{(-1)^{j+1}}{j^{2}}K_{2}(2jmL),
\end{equation}
whereas in the massless situation we have instead
\begin{equation}
	E_Z(L,0)=\frac{7\pi^2 A}{11520L^{3}}.
\end{equation}
The temperature dependent part of the free energy has the form
\begin{align}\nonumber
	&F^{l\not=0}_Z(L,m,\beta)=-\frac{AL}{2\pi^2\beta^{3}}\int_{m}^\infty dk\ \left(1-\tanh(kL)\right)\\\label{Free_energy_zaremba_3d}
	&\times \left( \left(\beta \sqrt{k^2-m^2}\ \text{Re}\left(\text{Li}_2\left(e^{-i\beta \sqrt{k^2-m^2}}\right)\right)+\text{Im}\left(\text{Li}_3\left(e^{-i\beta \sqrt{k^2-m^2}}\right)\right)\right)\right).
\end{align}
\end{enumerate}

\subsection{Asymptotic behaviour}
\begin{figure}[H]
	\centering
	\includegraphics[width=1\textwidth]{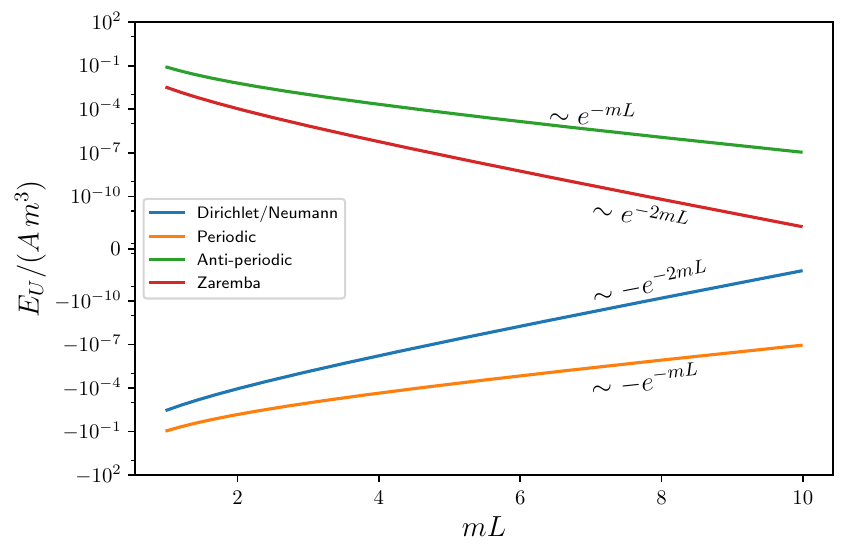}
	\caption{Exponential decay of the dimensionless Casimir energy (in logarithmic scale) as a function of the dimensionless distance $mL$ for different boundary conditions in 3+1 dimensions.}
	\label{fig:plot_Casimir_3d}
\end{figure}

Like in the $2+1$  dimensional case we compute the dependence on the distance between the plates $mL$ and the temperature $\beta/L$ of the Casimir energy, the free energy and their quotient to show their asymptotic behaviour. First, we display in \autoref{fig:plot_Casimir_3d} the Casimir energy for the five different boundary conditions. \autoref{fig:plot_Casimir_3d} shows the expected exponential behaviour of the Casimir energy, where \acrshort*{dbc}, \acrshort*{nbc} and \acrshort*{zbc} decay exponentially with $e^{-2mL}$ whereas for \acrshort*{pbc} and \acrshort*{abc} the decay is slower $e^{-mL}$. Also, as in the $2+1$ dimensional case there are boundaries conditions with negative and positive Casimir energy. This is actually the behaviour for any dimension as can be seen in the formulas we obtained for the Casimir energy in \autoref{sec:Particular_conditions}. \acrshort*{dbc}, \acrshort*{nbc} and \acrshort*{pbc} have a negative Casimir energy for every dimensional case, whereas \acrshort*{abc} and \acrshort*{zbc} have a positive Casimir energy.

\begin{figure}[H]
	\centering
	\includegraphics[width=1\textwidth]{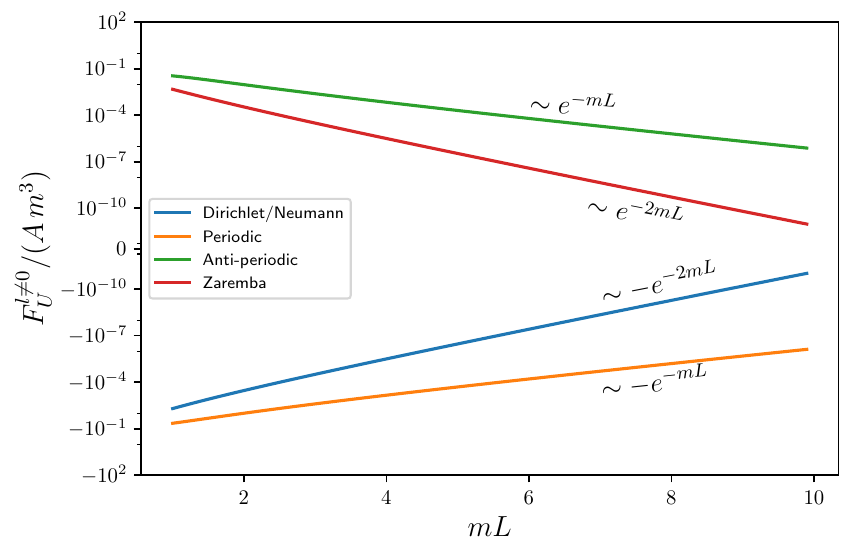}
	\caption{Exponential decay of the dimensionless free energy (in logarithmic scale) as a function of the dimensionless distance $mL$ for different boundary conditions with $m\beta=1$ in 3+1 dimensions.}
	\label{fig:plot_FreeEnergy_3d}
\end{figure}

\pagebreak
In a similar way we analyze the temperature dependent part of the free energy in \autoref{fig:plot_FreeEnergy_3d}. As can be seen the free energy follows the same asymptotic behaviour than the Casimir energy (as was hinted in \autoref{sec:Particular_conditions}). With the boundary conditions that impose independent conditions on each wall (tr$(U\sigma_1)=0$) there is the fast exponential decay $e^{-2mL}$, whereas those who involve relations between the values of the fields at  both walls (tr$(U\sigma_1)\not=0$) exhibit a slower decay $e^{-mL}$.

\begin{figure}[t]
	\centering
	\includegraphics[width=1\textwidth]{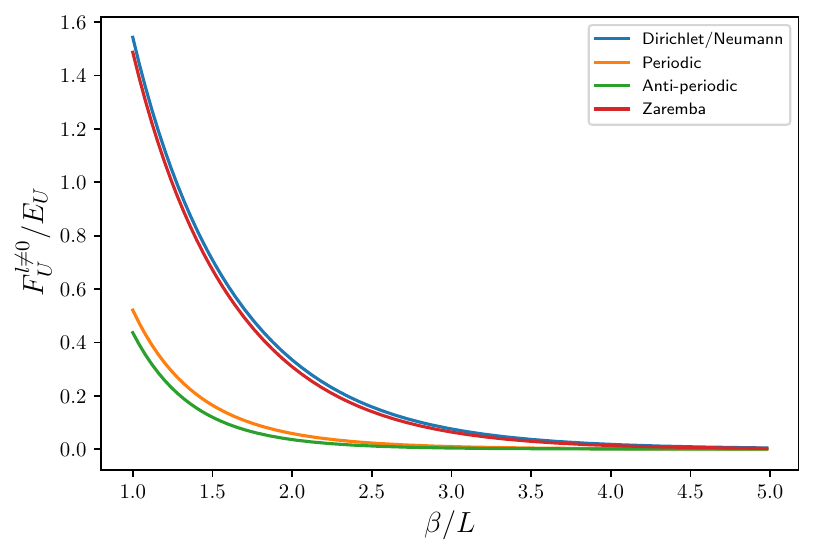}
	\caption{Quotient of the temperature dependent terms of the free energy and the Casimir energy as a function of $\beta/L$ for different boundary conditions with $mL=1$ in 3+1 dimensions.}
	\label{fig:plot_FreeEnergy__temp_3d}
\end{figure}

Finally, in \autoref{fig:plot_FreeEnergy__temp_3d} we show the quotient of the temperature dependent terms of the free energy and the Casimir energy to explore how it behaves when $\beta>L$ and check that the asymptotic behaviour is similar to that of the $2+1$  dimensional theory. Whenever the temperature becomes low with respect to the distance between the walls ($\beta \gg L$) the Casimir energy is much larger than the rest of the free energy terms. As was remarked in the $2+1$ dimensional theory. This will be very relevant for  lattice simulations since it means that when the size associated with the temperature is large enough with respect to the size of the transverse direction we obtain the Casimir energy because the rest of the temperature dependent terms of the free energy become negligible.

    \addtocontents{toc}{\protect\newpage}
    \chapter{Scalar fields on the lattice}\label{chp:scalar_lattice}
The goal of this work is to calculate the Casimir energy in  non-abelian gauge theories on the lattice.
But before doing that we shall test the method with the scalar field where we can compare directly with the continuum results we have in \autoref{chp:scalar_cont}. Thus, the main goal of this chapter is to compute the Casimir energy of a real massive scalar field using the lattice formalism for different spatial dimensions and boundary conditions, and check it with the results in the continuum.

In \autoref{sec:Lattice_form} we summarize the basic ingredients of  the lattice formalism and how we can calculate the expectation values of observables. In \autoref{sec:BC_scalar} we show how to implement the boundary conditions  and compute the partition function with those boundary conditions. In \autoref{sec:Energy_scalar} we shall analytically calculate the internal energy on the lattice. In \autoref{sec:Casimir_scalar} we compute the Casimir energy for the different boundary conditions and compare the results with the continuum values. Finally, in \autoref{sec:21_scalar} we will focus on the 2+1 dimensional case and analyze how the different sizes of the lattice affect the calculation of the Casimir energy.
\section{Lattice formalism}\label{sec:Lattice_form}
The lattice formalism is based on the connection between the functional integral \cite{feynman1948space,Feynman:100771} in Euclidean time and statistical mechanics. For a more extensive analysis about this, see for example \cite{gattringer2009quantum,roepstorff2012path,rivers1988path,rothe2012lattice,wipf2013statistical}. 

The continuum action of a free massive real scalar field in $D+1$ space-time dimensions is
\begin{equation}
	S_M[\psi]=\frac{1}{2}\int d\tau\  d^{D}x \left(\left(\frac{\partial \psi(\tau,\mathbf x)}{\partial \tau}\right)^2-\sum_{i=1}^D \left(\frac{\partial \psi(\tau,\mathbf x)}{\partial x_i}\right)^2- m^2 \psi^2(\tau,\mathbf x)\right),
\end{equation}
with $x=(\tau,x_1,x_2,\ldots,x_D)$ the space-time coordinates  in Minkowski space. With this action, we can construct the partition function in the functional integral formalism as
\begin{equation}
	Z=\int \mathcal D \psi\ e^{iS_M[\psi]}
\end{equation}
where the integral is over all possible values of the field at every space-time point.
Now, we do a Wick rotation in time $\tau=ix_0$ and transform the action into the Euclidean action $S_E[\psi(x_0,{\mathbf x})]=-iS_M[\psi(i x_0,{\mathbf x})]$, i.e.
\begin{equation}\label{eq:Euc_action_cont}
	S_E[\psi]=\frac{1}{2}\int d^{D+1}x \left(\sum_{\mu=0}^D \left(\frac{\partial \psi( x)}{\partial x_\mu}\right)^2+ m^2 \psi^2(x)\right),
\end{equation}
where  ${x}=(x_0,x_1,x_2,\ldots,x_D)$ are the space-time coordinates in the Euclidean space ($x_0\equiv t$)\footnote{Since we are working with natural units ($c=1$), we use $x_0$ and $t$ indistinctly.}. Using this Euclidean action the partition function becomes
\begin{equation}\label{eq:Z_cont}
	Z=\int \mathcal D \psi\ e^{-S_E[\psi]}
\end{equation}
which we can recognise as the partition function of a statistical mechanics system where the usual Boltzmann factor $e^{-\beta H}$ is replaced by $e^{-S_E}$ instead. Therefore, we can obtain the expectation value of an observable $O$ as is done in statistical mechanics, by computing
\begin{equation}
	\braket{O}=\frac{1}{Z}\int \mathcal D \psi\ O(\psi)\ e^{-S_E[\psi]}.
\end{equation}

\begin{figure}[h]
	\centering
	\includegraphics[width=.6\textwidth]{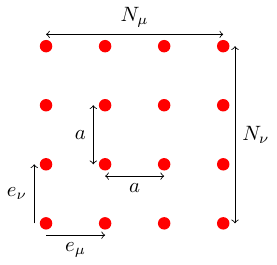}
	\caption{Schematic representation of the lattice.}
	\label{fig:lattice_representation}
\end{figure}

Once we have seen how we can use the functional integral to compute the expectation value of physical observables in the Euclidean space-time, we can proceed with the lattice formalism. First, we discretize the space-time into a lattice of $D+1$ dimensions $\Lambda$ with the same spacing $a$ for every direction of the lattice, $N_\mu$ the number of points in each direction and $N=N_0N_1\cdots N_{D-1}N_D$ the total number of sites on the lattice. The coordinates on the lattice are of the form ${n}=(n_0,n_1,n_2,\ldots,n_D)$ with $n_\mu=1,2,\ldots,N_\mu$, which gives the physical coordinates of the sites of the lattice ${x}=a\mathit{n}$, and in each point of the lattice the field takes a real value $\psi(an)$. Also, we define ${e_\mu}$ as the unitary vector in the direction $\mu$. In \autoref{fig:lattice_representation} we show a representation of a two dimensional lattice with the parameters we just mentioned.

Now, for discretizing the Euclidean action \eqref{eq:Euc_action_cont} we need to  re-write the integral as a Riemann sum over the lattice
\begin{equation}\label{eq:integral_sum}
	\int d^{D+1}x\ f( x) \rightarrow a^{D+1}\sum_{n\in \Lambda}f(a n)
\end{equation}
and the derivative as
\begin{equation} \label{eq:discrete_derivative}
	\frac{\partial \psi ( x)}{\partial x_\mu}\simeq\frac{\psi(a( n+ e_\mu))-\psi(a n)}{a},
\end{equation}
which leads to the Euclidean action on the lattice taking the form
\begin{equation}
	S_E[\psi]=\frac{a^{D+1}}{2}\sum_{n \in \Lambda} \left(\sum_{\mu=0}^{D}\frac{(\psi( a( n+ e_\mu))-\psi(a n))^2}{a^2}+ m^2 \psi(a n)^2 \right)
\end{equation}
where by $\sum\limits_{ n \in \Lambda}$ we indicate the sum over all the points of the lattice. It can be seen that by taking the limits $a\rightarrow 0$ and $N_\mu\rightarrow \infty$ we recover the continuum expression \eqref{eq:Euc_action_cont}. This formula can be re-arranged in a more advantageous way 
\begin{equation}\label{eq:Euc_action_lattice}
	S_E[\phi]=\frac{1}{2}\sum_{ n \in \Lambda} \left( \left(2(D+1)+M^2\right) \phi(n)^2-\sum_{\mu=0}^{D}\left(\phi( n+ e_\mu)+\phi( n- e_{\mu})\right)\phi( n)\right),
\end{equation}
where we have changed to the dimensionless variables $M=am$ and $\phi=\sqrt{a^{D-1}}\psi$, and taken advantage of the fact that when summing over all the lattice we have
\begin{equation}\label{eq:sum_relation}
	2\sum_{ n\in \Lambda}\phi( n+ e_\mu)\phi( n)=\sum_{ n\in \Lambda}\left(\phi( n+ e_\mu)+\phi( n- e_{\mu})\right)\phi( n).
\end{equation} 
An important feature of formula \eqref{eq:Euc_action_lattice} is that it can be interpreted as the discretization of the Laplacian, which is given by
\begin{equation}\label{eq:discrete_Laplacian}
\sum_{\mu=0}^{D}\frac{\partial^2 \phi(x)}{\partial x^2_\mu}\simeq \sum_{\mu=0}^{D}\frac{\phi(n+ e_\mu)+\phi( n- e_{\mu})-2\phi( n)}{a^2}.
\end{equation}

On the lattice formalism, the functional integral of the partition function \eqref{eq:Z_cont} becomes a product of the integrals over the possible values of the fields in each point of the lattice weighted by the corresponding action \eqref{eq:Euc_action_lattice}
\begin{equation}\label{eq:Zeta_lattice}
	Z=\int \prod_{ n \in \Lambda} d\phi_{n}\ e^{-S_E[\phi]}.
\end{equation}

Now, we can use another way of mapping the points on the lattice, instead of using the $D+1$ vector ${n}=(n_0,n_1,n_2,\ldots, n_D)$, the sites can be indexed by an integer in the following way $n=n_D+n_{D-1}N_D+\ldots+n_1N_2\cdots N_{D-1}N_{D}+n_0N_1N_2\cdots N_{D-1}N_{D}$. This allows us to rewrite the lattice action \eqref{eq:Euc_action_lattice} as
\begin{equation}
	S_E[\phi]=\frac{1}{2}\sum_{m,n=1}^{N}\phi_{n}\Sigma_{nm}\phi_m
\end{equation}
where we have defined $\phi_n\equiv\phi(n)$ and $\Sigma$ is a $N\times N$ symmetric matrix that describes the interactions between the different lattice points given by the action. This matrix has the following form:
\begin{equation}\label{mpbc}
	\Sigma_{nm}=(M^2+2(D+1))\delta_{nm}-\sum_{\mu=0}^{D}\left(\delta_{n+e_\mu,m}+\delta_{n-e_\mu,m}\right),
\end{equation}
except in the boundary sites of the lattice in which the form of the matrix elements depends on the chosen boundary condition. Since this matrix is symmetric, we can diagonalize it and transform the expression of the partition function \eqref{eq:Zeta_lattice} into a product of $N$ independent Gaussian integrals. Therefore, the partition function can be integrated as this product of Gaussian integrals 
\begin{equation}\label{partition_lattice}
	Z=\frac{(2\pi)^{N/2}}{\sqrt{|\Sigma|}},
\end{equation}
and we just need to obtain the determinant of $\Sigma$ for computing the partition function. As we just mentioned, the specific form of $\Sigma$ is highly dependent on the boundary conditions we impose on the lattice.

\section{Boundary conditions}\label{sec:BC_scalar}
In this section, we shall analyze the boundary conditions that are as close as possible  to the setting of those studied in \autoref{chp:scalar_cont}, i.e. two parallel homogeneous infinite walls where the fields satisfy the boundary conditions (\autoref{fig:Boundary_wall_representation}).  First, we consider \acrshort*{pbc} on every direction of the lattice (including Euclidean time) except the direction $D$. For \acrshort*{pbc} the two boundary sides of the lattice are identified giving rise to a torus. For example, in the time dimension the points $\phi(1,n_1,n_2,\ldots,n_D)$ and $\phi(N_0,n_1,n_2,\ldots,n_D)$ are  neighbours and interact with each other. 
The same is done for the space dimensions (apart from the $D$ direction), the boundary points in those directions $n_\mu=1$ and $n_\mu=N_\mu$ are also neighbours and interact, where $\mu=1,\ldots D-1$. 
 In other words, when we move to the left of the first point of the lattice in a certain direction we arrive at the last point of the lattice in the same direction and when move to the right of the last point of the lattice we arrive at the first point of the lattice. Now, we shall consider different boundary conditions on the  direction $D$ and we will calculate the corresponding 
 partition function.
\subsection{Periodic boundary conditions}
Let us consider first also \acrshort*{pbc} on the $D$ direction  as we did for the other directions. This implies that we are connecting the points with $n_D=1$ and $n_D=N_D$ like in the rest of the directions of the lattice.
Now that we have all the boundary conditions of the lattice,  we can determine the matrix $\Sigma$ that describes the action. If we have $D$ spatial dimensions the $N\times N$ matrix $\Sigma$ corresponding to \acrshort*{pbc} is of the form \eqref{mpbc}, but taking into account that  the sites with $n_\mu=1$ and $n_\mu=N_\mu$ 
also interact with each other.
The eigenvalues of this matrix  are
\begin{equation}
		\lambda^P_{n_0,n_1,\ldots,n_D}=2(D+1)-2\sum_{\mu=0}^D\cos\left(\frac{2\pi n_\mu}{N_\mu}\right)+M^2=4\sum_{\mu=0}^D \sin^2\left(\frac{\pi n_\mu}{N_\mu}\right)+M^2,
\end{equation}
with $n_\mu$ running from $1$ to $N_\mu$. Therefore,  the partition function \eqref{partition_lattice} is given by the product of these eigenvalues
\begin{equation}\label{eq:partition_per}
	Z_P=(2\pi)^{N/2}\left(\prod_{n_0,n_1,\ldots,n_D=1}^{N_i}\left(4\sum_{\mu=0}^D \sin^2\left(\frac{\pi n_\mu}{N_\mu}\right)+M^2\right)\right)^{-1/2}.
\end{equation}

Let us solve an explicit example to see more clearly the form of the matrix $\Sigma_P$. We consider a 2 dimensional lattice (1 time and 1 spatial dimension) with 3 points in each direction $N_0=N_1=3$ with the \acrshort*{pbc}s. Hence, the matrix $\Sigma_P$ has the form
\begin{align}\nonumber
&\Sigma_P=\\
&\begin{pmatrix}
	4 + M^2 & -1 & -1 & -1 & 0 & 0 & -1 & 0 & 0 \\
	-1 & 4 + M^2 & -1 & 0 & -1 & 0 & 0 & -1 & 0 \\
	-1 & -1 & 4 + M^2 & 0 & 0 & -1 & 0 & 0 & -1 \\
	-1 & 0 & 0 & 4 + M^2 & -1 & -1 & -1 & 0 & 0 \\
	0 & -1 & 0 & -1 & 4 + M^2 & -1 & 0 & -1 & 0 \\
	0 & 0 & -1 & -1 & -1 & 4 + M^2 & 0 & 0 & -1 \\
	-1 & 0 & 0 & -1 & 0 & 0 & 4 + M^2 & -1 & -1 \\
	0 & -1 & 0 & 0 & -1 & 0 & -1 & 4 + M^2 & -1 \\
	0 & 0 & -1 & 0 & 0 & -1 & -1 & -1 & 4 + M^2
\end{pmatrix}
\end{align}
and the partition function is
\begin{equation}
	Z_P=(2\pi)^{9/2}\left(\prod_{n_0,n_1=1}^{3}\left(4\sin^2\left(\frac{\pi n_0}{3}\right)+4\sin^2\left(\frac{\pi n_1}{3}\right)+M^2\right)\right)^{-1/2}.
\end{equation}
\subsection{Dirichlet boundary conditions}
With \acrshort*{dbc} the fields must vanish at the boundaries, this can be implemented on the lattice by fixing the fields to zero at one of the boundary points in the $D$ direction, i.e.
\begin{equation}\label{eq:boundary_dirichlet_lattice}
	\phi(n_0,n_1,n_2,\ldots,1)=0,
\end{equation}
and imposing \acrshort*{pbc} on the other side. In this way the points with $n_D=N_D$ have as neighbours the sites with $n_D=1$ in which the fields are fixed to zero. Since the fields at the points of the lattice with $n_D=1$ are zero we actually have less degrees of freedom, and this translates into the matrix that encodes the lattice action $\Sigma$ being smaller, in particular $N'\times N'$ dimensional ($N'=N(N_D-1)/N_D$) instead of $N\times N$. 
In this case the boundary points in the $D$ direction in the matrix $\Sigma$ are different than the rest of the directions. The points in the matrix with $n_D=1$ have a zero coupling to the left, i.e. 
\begin{equation}
	\Sigma_{nm}=(M^2+2(D+1))\delta_{nm}-\delta_{n+e_D,m}-\sum_{\mu=0}^{D-1}\left(\delta_{n+e_\mu,m}+\delta_{n-e_\mu,m}\right)
\end{equation}
	when $n=(n_0,n_1,\ldots,n_D=1)$. Whereas the points in the matrix with $n_D=N_D-1$\footnote{Since the matrix with \acrshort*{dbc} is smaller we shift the indices of the matrix, $n_D=N_D-1$ corresponds to the boundary points in the matrix. \label{fn:matrix_dir}} have a zero coupling to the right, i.e.
\begin{equation}
	\Sigma_{nm}=(M^2+2(D+1))\delta_{nm}-\delta_{n-e_D,m}-\sum_{\mu=0}^{D-1}\left(\delta_{n+e_\mu,m}+\delta_{n-e_\mu,m}\right)
\end{equation}
when $ n=(n_0,n_1,\ldots,n_D=N_D-1)$\footref{fn:matrix_dir}.

%
We can calculate the eigenvalues of this matrix
\begin{equation}
	\lambda^D_{n_0,n_1,\ldots,n_D}=4\sum_{i=0}^{D-1} \sin^2\left(\frac{\pi n_i}{N_i}\right)+4\sin^2\left(\frac{\pi n_D}{2N_D}\right)+M^2,
\end{equation}
with $n_i$ running from $1$ to $N_i$ and $n_D$ from $1$ to $N_D-1$. Therefore, we have $N'$ instead of $N$ eigenvalues which agrees with the loss of degrees of freedom due to the fixed value of the field at $n_D=1$. Taking all of this into account, the partition function is given by
\begin{equation}\label{eq:partition_dir}
	Z_D=(2\pi)^{N'/2}\left(\prod_{n_0,\ldots,n_{D-1}=1}^{N_i}\prod_{n_D=1}^{N_D-1}\left(4\sum_{\mu=0}^{D-1} \sin^2\left(\frac{\pi n_\mu}{N_\mu}\right)+4\sin^2\left(\frac{\pi n_D}{2N_D}\right)+M^2\right)\right)^{-1/2}.
\end{equation}

Let us consider to clarify this issue, a 2 dimensional lattice with 3 sites in each direction but with \acrshort*{dbc}. The matrix $\Sigma_D$ will be $6\times6$ with the following form
\begin{equation}
\Sigma_D=	\begin{pmatrix}
		4 + M^2 & -1 & -1 & 0 & -1 & 0 \\
		-1 & 4 + M^2 & 0 & -1 & 0 & -1 \\
		-1 & 0 & 4 + M^2 & -1 & -1 & 0 \\
		0 & -1 & -1 & 4 + M^2 & 0 & -1 \\
		-1 & 0 & -1 & 0 & 4 + M^2 & -1 \\
		0 & -1 & 0 & -1 & -1 & 4 + M^2
	\end{pmatrix},
\end{equation} 
notice how it has only 3 couplings per column or row due to the direction associated to the \acrshort*{dbc} being only 2 points, the points $n_1=1$ have left coupling always zero whereas the points with $n_1=2$ have the right coupling always zero. The partition function will be given by the determinant of this matrix
\begin{equation}
	Z_D=(2\pi)^{3}\left(\prod_{n_0=1}^{3}\prod_{n_1=1}^{2}\left(4\sin^2\left(\frac{\pi n_0}{3}\right)+4\sin^2\left(\frac{\pi n_1}{6}\right)+M^2\right)\right)^{-1/2}.
\end{equation}
\subsection{Anti-periodic boundary conditions}
These boundary conditions are similar to \acrshort*{pbc} but with a change of sign between the fields at the boundary terms. Therefore, we have the same interaction between the boundary points than in the \acrshort*{pbc} case, where the coupling to the left of the points with $n_D=1$ is with the points $n_D=N_D$ but with a minus sign and vice versa.  
In this case the coupling of the boundary points where $n_D=1$ have a change of sign in their left coupling
\begin{equation}
	\Sigma_{nm}=(M^2+2(D+1))\delta_{nm}+\delta_{n-e_D,m}-\delta_{n+e_D,m}-\sum_{\mu=0}^{D-1}\left(\delta_{n+e_\mu,m}+\delta_{n-e_\mu,m}\right)
\end{equation}
when $ n=(n_0,n_1,\ldots,n_D=1)$. The points in the matrix with $n_D=N_D$ have a positive coupling to the right, i.e.
\begin{equation}
	\Sigma_{nm}=(M^2+2(D+1))\delta_{nm}-\delta_{n-e_D,m}+\delta_{n+e_D,m}-\sum_{\mu=0}^{D-1}\left(\delta_{n+e_\mu,m}+\delta_{n-e_\mu,m}\right)
\end{equation}
when $ n=(n_0,n_1,\ldots,n_D=N_D)$. 
Hence, the eigenvalues are 
\begin{equation}
	\lambda^A_{n_0,n_1,\ldots,n_D}=4\sum_{\mu=0}^{D-1} \sin^2\left(\frac{\pi n_\mu}{N_\mu}\right)+4\sin^2\left(\frac{\pi (2n_D+1)}{2N_D}\right)+M^2,
\end{equation}
with $n_\mu$ running from $1$ to $N_\mu$. Thus, the product of the $N$ eigenvalues gives  the following partition function
\begin{equation}\label{eq:partition_anti}
	Z_A=(2\pi)^{N/2}\left(\prod_{n_0,n_1,\ldots,n_D=1}^{N_\mu}\left(4\sum_{\mu=0}^{D-1} \sin^2\left(\frac{\pi n_\mu}{N_\mu}\right)+4\sin^2\left(\frac{\pi( 2n_D+1)}{2N_D}\right)+M^2\right)\right)^{-1/2}.
\end{equation}
Now, we  can again consider  to illustrate the algebra the simpler example of a 2 dimensional lattice with $3$ sites on each direction. The matrix $\Sigma_A$ with the \acrshort*{abc} has the form
\begin{align}\nonumber
	&\Sigma_A=\\
	&\begin{pmatrix}
		4 + M^2 & -1 & 1 & -1 & 0 & 0 & -1 & 0 & 0 \\
		-1 & 4 + M^2 & -1 & 0 & -1 & 0 & 0 & -1 & 0 \\
		1 & -1 & 4 + M^2 & 0 & 0 & -1 & 0 & 0 & -1 \\
		-1 & 0 & 0 & 4 + M^2 & -1 & 1 & -1 & 0 & 0 \\
		0 & -1 & 0 & -1 & 4 + M^2 & -1 & 0 & -1 & 0 \\
		0 & 0 & -1 & 1 & -1 & 4 + M^2 & 0 & 0 & -1 \\
		-1 & 0 & 0 & -1 & 0 & 0 & 4 + M^2 & -1 & 1 \\
		0 & -1 & 0 & 0 & -1 & 0 & -1 & 4 + M^2 & -1 \\
		0 & 0 & -1 & 0 & 0 & -1 & 1 & -1 & 4 + M^2
	\end{pmatrix},
\end{align}
where we can see how there is a change of sign when we have the coupling between the points $n_1=1$ and $n_1=3$ which are the boundary ones on this lattice. Finally, from the determinant of this matrix the partition function is
\begin{equation}
	Z_A=(2\pi)^{9/2}\left(\prod_{n_0,n_1=1}^{3}\left(4\sin^2\left(\frac{\pi n_0}{3}\right)+4\sin^2\left(\frac{\pi (2n_1+1)}{6}\right)+M^2\right)\right)^{-1/2}.
\end{equation}
\subsection{Neumann boundary conditions}
\acrshort*{nbc} correspond to the case where the outwards normal derivatives of the fields at the boundaries vanish. Taking into account the form of the discretization of the derivative \eqref{eq:discrete_derivative}, setting to zero the normal derivative means that the point following the boundary site has the same value $\phi(n\mp e_D)=\phi( n)$. We can implement this condition by modifying the action \eqref{eq:Euc_action_lattice} of the boundary points in the following way:
\begin{equation}\label{eq:action_Neumann}
2	S_E[\phi( n)]=(2D+1+M^2)\phi^2(n)-\sum_{\mu=0}^{D-1}\left(\left(\phi(n+ e_\mu)+\phi( n-e_{\mu})\right)\phi( n)\right)+\phi(n+ e_D)\phi( n)
\end{equation}
for the sites with $n_D=1$, and 
\begin{equation}\label{eq:action_Neumann2}
	2	S_E[\phi( n)]=(2D+1+M^2)\phi^2( n)-\sum_{\mu=0}^{D-1}\left(\left(\phi( n+e_\mu)+\phi( n-e_{\mu})\right)\phi(n)\right)+\phi( n- e_D)\phi( n)
\end{equation}
for the points with $n_D=N_D$.
Thus, with these conditions the values on the diagonal of the $\Sigma$ matrix are $2D+1+M^2$ in the boundary points instead of the usual $2(D+1)+M^2$. Also, it has the same coupling to the neighbours of the boundary points than \acrshort*{dbc} where the coupling to the left of the points with $n_D=1$ is zero and to the right of the points $n_D=N_D$. Thus, the matrix elements of the boundary points are given by
\begin{equation}
		\Sigma_{nm}=(M^2+2D+1)\delta_{nm}-\delta_{n+e_D,m}-\sum_{\mu=0}^{D-1}\left(\delta_{n+e_\mu,m}+\delta_{n-e_\mu,m}\right)
\end{equation}
	when $ n=(n_0,n_1,\ldots,n_D=1)$ and
\begin{equation}
	\Sigma_{nm}=(M^2+2D+1)\delta_{nm}-\delta_{n-e_D,m}-\sum_{\mu=0}^{D-1}\left(\delta_{n+e_\mu,m}+\delta_{n-e_\mu,m}\right)
\end{equation}
when $ n=(n_0,n_1,\ldots,n_D=N_D)$.

%
The eigenvalues of this matrix are
\begin{equation}
	\lambda^N_{n_0,n_1,\ldots,n_D}=4\sum_{\mu=0}^{D-1} \sin^2\left(\frac{\pi n_\mu}{N_\mu}\right)+4\sin^2\left(\frac{\pi n_D}{2N_D}\right)+M^2,
\end{equation}
with $n_\mu$ running from $1$ to $N_\mu$ and $n_D$ from $0$ to $N_D-1$. Therefore, the partition function with this boundary condition is given by
\begin{equation}\label{eq:partition_neu}
	Z_N=(2\pi)^{N/2}\left(\prod_{n_0,\ldots,n_{D-1}=1}^{N_i}\prod_{n_D=0}^{N_D-1}\left(4\sum_{\mu=0}^{D-1} \sin^2\left(\frac{\pi n_\mu}{N_\mu}\right)+4\sin^2\left(\frac{\pi n_D}{2N_D}\right)+M^2\right)\right)^{-1/2}.
\end{equation}

This can be illustrated with the simple case of a 2 dimensional lattice with $N_0=N_1=3$. The matrix in this case is given by
\begin{align}\nonumber
	&\Sigma_N=\\
	&\begin{pmatrix}
		3 + M^2 & -1 & 0 & -1 & 0 & 0 & -1 & 0 & 0 \\
		-1 & 4 + M^2 & -1 & 0 & -1 & 0 & 0 & -1 & 0 \\
		0 & -1 & 3 + M^2 & 0 & 0 & -1 & 0 & 0 & -1 \\
		-1 & 0 & 0 & 3 + M^2 & -1 & 0 & -1 & 0 & 0 \\
		0 & -1 & 0 & -1 & 4 + M^2 & -1 & 0 & -1 & 0 \\
		0 & 0 & -1 & 0 & -1 & 3 + M^2 & 0 & 0 & -1 \\
		-1 & 0 & 0 & -1 & 0 & 0 & 3 + M^2 & -1 & 0 \\
		0 & -1 & 0 & 0 & -1 & 0 & -1 & 4 + M^2 & -1 \\
		0 & 0 & -1 & 0 & 0 & -1 & 0 & -1 & 3 + M^2
	\end{pmatrix},
\end{align}
where we can see the different factor for the coupling of the diagonal in the boundary points, and the null one between the points $n_1=1$ and $n_1=3$. Finally, we can compute the partition function 
\begin{equation}
	Z_N=(2\pi)^{9/2}\left(\prod_{n_0=1}^{3}\prod_{n_1=0}^{2}\left(4\sin^2\left(\frac{\pi n_0}{3}\right)+4\sin^2\left(\frac{\pi n_1}{6}\right)+M^2\right)\right)^{-1/2}.
\end{equation}
\subsection{Zaremba boundary conditions}
The \acrshort*{zbc} are a combination of \acrshort*{dbc} and \acrshort*{nbc}, in one boundary wall they satisfy \acrshort*{dbc} and in the opposite boundary \acrshort*{nbc}. Since it is irrelevant in which one we impose each condition, we choose \acrshort*{dbc} in the left side $n_D=1$ and \acrshort*{nbc} in the right side $n_D=N_D$. 
If we set \acrshort*{dbc} on the left side $n_D=1$ the $\Sigma$  matrix elements are given by
\begin{equation}
	\Sigma_{nm}=(M^2+2(D+1))\delta_{nm}-\delta_{n+e_D,m}-\sum_{\mu=0}^{D-1}\left(\delta_{n+e_\mu,m}+\delta_{n-e_\mu,m}\right)
\end{equation}
when $ n=(n_0,n_1,\ldots,n_D=1)$. On the other side we have \acrshort*{nbc} which implies that 
the right side elements of the matrix $n_D=N_D$ are
\begin{equation}
	\Sigma_{nm}=(M^2+2D+1)\delta_{nm}-\delta_{n-e_D,m}-\sum_{i=0}^{D-1}\left(\delta_{n+e_i,m}+\delta_{n-e_i,m}\right)
\end{equation}
when $ n=(n_0,n_1,\ldots,n_D=N_D)$.

The corresponding eigenvalues are 
\begin{equation}
	\lambda^Z_{n_0,n_1,\ldots,n_D}=4\sum_{\mu=0}^{D-1} \sin^2\left(\frac{\pi n_\mu}{N_\mu}\right)+4\sin^2\left(\frac{\pi (2n_D+1)}{4N_D-2}\right)+M^2,
\end{equation}
with $n_\mu$ running from $1$ to $N_i$ and $n_D$ from $0$ to $N_D-2$. By inspecting the term of the eigenvalues that corresponds to the \acrshort*{zbc} we can notice an important fact, the denominator in the sine is proportional to $N_D-1/2$ instead of the expected behaviour with $N_D$ like in the rest of boundary conditions. This is due to the lattice setup we use to implement \acrshort*{zbc}. Since we only have $N_D$ points the physical distance is $a(N_D-1)$ plus $a/2$ from considering the derivative when applying \acrshort*{nbc}. Thus, the total physical distance that the lattice involves is $a(N_D-1/2)$, which is the factor that appears in the denominator. Notice, that in \acrshort*{pbc}, \acrshort*{dbc} and \acrshort*{abc} although we also have $N_D$ points, connecting the points $n_D=N_D$ and $n_D=1$ makes the physical distance of the lattice $aN_D$. In \acrshort*{nbc} considering the derivative of the boundary points on both ends is equivalent of having the physical distance of $aN_D$. With these eigenvalues, the partition function is
\begin{equation}\label{eq:partition_zar}
	Z_Z\!=\!(2\pi)^{N'/2}\!\left(\prod_{n_0,\ldots,n_{D-1}=1}^{N_\mu}\prod_{n_D=0}^{N_D-2}\!\left(\!4\sum_{\mu=0}^{D-1} \sin^2\!\left(\frac{\pi n_\mu}{N_\mu}\right)\!+\!4\sin^2\left(\frac{\pi (2n_D+1)}{4N_D-2}\right)\!+\!M^2\right)\right)^{-1/2}.
\end{equation}
In our toy example of a 2 dimensional lattice with 3 points per side, the matrix has the form
\begin{equation}
	\Sigma_Z=	\begin{pmatrix}
		4 + M^2 & -1 & -1 & 0 & -1 & 0 \\
		-1 & 3 + M^2 & 0 & -1 & 0 & -1 \\
		-1 & 0 & 4 + M^2 & -1 & -1 & 0 \\
		0 & -1 & -1 & 3 + M^2 & 0 & -1 \\
		-1 & 0 & -1 & 0 & 4 + M^2 & -1 \\
		0 & -1 & 0 & -1 & -1 & 3 + M^2
	\end{pmatrix},
\end{equation} 
where we can see how it is a combination of the matrix of \acrshort*{dbc} and \acrshort*{nbc}. The partition in this particular case is
\begin{equation}
	Z_Z=(2\pi)^{3}\left(\prod_{n_0=1}^{3}\prod_{n_1=0}^{1}\left(4\sin^2\left(\frac{\pi n_0}{3}\right)+4\sin^2\left(\frac{\pi (2n_1+1)}{10}\right)+M^2\right)\right)^{-1/2}.
\end{equation}
\section{Vacuum Energy}\label{sec:Energy_scalar}
In order to calculate the Casimir energy we have to obtain first the vacuum energy on the lattice formalism. Unlike the continuum counterpart this will be a finite quantity. The Euclidean internal  energy density can be obtained with the first component of the energy-momentum tensor, which in Euclidean space is given by
\begin{equation}
	T_{00}=\frac{1}{2}\left(\sum_{i=1}^D \left(\frac{\partial \psi(x)}{\partial x_i}\right)^2-\left(\frac{\partial \psi( x)}{\partial x_0}\right)^2+ m^2 \psi^2( x)\right),
\end{equation}
and then integrate over the spatial dimensions to obtain the energy
\begin{equation}
	\mathcal E[\psi]=\frac{1}{2}\int d^D \mathbf x\left(\sum_{i=1}^D \left(\frac{\partial \psi(  x)}{\partial x_i}\right)^2-\left(\frac{\partial \psi( x)}{\partial x_0}\right)^2+ m^2 \psi^2( x)\right).
\end{equation}
This expression can be discretized in the same way as was done for the Euclidean action by using the discretization of the integral as a Riemann sum \eqref{eq:integral_sum} and the derivative \eqref{eq:discrete_derivative}, obtaining
\begin{align}\nonumber
	&\mathcal E[\psi]=\frac{a^{D}}{2N_0}\\
	&\times \sum_{n \in \Lambda} \left(\sum_{i=1}^{D}\frac{(\psi( a( n+ e_i))-\psi(a n))^2}{a^2}-\frac{(\psi( a( n+ e_0))-\psi(a n))^2}{a^2}+ m^2 \psi^2(a n) \right).
\end{align}
Notice that we are also averaging over the time direction, which will help in the analytical computation of the energy \footnote{Remember that since we are always using \acrshort*{pbc} on the time direction all the points that only differ in the time coordinate are equivalent.}. This formula can be transformed into a more convenient expression by using \eqref{eq:sum_relation} and the dimensionless variables
\begin{align}\nonumber
	\mathcal E[\phi]=&\frac{1}{2aN_0}\sum_{n \in \Lambda}\biggl( \left(2(D-1)+M^2\right) \phi^2( n)+\left(\phi(n+ e_0)+\phi( n- e_{0})\right)\phi( n)\\\label{eq:Euc_energy_lattice}
	&-\sum_{i=1}^{D}\left(\phi( n+ e_i)+\phi(n- e_{i})\right)\phi( n)\biggr),
\end{align}
where as in the action expression we obtain the discretized form of the second derivatives \eqref{eq:discrete_Laplacian}. Now, we have to compute the expectation value of this observable  as
\begin{equation}
	\braket{\mathcal E}=\frac{1}{Z}\int \prod_{ n \in \Lambda} d\phi_{n}\ \mathcal E[\phi ]\ e^{-S_E[\phi]}.
\end{equation}
Since we have already solved the partition function in \autoref{sec:BC_scalar}, the most natural way to compute this expectation value is by writing the energy in terms of the action. For this matter, first we add two parameters $\alpha$ and $\eta$ to the action \eqref{eq:Euc_action_lattice}
\begin{align}\nonumber
	S_E[\phi](\alpha,\eta)=&\frac{1}{2}\sum_{ n \in \Lambda} \Biggl( \left(2D\alpha+\alpha M^2+2\eta \right) \phi^2(n)-\eta\left(\phi(n+ e_0)+\phi( n- e_{0})\right)\phi( n)\\ \label{eq:Euc_action_lattice_mod}
	&-\alpha\sum_{i=1}^{D}\left(\phi(n+ e_i)+\phi( n- e_{i})\right)\phi( n)\Biggr),
\end{align}
where we recover the original action \eqref{eq:Euc_action_lattice} when $\alpha=\eta=1$. These parameters allow us to obtain the lattice energy \eqref{eq:Euc_energy_lattice} as combination of derivatives of this parametrized action:
\begin{equation}
	\mathcal E[\phi]=\frac{1}{aN_0}\left(\frac{\partial S_E[\phi](\alpha,\eta)}{\partial \alpha}-\frac{\partial S_E[\phi](\alpha,\eta)}{\partial \eta}\right).
\end{equation}
Therefore, we can use differentiation under the integral sign with this same combination of derivatives of the parameters $\alpha$ and $\eta$ to get the expectation value of the energy in the following way
\begin{align}\nonumber
\braket{\mathcal E}&=-\frac{1}{aN_0Z}\int \prod_{ n \in \Lambda}d\phi_{n}\ \left.\left(\frac{\partial }{\partial \alpha}-\frac{\partial }{\partial \eta}\right)\left(e^{-S_E[\phi](\alpha,\eta)}\right)\right|_{\alpha,\eta=1}\\ \label{eq:energy_lattice}
&=-\frac{1}{aN_0Z}\left.\left(\frac{\partial Z(\alpha,\eta) }{\partial \alpha}-\frac{\partial Z(\alpha,\eta) }{\partial \eta}\right)\right|_{\alpha,\eta=1}.
\end{align}
This means that to calculate the expectation value of the energy we just have to obtain the partition function with the extra dependence on the parameters $Z(\alpha,\eta)$ and differentiate it. As before, this will be given by the determinant of the matrix $\Sigma(\alpha,\eta)$ that will depend on the parameters $\alpha$ and $\eta$, and the boundary conditions.
\subsection{Periodic boundary conditions}
 The eigenvalues of the matrix $\Sigma_P(\alpha,\eta)$ with \acrshort*{pbc} have the form
\begin{equation}
	\lambda^P_{n_0,n_1,\ldots,n_D}(\alpha,\eta)=4\eta \sin^2\left(\frac{\pi n_0}{N_0}\right)+\alpha\left(\sum_{i=1}^D 4\sin^2\left(\frac{\pi n_i}{N_i}\right)+M^2\right),
\end{equation}
and the partition function is
	\begin{equation}
		Z_P(\alpha,\eta)=(2\pi)^{N/2}\left(\prod_{n_0,n_1,\ldots,n_D=1}^{N_i}\left(	\lambda^P_{n_0,n_1,\ldots,n_D}(\alpha,\eta)\right)\right)^{-1/2}.
	\end{equation}
	Now we can differentiate this expression
	\begin{align}\nonumber
		&\left.\left(\frac{\partial Z_P(\alpha,\eta) }{\partial \alpha}-\frac{\partial Z_P(\alpha,\eta)}{\partial \eta}\right)\right|_{\alpha,\eta=1}=-\frac{(2\pi)^{N/2}|\Sigma_P(\alpha,\eta)|}{2|\Sigma_P(\alpha,\eta)|^{3/2}}\\
		&\times \left.\sum_{n_0,n_1\ldots n_D=1}^{N_i}\frac{\lambda^P_{n_0,n_1,\ldots,n_D}(1,-1)}{\lambda^P_{n_0,n_1,\ldots,n_D}(\alpha,\eta)}\right|_{\alpha,\eta=1}
		=-\frac{Z}{2}\sum_{n_0,n_1\ldots n_D=1}^{N_i}\frac{\lambda^P_{n_0,n_1,\ldots,n_D}(1,-1)}{\lambda^P_{n_0,n_1,\ldots,n_D}(1,1)}.
	\end{align}
	Finally we insert this result into \eqref{eq:energy_lattice} to obtain the expectation value of the energy with \acrshort*{pbc}
	\begin{equation}
		\braket{\mathcal E_P}=\frac{1}{2aN_0}\sum_{n_0,n_1\ldots n_D=1}^{N_i}\frac{4\sum\limits^{D}_{i=1}\sin^2\left(\frac{\pi n_i}{N_i}\right)-4\sin^2\left(\frac{\pi n_0}{N_0}\right)+M^2}{4\sum\limits^{D}_{\mu=0}\sin^2\left(\frac{\pi n_\mu}{N_\mu}\right)+M^2},
	\end{equation}
     that we can simplify to
     	\begin{equation}\label{eq:energy_lscalar_per}
     	\braket{\mathcal E_P}=\frac{N}{2aN_0}-\frac{4}{aN_0}\sum_{n_0,n_1\ldots n_D=1}^{N_\mu}\frac{\sin^2\left(\frac{\pi n_0}{N_0}\right)}{4\sum\limits^{D}_{\mu=0}\sin^2\left(\frac{\pi n_\mu}{N_\mu}\right)+M^2}.
     \end{equation}
	We can see how in the continuum limit $a\to 0$ this energy is \acrshort*{uv} divergent, also when $N_i\to \infty$ the expression has an infrared divergence  proportional to the spatial volume of the system $N/N_0$.
	\subsection{Dirichlet boundary conditions}
 The eigenvalues of the matrix $\Sigma_D(\alpha,\eta)$ with \acrshort*{dbc} are given by
	\begin{equation}
		\lambda^D_{n_0,n_1,\ldots,n_D}(\alpha,\eta)=4\eta \sin^2\left(\frac{\pi n_0}{N_0}\right)+\alpha\left(4\sum\limits^{D-1}_{i=1}\sin^2\left(\frac{\pi n_i}{N_i}\right)+4\sin^2\left(\frac{\pi n_D}{2N_D}\right)+M^2\right),
	\end{equation}
and the partition function is
	\begin{equation}
	Z_D(\alpha,\eta)=(2\pi)^{N'/2}\left(\prod_{n_0,n_1,\ldots,n_{D-1}=1}^{N_i}\prod_{
		n_{D}=1}^{N_D-1}\left(	\lambda^D_{n_0,n_1,\ldots,n_D}(\alpha,\eta)\right)\right)^{-1/2}.
\end{equation}
By  differentiating  this expression we obtain the expectation value of the energy
	\begin{equation}
	\braket{\mathcal E_D}=\frac{1}{2aN_0}\sum_{n_0,n_1\ldots n_{D-1}=1}^{N_\mu}\sum_{n_D=1}^{N_D-1}\frac{4\sum\limits^{D-1}_{i=1}\sin^2\left(\frac{\pi n_i}{N_i}\right)+4\sin^2\left(\frac{\pi n_D}{2N_D}\right)-4\sin^2\left(\frac{\pi n_0}{N_0}\right)+M^2}{4\sum\limits^{D-1}_{\mu=0}\sin^2\left(\frac{\pi n_\mu}{N_\mu}\right)+4\sin^2\left(\frac{\pi n_D}{2N_D}\right)+M^2},
\end{equation}
which we can again simplify to
\begin{equation}\label{eq:energy_lscalar_dir}
	\braket{\mathcal E_D}=\frac{N(N_D-1)}{2aN_0N_D}-\frac{4}{aN_0}\sum_{n_0,n_1\ldots n_{D-1}=1}^{N_\mu}\sum_{n_D=1}^{N_D-1}\frac{\sin^2\left(\frac{\pi n_0}{N_0}\right)}{4\sum\limits^{D-1}_{\mu=0}\sin^2\left(\frac{\pi n_\mu}{N_\mu}\right)+4\sin^2\left(\frac{\pi n_D}{2N_D}\right)+M^2}.
\end{equation}
\subsection{Anti-periodic boundary conditions}
In this case the eigenvalues of the modified matrix $\Sigma_A(\alpha,\eta)$ are given by
\begin{equation}
	\lambda^A_{n_0,n_1,\ldots,n_D}(\alpha,\eta)=4\eta \sin^2\left(\frac{\pi n_0}{N_0}\right)+\alpha\!\left(\!4\sum\limits^{D-1}_{i=1}\sin^2\left(\frac{\pi n_i}{N_i}\right)\!+\!4\sin^2\!\left(\frac{\pi (2n_D+1)}{2N_D}\right)+M^2\right),
\end{equation}
and the partition function is
\begin{equation}
	Z_A(\alpha,\eta)=(2\pi)^{N/2}\left(\prod_{n_0,n_1,\ldots,n_{D}=1}^{N_i}\left(	\lambda^A_{n_0,n_1,\ldots,n_D}(\alpha,\eta)\right)\right)^{-1/2}.
\end{equation}
As the previous cases we can use \eqref{eq:energy_lattice} and by differentiating this partition function obtain the expectation value of the energy as
\begin{equation}
	\braket{\mathcal E_A}=\frac{1}{2aN_0}\sum_{n_0,n_1\ldots n_{D}=1}^{N_\mu}\frac{4\sum\limits^{D-1}_{i=1}\sin^2\left(\frac{\pi n_i}{N_i}\right)+4\sin^2\left(\frac{\pi (2n_D+1)}{2N_D}\right)-4\sin^2\left(\frac{\pi n_0}{N_0}\right)+M^2}{4\sum\limits^{D-1}_{\mu=0}\sin^2\left(\frac{\pi n_\mu}{N_\mu}\right)+4\sin^2\left(\frac{\pi (2n_D+1)}{2N_D}\right)+M^2},
\end{equation}
which can be rewritten as
\begin{equation} \label{eq:energy_lscalar_anti}
	\braket{\mathcal E_A}=\frac{N}{2aN_0}-\frac{4}{aN_0}\sum_{n_0,n_1\ldots n_{D}=1}^{N_\mu}\frac{\sin^2\left(\frac{\pi n_0}{N_0}\right)}{4\sum\limits^{D-1}_{\mu=0}\sin^2\left(\frac{\pi n_\mu}{N_\mu}\right)+4\sin^2\left(\frac{\pi (2n_D+1)}{2N_D}\right)+M^2}.
\end{equation}
	\subsection{Neumann boundary conditions}
The eigenvalues of the modified matrix $\Sigma_N(\alpha,\eta)$ for \acrshort*{nbc} are
\begin{equation}
	\lambda^N_{n_0,n_1,\ldots,n_D}(\alpha,\eta)=4\eta \sin^2\left(\frac{\pi n_0}{N_0}\right)+\alpha\left(4\sum\limits^{D-1}_{i=1}\sin^2\left(\frac{\pi n_i}{N_i}\right)+4\sin^2\left(\frac{\pi n_D}{2N_D}\right)+M^2\right),
\end{equation}
and the partition function is
\begin{equation}
	Z_N(\alpha,\eta)=(2\pi)^{N/2}\left(\prod_{n_0,n_1,\ldots,n_{D-1}=1}^{N_i}\prod_{
		n_{D}=0}^{N_D-1}\left(	\lambda^N_{n_0,n_1,\ldots,n_D}(\alpha,\eta)\right)\right)^{-1/2}.
\end{equation}
By differentiating the parameters we can obtain the expectation value of the energy
\begin{equation}
	\braket{\mathcal E_N}=\frac{1}{2aN_0}\sum_{n_0,n_1\ldots n_{D-1}=1}^{N_\mu}\sum_{n_D=0}^{N_D-1}\frac{4\sum\limits^{D-1}_{i=1}\sin^2\left(\frac{\pi n_i}{N_i}\right)+4\sin^2\left(\frac{\pi n_D}{2N_D}\right)-4\sin^2\left(\frac{\pi n_0}{N_0}\right)+M^2}{4\sum\limits^{D-1}_{\mu=0}\sin^2\left(\frac{\pi n_\mu}{N_\mu}\right)+4\sin^2\left(\frac{\pi n_D}{2N_D}\right)+M^2},
\end{equation}
or
\begin{equation}\label{eq:energy_lscalar_neu}
	\braket{\mathcal E_N}=\frac{N}{2aN_0}-\frac{4}{aN_0}\sum_{n_0,n_1\ldots n_{D-1}=1}^{N_\mu}\sum_{n_D=0}^{N_D-1}\frac{\sin^2\left(\frac{\pi n_0}{N_0}\right)}{4\sum\limits^{D-1}_{\mu=0}\sin^2\left(\frac{\pi n_\mu}{N_\mu}\right)+4\sin^2\left(\frac{\pi n_D}{2N_D}\right)+M^2}.
\end{equation}
	\subsection{Zaremba boundary conditions}
In the case of \acrshort*{zbc} the  modified matrix $\Sigma_Z(\alpha,\eta)$ has the following eigenvalues
\begin{equation}
	\lambda^Z_{n_0,n_1,\ldots,n_D}(\alpha,\eta)=4\eta \sin^2\left(\frac{\pi n_0}{N_0}\right)+\alpha\left(4\sum\limits^{D-1}_{i=1}\sin^2\left(\frac{\pi n_i}{N_i}\right)+4\sin^2\left(\frac{\pi (2n_D+1)}{4N_D-2}\right)+M^2\right),
\end{equation}
and the corresponding partition function is given by
\begin{equation}
	Z_Z(\alpha,\eta)=(2\pi)^{N'/2}\left(\prod_{n_0,n_1,\ldots,n_{D-1}=1}^{N_i}\prod_{
		n_{D}=0}^{N_D-2}\left(	\lambda^Z_{n_0,n_1,\ldots,n_D}(\alpha,\eta)\right)\right)^{-1/2}.
\end{equation}
The expectation value of energy can be obtained by using \eqref{eq:energy_lattice}
\begin{equation}
	\braket{\mathcal E_Z}=\frac{1}{2aN_0}\sum_{n_0,n_1\ldots n_{D-1}=1}^{N_\mu}\sum_{n_D=0}^{N_D-2}\frac{4\sum\limits^{D-1}_{i=1}\sin^2\left(\frac{\pi n_i}{N_i}\right)+4\sin^2\left(\frac{\pi (2n_D+1)}{4N_D-2}\right)-4\sin^2\left(\frac{\pi n_0}{N_0}\right)+M^2}{4\sum\limits^{D-1}_{\mu=0}\sin^2\left(\frac{\pi n_\mu}{N_\mu}\right)+4\sin^2\left(\frac{\pi (2n_D+1)}{4N_D-2}\right)+M^2},
\end{equation}
which can be rewritten as
\begin{equation}\label{eq:energy_lscalar_zar}
	\braket{\mathcal E_Z}=\frac{N(N_D-1)}{2aN_0N_D}-\frac{4}{aN_0}\sum_{n_0,n_1\ldots n_{D-1}=1}^{N_\mu}\sum_{n_D=0}^{N_D-2}\frac{\sin^2\left(\frac{\pi n_0}{N_0}\right)}{4\sum\limits^{D-1}_{\mu=0}\sin^2\left(\frac{\pi n_\mu}{N_\mu}\right)+4\sin^2\left(\frac{\pi (2n_D+1)}{4N_D-2}\right)+M^2}.
\end{equation}
\pagebreak
\section{Casimir Energy}\label{sec:Casimir_scalar}
Once we have derived the vacuum energy we can compute the Casimir energy. As in the continuum case, we have to subtract the contributions that do not depend on the distance between the boundary walls ($N_D$ on the lattice) or are linear with it,  which were divergent in the continuum but are going to be finite on the lattice. The Casimir energy can be calculated following the same strategy as in
the continuum case. We consider lattice sizes  $N_0,N_1,\ldots N_{D-1}\gg N_D$ much larger in all directions $i=0, \cdots D-1$ than in the $D$ direction as required by the framework of two infinite homogeneous walls (\autoref{fig:Boundary_wall_representation}) and the low temperature limit.

We can now subtract the vacuum energy contributions we are not interested in as in the continuum case \eqref{eq:zeta_re}.
In particular, we shall use the combination
\begin{equation}
	E_U(N_L,M)=\lim_{N_{L_0}\rightarrow \infty}\left(\braket{\mathcal E_U(N_L,M)}+\braket{\mathcal E_U(2{N_{L_0}}+N_L,M)}-2\braket{\mathcal E_U({N_{L_0}}+N_L,M)}\right),
\end{equation} 
that as was argued in the continuum case cancels the volume and surface terms of the energy. In this expression $\braket{\mathcal E_U(N_L,M)}$ is  the vacuum energy of a field with mass $M$ on a lattice with lattice sizes  $N_0,N_1,\ldots N_{D-1}\gg N_D=N_L$ much larger in all directions $i=0, \cdots D-1$ than in the $D$ direction $N_D=N_L$,  and boundary conditions  on the wall given by the unitary $2\times 2$ matrix $U$ \footnote{This expression is analogous to \eqref{eq:zeta_re} by recovering the physical quantities $L=aN_L$, $A=a^{D-1}N_1\cdots N_{D-1}$, $\beta=aN_0$ and $m=M/a$.}. 

\begin{figure}[H]
	\centering
	\includegraphics[width=0.9\textwidth]{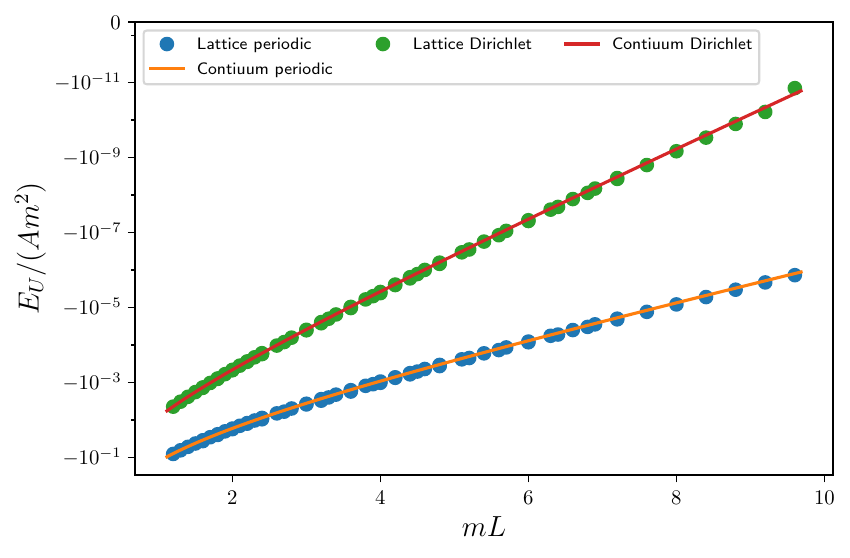}
	\caption{Comparison of the dimensionless Casimir energy of a $2+1$ dimensional massive scalar field  calculated on the lattice and continuum formalism  for \acrshort*{pbc} and \acrshort*{dbc}  in logarithmic scale, for a lattice with parameters $N_0=96$, $N_1=96$, $N_{L_0}=48$, $N_{L}=8-24$ and $M=0.1-0.4$.}
	\label{fig:Casimir_renormalization_lattice}
\end{figure}

We can check if this method works on the lattice by comparing with the continuum results. Let us consider a $2+1$ dimensional theory with \acrshort*{pbc} and \acrshort*{dbc}, which will correspond to a three dimensional lattice. We plot in \autoref{fig:Casimir_renormalization_lattice} the dimensionless Casimir energy $E_U/Am^2$ for \acrshort*{pbc} and \acrshort*{dbc} and compare with the continuum, where $A=a^{D-1}N_1\cdots N_{D-1}$ on the lattice. In \autoref{fig:Casimir_renormalization_lattice} it can be seen how the lattice computed Casimir energy fits well with the continuum prediction given by \eqref{eq:Cas_per_2d} and \eqref{eq:Cas_dir_2d} in a wide range of values for the effective distance $mL$. We observe the expected asymptotic behaviour \eqref{eq:rate} of the Casimir energy, where \acrshort*{dbc} decay as $e^{-2mL}$ whereas with \acrshort*{pbc} it does as $e^{-mL}$.

However, this method of implementing the renormalization prescription has a big disadvantage. For every distance $L$ we want to compute we need to obtain the energy for three different lattices (with $N_D=N_{L}$, $2{N_{L_0}}+L$ and ${N_{L_0}}+L$) and this would be very expensive in computing time in chapters \ref{chp:su2_21} and \ref{chp:su2_31} when we compute the Casimir energy in the non-abelian gauge theories case. The point of this renormalization prescription is to cancel the contributions to the energy that do not go to zero as the distance between the boundary walls goes to infinity, which on the lattice we expect should behave as

\begin{equation}\label{eq:asymptotic_energy_lattice}
	\braket{\mathcal E_U(N_L,M)}=C_0(M) N_1\cdots N_{D-1}N_L+C_1(M) N_1\cdots N_{D-1}+E_U(N_L,M)+\ldots
\end{equation}
where $C_0(M)$ is the term associated to the bulk energy density, $C_1(M)$ is the energy on the boundary walls and $E_U(N_L,M)$ the Casimir energy we want to calculate.

\subsection{Periodic boundary conditions}

We can take advantage of these dependencies by using the fact that when we have \acrshort*{pbc}, the physical set-up is equivalent to not having a boundary wall because both boundaries are connected forming a torus. This implies that the contributions associated to the existence of the boundary walls should cancel out $C_1(M)=0$. We can actually test this by plotting the energy density for different values of $N_D$ and checking if it is independent of $N_D$ (apart from the small fluctuations giving rise to the Casimir energy), since any non-trivial  contribution from the $C_1(M)$ term should imply the appearance of a large $1/N_L$ dependence.

\begin{figure}[h]
	\centering
	\includegraphics[width=0.8\textwidth]{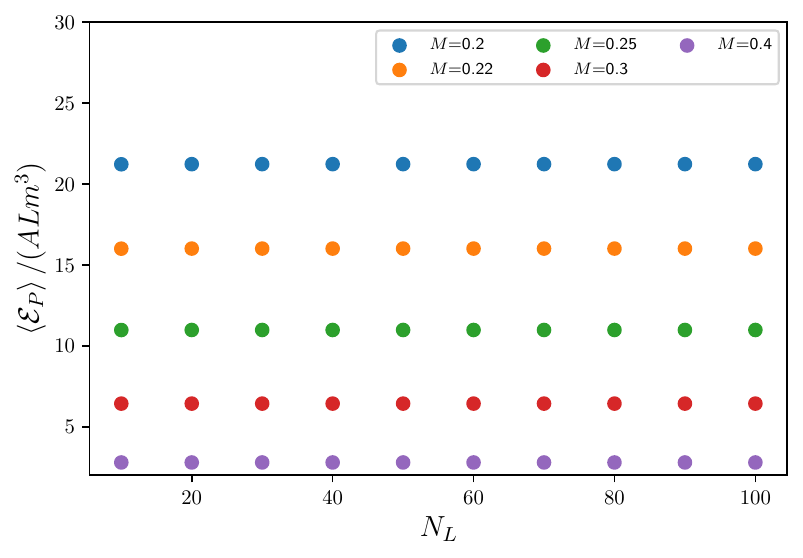}
	\caption{Dimensionless energy density in a three dimensional lattice with \acrshort*{pbc} for different lattice sizes and masses. The chosen parameters of the lattice are $N_0=100$, $N_1=100$, $N_{L}=10-100$ and $M=0.2-0.4$.}
	\label{fig:Energy_periodic_lattice}
\end{figure}

In \autoref{fig:Energy_periodic_lattice} we can see the predicted behaviour, where the energy density with \acrshort*{pbc} is mostly independent of $N_L$ and it does not show any contribution from the boundary walls (that would be noticeable due to its $1/N_L$ dependence). Hence, we can compute this bulk energy contribution as
\begin{equation}\label{eq:bulk_periodic_scalar}
	C_0(M)=\lim_{N_{L_{0}}\rightarrow \infty}\braket{\mathcal E_P(N_{L_{0}},M)}/(N_1\cdots N_{D-1} N_{L_0})
\end{equation}
and obtain the Casimir energy with \acrshort*{pbc} with the following subtraction\footnote{When we make explicit the dependence of the expectation value of the energy on $N_D$  we are taking the size of the rest of the directions of the lattice  very large in comparison with $N_D$.}
\begin{equation}\label{eq:Casimir_periodic_lattice}
	E_P(N_L,M)=\lim_{N_{L_{0}}\rightarrow \infty}\left(\braket{\mathcal E_P(N_{L},M)}-\frac{N_L}{N_{L_0}}\braket{\mathcal E_P(N_{L_{0}},M)}\right),
\end{equation}
where we are setting $N_0,N_1,\ldots N_{D-1}$ with the same size in both lattices and with the factor $N_L/N_{L_0}$ we compensate the different sizes of the lattices. Notice, that by taking the $N_L{_0}\rightarrow \infty$ limit we are suppressing the Casimir energy of the second subtracting term. Now, we can test if this method of computing the Casimir energy with \acrshort*{pbc} gives a  result compatible with the continuum formula \eqref{eq:Casimir_periodic}.
\begin{figure}[H]
	\centering
	\includegraphics[width=1\textwidth]{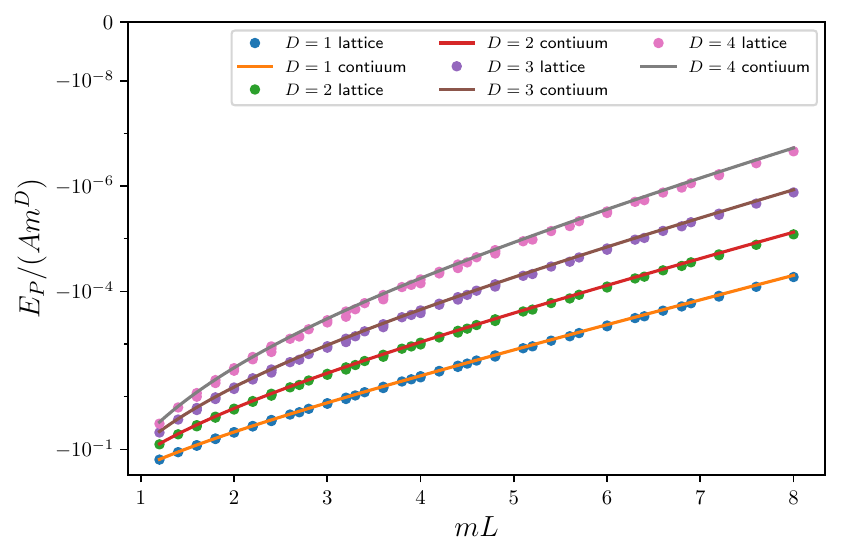}
	\caption{Dimensionless Casimir energy in logarithmic scale with \acrshort*{pbc} for different dimensions and lattice sizes with parameters $N_0=96$, $N_{i\not = D}=96$, $N_{L_0}=96$, $N_{L}=8-24$ and $M=0.1-0.4$ for $D=1-3$. For $D=4$ we use instead  $N_0=48$, $N_{1,2,3}=48$, $N_{L_0}=48$, $N_{L}=8-24$ and $M=0.1-0.4$.}
	\label{fig:Casimir_periodic_lattice}
\end{figure}
In \autoref{fig:Casimir_periodic_lattice} we plot the results we obtain with this formula and the continuum one for different dimensions and effective dimensionless distances $mL$. It clearly shows how this method \eqref{eq:Casimir_periodic_lattice} for obtaining the Casimir energy with \acrshort*{pbc} works, since the lattice points fit the continuum formula. Also, as it was mentioned before this method is much more efficient, for each value of the mass we just have to compute one reference value to subtract instead of two more lattices for each individual point in the plot.

\subsection{Dirichlet boundary conditions}
Now, we can try to use a similar method for \acrshort*{dbc}. First, for the term $C_0(M)$ we can apply that it should be independent of the boundary condition, since it is associated with the bulk energy, and use the value computed for \acrshort*{pbc}. The contribution $C_1(M)$ can be obtained by using a similar strategy, i.e. subtracting the energy of another lattice with \acrshort*{dbc} (where we also subtract the bulk energy to this energy) with the size $N_D$ big enough so the rest of the Casimir energy contributions are negligible. Like in the \acrshort*{pbc} case, we can test if the energy has a $N_L$ dependence given by \eqref{eq:asymptotic_energy_lattice} by plotting the difference between the energy of a lattice with \acrshort*{dbc} and the same lattice with \acrshort*{pbc} but with the $D$ direction fixed to a large $N_L$ size. If we find a constant term then we know we have subtracted the bulk contributions and we are left with just the boundary term $C_1(M)$ that does not depend on $N_L$.
\begin{figure}[h]
	\centering
	\includegraphics[width=0.8\textwidth]{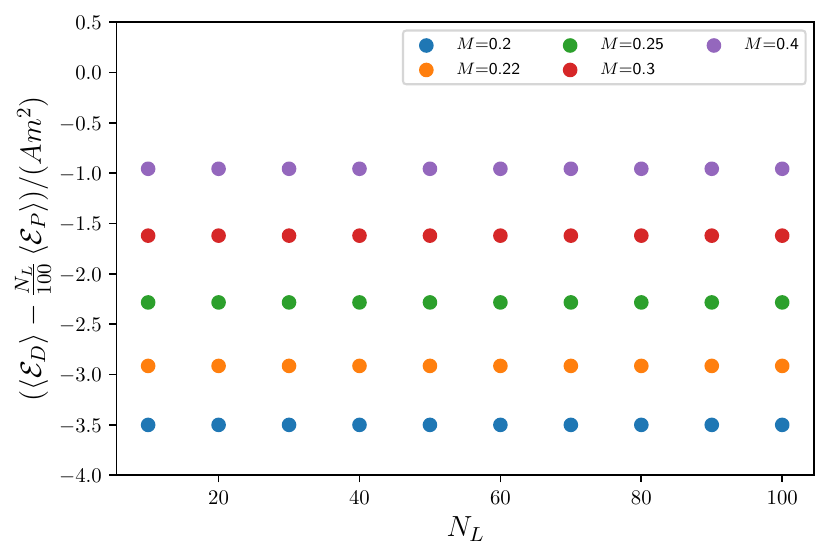}
	\caption{Difference between the dimensionless energy with \acrshort*{dbc} and \acrshort*{pbc} in a three dimensional lattice for different lattice sizes and masses. The chosen parameters of the lattice are $N_0=100$, $N_1=100$, $N_{L}=10-100$ (\acrshort*{dbc}), $N_L=100$ (\acrshort*{pbc}) and $M=0.2-0.4$.}
	\label{fig:Energy_Dirichlet_lattice}
\end{figure}

In \autoref{fig:Energy_Dirichlet_lattice} we show the expected behaviour, where after subtracting the bulk term $C_0(M)$ to the \acrshort*{dbc} we are left with the boundary term $C_1(M)$ which is independent of $N_D$ and the small fluctuations associated to the Casimir energy. Hence, we can calculate this boundary term $C_1(M)$ as
\begin{equation}
	C_1(M)=\lim_{N_{L_0}\rightarrow \infty}\left(\braket{\mathcal E_D(N_{L_0},M)}-\braket{\mathcal E_P(N_{L_{0}},M)}\right)/(N_1\cdots N_{D-1}),
\end{equation}
We can define the Casimir energy for \acrshort*{dbc} as
\begin{align}\nonumber
	&E_D(N_L,M)=\lim_{N_{L_0}\rightarrow \infty}\left(\braket{\mathcal E_D(N_{L},M)}-\frac{N_L}{N_{L_0}}\braket{\mathcal E_P(N_{L_{0}},M)}-\braket{\mathcal E_D(N_{L_0},M)}\right.\\\label{eq:Casimir_Dirichlet_lattice}
	&\left.+\braket{\mathcal E_P(N_{L_{0}},M)}\right)
	=\!\!\lim_{N_{L_0}\rightarrow \infty}\!\!\left(\braket{\mathcal E_D(N_{L},M)}\!-\braket{\mathcal E_D(N_{L_0},M)}\!-\frac{N_L-N_{L_0}}{N_{L_0}}\braket{\mathcal E_P(N_{L_{0}},M)}\right),
\end{align}
where we are also fixing  the other sizes of the lattice   $N_0=N_1=\ldots =N_{D-1}$ equal. In the limit  $N_{L_0}\rightarrow \infty$ we have the cancellation of  the Casimir energy on the lattices with $N_D=L_0$. Thus, we can compare with the continuum  \eqref{eq:Casimir_Dirichlet} for different dimensions.
\begin{figure}[H]
	\centering
	\includegraphics[width=1\textwidth]{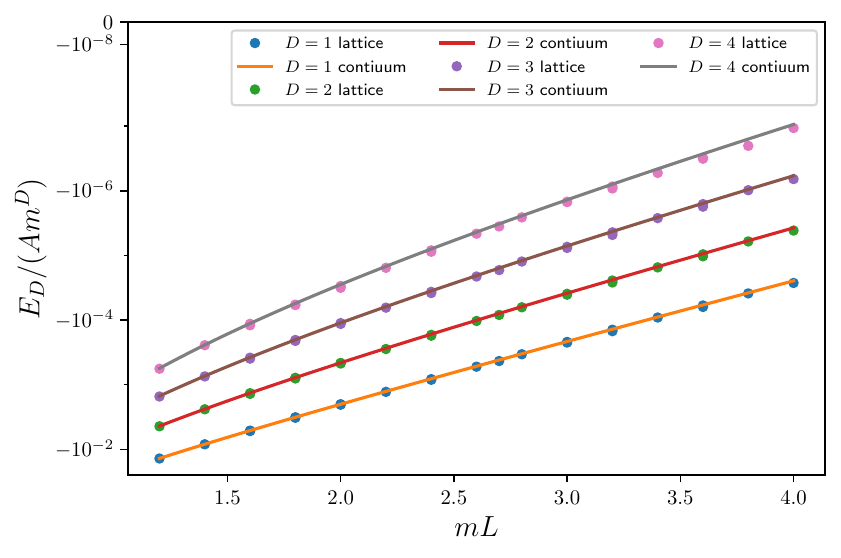}
	\caption{Dimensionless Casimir energy in logarithmic scale with \acrshort*{dbc} for different dimensions and lattice sizes with parameters $N_0=96$, $N_{i\not = D}=96$, $N_{L_0}=96$, $N_{L}=8-24$ and $M=0.1-0.4$ for $D=1-3$. For $D=4$ we use instead  $N_0=48$, $N_{1,2,3}=48$, $N_{L_0}=48$, $N_{L}=8-20$ and $M=0.1-0.4$.}
	\label{fig:Casimir_Dirichlet_lattice}
\end{figure}
In \autoref{fig:Casimir_Dirichlet_lattice} we can see how the lattice Casimir energy obtained by using the renormalization prescription given by \eqref{eq:Casimir_Dirichlet_lattice} fits perfectly with the Casimir energy in the continuum for a wide range of distances. This proves that we can use the formula \eqref{eq:Casimir_Dirichlet_lattice} to compute the Casimir energy with \acrshort*{dbc}, which only needs one lattice with \acrshort*{pbc} and another with \acrshort*{dbc} to obtain the Casimir energy for any value of $N_L$ and each mass. This means a great advantage in comparison with the continuum prescription used in \autoref{chp:scalar_cont} where we have to compute three different energies for each point.

\subsection{Anti-periodic boundary conditions}
\begin{figure}[h]
	\centering
	\includegraphics[width=0.8\textwidth]{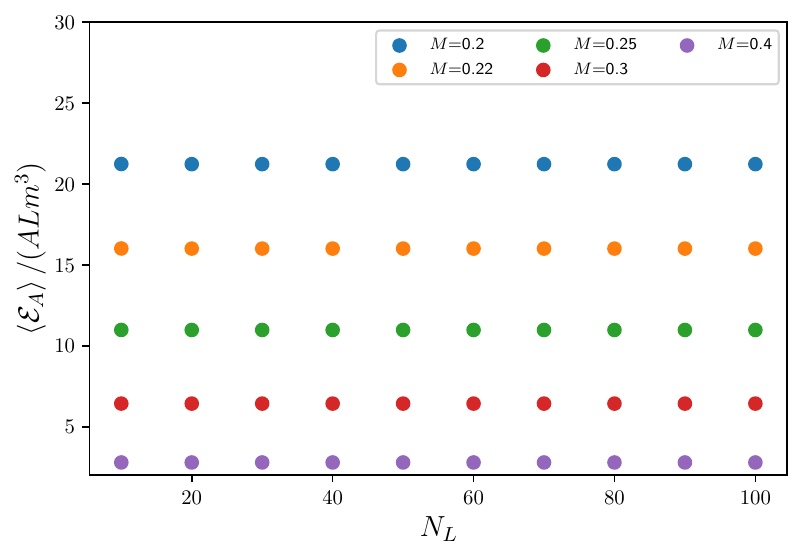}
	\caption{Dimensionless energy density in a three dimensional lattice with \acrshort*{abc} for different lattice sizes and masses. The chosen parameters of the lattice   are $N_0=100$, $N_1=100$, $N_{L}=10-100$ and $M=0.2-0.4$.}
	\label{fig:Energy_antiperiodic_lattice}
\end{figure}

With \acrshort*{abc} we expect the energy to have a similar behaviour to the \acrshort*{pbc} case, where the boundary term $C_1(M)$ also vanishes for similar reasons. Thus, the energy has a dominant bulk term $C_0(M)$ that should have the same value as for \acrshort*{pbc}, since it is independent of the boundary condition, and the small Casimir energy contribution. We can test this behaviour by plotting the energy density for different sizes $N_{L}$ of the lattice and masses in \autoref{fig:Energy_antiperiodic_lattice}. We check this behaviour where we have the dominant term $C_0(M)$ that only depends on the mass. The results (see \autoref{fig:Energy_periodic_lattice}) confirm these  expectations, i.e. $C_0(M)$ is the same for both boundary conditions.

\begin{figure}[H]
	\centering
	\includegraphics[width=1\textwidth]{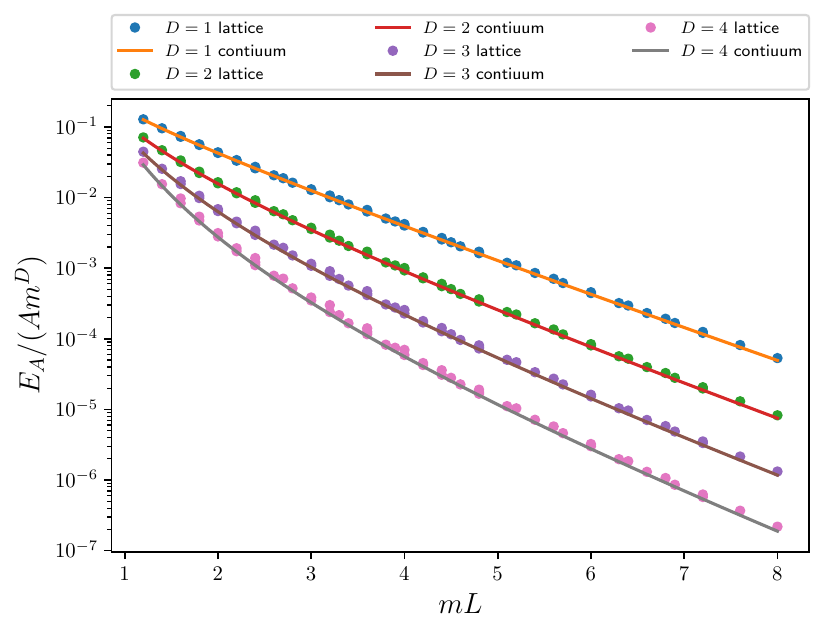}
	\caption{Dimensionless Casimir energy in logarithmic scale with \acrshort*{abc} for different dimensions and lattice sizes with parameters $N_0=96$, $N_{i\not = D}=96$, $N_{L_0}=96$, $N_{L}=8-24$ and $M=0.1-0.4$ for $D=1-3$. For $D=4$ we use instead  $N_0=48$, $N_{1,2,3}=48$, $N_{L_0}=48$, $N_{L}=8-20$ and $M=0.1-0.4$.}
	\label{fig:Casimir_antiperiodic_lattice}
\end{figure}

Thus, this boundary condition can also be used to obtain the value of $C_0(M)$
\begin{equation}
	C_0(M)=\lim_{N_{L_{0}}\rightarrow \infty}\braket{\mathcal E_A(N_{L_{0}},M)}/(N_1\cdots N_{D-1} N_{L_0}),
\end{equation}
and we can obtain  the Casimir energy with \acrshort*{abc} as
\begin{equation}\label{eq:Casimir_antiperiodic_lattice}
	E_A(N_L,M)=\lim_{N_{L_{0}}\rightarrow \infty}\left(\braket{\mathcal E_A(N_{L},M)}-\frac{N_L}{N_{L_0}}\braket{\mathcal E_A(N_{L_{0}},M)}\right),
\end{equation}
where we set $N_0=\ldots =N_{D-1}\gg N_L$. Finally, we plot the Casimir energy we obtain with this method and compare it with the continuum value \eqref{eq:Casimir_anti} in \autoref{fig:Casimir_antiperiodic_lattice}. We can see how well it fits the continuum curve for every dimension. We also obtain the expected asymptotic decay with $e^{-mL}$ and the correct sign since with \acrshort*{abc} the Casimir energy is positive.

\subsection{Neumann boundary conditions}
\begin{figure}[H]
	\centering
	\includegraphics[width=1\textwidth]{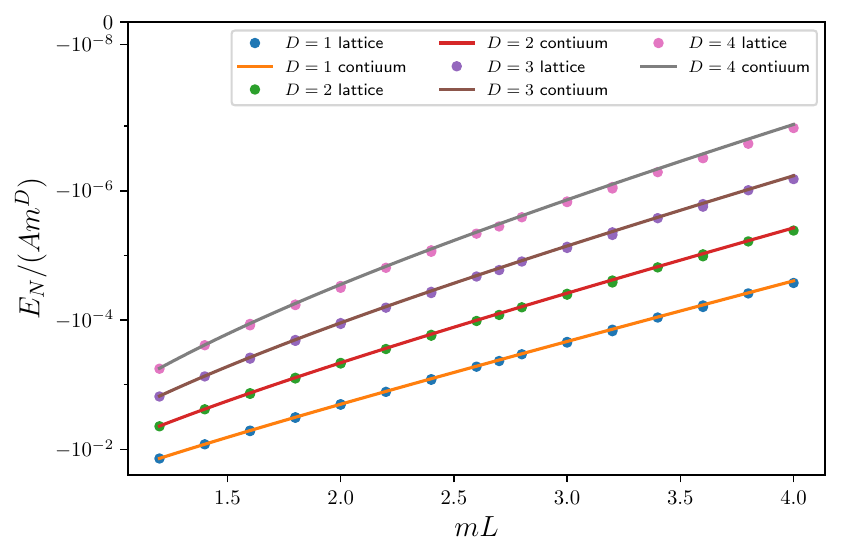}
	\caption{Dimensionless Casimir energy in logarithmic scale with \acrshort*{nbc} for different dimensions and lattice sizes with parameters $N_0=96$, $N_{i\not = D}=96$, $N_{L_0}=96$, $N_{L}=8-24$ and $M=0.1-0.4$ for $D=1-3$. For $D=4$ we use instead  $N_0=48$, $N_{1,2,3}=48$, $N_{L_0}=48$, $N_{L}=8-20$ and $M=0.1-0.4$.}
	\label{fig:Casimir_Neumann_lattice}
\end{figure}

\acrshort*{nbc} are expected to have the same behaviour of the Casimir energy than for \acrshort*{dbc}, i.e. the existence of  a non-vanishing  boundary contribution $C_1(M)$. We can test this fact by plotting the difference between the energy density with \acrshort*{nbc} and \acrshort*{pbc}, which should be given by the factor $C_1(M)$ that is independent of $N_L$ and an extra small contribution due to the Casimir energy.
\begin{figure}[h]
	\centering
	\includegraphics[width=0.8\textwidth]{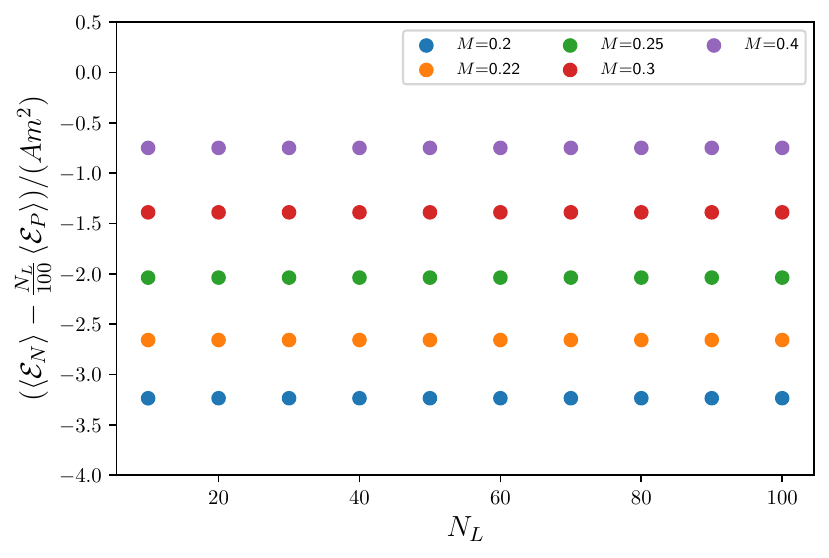}
	\caption{Difference between the dimensionless vacuum  energy with \acrshort*{nbc} and \acrshort*{pbc} in a three dimensional lattice for different lattice sizes. The chosen parameters of the lattice are $N_0=100$, $N_1=100$, $N_{L}=10-100$ (\acrshort*{nbc}), $N_L=100$ (\acrshort*{pbc}) and $M=0.2-0.4$.}
	\label{fig:Energy_Neumann_lattice}
\end{figure}
In \autoref{fig:Energy_Neumann_lattice} we can see the expected behaviour of the vacuum energy where after subtracting the bulk density contribution $C_0(M)$ 
we obtain a value that is  independent of $N_L$ and corresponds to the boundary energy contribution $C_1(M)$. Although, if we compare with the same plot for \acrshort*{dbc} (\autoref{fig:Energy_Dirichlet_lattice}), it can bee seen that we obtain different values for the same masses. Therefore, this implies that we have different values of $C_1(M)$ for each boundary condition and we have to compute it independently. From now on we will indicate to which  boundary condition it corresponds, i.e. $C_1^U(M)$. We can obtain it in the same manner than with \acrshort*{dbc} but using this condition
\begin{equation}
	C^N_1(M)=\lim_{N_{L_0}\rightarrow \infty}\left(\braket{\mathcal E_N(N_{L_0},M)}-\braket{\mathcal E_P(N_{L_{0}},M)}\right)/(N_1\cdots N_{D-1}),
\end{equation}
and define the Casimir energy for \acrshort*{nbc} as
\begin{equation}\label{eq:Casimir_Neumann_lattice}
	E_N(N_L,M)=\lim_{N_{L_0}\rightarrow \infty}\left(\braket{\mathcal E_N(N_{L},M)}-\braket{\mathcal E_N(N_{L_0},M)}-\frac{N_L-N_{L_0}}{N_{L_0}}\braket{\mathcal E_P(N_{L_{0}},M)}\right).
\end{equation}
This combination allows us to cancel the bulk density $C_0(M)$ and the boundary term $C_1^N(M)$ contributions, so we are left with just the Casimir energy. Now, we test this formula by comparing with the continuum result \eqref{eq:Casimir_Dirichlet} \footnote{In \autoref{chp:scalar_cont} it was shown that \acrshort*{dbc} and \acrshort*{nbc} share the same exact formula for the Casimir energy in the continuum.} for different dimensions in \autoref{fig:Casimir_Neumann_lattice}. We find an excellent  fit with the continuum expression for every dimension, including the correct asymptotic behaviour $e^{-2mL}$. We also recover on the lattice the continuum property that \acrshort*{nbc} and \acrshort*{dbc} have the same Casimir energy.

\subsection{Zaremba boundary conditions}
\begin{figure}[H]
	\centering
	\includegraphics[width=0.8\textwidth]{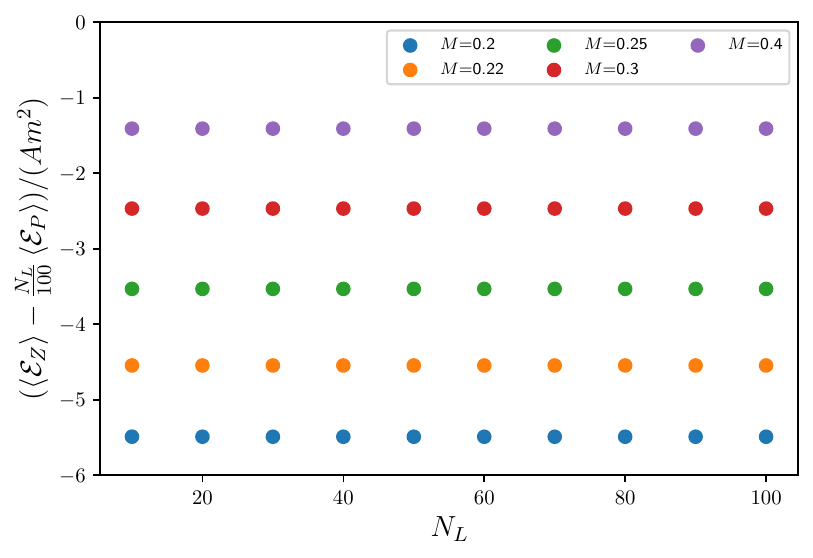}
	\caption{Difference between the dimensionless energy with \acrshort*{zbc} and \acrshort*{pbc} in a three dimensional lattice for different lattice sizes and masses. The chosen parameters of the lattice are $N_0=100$, $N_1=100$, $N_{L}=10-100$ (\acrshort*{nbc}), $N_L=100$ (\acrshort*{pbc}) and $M=0.2-0.4$.}
	\label{fig:Energy_Zaremba_lattice}
\end{figure}

Let us conclude by  analyzing the \acrshort*{zbc} case. Since these conditions are  a combination of \acrshort*{dbc} and \acrshort*{nbc} they are expected to have  the same behaviour, i.e. we have a boundary contribution $C_1^Z(M)$ that has to be subtracted at the same time than the bulk contribution $C_0(M)$. First, we plot the vacuum energy after subtracting the energy density with \acrshort*{pbc} to obtain $C^Z_1(M)$ plus a small Casimir energy contribution  in \autoref{fig:Energy_Zaremba_lattice}, and we observe the expected behaviour: this quantity is independent of $N_L$. Also, we can see how it is a very different value from the boundary term of \acrshort*{dbc} or \acrshort*{nbc}. Thus, we can not use any of those boundary conditions to subtract it.
\vspace{1cm}
\vspace{1cm}
\pagebreak

\begin{figure}[H]
	\centering
	\includegraphics[width=1\textwidth]{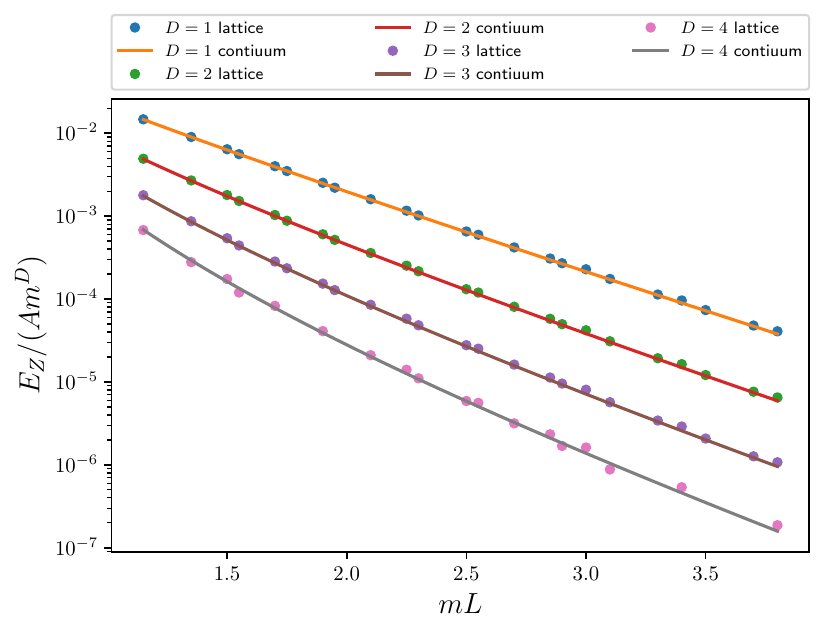}
	\caption{Dimensionless Casimir energy in logarithmic scale with \acrshort*{zbc} for different dimensions and lattice sizes with parameters $N_0=96$, $N_{i\not = D}=96$, $N_{L_0}=96$, $N_{L}=8-24$ and $M=0.1-0.4$ for $D=1-3$. For $D=4$ we use instead  $N_0=48$, $N_{1,2,3}=48$, $N_{L_0}=48$, $N_{L}=8-18$ and $M=0.1-0.2$.}
	\label{fig:Casimir_Zaremba_lattice}
\end{figure}

Therefore, we compute the boundary term $C^Z_1(M)$ contribution as
\begin{equation}
	C^Z_1(M)=\lim_{N_{L_0}\rightarrow \infty}\left(\braket{\mathcal E_Z(N_{L_0},M)}-\braket{\mathcal E_P(N_{L_{0}},M)}\right)/(N_1\cdots N_{D-1}),
\end{equation}
and the Casimir energy with the same combination we used for \acrshort*{dbc} and \acrshort*{nbc} but with \acrshort*{zbc}
\begin{equation}\label{eq:Casimir_Zaremba_lattice}
	E_Z(N_L,M)=\lim_{N_{L_0}\rightarrow \infty}\left(\braket{\mathcal E_Z(N_{L},M)}-\braket{\mathcal E_Z(N_{L_0},M)}-\frac{N_L-N_{L_0}}{N_{L_0}}\braket{\mathcal E_P(N_{L_{0}},M)}\right).
\end{equation}
Again, with this subtraction we cancel the $C_0(M)$ and $C_1^Z(M)$ terms, and only the Casimir energy is left. Finally, we test this formula giving the Casimir energy in the plots of \autoref{fig:Casimir_Zaremba_lattice} for different dimensions. Here, we have to take into account what we discussed when computing the eigenvalues, the physical distance in the $n_D$ direction is $a(N_D-1/2)$. Hence, the effective distance that we have to use when comparing with the continuum result is $mL=M(N_L-1/2)$. We also get  the correct sign since it is positive and exponentially decays with distance $L$ between the plates  as $e^{-2mL}$ matching   the continuum curve.

\pagebreak
\section{2+1 dimensional case}\label{sec:21_scalar}
  In this section we analyze the 2+1 dimensional case in more detail. We restrict the general formulas previously obtained for a generic space dimension $D$ for $D=2$. In particular, we have that $n_0$ is the time coordinate, $n_1$ the longitudinal spatial direction where we fix \acrshort*{pbc} and $n_2$ the transversal spatial direction in which we consider a variety of  boundary conditions to compute the Casimir energy. For \acrshort*{pbc} in the transverse direction the formula for Casimir energy \eqref{eq:Casimir_periodic_lattice} reduces to
    \begin{align}\nonumber
	E_P(N_L,M)=&-\frac{1}{aN_0}\sum_{n_0,n_1=1}^{N_A}\frac{\sin^2\left(\frac{\pi n_0}{N_A}\right)}{S(n_0,n_1,N_A,M)}\times\left(\sum_{n_2=1}^{N_L}\left(1+\frac{\sin^2\left(\frac{\pi n_2}{N_L}\right)}{S(n_0,n_1,N_A,M)}\right)^{-1}\right.\\
	&-\left.\sum_{n_2=1}^{N_{L_0}}\left(1+\frac{\sin^2\left(\frac{\pi n_2}{N_{L_0}}\right)}{S(n_0,n_1,N_A,M)}\right)^{-1}\right),
\end{align}
 \acrshort*{dbc} \eqref{eq:Casimir_Dirichlet_lattice} give instead
  \begin{align}\nonumber 
  	&E_D(N_L,M)=-\frac{1}{aN_0}\sum_{n_0,n_1=1}^{N_A}\frac{\sin^2\left(\frac{\pi n_0}{N_A}\right)}{S(n_0,n_1,N_A,M)}\times\left(\sum_{n_2=1}^{N_L-1}\left(1+\frac{\sin^2\left(\frac{\pi n_2}{2N_L}\right)}{S(n_0,n_1,N_A,M)}\right)^{-1}\right.\\ 
  	&\left.-\sum_{n_2=1}^{N_{L_0}-1}\left(1+\frac{\sin^2\left(\frac{\pi n_2}{2N_{L_0}}\right)}{S(n_0,n_1,N_A,M)}\right)^{-1}-\frac{N_L-N_{L_0}}{N_{L_0}}\sum_{n_2=1}^{N_{L_0}}\left(1+\frac{\sin^2\left(\frac{\pi n_2}{N_{L_0}}\right)}{S(n_0,n_1,N_A,M)}\right)^{-1}\right),
  \end{align}
  whereas for \acrshort*{abc} \eqref{eq:Casimir_antiperiodic_lattice} we get
     \begin{align}\nonumber
  	E_A(N_L,M)=&-\frac{1}{aN_0}\sum_{n_0,n_1=1}^{N_A}\frac{\sin^2\left(\frac{\pi n_0}{N_A}\right)}{S(n_0,n_1,N_A,M)}\times\left(\sum_{n_2=1}^{N_L}\left(1+\frac{\sin^2\left(\frac{\pi (2n_2+1)}{2N_L}\right)}{S(n_0,n_1,N_A,M)}\right)^{-1}\right.\\
  	&-\left.\sum_{n_2=1}^{N_{L_0}}\left(1+\frac{\sin^2\left(\frac{\pi (2n_2+1)}{2N_{L_0}}\right)}{S(n_0,n_1,N_A,M)}\right)^{-1}\right).
  \end{align}
  For \acrshort*{nbc} \eqref{eq:Casimir_Neumann_lattice} we have
    \begin{align}\nonumber 
  	&E_N(N_L,M)=-\frac{1}{aN_0}\sum_{n_0,n_1=1}^{N_A}\frac{\sin^2\left(\frac{\pi n_0}{N_A}\right)}{S(n_0,n_1,N_A,M)}\times\left(\sum_{n_2=0}^{N_L-1}\left(1+\frac{\sin^2\left(\frac{\pi n_2}{2N_L}\right)}{S(n_0,n_1,N_A,M)}\right)^{-1}\right.\\ 
  	&\left.-\sum_{n_2=0}^{N_{L_0}-1}\left(1+\frac{\sin^2\left(\frac{\pi n_2}{2N_{L_0}}\right)}{S(n_0,n_1,N_A,M)}\right)^{-1}-\frac{N_L-N_{L_0}}{N_{L_0}}\sum_{n_2=1}^{N_{L_0}}\left(1+\frac{\sin^2\left(\frac{\pi n_2}{N_{L_0}}\right)}{S(n_0,n_1,N_A,M)}\right)^{-1}\right),
  \end{align}
  and finally with \acrshort*{zbc} \eqref{eq:Casimir_Zaremba_lattice} we arrive at 
    \begin{align}\nonumber 
  	&E_Z(N_L,M)=-\frac{1}{aN_0}\sum_{n_0,n_1=1}^{N_A}\frac{\sin^2\left(\frac{\pi n_0}{N_A}\right)}{S(n_0,n_1,N_A,M)}\times\left(\sum_{n_2=0}^{N_L-2}\left(1+\frac{\sin^2\left(\frac{\pi (2n_2+1)}{4N_L-2}\right)}{S(n_0,n_1,N_A,M)}\right)^{-1}\right.\\ 
  	&\left.-\sum_{n_2=1}^{N_{L_0}-1}\left(1+\frac{\sin^2\left(\frac{\pi (2n_2+1)}{4N_{L_0}-2}\right)}{S(n_0,n_1,N_A,M)}\right)^{-1}-\frac{N_L-N_{L_0}}{N_{L_0}}\sum_{n_2=1}^{N_{L_0}}\left(1+\frac{\sin^2\left(\frac{\pi n_2}{N_{L_0}}\right)}{S(n_0,n_1,N_A,M)}\right)^{-1}\right),
  \end{align}
  where
  \begin{equation}
  	S(n_0,n_1,N_A,M)=\sin^2\left(\frac{\pi n_0}{N_A}\right)+\sin^2\left(\frac{\pi n_1}{N_A}\right)+(M/2)^2,
  \end{equation}
  is a common factor  to all the formulas that does not depend on the boundary conditions since it only involves the other directions of the lattice and the mass. We have considered  $N_0=N_1=N_A\gg N_L$ in order to make every direction much larger than the one we want to compute the Casimir energy $N_L$.

\begin{figure}[H]
	\centering
	\includegraphics[width=1\textwidth]{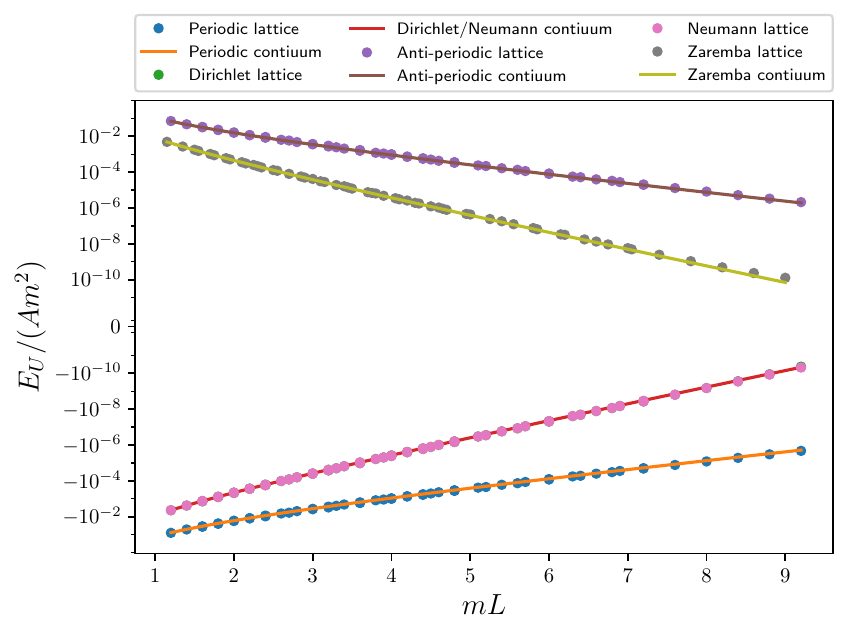}
	\caption{Dimensionless Casimir energy in logarithmic scale of a three dimensional lattice for various boundary conditions with $N_A=96$, $N_{L_0}=96$, $N_{L}=8-24$ and $M=0.1-0.4$.}
	\label{fig:Casimir_lattice_2d}
\end{figure}

The Casimir energy  is plotted in \autoref{fig:Casimir_lattice_2d} for all these different boundary conditions. First, we observe the expected asymptotic behaviour for \acrshort*{dbc}, \acrshort*{nbc} and \acrshort*{zbc} since they decay twice faster ($e^{-2mL}$) than for \acrshort*{pbc} and \acrshort*{abc} ($e^{-mL}$), following the continuum formulas. We also obtain the correct sign for the Casimir energy, \acrshort*{abc} and \acrshort*{zbc} have a positive Casimir energy whereas we get a negative Casimir energy for the other three boundary conditions. Moreover, the equivalence between the Casimir energy of \acrshort*{dbc} and \acrshort*{nbc} is recovered since their points can barely be distinguished in the plot. Notice that the points in \acrshort*{zbc} are located at different values of $mL$ than the rest of the boundary condition, as was explained before, this is due to our lattice set-up for imposing these boundary conditions corresponding to a physical system of $a(N_L-1/2)$ length instead of $aN_L$. Finally, we can see how the plot for \acrshort*{dbc} and \acrshort*{pbc} is roughly the same as in \autoref{fig:Casimir_renormalization_lattice} where the values for the Casimir energy were computed using the continuum renormalization prescription. Thus, both different methods for obtaining the Casimir energy on the lattice are equivalent.
\subsection{Finite volume errors}
Let us now test how the finite volume of the lattice affects the calculation of the Casimir energy. We choose one boundary condition that does not have the boundary contribution $C^U_1(M)$ and one that actually does have it, this way we can test how the errors due to the finite volume of the two different combinations of energies that we are using to obtain the Casimir energy behave. In particular, we shall analyze \acrshort*{pbc} and \acrshort*{dbc} to show one case (\acrshort*{pbc}) with the slow asymptotic decay of the Casimir energy and another case (\acrshort*{dbc}) with faster decay.

\begin{figure}[H]
 	\centering\includegraphics[width=0.84\textwidth]{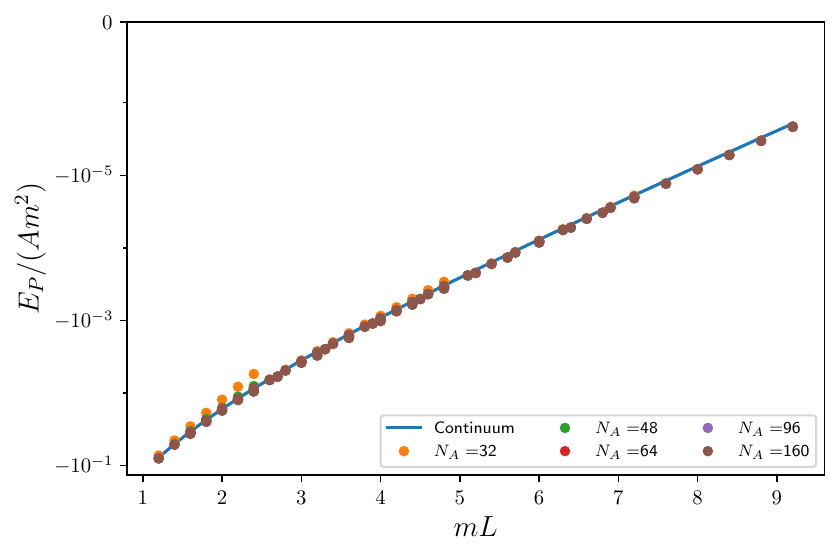} 
 	\caption{Dimensionless Casimir energy with \acrshort*{pbc} in logarithmic scale of a three dimensional lattice for different values of $N_A=\{32,48,64,96,160\}$. The rest of the lattice parameters are fixed at $N_{L_0}=96$, $N_{L}=8-24$ and $M=0.1-0.4$.}
 	\label{fig:Casimir_lattice_2d_sizeA_periodic}
 \end{figure}

 First, we study the effect of the lattice size in the time and longitudinal spatial direction $N_A$. By fixing all the other lattice parameters $N_L$ and $N_{L_0}$ we can compute the Casimir energy for different values of $N_A$.
We plot in \autoref{fig:Casimir_lattice_2d_sizeA_periodic} the results for \acrshort*{pbc}. We see how when $N_A\geq 48$ the results for the Casimir energy in \acrshort*{pbc} are very similar and fit well with the continuum formula for a wide range of distances $mL$. This is obviously compatible with the condition we imposed of $N_A\gg N_L$ since we are taking values at a maximum of $N_L=24$.

\begin{figure}[H]
	\centering\includegraphics[width=0.84\textwidth]{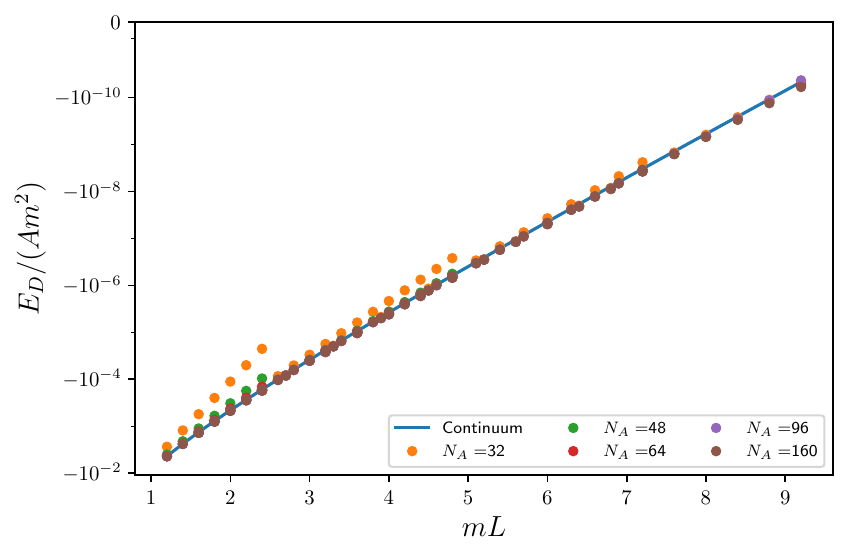}  
	\caption{Dimensionless Casimir energy with \acrshort*{dbc} in logarithmic scale of a three dimensional lattice for different values of $N_A=\{32,48,64,96,160\}$. The rest of the lattice parameters are fixed at $N_{L_0}=96$, $N_{L}=8-24$ and $M=0.1-0.4$.}
	\label{fig:Casimir_lattice_2d_sizeA_Dir}
\end{figure}

 We do the same plot but for \acrshort*{dbc} in \autoref{fig:Casimir_lattice_2d_sizeA_Dir}. The error associated to having smaller lattices in this case is larger than for \acrshort*{pbc} case, and having a size of $N_A\geq 64$ seems to be enough to fit the continuum values.

\begin{figure}[H]
\centering\includegraphics[width=1\textwidth]{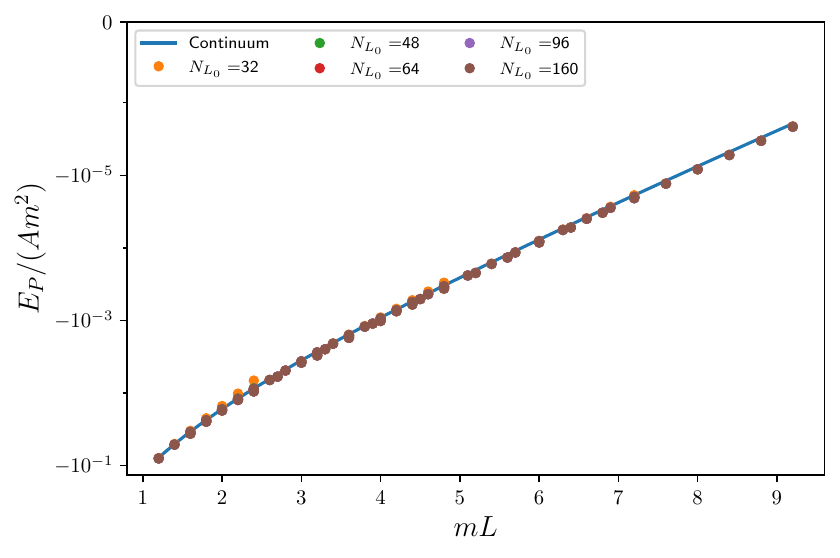}
	\caption{Dimensionless Casimir energy with \acrshort*{pbc} in logarithmic scale of a three dimensional lattice for different values of $N_{L_0}=\{32,48,64,96,160\}$. The other lattice parameters are fixed at $N_{A}=96$, $N_{L}=8-24$ and $M=0.1-0.4$.}
	\label{fig:Casimir_lattice_2d_sizeL0_periodic}
\end{figure}

Now, we repeat the analysis but changing the size of the reference value of the transverse spatial direction $N_{L_0}$ that we use to subtract the $C_0(M)$ and $C_1^U(M)$ contributions, and fix the rest of the parameters. First, we plot the \acrshort*{pbc} case in \autoref{fig:Casimir_lattice_2d_sizeL0_periodic}. It can be seen that the size errors are very similar to the previous case when varying $N_A$. Having $N_{L_0}\geq 48$ seems to be sufficient to make the errors small enough.

\begin{figure}[H]
	\includegraphics[width=1\textwidth]{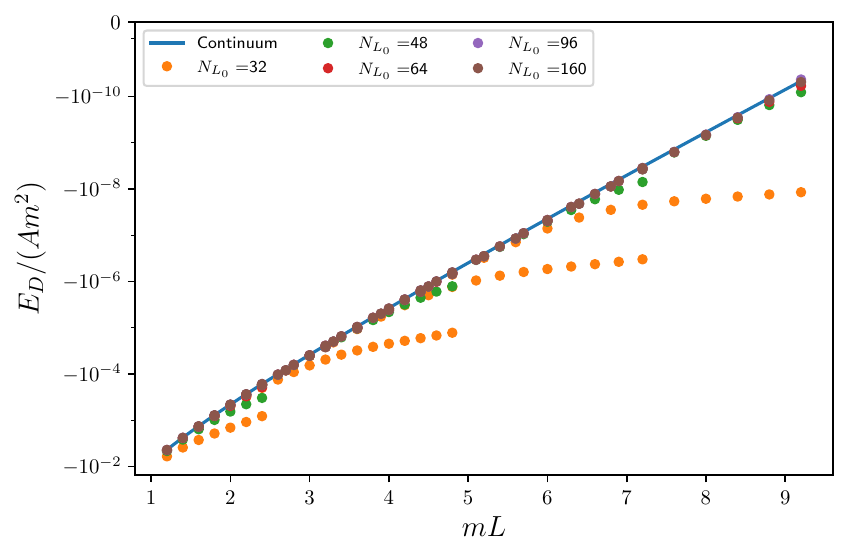} 
	\caption{Dimensionless Casimir energy with \acrshort*{dbc} in logarithmic scale of a three dimensional lattice for different values of $N_{L_0}=\{32,48,64,96,160\}$. The rest of the lattice parameters are fixed at $N_{A}=96$, $N_{L}=8-24$ and $M=0.1-0.4$.}
	\label{fig:Casimir_lattice_2d_sizeL0_Dir}
\end{figure}

For \acrshort*{dbc} the results are displayed in \autoref{fig:Casimir_lattice_2d_sizeL0_Dir}. We can clearly see how in this case the error is much larger than in all the previous cases. With $N_{L_0}=32$ the results are not even at the correct order of magnitude and even with $N_{L_0}=64$ there is some appreciable difference in some points of the plot. Although one may think the main contribution to this error is the term with \acrshort*{dbc} of size $N_{L_0}$ in \eqref{eq:Casimir_Dirichlet_lattice}, this is not true. The main contribution is the term with periodic boundary conditions of size $N_{L_0}$. This is due to the different rate of the exponential decay of the Casimir energy of both boundary conditions, when $N_{L_0}< 2N_L$ the Casimir contribution for the extra term with \acrshort*{pbc} is the dominant term for the Casimir energy of the \acrshort*{dbc}. This can be clearly seen in the plot, since when $N_{L_0}=32$ the Casimir energy for the higher values of $N_L$ is roughly independent of $N_L$. Thus, in those points we are actually seeing the Casimir energy of the lattice with \acrshort*{pbc} instead of the one with \acrshort*{dbc}. For this reason, the condition $N_{L_0}\gg N_L$ is very important to obtain the correct value of the Casimir energy. In particular we need to take values of the order of $N_{L_0}=96$ if we are using $N_L$ of the order of $20$. 

Finally, we analyze how the lattice size $N_L$ affects the result. Although  we have already pointed out  for all the previous plots we use values of $N_L$ from $8$ to $24$ and we have shown how well they fit with the continuum curve. Now we shall be more precise about these features. We shall now compute the Casimir energy for different values of $N_L$ and adjust the value of the mass $M$, so the dimensionless distance $mL$ is the same for all $N_L$. This allows us to compare the differences more clearly since in the continuum limit they should tend to the same value of the Casimir energy. 

\begin{figure}[H]
	\includegraphics[width=1\textwidth]{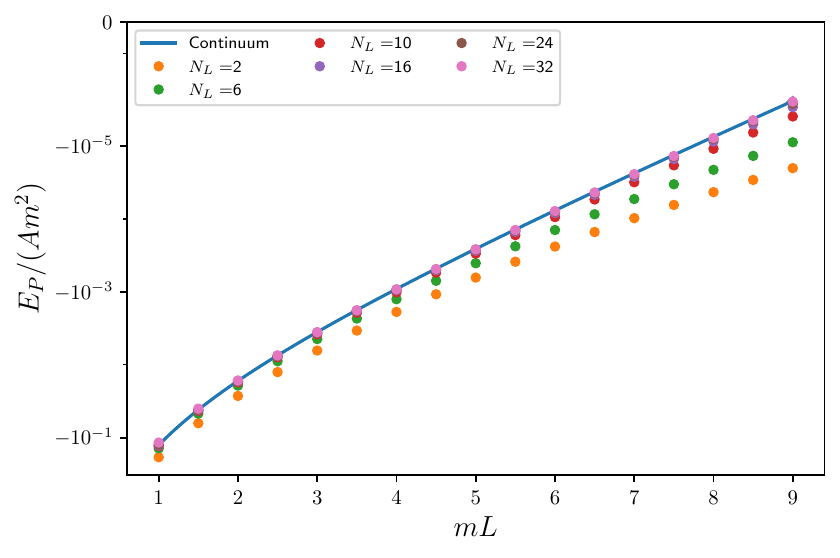} 
	\caption{Dimensionless Casimir energy with \acrshort*{pbc} in logarithmic scale of a three dimensional lattice for different values of $N_{L}=\{2,6,10,16,24,32\}$ and $mL=1-9$. The other lattice parameters are fixed at $N_{A}=96$ and $N_{L_0}=96$.}
	\label{fig:Casimir_lattice_2d_sizeL_per}
\end{figure}

In \autoref{fig:Casimir_lattice_2d_sizeL_per} we display the Casimir energy for different values of $N_L$ at fixed $mL$ for \acrshort*{pbc}. We can see how the results in the higher masses ($M=mL/N_L$) for the smaller lattices are very far from the expected one but when we have values $N_L>10$ it gets reasonable close. Also, it is important to notice that in the scalar field lattice the continuum limit is $am=M\rightarrow 0$ ($a\rightarrow 0$ at a fix physical mass $m$), thus, the points where $mL$ is larger in the smaller lattices suffer from having a mass $M$ too large that adds into more systematic errors than just $N_L$ being small. A clear example of this is the case with $N_L=10$, when $mL>5$ it starts to clearly deviate from the continuum curve but for smaller values of the mass it fits well. This fact also explains why in all the previous plots, the values of $N_L=8$ fit well the continuum since we are using smaller values of the masses (between $M=0.1$ and $M=0.4$), whereas in this plot we can reach $M=0.9$ for $N_L=10$.

\begin{figure}[h]
	\includegraphics[width=1\textwidth]{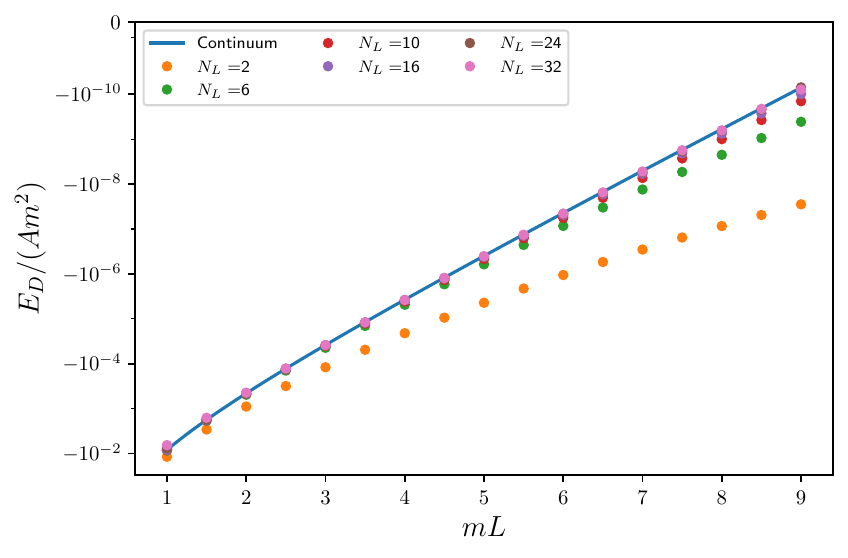} 
	\caption{Dimensionless Casimir energy with \acrshort*{dbc} in logarithmic scale of a three dimensional lattice with different values of $N_{L}=\{2,6,10,16,24,32\}$ and $mL=1-9$. The other lattice parameters are fixed at $N_{A}=96$ and $N_{L_0}=96$.}
	\label{fig:Casimir_lattice_2d_sizeL_Dir}
\end{figure}

Plotting again the Casimir energy with different sizes $N_L$ at fixed $mL$ for \acrshort*{dbc} in \autoref{fig:Casimir_lattice_2d_sizeL_Dir} we observe
the same behaviour than for \acrshort*{pbc}. The difference in the Casimir energy with the different sizes increases as the dimensionless effective mass $mL$ grows for the same reason as in  \acrshort*{pbc}, and the smaller sizes of $N_L$ adjust very poorly to the continuum behaviour. The quality of the fit improves notably as $N_L$ grows, getting a good fit for the larger values.

     \chapter{Lattice gauge theories}\label{chp:gauge_theory}
     In this chapter we shall introduce  gauge theories regularized on a space-time lattice and the most relevant aspects that are necessary for the numerical simulations that will be done in chapters \ref{chp:su2_21} and \ref{chp:su2_31}.

In \autoref{sec:Nair_karabali} we introduce the Karabali-Nair parametrization (\acrshort*{nkp}), where in 2+1 dimensions the gauge invariant degrees of freedom are expressed in terms of massive scalar fields with a mass value which is much lower than the glueball. In \autoref{sec:Wilson_action} we present the lattice version of the pure gauge theories by using the Wilson action. Then, in \autoref{sec:MC_sim} we introduce the \acrshort*{mc} method that will be used to do the numerical simulations. In \autoref{sec:BC} we show how to introduce new boundary conditions on the lattice. Finally, in sections \ref{sec:Run_sim} and \ref{sec:statistical_erro} we include some key details of the numerical simulations, in particular, the parallelization for the Wilson action and how to obtain the statistical errors from the \acrshort*{mc} results.

Gauge theories are  described by Yang-Mills theories in a continuum space-time, where the action is given by
\begin{equation}\label{eq:YM_cont_action}
	S_{YM}=-\frac{1}{2g^2}\int dx^{D+1}\ \text{tr}\left(F_{\mu\nu}F^{\mu\nu}\right)
\end{equation}
where $F_{\mu\nu}$ is the field strength tensor 
\begin{equation}
	F_{\mu\nu}=\partial_\mu A_\nu -\partial_\nu A_\mu -[A_\mu,A_\nu],
\end{equation}
$A_\mu=A^a_\mu t^a$ are the gauge fields and $t^a$ the generators of the Lie algebra of the gauge group $G$. The gauge transformations of the gauge fields are given by
\begin{equation}
	A_\mu\rightarrow \Omega\ A_\mu \Omega^{-1}+i\ \Omega\ \partial_\mu \Omega^{-1},
\end{equation}
with $\Omega\in G$. This gauge transformation translates into the field strength tensor as
\begin{equation}
	F_{\mu\nu}\rightarrow \Omega\ F_{\mu\nu}\Omega^{-1},
\end{equation}
and  the action \eqref{eq:YM_cont_action} is invariant under such transformations.
\section{Karabali-Nair parametrization}\label{sec:Nair_karabali}
Let us  introduce an approximation for $SU(N_c)$ gauge theories in 2+1 dimensions that was developed by D. Karabali and V. P. Nair \cite{karabali1996gauge,karabali1998planar}, where the invariant degrees of freedom of the gauge fields are parametrized in terms of massive scalar fields. This will not only justify the comparison between the Casimir energy that we obtain on the numerical simulations of the gauge theory with the massive scalar field expressions, but will also allow us to check the validity of this parametrization \cite{karabali2018casimir} by comparing the results of the Casimir energy for different boundary conditions.

Let us consider the Hamiltonian approach and the $A_0=0$ gauge. The spatial gauge fields $A_1$ and $A_2$ define  two complex conjugate combinations of these spatial gauge fields
\begin{equation}
	A_z=\frac{1}{2}\left(A_1+iA_2\right),\hspace{1cm}A_{\bar z}=\frac{1}{2}\left(A_1-iA_2\right).
\end{equation}
We can parametrize them in terms of complex $SL(N,C)$ matrices $M$ and $M^\dagger$ as
\begin{equation}
	A_z=-\partial_z M M^{-1},\hspace{1cm}A_{\bar z}=-\left(M^{\dagger}\right)^{-1}\partial_{\bar z}M^{\dagger},
\end{equation}
where $\partial_z=\frac{1}{2}(\partial_1+i\partial_2)$ and $\partial_{\bar z}=\frac{1}{2}(\partial_1-i\partial_2)$. This matrices transform under gauge transformations as 
\begin{equation}
	M\rightarrow \Omega M\hspace{1cm} M^\dagger\rightarrow M^\dagger \Omega^{-1}
\end{equation}
with $\Omega\in SU(N_c)$. We can actually construct a gauge invariant field as $H=M^\dagger M$, which belongs to $SL(N_c,C)/SU(N_c)$. This gauge invariant field can be used to write the inner product of physical states as 
\begin{equation}
	\braket{1|2}=\int d\mu(H)\ e^{2c_A\mathcal S(H)}\ \Psi^*_1(H)\Psi_2(H)
\end{equation}
where $S(H)$ is the Wess-Zumino-Witten (\acrshort*{wzw}) action for the matrix $H$, $d\mu(H)$ is the Haar measure and $c_A$ the quadratic Casimir operator in the adjoint representation ($c_A=N_c$ in $SU(N_c)$). Thus, the Hamiltonian and other observables can be expressed in terms of the current (see \cite{karabali1996gauge,karabali1998planar} for the explicit formula)
\begin{equation}
	J_a=\frac{c_A }{\pi}  \left(\partial H H^{-1}\right)_a.
\end{equation}
Now, we write  $H=e^{t_a\varphi_a}$ and  expand in powers of this field $\varphi$ for the Hamiltonian and the \acrshort*{wzw} action. But we retain the exponential factor
\begin{equation}\label{eq:nahir_exp_factor}
e^{2c_AS(H)}\simeq e^{-\frac{c_A}{2\pi}\int d^2x\ \partial \varphi \overline\partial \varphi}
\end{equation}
in the functional integral
instead of expanding the exponential. Also, by absorbing this factor in \eqref{eq:nahir_exp_factor} into the wave functions $\Phi=e^{-\frac{c_A}{4\pi}\int \partial \varphi \overline\partial \varphi} \Psi$, the inner product with these expansions simplifies to
\begin{equation}
	\braket{1|2}=\int [d\varphi]\ \Phi^*_1\Phi_2.
\end{equation}
Using this wave function the Hamiltonian takes the form
\begin{equation}
\mathcal H=\frac{m\pi}{c_A}\!\!\int\!\! d^2\mathbf xd^2\mathbf y\ C(\mathbf x,\mathbf y)\frac{\delta}{\delta \varphi_a(\mathbf x)}\frac{\delta}{\delta \varphi_a(\mathbf y)}+\frac{mc_A}{4\pi}\!\!\int\!\! d^2\mathbf x\!\left(\! \partial \varphi_a \overline\partial \varphi_a+\frac{4}{m^2} \partial\varphi_a(-\partial \overline \partial )\overline \partial \varphi_a\!\right)\!+\ldots,
\end{equation}
where we have defined the mass $m=g^2c_A/(2\pi)$ and
\begin{equation}
	 C(\mathbf x,\mathbf y)=-\frac{1}{(2\pi)^2}\int d^2\mathbf k\frac{e^{i\mathbf k(\mathbf x-\mathbf y)}}{k\bar k}.
\end{equation}
Finally, by defining new fields as $\phi_a(\mathbf k)=\sqrt{c_Ak\overline k/(2\pi m)}\ \varphi_a(\mathbf k)$, we obtain the Hamiltonian of $N_c^2-1$ massive scalar fields
\begin{equation}
	\mathcal{H}\simeq\frac{1}{2}\int d^2\mathbf x\left( -\frac{\delta^2}{\delta \phi^2_a(\mathbf x)}+\phi_a(\mathbf x)\left(m^2-\nabla^2\right)\phi_a(\mathbf x)+\ldots\right),
\end{equation}
with the following action
\begin{equation}
	S_\phi=\frac{1}{2}\int d^3x\left(\dot \phi^2_a-(\nabla \phi_a)^2-m^2\phi^2_a\right).
\end{equation}
We can see how the propagator of the gauge  degrees of freedom takes the form of a massive adjoint scalar field with mass $m_{\mathrm{K}}=g^2c_A/(2\pi)$. Thus, one would expect that the Casimir energy behaves similarly to the one given by a massive scalar field with this mass $m$ predicted by this approximation. Moreover, the relation between the gauge fields and the adjoint scalar fields in this small $\varphi$ expansion is given by
\begin{equation}\label{eq:relation_NH}
	A_\mu^a\simeq\frac{1}{2}\epsilon_{\mu\nu} \partial_\nu \varphi^a+\ldots,
\end{equation}
from which we can derive what boundary conditions will emerge on the massive scalar field $\phi$ when imposing a certain boundary condition on the gauge fields $A_\mu$.
\section{Wilson action}\label{sec:Wilson_action}
\begin{figure}[H]
	\centering
	\includegraphics[width=.8\textwidth]{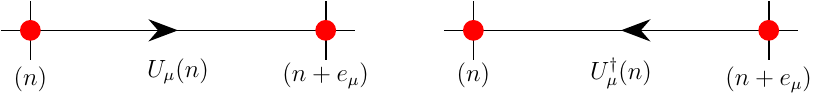}
	\caption{Representation of the link variables $U_\mu(n)$ and $U_\mu^\dagger (n)$ on the lattice. }
	\label{fig:link_representation}
\end{figure}

On the lattice regularization of pure Yang-Mills theories it is standard to use the Wilson action \cite{wilson1974confinement}. We recover the notation used in \autoref{chp:scalar_lattice}, where $n=(n_0,n_1,\ldots,n_D)$ are the coordinates of the lattice sites, $a$ is the distance between two consecutive lattice points and $ e_\mu$ is the unitary vector in the direction $\mu$ of the lattice. In this formulation the variables are associated with the links between the sites of the lattice, which are defined in the continuum as
\begin{equation}
	U_{\mu}( x)=\mathcal P \exp \left({i\int_{ x}^{ x+a e_\mu}}A_\mu(y)\  dy\right),
\end{equation}
and can be approximated for smooth fields on the lattice by
\begin{equation}\label{eq:link_matrix}
	U_\mu ( n)=e^{iA_\mu(a n)}
\end{equation}
up to $O(a)$. These link variables connect consecutive points on the lattice and are oriented, $U_\mu( n)$ links the point $ n$ with $ n+ e_\mu$ whereas $U_{-\mu}( n)$ links the point $ n$ with $ n- e_\mu$, see \autoref{fig:link_representation}. The positive and negative oriented link variables are actually related: $U_{-\mu}( n)=U^\dagger_{\mu}( n-e_\mu)$. This is consistent with the fact that the links variable are elements of the gauge group $SU(N_c)$, and $U_\mu^\dagger( n)$ is the inverse of $U_\mu( n)$. These link variables transform under a gauge transformation as
\begin{equation}
		U_\mu( x)\rightarrow \Omega( x)U_\mu( x)\Omega^\dagger( x+ae_\mu),
\end{equation}
which depends on the starting and endpoint of the link.

\begin{figure}[H]
	\centering
	\includegraphics[width=.6\textwidth]{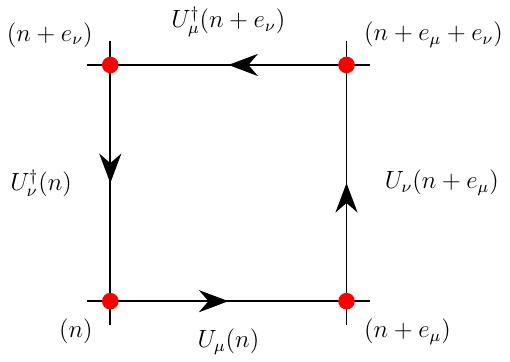}
	\caption{Representation of a plaquette $P_{\mu\nu}( n)$ on the lattice. }
	\label{fig:plaquette_representation}
\end{figure}

 The Wilson action is constructed with these link variables as the sum of all the different possible smallest closed loops $P_{\mu\nu}( n)$ on the lattice
\begin{equation}\label{eq:wilson_action}
	S_W=\beta \sum_{ n\in \Lambda}\sum_{\mu <\nu} \left(1-\frac{1}{N_c}\text{Re tr }P_{\mu\nu}( n)\right),
\end{equation}
where $\beta$ is the coupling constant associated to $g^2 $ in \eqref{eq:YM_cont_action}. The smallest closed loops are called \textit{plaquettes} $P_{\mu\nu}$ and are given by
\begin{equation}\label{eq:plaquette}
	P_{\mu\nu}(n)=U_\mu( n)U_\nu( n+e_\mu)U^\dagger_\mu( n
	+ e_\nu)U^\dagger_\nu( n),
\end{equation}
i.e. a square on the lattice of size $a$ in the directions $\mu$ and $\nu$ (see \autoref{fig:plaquette_representation}).

In order to have a well defined large $N_c$ limit it is convenient to  use a $N_c$ depending coupling constant \footnote{Notice how in the 3+1 dimensional case there is no $a$ in the definition of $\beta$ since $g$ is dimensionless.}
\begin{equation}
	\beta(N_c,D)=\frac{2 N_c}{a^{3-D}g^2}.
\end{equation}
Obviously, this action is a gauge invariant quantity since under a gauge transformation the plaquette behaves as
\begin{equation}
	P_{\mu\nu}(n)\rightarrow \Omega( n) P_{\mu\nu}( n) \Omega^\dagger(  n),
\end{equation} 
and by taking the trace the remaining transformations cancel.

One can see that the Wilson action converges to the continuum action \eqref{eq:YM_cont_action} in the limit when the lattice spacing goes to zero $a\rightarrow 0$. This can be shown by inserting the lattice definition of the link variables \eqref{eq:link_matrix} into the formula of the plaquette \eqref{eq:plaquette} and using the Baker–Campbell–Hausdorff formula \cite{hausdorff1906symbolische}
\begin{equation}
	e^{A}e^{B}=A^{A+B+\frac{1}{2}[A,B]+\ldots}
\end{equation}
for the product of link variables. The result is (see Chapter 2 in Ref. \cite{gattringer2009quantum} for an extended development) that the plaquette can be written in terms of the field strength tensor as
\begin{equation}\label{eq:plaqutte_exp}
	P_{\mu\nu}( n)=e^{ia^2 F_{\mu\nu}(a n)+O(a^3)},
\end{equation}
where we have used a Taylor expansion for the gauge fields
\begin{equation}
	A_\mu(a n+ae_\nu)=A_\mu(an)+a\ \partial_\nu A_\mu(a n)+O(a^2).
\end{equation}
Now, we can expand the exponential in \eqref{eq:plaqutte_exp} and take the trace, obtaining the following relation between the plaquette and the field strength tensor
\begin{equation}\label{eq:plaquette_F}
	\text{Re tr}\left(P_{\mu\nu}( n)\right)=N_c-\frac{a^4}{2}\text{ tr }\left(F^2_{\mu\nu}(a n)\right)+ O(a^6),
\end{equation}
where due to the fact that it is the real part only the leading term with $a^2$ contributes to the expansion.
Finally, we can insert this relation \eqref{eq:plaquette_F} into the Wilson action \eqref{eq:wilson_action}
\begin{equation}
	S_W=\frac{a^{D+1}}{g^2}\sum_{ n\in \Lambda}\sum_{\mu <\nu}\ \text{tr}\ (F_{\mu\nu}^2(a n))+O(a^2)=\frac{1}{2g^2}\int dx^{D+1}\ \text{tr}\left(F_{\mu\nu}F^{\mu\nu}\right)+O(a^2)
\end{equation}
which shows that in the continuum limit $a\rightarrow 0$ we recover the Euclidean version of the action \eqref{eq:YM_cont_action}. We also have used \eqref{eq:integral_sum} to convert the sum on the lattice into a continuum integral and  that the sum without repetition of the square strength tensor is half of the fully contracted strength tensor (in Euclidean space) $\sum\limits_{u<\mu}F_{\mu\nu}^2=\frac{1}{2}F_{\mu\nu}F^{\mu\nu}$. 

In this work, we shall  focus on the case of $SU(2)$, thereby, the Wilson action reduces to
\begin{equation}\label{eq:wilson_action_su2}
	S_W[U]=\frac{4}{a^{3-D}g^2}\sum_{ n\in \Lambda}\sum_{\mu <\nu} \left(1-\frac{1}{2}\text{tr }(P_{\mu\nu}( n))\right),
\end{equation}
where we have omitted the projection to the real part since the trace of the $SU(2)$ matrices is always real.
\section{Monte Carlo simulations}\label{sec:MC_sim}
With this  Euclidean   lattice action that converges to the continuum, we can compute the expected values of an observable $O[U]$ analogously to what was done for the scalar field
\begin{equation}\label{eq:expected_YM}
\braket{O}=\frac{1}{Z}\int \prod_{ n\in \Lambda }\prod_{\mu=0}^{D}dU_\mu ( n)\ O[U]\ e^{-S_W[U]}
\end{equation}
where $Z$ is the partition function
\begin{equation}
	Z=\int \prod_{ n\in \Lambda }\prod_{\mu=0}^{D} dU_\mu ( n)\ e^{-S_W[U]}
\end{equation}
and the integral is over the Haar-measure of $SU(N_c)$.

Unfortunately, the partition function can not be solved analytically like in the free scalar case or even by direct numerical integration since the number of variables to be integrated is too large, for example in a small $10^4$ lattice with SU(2) pure gauge an integral of 120000 variables is required. Therefore, we have to use statistical methods to compute the expected value \eqref{eq:expected_YM}. In particular, we use the \acrshort*{mc} method, that was first introduced in 1949 \cite{metropolis1949monte} to solve the problem of neutron diffusion. In our case, we generate random configurations of the link variables $\{U\}$ distributed according to the probability driven by the action $e^{-S_W[U]}$, where configurations with the lowest action values are the most relevant. This is the central idea of the \textit{importance sampling} in these \acrshort*{mc} methods, the configurations which are used for computing the average of observables are sampled according to the probability factor given by the action. With these configurations, we can approximate the expected value of an observable by taking the mean value 
\begin{equation}
	\braket{O}\simeq\frac{1}{N}\sum_{i=1}^N O(\{U\}_i)
\end{equation}
where the sum is over the configurations $\{U\}_i$ generated, and this expected value will have a statistical error of order  $1/\sqrt{N}$\footnote{This is true if the measurements are statistically independent, this issue will be discussed in \autoref{sec:statistical_erro}.}.

Now, the question is how we obtain the configurations $\{U\}_i$ that follow the adequate probability. We have to start with an arbitrary configuration and follow a stochastic sequence that will converge to an equilibrium state with the desired probability distribution $e^{-S_W[U]}$. This sequential process is done with a Markov process, where the result of every step $\{U\}_n$ only depends on the previous configuration $\{U\}_{n-1}$, and the transition probability between the new configuration $\{U'\}$ and the old one $\{U\}$ has to obey
\begin{equation}
	0\leq T(\{U'\}|\{U\})\leq 1,\hspace{1cm} \sum_{\{U'\}} T(\{U'\}|\{U\})=1,
\end{equation}
where the sum in $\{U'\}$ also includes the old configuration $\{U\}$. When the equilibrium is reached, the Markov process cannot have sources or sinks of probability which translates into the balance equation
\begin{equation}
	\sum_{\{U\}} T(\{U'\}|\{U\})P(\{U\})=\sum_{\{U'\}} T(\{U\}|\{U'\})P(\{U'\}).
\end{equation}
A solution to this equation is known as the \textit{detailed balanced condition} where the previous equation is satisfied term by term
\begin{equation}
	 T(\{U'\}|\{U\})P(\{U\})= T(\{U\}|\{U'\})P(\{U'\}).
\end{equation}
Another important property that this Markov process has to fulfill is that it must be possible to reach any configuration in a finite number of steps, this property is called \textit{ergodicity}.

Now, we shall  explicitly analyze some algorithms that fulfill these conditions and allow us to perform \acrshort*{mc} simulations in the particular case we are interested, i.e  a pure $SU(2)$ gauge theory.
\subsection{Metropolis algorithm}
The Metropolis algorithm, which was introduced in 1953 \cite{metropolis1953equation}, is one of the most general and simplest methods of applying a Markov process with the properties described above.

Let us to focus on the specific case we are interested in of a pure $SU(2)$ gauge theory. The link matrices $U_\mu( n)$ are elements of $SU(2)$ and can be parametrized by four real numbers in the following way
\begin{equation}\label{eq:su2_representation}
	U_\mu( n)=x_0\ \mathbb I+i\mathbf x\cdot \boldsymbol \sigma \hspace{1cm}\text{with   } x_0^2+\|\mathbf x\|^2=1
\end{equation}
where $\boldsymbol \sigma$ are the Pauli matrices and $x_i$ are real numbers. Although the $SU(2)$ elements only have three degrees of freedom it is more efficient computationally to use this redundant representation with four real numbers and the extra condition of fixing the determinant.

The first step of the algorithm is to  generate a new candidate $U'_\mu( n)$ for the link $U_\mu( n)$ we are updating with some probability $T(U'|U)$. This has to be done in a way that the probability of the change being accepted is not too low, thus, we need that the change in the action due to the new candidate variable is small. A method for achieving this small change in the action is computing the new candidate variable as
\begin{equation}
	U'_\mu(n)=M U_\mu( n)
\end{equation}
where $M$ is an element of $SU(2)$ close to the identity. If one wants the transition probability $T(U'|U)$ to be symmetric, $M$ has to fulfill that the probability for obtaining $M$ and $M^{-1}$ is the same. A method for generating the random matrix $M$ with these conditions is by generating three random numbers $r_i$ between $(-1/2,1/2)$ and computing the matrix $M$ as
\begin{equation}
	\mathbf m=\epsilon \frac{\mathbf r}{\|\mathbf r\|},\hspace{1cm} m_0=\sqrt{1-\epsilon^2},
\end{equation}
where $(m_0,\mathbf m)$ is the representation of $M$ with \eqref{eq:su2_representation} and the parameter $\epsilon$ allow us to control how close to the identity the matrix $M$ is.

After generating the candidate link $U_\mu'( n)$, we compute how much the action would vary if we accept the change. The local contribution of a single link to the Wilson action \eqref{eq:wilson_action_su2} is
\begin{equation}\label{eq:local_wilson_action}
	S_W[U'_\mu( n)]=\beta\left(1-\frac{1}{2} \text{tr}\left(U_\mu'( n) \Gamma_\mu( n)\right)\right)
\end{equation}
where
\begin{align}\nonumber
	\Gamma_\mu(\ n)&=\sum_{\nu\not =\mu}\left(U_\nu( n+ e_\mu)U_\mu^\dagger( n+ e_\nu)U^\dagger_\nu(n )\right.\\ \label{eq:staple_2d}
	&\left.+U^\dagger_\nu( n+ e_\mu- e_\nu)U_\mu^\dagger( n- e_\nu)U_\nu( n- e_\nu )\right)
\end{align}
are the terms that appear in the action related to $U_\mu( n)$ which are usually named \textit{staples}. Once we have this quantity computed, it is easy to see that the variation of the action due to changing the link is given by
\begin{equation}
	\Delta S=S_W[U'_\mu( n)]-S_W[U_\mu( n)]=-\frac{\beta}{2}\text{tr}\left((U'_\mu( n)-U_\mu( n))\Gamma_\mu( n)\right).
\end{equation}

The acceptance probability of the candidate variable is given by $\min \{1,e^{-\Delta S}\}$, which means that when the action decreases with the candidate variable the change is always accepted, whereas if the action increases we generate an uniform random number $r$ between $(0,1)$ and accept the change if $e^{-\Delta S}>r$. Also, it is important to realize that when the change is not accepted the ``new'' link is the old one and we measure with this link again.
\subsection{Heat-bath algorithm}
In this algorithm the new value of the variable link is computed according to the local distribution of probability given by the local action of that link. This is equivalent to iterating the Metropolis algorithm on one link variable an infinite amount of times, thermalizing it with its local action. The method for applying this algorithm to the $SU(2)$ pure gauge case was first proposed in 1980 \cite{creutz1980monte}. The local probability distribution is given by
\begin{equation}\label{eq:su2_local_prob}
	dP(U)=dU \exp\left(\frac{\beta}{2} \text{tr} (U_\mu \Gamma_\mu)\right)
\end{equation}
where $\Gamma_\mu$ was defined previously in \eqref{eq:staple_2d} and $dU$ is the Haar integration measure of the group. This case of $SU(2)$ has a great advantage since the sum of elements of the group is proportional to an element of the group. In particular, from the sum of staples $\Gamma_\mu$ we can construct
\begin{equation}\label{eq:norm_staple}
	V_\mu=\Gamma_\mu/v \hspace{1cm}\text{where }\hspace{0.5cm}v=\sqrt{\text{det}(\Gamma_\mu)},
\end{equation}
and $V_\mu$ belongs to $SU(2)$. Since the product $W=U_\mu V_\mu$ is also an element of the group $SU(2)$ we can transform the local probability distribution \eqref{eq:su2_local_prob} in terms of this new matrix $W$
\begin{equation}\label{eq:su2_local_prob_W}
	dP(W)=dW \exp\left(\frac{v\beta}{2} \text{tr} (W)\right),
\end{equation}
and once we have generated the matrix $W$ with this probability distribution, we can undo the change to obtain the new variable link as
\begin{equation}\label{eq:new_link_hb}
	U'_\mu( n)=\frac{1}{v}W\Gamma^\dagger_\mu( n).
\end{equation}
Finally, we have to deal with the Haar measure of the group to generate $W$ with the correct probability. In the representation \eqref{eq:su2_representation} this Haar measure simplifies to generating random points in a four dimensional sphere
\begin{equation}
	dW=\frac{1}{\pi^2}\ d^4x\  \delta(x^2-1).
\end{equation}
Using spherical coordinates for the $\mathbf x$ vector ($d^2\Omega=d\cos\theta\ d\psi$), the Dirac delta to integrate the norm $\|\mathbf x\|$ and that the trace in this representation is tr$(W)=2x_0$ the local probability distribution \eqref{eq:su2_local_prob_W} reduces to
\begin{equation}
	dP(W)=\frac{1}{2\pi^2}d\cos\theta\ d \psi\ dx_0\ \sqrt{1-x_0^2}\  e^{v\beta x_0},
\end{equation}
with $x_0\in[-1,1]$, $\cos\theta\in[-1,1]$ and $\psi\in [0,2\pi)$. Thus, we just have to generate the direction of $\mathbf x$ randomly and $x_0$ according to the probability
\begin{equation}
	P(x_0)=\sqrt{1-x_0^2}\ e^{v\beta x_0}.
\end{equation}
An efficient method for finding this variable (see \cite{fabricius1984heat,kennedy1985improved} for a detailed explanation) is introducing an auxiliary variable $2\zeta=1-x_0$ and generating three uniform random numbers $r_i$ from $(0,1]$ that define the auxiliary variable
\begin{equation}
	\zeta=-\frac{1}{2a\beta}\left(\ln(r_1)+\cos^2(2\pi r_2)\ln(r_3)\right).
\end{equation}
Now, we generate another uniform random number $r_4$ between $[0,1)$ and if $r_4^2\leq 1-\zeta$ we accept this value of $\zeta$ from which we obtain $x_0$ as $x_0=1-2\zeta$. In the opposite case where $r_4^2> 1-\zeta$, we start again the process until we generate an accepted value of $\zeta$.

Finally we just have to generate $\mathbf x$, this will be a three dimensional vector with module $\|\mathbf x\|=\sqrt{1-x_0^2}$ and random direction. We can generate two uniform random numbers $r_i$ from $[0,1]$ and compute the angular parameters for the vector, $\psi=2\pi r_1$ and $\cos\theta=2r_2-1$. Hence, we can compute the three components of the vector with these angular variables as
\begin{equation}
	x_1=\|\mathbf x\|\cos\theta\hspace{0.5cm} x_2=\|\mathbf x\|\sqrt{1-\cos^2\theta}\cos \psi \hspace{0.5cm} x_3=\|\mathbf x\|\sqrt{1-\cos^2\theta}\sin \psi.
\end{equation} 
With the four components of $x$ we have fully characterized $W$ 
\begin{equation}
	W=x_0\ \mathbb{I}+\mathbf x \cdot \boldsymbol \sigma,
\end{equation}
and we can compute the new link $U'_\mu(n)$ from this $W$ by using \eqref{eq:new_link_hb}.
\subsection{Overrelaxation}
The last algorithm we shall consider   is overrelaxation \cite{adler1988overrelaxation}. This method has the advantage that new link variable is much farther away from the old one in configuration space than the variable generated with the Metropolis or heat-bath algorithms. The main idea is to find a candidate link $U_\mu'(n)$ that does not change the action and, therefore, has the same probability weight as the old link $U_\mu(n)$. As we mentioned for the heath bath algorithm, the fact that a sum of elements of the group is proportional to an element of the group in $SU(2)$ makes this case very simple, we can just compute the new link as
\begin{equation}
	U'_\mu( n)=V_\mu^\dagger (n)U^\dagger_\mu( n)V_\mu^\dagger ( n)
\end{equation}
where $V_\mu^\dagger ( n)$ is the normalized sum of staples  defined in \eqref{eq:norm_staple} which is also an element of $SU(2)$. We can check how the factor involving this new link in the action \eqref{eq:local_wilson_action} is invariant (we omit the dependence with the point of the lattice):
\begin{equation}
	\text{tr}(U'_\mu\ \Gamma_\mu )=	v\ \text{tr}(V_\mu^\dagger U^\dagger_\mu V_\mu^\dagger V_\mu  )=\text{tr}(\Gamma_\mu^\dagger U^\dagger_\mu   )=\text{tr}(U_\mu \Gamma_\mu)
\end{equation}
where in the last step we have used that the trace of $SU(2)$ is real. We see how the new action will be equal to the old one since the trace factor in the plaquette is invariant.

It is important to mention that this algorithm is not ergodic since it only moves the configurations on a subspace of constant action. Thus, we have to combine this method with another algorithm like Metropolis or heat-bath to move around the whole configuration space.

\section{Boundary conditions}\label{sec:BC}
In this section we  analyze the boundary conditions  of  the pure Yang-Mills theory on the lattice. In particular, we shall focus on the 2+1 and 3+1 dimensional cases. Like in the scalar field theory, we shall consider \acrshort*{pbc} on the directions that are parallel to the boundary wall (i.e. perpendicular to direction $D$) and the time direction. We also use the same notation,  the lattice points in any direction $i=0,1,2,3$ go from $n_i=1$ till $n_i=N_i$. Similarly to the scalar theory, the \acrshort*{pbc} are defined by connecting the boundary points of the lattice with each other. In terms of the links variables this means that, for example, the link $U_0(N_0,n_1,n_2$) connects the sites $(N_0,n_1,n_2)$ and $(1,n_1,n_2)$. 
Once we have fixed the boundary conditions in these directions, we have to deal with the transverse spatial direction. We shall use two different boundary conditions in the transverse direction $D$ to compute the Casimir energy in chapters \ref{chp:su2_21} and \ref{chp:su2_31}. First, we introduce \acrshort*{pbc}, and later on the so called perfect colour conductor boundary conditions (\acrshort*{pccbc}).
\subsection{Periodic boundary conditions}
As we just mentioned, we consider also \acrshort*{pbc} in the transverse spatial dimension. We implement this boundary condition in the same way than the other directions by connecting the boundary points in the $D$ direction, i.e in 2+1 dimensions the link $U_2(n_0,n_1,N_2)$ goes from the site $(n_0,n_1,N_2)$ to $(n_0,n_1,1)$. Equivalently in 3+1 dimensions we have that the link $U_3(n_0,n_1,n_2,N_3)$ connects the point  $(n_0,n_1,n_2,N_3)$ to $(n_0,n_1,n_2,1)$.
 When we use these boundary conditions in the transverse direction, all the directions of the lattice have the same boundary conditions and if they have the same number of links they are equivalent.
\subsection{Perfect colour conductor boundary conditions}\label{sec:perfect_conductor}
The second type of boundary conditions we will consider is the equivalent to the perfect  conductor boundary condition in electromagnetism for the non-abelian fields. In electromagnetism, in 3+1 dimensions this condition consists in imposing that the parallel electric fields and normal magnetic field cancel on the boundary, i.e. $E_1=E_2=B_3=0$. This condition can be written in terms of the field strength tensor as $F_{01}=F_{02}=F_{12}=0$, which is the condition we shall consider for the non-abelian case.

Now, we have to see what is the lattice equivalent of this continuum condition. We can use the relationship in \eqref{eq:plaquette_F} between the plaquette and the field strength tensor to arrive at 
\begin{equation}
	\text{tr }(P_{\mu\nu}( n))\simeq 2 -\frac{a^4}{2}\ \text{tr }(F^2_{\mu\nu}( n)),
\end{equation}
where we have used that the trace in $SU(2)$ is real. Thus, imposing $P_{\mu\nu}=\mathbb I$ on the lattice is the analogous condition to canceling the field strength tensor. Therefore, the \acrshort*{pccbc} on the lattice are given by 
\begin{align}\label{eq:perfect_conductor_3d}
	&P_{01}(n_0,n_1,n_2,1)=P_{02}(n_0,n_1,n_2,1)=P_{12}(n_0,n_1,n_2,1)=\mathbb I
\end{align}
The simplest way to actually implement this condition is fixing all the link variables in the boundary to the identity
\begin{equation}\label{eq:bound_ident_3}
	U_0(n_0,n_1,n_2,1)=U_1(n_0,n_1,n_2,1)=U_2(n_0,n_1,n_2,1)=\mathbb I
\end{equation}
and imposing  \acrshort*{pbc} on the other side on the lattice in the same way we did in \acrshort*{dbc} in the scalar field,
i.e.
the link variable $U_3(n_0,n_1,n_2,N_3)$ connects the sites $(n_0,n_1,n_2,N_3)$ and $(n_0,n_1,n_2,1)$. Clearly, the easiest way for implementing \eqref{eq:bound_ident_3}, is to set these links to the identity in the initial configuration and not evolve them in the \acrshort*{mc} steps. This procedure ensures that every configuration on the simulation fulfills the \acrshort*{pccbc}  \eqref{eq:perfect_conductor_3d}.

For the 2+1 dimensional case we apply the analogous condition but taking into account the fewer dimensions. In two spatial dimensions, the electromagnetic fields reduce to just two electric fields $E_1$ and $E_2$, and a magnetic field $B$ that is a scalar (see for example \cite{maggi2022dimensional} for an analysis about this dimensional reduction in electromagnetism). Thus, the \acrshort*{pccbc} just imposes $E_1=0$ as a condition, which translates into the non-abelian case as $F_{01}=0$. Following the same arguments as in the 3+1 dimensional case this condition on the lattice is given by
\begin{equation}\label{eq:perfect_conductor_21}
	P_{01}(n_0,n_1,1)=\mathbb I.
\end{equation}
We implement this condition in the same way by fixing the links variables at the boundary to the identity
\begin{equation}
	U_0(n_0,n_1,1)=U_1(n_0,n_1,1)=\mathbb I
\end{equation}
and applying \acrshort*{pbc} on the other side of the lattice.
Therefore, the link variable $U_2(n_0,n_1,N_2)$ connects the points $(n_0,n_1,N_2)$ and $(n_0,n_1,1)$.

Notice how these \acrshort*{pccbc} are equivalent to \acrshort*{dbc} in the scalar field since in both it is imposed that the action at the boundary points is zero.

\section{Numerical  simulations}\label{sec:Run_sim}
Now, we  analyze some technical details on how  to run the \acrshort*{mc} simulations to obtain the results that will be discussed in chapters \ref{chp:su2_21} and \ref{chp:su2_31}. Systems with local action (like in this case) are perfect candidates for being parallelized, in the sense, that instead of applying sequentially an update for each link by using one of the algorithms described before, one can update several links at the same time as long as they directly do not interact with each other. This allows to substantially speed up the simulations. The simulations will be run in two different systems, using a graphics processing unit (\acrshort*{gpu}) and also by using several central processing units (\acrshort*{cpu}s) in parallel.

\subsection{Simulating on GPU}  \label{sec:run_sim}

\begin{figure}[h]
	\centering
	\includegraphics[width=.6\textwidth]{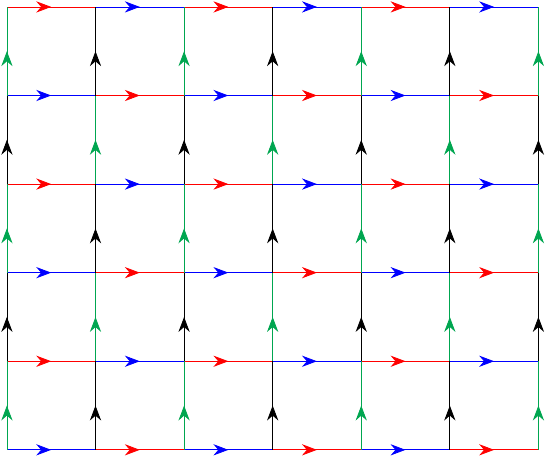}
	\caption{Representation of the parallelization for \acrshort*{gpu} running using the chessboard pattern, where the links of the same colour are updated simultaneously.}
	\label{fig:lattice_div}
\end{figure}

The \acrshort*{gpu}s are specialized in vectorized float operations, which makes them perfectly suited for these type of parallel calculations. For coding in these \acrshort*{gpu}s we shall use CUDA \cite{cuda}, that is a programming platform based in C++ that allows to easily program the parallel computation on NVIDIA's \acrshort*{gpu}s (see e.g.\cite{cardoso20112} for an implementation of $SU(2)$ lattice gauge theory using CUDA). A great advantage that CUDA offers is that it manages the threads\footnote{A thread is each of the sequential processes we send in parallel.} distribution in the \acrshort*{gpu}, which allow us to send a huge number of threads without having to worry if it is larger than the number of processors of the \acrshort*{gpu}. Thus, we can parallelize by running at the same time half of the lattice points and one direction at a time, where we divide the lattice points by the parity of the sum of their lattice indexes (like a chessboard pattern), see \autoref{fig:lattice_div} where each colour represents the links that are updated in parallel.  

This is actually the maximum parallelization the Wilson action \eqref{eq:wilson_action} allows. Consequently, to perform a full sweep of updates on the lattice six of these parallel updates where we update $N/2$ links at a time are required in the 2+1 dimensional case, whereas  eight full updates are necessary in the 3+1 dimensional case. This method has a disadvantage. For  \acrshort*{pbc} it is necessary  that every direction on the lattice has an even number of links. If not, when applying this chessboard pattern for updating the links we encounter that the boundary links have be updated simultaneously but they are first neighbours and have common plaquettes. For example, if in the transverse spatial direction $ e_2$ we impose \acrshort*{pbc} and size of $N_2=5$ we would have to update at the same time the links $U_1(n_0,n_1,1)$ and $U_1(n_0,n_1,5)$ which are neighbours (the same happens for the time links). This can also be seen in \autoref{fig:lattice_div} where if we have \acrshort*{pbc} in the vertical direction the horizontal links in the first and last row would coincide but they have to be updated at different times. Usually this is not a problem since it is enough to use only even lattice sizes, but since we are interested on the behaviour of the Casimir energy with the distance between the boundary walls, which is proportional to the number of links in the transverse spatial direction, having an odd number allows us to double the statistics. This problem can be solved by using \acrshort*{cpu}s where we do not parallelize to this maximum extent.

For the \acrshort*{gpu} simulations a NVIDIA GeForce RTX 3080 is used, which is a medium range \acrshort*{gpu} for gaming in personal computers. Being able to do some of the hard simulations of this work in  this \acrshort*{gpu}, shows the power of this type of hardware for easily parallelizable lattice problems.

\subsection{Simulating on CPU}

\begin{figure}[h]
	\centering
	\includegraphics[width=.6\textwidth]{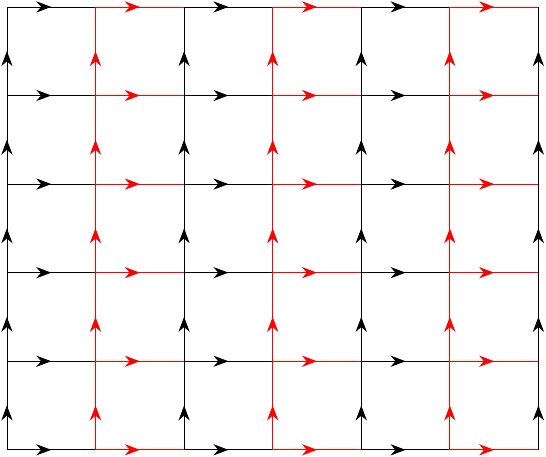}
	\caption{Representation of the parallelization for \acrshort*{cpu} running dividing the horizontal direction in even and odd sites, where the links of the same colour are updated simultaneously but each column is done sequentially by the same thread.}
	\label{fig:lattice_div_cpu}
\end{figure}

But we also use \acrshort*{cpu}s for the simulations and parallelize the code by using more than one processor simultaneously for the execution, although not to the extent of the \acrshort*{gpu} case. Our code was written in C/C++ language and for the parallelization the protocol OpenMP API \cite{omp} is used. We will not use the same maximal parallelization of the \acrshort*{gpu} case, since we would need $N/2$ threads available. Instead, we take one direction of the lattice (for example the time direction), and evolve in parallel all the links variables with an odd numeration in that direction, where each thread does sequentially all the links with that specific time value. After all the threads finish, we repeat the same process but with the links with an even numeration in the time direction. For example in a $8^3$ lattice, we would use four threads indexed as $c=0,1,2,3$, and each would update sequentially the links given by $U_\mu(2c+1,n_1,n_2)$ for all $\mu,n_1,n_2$. Once all the threads have finished, each one starts updating the rest of the links, i.e. $U_\mu(2c+2,n_1,n_2)$ for all $\mu,n_1,n_2$. After this second update is complete, a full sweep of updates over the lattice has been done. It is easy to notice how with this algorithm there is no risk of interference in the updates since the links that are updated in parallel are always one link away from each other (see \autoref{fig:lattice_div_cpu}).

This method can also be applied to the case where we have some direction of the lattice with an odd number of sites as long as is not the direction we use for the parallelization (in the previous example the time direction). Thus, we will use this algorithm for dealing with the cases when the number of links in the transverse spatial direction is odd.

For these \acrshort*{cpu} simulations we mainly use the Supercomputer Cluster CESAR. Although, we will not use that many cores for a particular simulation (at maximum $N_0/2$), since we need to obtain the energy for different lattice sizes and coupling constants. This cluster allowed us to also run in parallel different simulations.

\section{Statistical error}\label{sec:statistical_erro}
In this section we shall analyze some key aspects of the \acrshort*{mc} simulations. Once we have a sequence of $N$ measures of a certain observable $O$ from the configurations given by the \acrshort*{mc} process, these will be random variables $O_i$ with the appropriate distribution. Therefore, we need to use unbiased estimators to obtain from this sequence the expectation value and variance of the observable that we want to compute. For the expected value, the estimator is given by
\begin{equation}
	\widehat {O}=\frac{1}{N}\sum_{i=1}^N O_i,
\end{equation}
whereas for the variance of the measures $O_i$ the unbiased estimator is of the form
\begin{equation}
	\widehat{\sigma}^2_O=\frac{N}{N-1}\left(\frac{1}{N}\sum_{i=1}^N O_i^2-\widehat O^2\right).
\end{equation}
A key quantity for computing the statistical error of the expected value we obtain from the measures is the variance of the estimator of the mean, that has the form
\begin{equation} \label{eq:var_est_mean}
	\sigma^2_{\widehat O}=\Big\langle\left(\widehat O-\braket O\right)^2\Big\rangle=\frac{1}{N}\braket{O^2}-\braket O^2+\frac{1}{N^2}\sum_{i\not =j}\braket{O_iO_j},
\end{equation} 
and when the measures $O_i$ of the \acrshort*{mc} are uncorrelated, it simplifies to
\begin{equation}
	\sigma^2_{\widehat O}=\frac{1}{N}\left(\braket{O^2}-\braket{O}^2\right)=\frac{\sigma^2_O}{N}.
\end{equation}
Thus, the estimator of the variance of the estimator of the expected value with $N$ uncorrelated measures is just the estimator of the variance divided by the number of measures, i.e.
\begin{equation}
	\sigma^2\equiv \widehat \sigma^2_{\widehat O}=\frac{\widehat{\sigma}^2_O}{N}.
\end{equation}
This quantity $\sigma$ will give the statistical error of the expected value we obtain in the \acrshort*{mc} simulations
\begin{equation}
	\braket O=\widehat O \pm \sigma,
\end{equation}
in which as we previously mentioned, the statistical error decreases with the square root of the number of measures $\sigma \sim 1/\sqrt{N}$.
\subsection{Autocorrelation}
Since in the computer simulations there will always be some correlation between the different measurements in the \acrshort*{mc} steps, we need to take this into account. The correlation between the measures is usually referred to as \textit{autocorrelation} and can be defined as
\begin{equation}\label{eq:autocorrelation}
	C_O(t)=\braket{O_iO_{i+t}}-\braket{O}^2,
\end{equation}
where $C_O(0)=\sigma_O^2$. It is more useful to use the normalized autocorrelation function
\begin{equation}\label{eq:norm_corr}
	\rho_O(t)=C_O(t)/C_O(0),
\end{equation}
from which we can recover formula \eqref{eq:var_est_mean} for the variance of the estimator of the mean and compute it when there is correlations between the measures
\begin{equation}
	\sigma^2_{\widehat O}=\frac{\sigma^2_O}{N}\sum_{t=-N}^N\rho_{O}(\lvert t\rvert)\left(1-\frac{\lvert t\rvert}{N}\right)\simeq 2\frac{\sigma_O^2}{N}\left(\frac{1}{2}+\sum_{t=1}^N\rho_{O}( t)\right)=2\frac{\sigma_O^2}{N}\tau_{O,\text{int}}
\end{equation}
where we have defined the \textit{integrated correlation time}
\begin{equation}
	\tau_{O,\text{int}}=\frac{1}{2}+\sum_{t=1}^N\rho_{O}( t).
\end{equation}
Therefore, when we have autocorrelation, the variance of the estimator of the expected value of an observable is given by
\begin{equation}
	\sigma^2_{\widehat O,\text{auto}}=2\tau_{O,\text{int}}\ 	\sigma^2_{\widehat O}
\end{equation} 
which is $2\tau_{O,\text{int}}$ times larger that in the uncorrelated case. This can be understood as having $N/(2\tau_{O,\text{int}})$ independent measurements instead of $N$.

The estimators for these quantities \cite{wolff2004monte,madras1988pivot,sokal1997monte} are given by 
\begin{equation}
	\widehat C_O(t)=\frac{1}{N-\lvert t\rvert }\sum_{i=1}^{N-\lvert t\rvert}\left(O_i-\widehat {O}\right)\left(O_{i+\lvert t\rvert}-\widehat {O}\right),
\end{equation}
 for the autocorrelation function, the normalized version is just the quotient of the previous estimators
\begin{equation}
	\widehat{\rho}_O(t)=\widehat C_O(t)/\widehat C_O(0),
\end{equation}
and finally the integrated correlation time has the following estimator
\begin{equation}\label{eq:integrated_auto}
	\widehat \tau_{O,\text{int}}=\frac{1}{2}+\sum_{t=1}^W	\widehat{\rho}_O(t).
\end{equation}
Notice that we cutoff the sum at $W$ instead of $N$, this is the main difference from what one would expect as the more natural estimator. This is due to the ratio of noise and signal as $t$ grows, for larger values of $t$ the autocorrelation $\widehat{\rho}_O(t)$ becomes so small that the variance of the estimator $\widehat{\rho}_O(t)$ becomes much larger than the value and this makes the variance of	$\widehat \tau_{O,\text{int}}$ when $W=N$ not vanish in the $N\rightarrow \infty$ limit. Thus, the key point is finding a reasonable value of $W$ (see \cite{wolff2004monte,sokal1997monte} where some methods to estimate $W$ are presented) that allows us to obtain a good enough value for the estimator of the integrated correlation time. Then, we can compute the error of the estimator of the mean when there is correlation between the measures as
\begin{equation}
	\sigma_{\text{auto}}=\sigma\sqrt{2\widehat \tau_{O,\text{int}}}.
\end{equation}
\subsection{Thermalization}
Before we start to measure on the configurations, first we have to be sure we have arrived at an equilibrium configuration. Usually, one considers two types of initial configurations: the \textit{hot start} where the links are set as random unitary matrices and the \textit{cold start} where all the links are set to the identity matrix $U_\mu( n)=\mathbb I$. A good method for checking if we have arrived at an equilibrium configuration is starting from both the cold and hot start, and seeing how the observables evolve until they approach the same values from the two different initial configurations. Then, one can think that we are close to an equilibrium configuration.

From the expected exponential behaviour of the normalized correlation function $\rho_O(t)$ \eqref{eq:norm_corr}
\begin{equation}\label{eq:exp_correlated}
	\rho_O(t)\simeq e^{-t/\tau_{O,\text{exp}}},
\end{equation} 
one can obtain the \textit{exponential correlation time} $\tau_{O,\text{exp}}$. This quantity is related with the number of \acrshort*{mc} steps needed to reach an equilibrium configuration, at least $20\tau_{O,\text{exp}}$ steps should be discarded but usually using an order of $1000-10000\tau_{O,\text{exp}}$ \acrshort*{mc} steps for thermalization is a good measure. 

\subsection{Random number generator}
The choice of what pseudo-random number generator we use is an important factor in the \acrshort*{mc} simulations. It needs to have a low enough correlation, a long period so that the sequence of random numbers does not repeat during the simulation and has to be fast producing these random numbers since one of the most consuming time factors of the simulations is precisely the generation of the random numbers. Also, when doing parallel calculations one has to be extremely cautious since each parallel thread needs to execute the pseudo-random number generator independently. The best method to avoid correlation between different sequences or even overlapping, is to have all the threads executing the same sequence of random numbers but separated a distance long enough that they will not overlap during the simulation. 

For our simulations in \acrshort*{gpu} we used the cuRAND implementation \cite{cuRAND} of the \newline \textit{MRG32k3a} generator \cite{l1999good,l2002object}. In this method, each CUDA thread is assigned a part of the same sequence generated with a certain seed and separated by $2^{67}$ numbers. We test this generator in the scalar case where we can compare with the analytic results from \autoref{chp:scalar_lattice}, and check how it gives matching results and a low autocorrelation.\footnote{Doing the same test with the \textit{XORWOW} implementation in cuRAND, we observed large autocorrelation in our measures of the energy and results that did not agree with the analytic prediction.}

In the \acrshort*{cpu} simulations we used the C implementation of the xoshiro256+ random number generator \cite{blackman2021scrambled,prng}. This has a period of $2^{256-1}$ and allows us to select $2^{64}$ starting points from the same sequence separated by $2^{192}$ random numbers. Thus, it is a perfect candidate for the parallelization. Again, we also test the results with this generator and see that it gives very low autocorrelation for the energy.

     \chapter{Casimir energy of gauge fields in 2+1 dimensions}\label{chp:su2_21}
     In this chapter we address the calculation of the Casimir energy of a pure gauge $SU(2)$ theory in 2+1 dimensions by \acrshort*{mc} lattice simulations. We consider two different boundary conditions belonging to each of the families  classified according to their different exponential decay \eqref{eq:rate} given in \autoref{chp:scalar_cont}. Our goal is to  obtain the value of the mass that drives the exponential decay of this Casimir energy with the distance between the two boundary walls.

In \autoref{sec:energy_21} we introduce the observable used to compute the internal energy in its  lattice version. In \autoref{sec:therm_21} we analyze the thermalization that will be used in the simulations and the autocorrelation of the energy observable. In \autoref{sec:reno_21} we present and test the renormalization procedure we use to extract the Casimir energy from the vacuum energy.\renewcommand\sectionautorefname{Section} \autoref{sec:Casimir_21} is devoted to compute the Casimir energy and compare its behaviour with the exponential decay in the Casimir energy of massive scalar fields. \renewcommand\sectionautorefname{section}  Finally, in \autoref{sec:su21_lattice_effects} we analyze the finite volume errors in our computation of the Casimir energy.
\section{The internal energy}\label{sec:energy_21}
Following the successful method used to  obtain the Casimir energy  for the scalar field on the lattice, we shall use the first component $T_{00}$ of the energy-momentum tensor to measure the energy on the lattice. On gauge theories in 2+1 dimensions the Euclidean version is given by
\begin{equation}
	T_{00}=\frac{1}{g^2}\text{ tr}\left(F^2_{12}(x)-F^2_{01}( x)-F^2_{02}( x)\right),
\end{equation}
that  can be integrated in the spatial dimensions to obtain the total energy
\begin{equation}
	\mathcal E=\frac{1}{g^2}\int d^2 \mathbf x \text{ tr} \left(F^2_{12}( x)-F^2_{01}( x)-F^2_{02}(x)\right).
\end{equation}
From the relationship between the plaquettes and the field strength tensor \eqref{eq:plaquette_F}, and replacing the integral by the sum over the lattice \eqref{eq:integral_sum}, the vacuum energy on the lattice has the form
\begin{equation}
	\mathcal E=-\frac{\beta}{aN_0}\sum_{ n \in \Lambda}\left(1+\frac{1}{2}\text{tr}\left(P_{12}( n)-P_{01}( n)-P_{02}( n)\right)\right),
\end{equation} 
where we are also averaging over the temporal side of the lattice. The dependence of the energy with $a^{-4}$ \eqref{eq:plaquette_F} gives rise to multiplicative radiative corrections. We can smooth this behaviour by using the mean-field improved coupling \cite{parisi1980recent,teper1998n} in which the mean value of the plaquette 
\begin{equation}
	u=\frac{1}{2}\braket{\text{tr} P_{\mu\nu}},
\end{equation}
is used to redefine the energy on the lattice as
\begin{equation}\label{eq:Energy_su2_21}
	\mathcal E=-\frac{u^4\beta}{aN_0}\sum_{ n \in \Lambda}\left(1+\frac{1}{2}\text{tr}\left(P_{12}(n)-P_{01}( n)-P_{02}( n)\right)\right).
\end{equation} 
This extra factor can also be understood as using an improved coupling 
\begin{equation}\label{eq:improved_coupling}
	\beta_I=\frac{1}{2}\braket{\text{tr} P_{\mu\nu}}\beta,
\end{equation}
that reduces discretization effects. 

Since our interest is to compute the Casimir energy, we need to have the temporal and longitudinal spatial sides as large as possible in comparison with the transverse spatial dimension where  the different boundary conditions are satisfied. The specific values of those lattice sizes are not relevant as long as they are large enough. For such a reason, it is convenient to  give the same value to the lattice sizes on these two directions $N_0=N_1\equiv N_A$. Thus, the expected value of the plaquettes  $P_{12}$ and $P_{02}$ should be equal $\braket{P_{12}}=\braket{P_{02}}$, this allows us to simplify the expected value of the energy to
\begin{equation}\label{eq:energy_lattice_21}
	\braket{\mathcal E(N_A,N_L,\beta)}=-\frac{u^4\beta}{aN_A}\left(1-\frac{1}{2}\braket{\sum_{ n \in \Lambda}\text{tr}\left(P_{01}( n)\right)}\right).
\end{equation}
This is very convenient because it not only simplifies the computation  of the expectation values of the difference between the three plaquettes, but also suppresses the statistical fluctuations of the two plaquettes which cancel and thus, the statistical error will be lower (it reduces the error a factor of order $1/\sqrt{3}$ with respect to the original one).

In the particular situation where we have a cubic lattice with only \acrshort*{pbc} we can actually use the plaquettes in all directions to compute the energy. Therefore, instead of using \eqref{eq:energy_lattice_21} the energy can be obtained as
\begin{equation}\label{eq:periodic_square_lattice_21}
	\braket{\mathcal E_P(N_A,\beta)}=-\frac{u^4\beta}{aN_A}\left(1-\frac{1}{6}\braket{\sum_{ n \in \Lambda}\text{tr}\left(P_{01}(n)+P_{02}( n)+P_{12}( n)\right)}\right),
\end{equation}
by using the fact that under \acrshort*{pbc} and all the directions being the same size they have the same expected value in every direction, i.e. $\braket{P_{01}}=\braket{P_{02}}=\braket{P_{12}}$. Averaging over the three different plaquettes will reduce the statistical error (approximately by a factor $1/\sqrt{3}$) without doing any extra calculation.
\section{Thermalization and autocorrelation}\label{sec:therm_21}
\begin{figure}[H]
	\centering
	\includegraphics[width=0.99\textwidth]{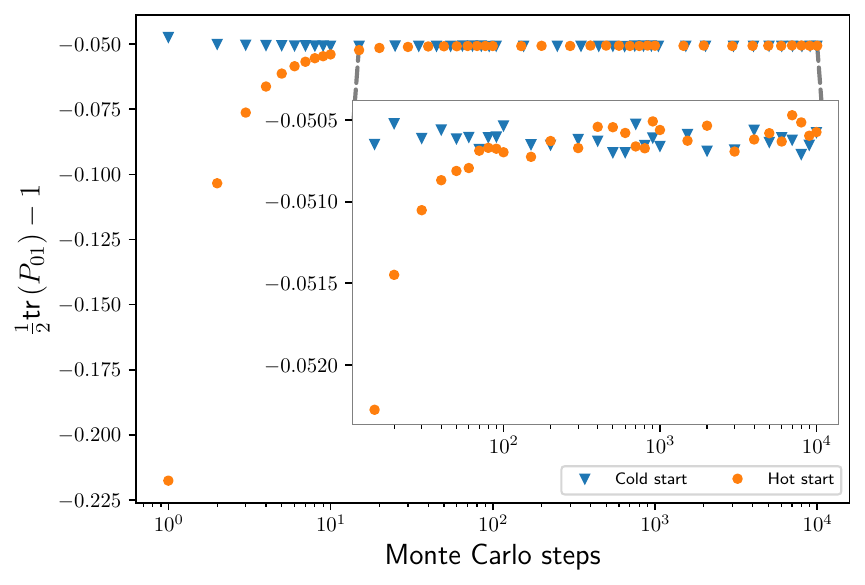}
	\caption{Evolution of the average over the lattice of the plaquette $P_{01}$ with the \acrshort*{mc} steps in logarithmic scale from hot and cold initial configurations for a lattice where $N_A=96$, $N_{L}=96$ and $\beta=20$, and \acrshort*{pbc}.}
	\label{fig:Energy_therm_ini_per}
\end{figure}

Let us start by analyzing the thermalization and autocorrelation of the energy. First, we check the thermalization by starting  independent simulations with hot and cold initial configurations. If the results converge we have a good indicator that we have done enough steps for the thermalization. In \autoref{fig:Energy_therm_ini_per} we plot the evolution of the plaquette $P_{01}$ (which is the one that defines the energy \eqref{eq:Energy_su2_21}) with the \acrshort*{mc} steps using a cold and hot starting configurations for a lattice with \acrshort*{pbc}. We can see how with these two starting configurations the initial values of the plaquettes are very different but after of the order of 100 \acrshort*{mc} steps both configurations seem to converge to the same expected value. In the zoomed zone it can be seen more clearly how after the 100 steps the average over the lattice of the plaquette starts to fluctuate around the same value and stays the same for the following thousands of \acrshort*{mc} steps.

\begin{figure}[H]
	\centering
	\includegraphics[width=1\textwidth]{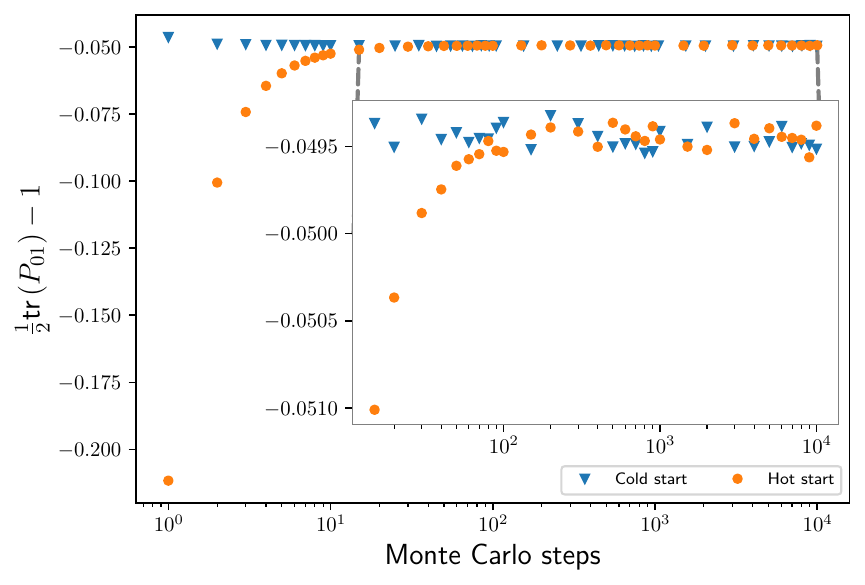}
	\caption{Evolution of the average of the plaquette $P_{01}$ over the lattice  with the \acrshort*{mc} steps in logarithmic scale from hot and cold initial configurations for a lattice where $N_A=96$, $N_{L}=48$ and $\beta=20$, and \acrshort*{pccbc} in the transverse spatial dimension.}
	\label{fig:Energy_therm_ini_cond}
\end{figure}

In \autoref{fig:Energy_therm_ini_cond}, we repeat the same plots but for a lattice with \acrshort*{pccbc} on the transverse spatial dimension. The plot shows the same behaviour than in the \acrshort*{pbc} case, after 100 \acrshort*{mc} (zoom in) steps the two configurations seem to converge to the same value and after these hundred steps the values start to fluctuate around it.
Consequently we can argue that after 1000 steps the configurations are already thermalized. We will use at least 10000 steps for thermalization in most lattices, and for the larger ones we will use 50000 steps for thermalization  to be sure we are thermalized (see \autoref{ch:21_values} where the results and thermalization values used for the simulations are listed).

\begin{figure}[H]
	\centering
	\includegraphics[width=0.85\textwidth]{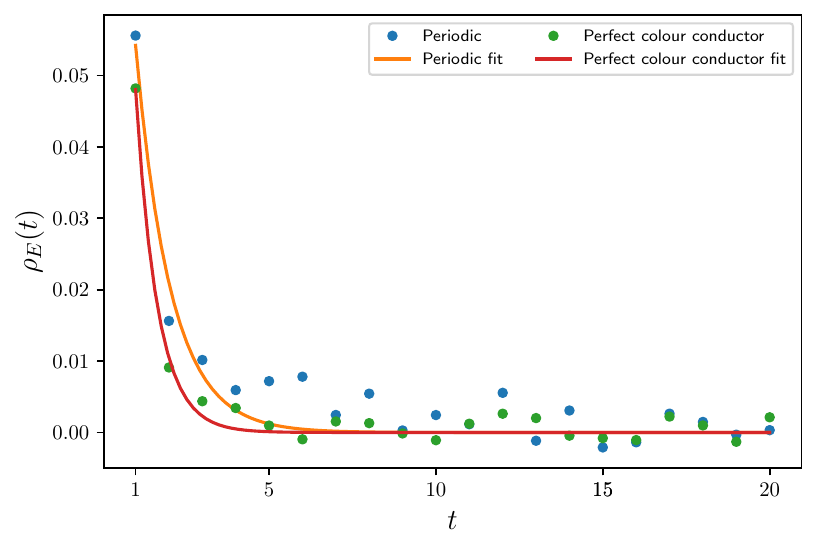}
	\caption{Normalized autocorrelation  of the mean value $\braket{P_{01}}$ for a lattice where $N_A=96$, $N_{L}=8$ and $\beta=20$ for \acrshort*{pbc} or \acrshort*{pccbc} in the transverse spatial dimension.}
	\label{fig:Energy_auto}
\end{figure}

Next, let us analyze the autocorrelation \eqref{eq:autocorrelation} of the energy observable after applying a thermalization of 10000 \acrshort*{mc} Steps. In \autoref{fig:Energy_auto} we plot the normalized autocorrelation for a lattice with \acrshort*{pbc} and another with \acrshort*{pccbc}. We can see how the autocorrelation is very small, in both cases in consecutive measures the value is $\rho_E(1)\simeq 0.05$ and at the following measures it becomes close to zero very fast. From these measures, we can compute the integrated correlation time \eqref{eq:integrated_auto}, obtaining $\tau_{E,\text{int}}\simeq 0.62$ and $\tau_{E,\text{int}}\simeq 0.57$ for \acrshort*{pbc} and \acrshort*{pccbc}, respectively. These values also show how small is the correlation between the measures of the energy on the lattice and that we can consider consecutive measures independent. 

In order to compute the exponential correlation time \eqref{eq:exp_correlated} we can also fit these data to an exponential decay, obtaining $\tau_{E,\text{exp}}\simeq 1.0$ for \acrshort*{pbc} and $\tau_{E,\text{exp}}\simeq 0.65$ for the lattice with \acrshort*{pccbc}. These results agree with those of the previous thermalization analysis, where after $1000\tau_{E,\text{exp}}\simeq 1000$ steps the energy seems to be thermalized, and using $10000\tau_{E,\text{exp}}\simeq 10000$ is a good reference for being safe in this regard.

All these results  have been obtained by means of the heat-bath algorithm for the \acrshort*{mc} updates. Since we obtain a small enough correlation and a successful thermalization, it is the algorithm we choose for the rest of the results we present in this chapter. We also tested  the Metropolis algorithm obtaining the same results, but with a higher correlation between the measures. Thus, we need to add extra \acrshort*{mc} steps between measures to reduce this correlation, and this makes using the Metropolis algorithm slower than heat-bath.

\section{Renormalization}\label{sec:reno_21}
In order to compute the Casimir energy from the vacuum energy, we have to deal with the continuum divergences of the energy (that are finite on the lattice) as was done in the scalar case. We expect that these contributions follow the same structure than the scalar case, i.e.
\begin{equation}\label{eq:energy_su2_21}
	\braket{\mathcal E_U(N_L,\beta)}=C_0(\beta) N_AN_L+C_1(\beta) N_A+E_U(N_L,\beta)+\ldots
\end{equation}
where the first term $C_0(\beta)$ is the one associated with the bulk energy density, $C_1(\beta)$ is associated with the vacuum energy of  the boundary walls and $E_U(N_L,\beta)$ is the Casimir energy.
\subsection{Periodic boundary conditions}
As in the scalar case, we expect that the term $C_1(\beta)$ associated to the boundary walls vanishes in the case of \acrshort*{pbc}, where there are no boundaries, i.e. $C_1(\beta)=0$. We can check this feature just by plotting the energy density for different sizes ($N_L$) of the transverse spatial direction .  

\begin{figure}[H]
	\centering
	\includegraphics[width=0.84\textwidth]{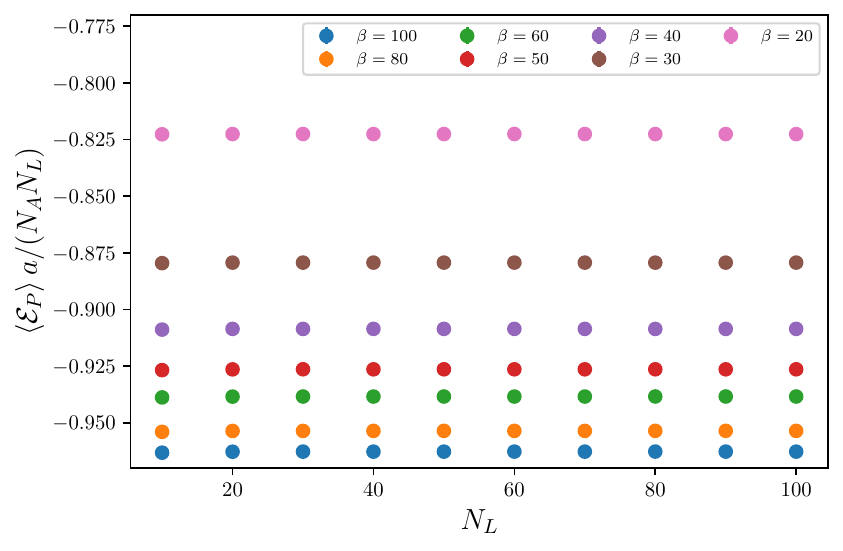}
	\caption{Energy density in a pure gauge $SU(2)$ three dimensional lattice with \acrshort*{pbc} for different lattices and coupling values. The chosen lattice parameters  are $N_A=96$, $N_{L}=10-100$ and $\beta=20-100$.}
	\label{fig:Energy_periodic_su21}
\end{figure}

\autoref{fig:Energy_periodic_su21} confirms this expected behaviour. We see that the vacuum energy density is essentially independent of the transversal spatial size 
$N_L$ for every value of the coupling constant $\beta$. The small fluctuations are due to the Casimir energy contribution and the statistical error of the \acrshort*{mc} simulation, in \autoref{tab:Energy_renom_periodic_1}, \autoref{tab:Energy_renom_periodic_2} and \autoref{tab:Energy_renom_periodic_3} the values used for this plot are shown.

The fact that the bulk contribution $C_0(\beta)$ should be independent of the boundary conditions we choose, allow us to use the energy with \acrshort*{pbc} to define this value as
\begin{equation}
	C_0(\beta)=\lim_{N_{L_0}\rightarrow \infty} \braket{\mathcal E_P(N_{L_0},\beta)}/(N_AN_{L_0}),
\end{equation}
where when we do not make explicit the dependence of the energy with $N_A$ it means we are considering a suitable large enough value.

\subsection{Perfect colour conductor boundary conditions}
When we defined the \acrshort*{pccbc} in \autoref{sec:perfect_conductor} we already pointed out that they are equivalent to \acrshort*{dbc} for the scalar field theory since  both  require the vanishing of  the action  at the boundary. Thus, we expect a similar behaviour where the boundary term $C_1(\beta)$ contributes to the vacuum energy. Again, we can check this behaviour by plotting the energy with \acrshort*{pccbc} for different sizes  $N_L$ of the transverse spatial direction.

\begin{figure}[H]
	\centering
	\includegraphics[width=0.84\textwidth]{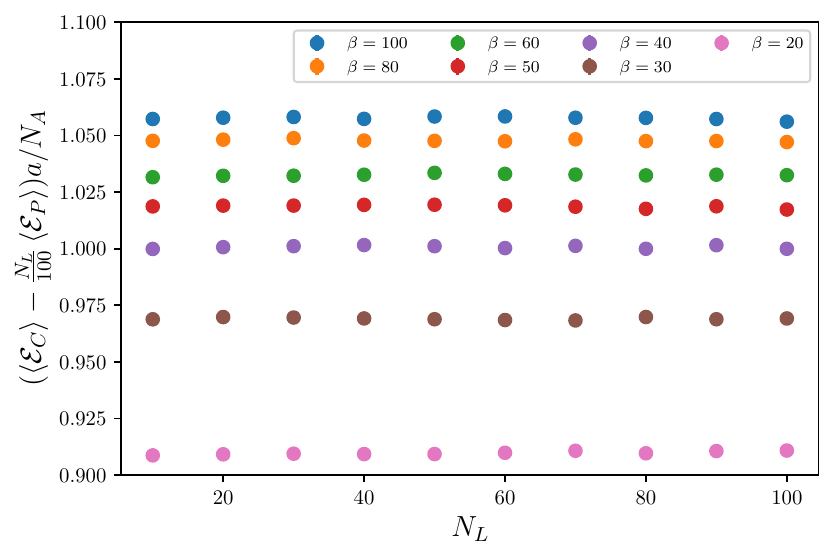}
	\caption{Energy in a pure gauge $SU(2)$ three dimensional lattice with \acrshort*{pccbc} after subtracting the bulk contribution $C_0(\beta$) for different lattices and coupling values. The chosen lattice parameters are $N_A=96$, $N_{L}=10-100$, $N_{L_0}=100$ (in \acrshort*{pbc}) and $\beta=20-100$.}
	\label{fig:Energy_conductor_su21}
\end{figure}

Precisely, we observe this behaviour in \autoref{fig:Energy_conductor_su21} where after subtracting the bulk term $C_0(\beta)$ the energy is independent of $N_L$ apart from small fluctuations due to the Casimir contribution and statistical error. The values used for this plot are shown in \autoref{tab:Energy_renom_cond_1}, \autoref{tab:Energy_renom_cond_2} and \autoref{tab:Energy_renom_cond_3}.

Therefore, we can obtain this boundary term $C_1(\beta)$ as
\begin{equation}
	C_1(\beta)=\lim_{N_{L_0}\rightarrow \infty}\left(\braket{\mathcal E_C(N_{L_0},\beta)}-\braket{\mathcal E_P(N_{L_0},\beta)}\right)/N_A
\end{equation} 
where we also set $N_A$ large.
\vspace{1.5cm}

\section{Casimir energy}\label{sec:Casimir_21}
Once we have analyzed how to compute the terms in the energy that we are not interested in, we can compute the Casimir energy by canceling all these divergent contributions. In order to compare the results with different coupling constants $\beta$ and give the results in physical units we need to analyze another observable. In our case we choose the \textit{string tension} $\sigma$, which is the coefficient of the linear term in the static quark-antiquark potential \cite{creutz1980monte} and drives the exponential decay of large Wilson loops with the area they enclose, i.e.
\begin{equation}
	\ln W=-(a^2\sigma )A+O(P),
\end{equation} 
with $A$ the area enclosed by a planar Wilson loop, $P$ its perimeter 
and $\sigma$ the string tension. Thus, the dimensionless factor $a\sqrt{\sigma}$ 
allows us to compute a scaling for the physical lattice size with the coupling $\beta$. 

In this particular case of 2+1 dimensions with $SU(2)$, the following phenomenological relation between the 
dimensionless factor $a\sqrt{\sigma}$ and the improved coupling $\beta_I$ \eqref{eq:improved_coupling} was
 found \cite{teper1998n,athenodorou2017n} 
\begin{equation}\label{eq:string_tension}
	\beta_I a\sqrt{\sigma}=1.341(7)-\frac{0.421(50)}{\beta_I},
\end{equation}
that allows us to have a precise estimate of the string tension for higher values of $\beta$.
 Hence, we can compute a dimensionless physical distance $\sqrt{\sigma}L=N_La\sqrt{\sigma}$ and obtain the mass factor that drives the exponential decay of the Casimir energy in terms of this string tension. Also, we will be able to compare with the mass of the scalar fields that emerge in the \acrshort*{nkp} since in terms of this string tension \cite{karabali1998planar} it has the value $m_{\mathrm{K}}\simeq 0.92 \sqrt{\sigma}$. Notice how this mass value is considerably smaller than the lightest glueball in $SU(2)$ which is given by $m_g\simeq 4.7\sqrt{\sigma}$ \cite{teper1998n,lucini2002n,athenodorou2017n}.
 
\subsection{Perfect colour conductor boundary conditions}
In order to compute the Casimir energy for \acrshort*{pccbc} we have to cancel the contribution $C_0(\beta)$ associated to the bulk and $C_1(\beta)$ to the boundary energy densities. The second contribution can be eliminated by subtracting the energy density of another lattice with the same parameters and boundary conditions but  transverse spatial dimension much larger. The \acrshort*{uv} divergence associated to the bulk term can be canceled by subtracting to both lattices with \acrshort*{pccbc} the energy of an equivalent lattice but with \acrshort*{pbc}. Hence, the formula for computing the Casimir energy is given by
\begin{align}\nonumber
	E_C(N_L,\beta)&=\lim_{N_{L_0}\rightarrow \infty}\left(\braket{\mathcal E_C(N_L,\beta)}-\frac{N_L}{N_{L_0}}\braket{\mathcal E_P(N_{L_0},\beta)}-\braket{\mathcal E_C(N_{L_0},\beta)}+\braket{\mathcal E_P(N_{L_0},\beta)}\right)\\ \label{eq:Casimir_conductor_21}
	&=\lim_{N_{L_0}\rightarrow \infty}\left(\braket{\mathcal E_C(N_L,\beta)}-\braket{\mathcal E_C(N_{L_0},\beta)}-\frac{N_L-N_{L_0}}{N_{L_0}}\braket{\mathcal E_P(N_{L_0},\beta)}\right).
\end{align}
This means that for each value of $\beta$ we just need to compute one large lattice with \acrshort*{pccbc} and another with \acrshort*{pbc} (that will also be used for computing the Casimir energy with \acrshort*{pbc} in \autoref{sec:Periodic_21}). Although ideally we would use the same large value for the lattice size in the transverse spatial dimension $N_{L_0}$ in both boundary conditions, it is more computationally effective to have a smaller one in the \acrshort*{pccbc}. This is due to the dependence of the $C_0(\beta)$ and $C_1(\beta)$ in \eqref{eq:energy_su2_21} with $N_L$. The bulk contribution is linear with $N_L$, and thus, when averaging over a lattice with larger $N_L$ there are more points to average from and the measures over this lattice will have less statistical error than with a smaller one. On the contrary, the boundary term $C_1(\beta)$ is not equal for every point in the $N_L$ direction (when we impose the \acrshort*{pccbc} we break the equivalence between the points in the transverse spatial dimension), hence, by making $N_L$ larger we do not improve the statistical error and just make the simulation slower. Taking this into account, it is clearly more convenient to optimize the simulation time taking a smaller value for $N_{L_0}\rightarrow N_{L_C}$ in the \acrshort*{pccbc} value in \eqref{eq:Casimir_conductor_21} than on the lattice with \acrshort*{pbc} $N_{L_0}\rightarrow N_{L_P}$, but large enough to make  the Casimir energy contribution with $N_{L_C}$ negligible. Also, setting these two lattices sizes independently allows us to fix the size $N_{L_P}$ of the lattice with \acrshort*{pbc} to the same value than the rest of the directions $N_{L_P}=N_A$ and use formula \eqref{eq:periodic_square_lattice_21}, which as we mentioned before reduces the statistical error of this contribution.

Another important aspect to analyze is to compare the vacuum energy of  the gauge field with \acrshort*{pccbc} with that of the scalar field analyzed in \autoref{sec:Nair_karabali} from the \acrshort*{nkp}. From the relation \eqref{eq:relation_NH}, we can see that \acrshort*{pccbc} on $F_{01}=0$ translates as \cite{karabali2018casimir}
\begin{equation}
\partial_t	\partial_2  \varphi^a=0
\end{equation}
for the scalar field.  Thus, imposing the \acrshort*{pccbc} on the gauge fields should force the emerging scalar field to satisfy \acrshort*{nbc}.

\begin{figure}[h]
	\centering
	\includegraphics[width=1\textwidth]{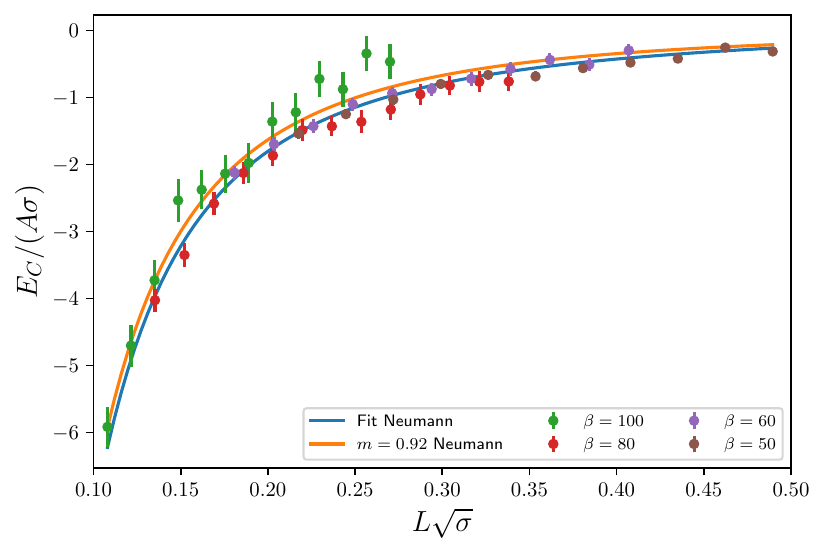}
	\caption{Dimensionless Casimir energy with \acrshort*{pccbc} in a three dimensional lattice for different couplings and lattices sizes with parameters $N_A=96$, $N_{L_P}=96$, $N_{L_C}=48$, $N_{L}=8-20$ and $\beta=50-100$. The curves are given by the formula for the Casimir energy of a massive scalar field with Neumann boundary conditions \eqref{eq:su2_neumann_fit}.}
	\label{fig:Casimir_perfect_cond}
\end{figure}

\begin{figure}[h]
	\centering
	\includegraphics[width=1\textwidth]{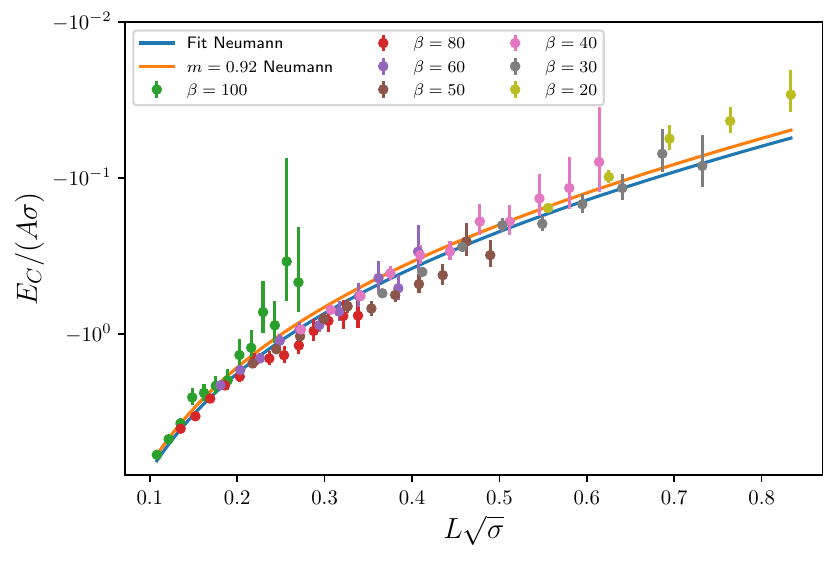}
	\caption{Dimensionless Casimir energy in logarithmic scale with \acrshort*{pccbc} in a three dimensional lattice for different couplings and lattice sizes with parameters $N_A=96$, $N_{L_P}=96$, $N_{L_C}=48$, $N_{L}=8-20$ and $\beta=20-100$. The curves are given by the formula for the Casimir energy of a massive scalar field with \acrshort*{nbc} \eqref{eq:su2_neumann_fit}.}
	\label{fig:Casimir_perfect_cond_log}
\end{figure}

In \autoref{fig:Casimir_perfect_cond} (linear scale) and \autoref{fig:Casimir_perfect_cond_log} (logarithmic scale) we plot the results we obtain for the Casimir energy with the \acrshort*{pccbc} by using formula \eqref{eq:Casimir_conductor_21}. The plaquette expectation values that are used for computing these data
 are displayed in \autoref{ch:21_values}: \autoref{tab:Casimir_21_cond_1}, \autoref{tab:Casimir_21_cond_2}, \autoref{tab:Casimir_21_periodic_1}, \autoref{tab:Casimir_21_periodic_2} and \autoref{tab:Casimir_21_periodic_3}. First, we can clearly see an exponential decay of the Casimir energy with the distance between walls  where the boundary conditions constrain the fields that is characteristic of massive theories, like the massive scalar field theory that was discussed in \autoref{chp:scalar_lattice}. Moreover, the results with different couplings $\beta$ seem to fit to the same behaviour which is a good indicator that there are not large discretization errors and we are close enough to the continuum limit. It is important to notice that the larger contribution to the error of the points is the factor $\braket{\mathcal E_C(N_{L_C},\beta)}$ in the Casimir energy formula \eqref{eq:Casimir_conductor_21} that affects all values with the same coupling constant. This is the reason why the points with the same $\beta$ seem to be biased in the same direction instead of being randomly distributed with the error around the curve as one would expect.

Let us now fit these data with the formula of a massive scalar field with \acrshort*{nbc} already computed in \autoref{chp:scalar_cont} for the 2+1 dimensional case \eqref{eq:Cas_dir_2d}. Since we use the square root of the string tension to give physical units, the mass that is obtained from the fit to an exponential decay will be given in terms of the fitting formula 
\begin{equation}\label{eq:su2_neumann_fit}
	\frac{E_C(L\sqrt{\sigma})}{A\sigma}=-\frac{G}{16\pi L^2\sigma}\left(2mL\sqrt{\sigma}\ \text{Li}_2\left(e^{-2mL\sqrt{\sigma}}\right)+\text{Li}_3\left(e^{-2mL\sqrt{\sigma}}\right)\right),
\end{equation}
where the parameters are the multiplicative factor $G$ and the mass $m$. This fit is labeled in the plots as ``Fit Neumann''. Apart from plotting the best fit with these two parameters free, we also plot the curve when fixing these parameters to $G=3$ and $m=0.92$ which are the values we expect from the \acrshort*{nkp}, this fit is labeled as ``$m=0.92$ Neumann'' on the plots. The value $G=3$ is due to the three degrees of freedom in $SU(2)$.

The method used to fit the parameters is  a non-linear least square method that minimizes the following $\chi^2$
\begin{equation}\label{eq:xi_square}
	\chi ^2=\sum_{i,j=1}^n\text{cov}(E)^{-1}_{ij}(E_i-f_i)(E_j-f_j)/(n-p),
\end{equation} 
where $\text{cov}(E)_{ij}$ is the covariance matrix between the $n$ Casimir energy results $E_i \  (i=1,\dots, n)$ and  the corresponding points $f_i$  of the fitting function, and $p$ the number of parameters that are fitted. In this case taking into account the covariance is necessary since the points with the same $\beta$ are obtained with the same reference energies used for subtracting the bulk and boundary terms, thus, there is a high correlation between the propagated statistical error of the Casimir energy values with the same $\beta$.

 The best fit parameters found are $$G=3.26\pm 0.06 \ \mathrm{and}\  m=0.89\pm 0.07,$$ which give $\chi^2=1.8$. The value of the mass is actually compatible with the expected result in the \acrshort*{nkp} of $m_{\mathrm{K}}=0.92 \sqrt{\sigma}$ and the coefficient is close to the most reasonable value of having the three scalar fields. This is consistent with the fact that in the plots, the best fit and the curve where we fix the parameters to the expected values in the \acrshort*{nkp} are very close and both provide a decent fit of the points. In particular, when we set the parameters to $G=3$ and $m=0.92$ the fit gives $\chi^2=2.3$, that is close to the value given in the best fit. In the linear plot \autoref{fig:Casimir_perfect_cond}, where only the higher values of $\beta$ are included due to the exponential behaviour of the Casimir energy, we can clearly see how considering the statistical error of the points both curves agree well with them. In the logarithmic plot \autoref{fig:Casimir_perfect_cond_log} we can clearly see the exponential decay that the Casimir presents with the distance between the walls. Also, both curves seem to adjust well to this exponential decay in the different ranges of $\beta$ values and   distance between the walls $L\sqrt{\sigma}$.

\pagebreak

Let us now consider an alternative fit to the same data where instead of using the formula of the Casimir energy of a massive scalar field with \acrshort*{dbc}/\acrshort*{nbc} we use directly an exponential function. This will allow us to compare better with the 3+1 dimensional case. The fitting function in this case will be
\begin{equation}\label{eq:exp_fit_21_perf}
	\frac{E_C(L\sqrt{\sigma})}{A\sigma}=-\frac{C}{(L\sqrt{\sigma})^\nu}e^{-2\mu L\sqrt{\sigma}},
\end{equation}
where the fitting parameters are the constant $C$, the mass $\mu$ and the $\nu$ exponent of the dimensionless distance. Notice that if we set $\nu=1$ we obtain the formula for the asymptotic behaviour \eqref{eq:Casimir_asymptotic_Dirichlet} of the \acrshort*{dbc}/\acrshort*{nbc} formula for the massive scalar case. The best fitting parameters are $$C=0.23\pm 0.04, \mu=1.14\pm 0.15 \ \mathrm{and}\  \nu=1.58\pm 0.07,$$ which give $\chi^2=1.5$.
\vspace{1cm}
\begin{figure}[H]
	\centering
	\includegraphics[width=1\textwidth]{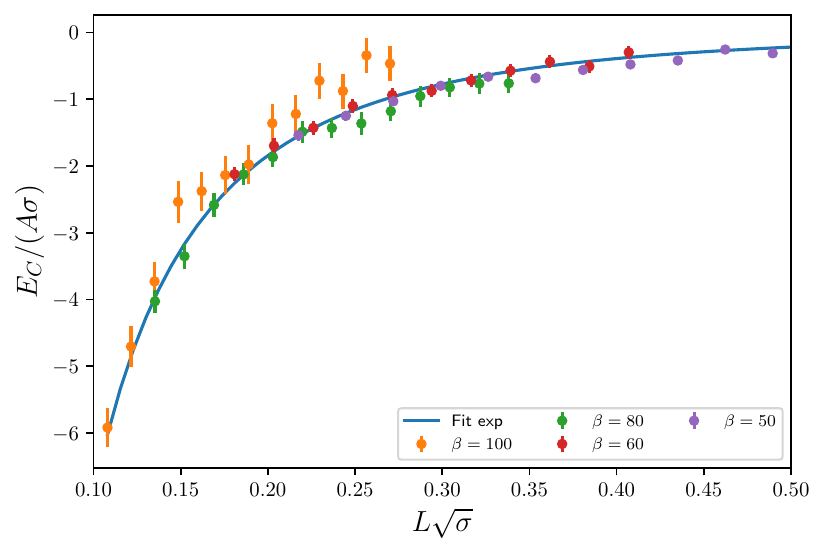}
	\caption{Dimensionless Casimir energy with \acrshort*{pccbc} in a three dimensional lattice for different couplings and lattices values with parameters $N_A=96$, $N_{L_P}=96$, $N_{L_C}=48$, $N_{L}=8-20$ and $\beta=50-100$. The fit is given by the formula \eqref{eq:exp_fit_21_perf}.}
	\label{fig:Casimir_perfect_cond_exp_lin}
\end{figure}

\begin{figure}[H]
	\centering
	\includegraphics[width=1\textwidth]{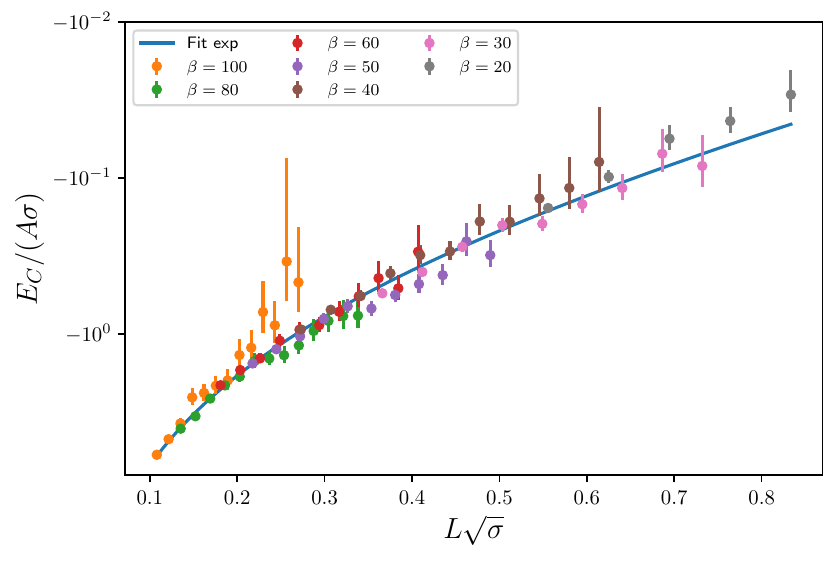}
	\caption{Dimensionless Casimir energy in logarithmic scale with \acrshort*{pccbc} in a three dimensional lattice for different couplings and lattice sizes  with parameters $N_A=96$, $N_{L_P}=96$, $N_{L_C}=48$, $N_{L}=8-20$ and $\beta=20-100$. The fit is obtained from the formula \eqref{eq:exp_fit_21_perf}.}
	\label{fig:Casimir_perfect_cond_exp}
\end{figure}
 In the plots \autoref{fig:Casimir_perfect_cond_exp_lin} (linear scale) and \autoref{fig:Casimir_perfect_cond_exp} (logarithmic scale) we can see how the exponential formula also fits the simulation results in both scales, and it has a similar but lower value of $\chi^2$, which is expected since this fit has an extra parameter in comparison with the previous fit. The obtained value for the mass $\mu=1.14\pm 0.15$ is within two sigmas from the \acrshort*{nkp} value and the mass obtained by fitting with the massive scalar field expression. Moreover, the exponent parameter $\nu=1.58\pm 0.07$ is an intermediate value between the asymptotic formula \eqref{eq:Casimir_asymptotic_Dirichlet}  where $\nu=1$ and the massless formula \eqref{eq:Casimir_massless_dir_2d} where $\nu=2$. This is consistent with the fact that the points are in an intermediate range of distance $\sqrt{\sigma}L$, and therefore, 
neither limit is a good enough approximation.

In summary, we find that the Casimir energy of a pure gauge $SU(2)$ theory in 2+1 dimensions under \acrshort*{pccbc} shows a massive exponential decay and the results are in agreement with the Casimir energy of  massive scalar fields with a mass of the order of $0.92\sqrt{\sigma}$ and \acrshort*{dbc}/\acrshort*{nbc}. This is compatible with the \acrshort*{nkp} of the gauge fields in terms of scalar fields we introduced previously, and most importantly with the mass predicted by this parametrization $m_{\mathrm{K}}=0.92\sqrt{\sigma}$. Moreover, the value of mass we obtain is much smaller than the lightest glueball $m_g\simeq 4.7\sqrt{\sigma}$, which is supposed to be the smallest mass in the theory and the one that drives the exponential decay of the Casimir energy. 

\textcolor{black}{  We can compare these results with the ones obtained in Ref. \cite{chernodub2018casimir} by M. N. Chernodub et al. where the Casimir energy was also computed for \acrshort*{pccbc} boundary conditions but with a different set up and implementation of the boundary conditions. In that case, the Casimir energy is fitted by an exponential with a fixed constant obtained from the massless limit of the Casimir energy for a scalar field \eqref{eq:Casimir_massless_dir_2d}. The value obtained in \cite{chernodub2018casimir} is $m_C=0.69\sqrt{\sigma}$, and does not seem compatible with our results $m\simeq 0.9\sqrt{\sigma}$ or the mass given by the \acrshort*{nkp} $m_{\mathrm{K}}=0.92\sqrt{\sigma}$. Although again, this mass in \cite{chernodub2018casimir}  is much smaller than the lightest glueball $m_g\simeq 4.7\sqrt{\sigma}$. A later analysis carried out by D. Karabali and V. P. Nair in Ref. \cite{karabali2018casimir} by fitting the data in Ref. \cite{chernodub2018casimir} to the formula of the massive scalar field \eqref{eq:su2_neumann_fit} with a fixed mass given by the one of the \acrshort*{nkp} ($0.92\sqrt\sigma$) showed agreement between both approaches. Thus, showing that when using a similar fit than the ones used in this work, our results with the \acrshort*{pccbc} are also compatible with these from Chernodub et al. \cite{chernodub2018casimir}. }
\subsection{Periodic boundary conditions}\label{sec:Periodic_21}
In \acrshort*{pbc} we only have to cancel the bulk contribution $C_0(\beta)$ to obtain the Casimir energy. Thus, by just subtracting the energy of a lattice with \acrshort*{pbc} and considering the transverse spatial dimension much larger  we can obtain the Casimir energy with \acrshort*{pbc}. We define this Casimir energy as
\begin{equation}\label{eq:Casimir_periodic_21}
	E_P(N_L,\beta)=\lim_{N_{L_P}\rightarrow \infty}\left(\braket{\mathcal E_P(N_L,\beta)}-\frac{N_L}{N_{L_P}}\braket{\mathcal E_P(N_{L_P},\beta)}\right).
\end{equation}
Hence, in the computation of the values at a certain $\beta$ we just need to calculate one extra value of the energy with that $\beta$. As in the \acrshort*{pccbc} case, the best strategy for computing this large lattice energy is setting the transverse spatial dimension with the same size than the time and other spatial dimension, i.e. $N_{L_P}=N_A$. In this way we can use formula \eqref{eq:periodic_square_lattice_21} to compute the energy that gives less statistical error than the general formula for the energy \eqref{eq:energy_lattice_21}.

\begin{figure}[H]
	\centering
	\includegraphics[width=1\textwidth]{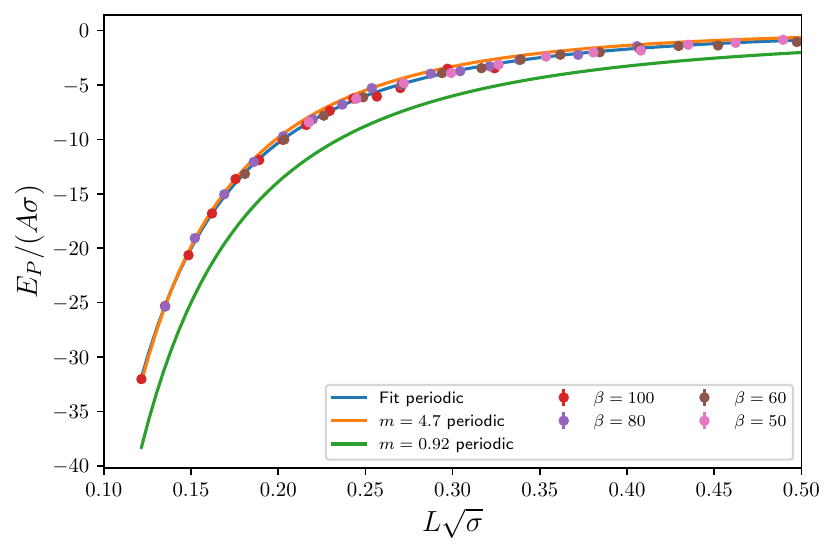}
	\caption{Dimensionless Casimir energy with \acrshort*{pbc} in a three dimensional lattice for different couplings and lattice sizes with parameters $N_A=96$, $N_{L_P}=96$, $N_{L}=8-24$ and $\beta=20-100$. The curves are given by  the Casimir energy formula of a massive scalar field with \acrshort*{pccbc} \eqref{eq:exp_fit_21_periodic}.}
	\label{fig:Casimir_periodic_cond}
\end{figure}

\begin{figure}[h]
	\centering
	\includegraphics[width=1\textwidth]{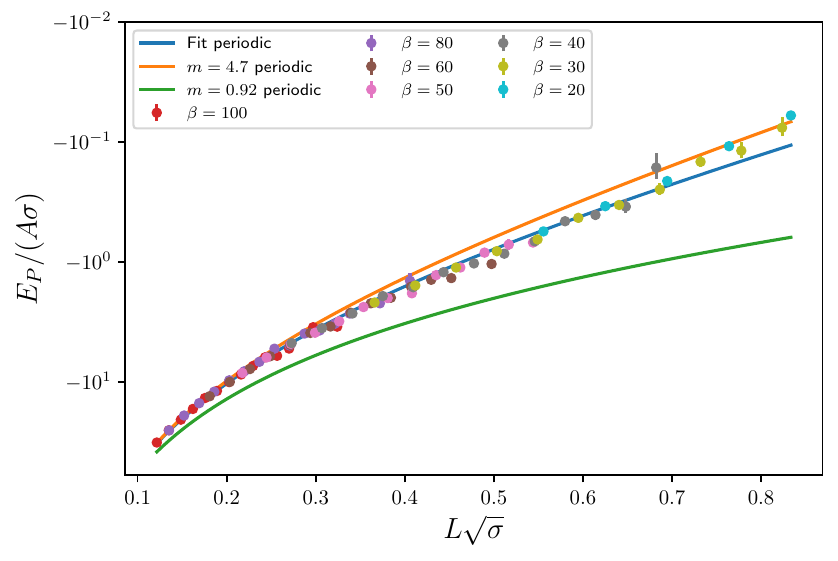}
	\caption{Dimensionless Casimir energy in logarithmic scale with \acrshort*{pbc} in a three dimensional lattice for different couplings and lattice sizes with parameters $N_A=96$, $N_{L_P}=96$, $N_{L}=8-24$ and $\beta=20-100$. The curves are given by  the Casimir energy formula for  massive scalar fields with \acrshort*{pbc} \eqref{eq:exp_fit_21_periodic}.}
	\label{fig:Casimir_periodic_cond_log}
\end{figure}

We display the results of the Casimir energy with \acrshort*{pbc} in \autoref{fig:Casimir_periodic_cond} (linear scale) and \autoref{fig:Casimir_periodic_cond_log} (logarithmic scale). The numerical values of the average plaquettes are explicitly shown in \autoref{ch:21_values}: \autoref{tab:Casimir_21_periodic_1}, \autoref{tab:Casimir_21_periodic_2} and \autoref{tab:Casimir_21_periodic_3}. The results show again a clear exponential decay in the Casimir energy with the distance between the walls, and for the different $\beta$ values they seem to share the same behaviour which suggests that the discretization errors are under control and we are close enough to the continuum limit. Unlike the \acrshort*{pccbc} case, the largest contribution to the statistical error is the term with the smaller spatial dimension $N_L$. Thus, we do not see a biased pattern where for each $\beta$ the points are either above or under the fits, they are randomly spread around it.

In this case we   fit the data with the formula of a massive scalar field with \acrshort*{pbc} that we computed in \autoref{chp:scalar_cont} for the two spatial dimensional case \eqref{eq:Cas_per_2d}. Thus the fitting expression is 
\begin{equation}\label{eq:exp_fit_21_periodic}
	\frac{E_P(L\sqrt{\sigma})}{A\sigma}=-\frac{G}{2\pi L^2\sigma}\left(mL\sqrt{\sigma}\ \text{Li}_2\left(e^{-mL\sqrt{\sigma}}\right)+\text{Li}_3\left(e^{-mL\sqrt{\sigma}}\right)\right),
\end{equation}
where we use the same parameters with  a multiplicative factor $G$ and  mass $m$. The fit is labeled as ``Fit periodic'' in  \autoref{fig:Casimir_periodic_cond} and \autoref{fig:Casimir_periodic_cond_log}. We also plot two more curves, one where we fix $G=3$ and $m=0.92$ that are the expected values from the \acrshort*{nkp}, and another with $G=3$ and $m=4.7$ which is the mass of the lightest glueball in the theory. These plots are labeled as ``$m=0.92$ periodic'' and ``$m=4.7$ periodic'' respectively.
Like in the \acrshort*{pccbc} case, we use a non-linear least square method for fitting the function that minimizes the $\chi^2$ defined previously \eqref{eq:xi_square} taking into account the correlation between the Casimir energy values with the same $\beta$, although in this case it is less relevant since the common error contribution is smaller than the individual one. The best fit parameters found are $$G=2.90\pm 0.02 \ \mathrm{and}\  m=3.96\pm 0.03,$$ with $\chi^2=4.5$. This value for the mass obtained from the fit is much higher than  expected from the \acrshort*{nkp} and much closer to the value of the mass of the lightest glueball. This can be seen clearly in \autoref{fig:Casimir_periodic_cond} where the fit and the curve with $m=4.7$ are very close with both fitting the simulation results  ($\chi^2=13$) whereas the curve with $m=0.92$ is very far from the results ($\chi^2=695$) . Since the value of the constant in the fit is close to the one with the fixed values ($G=3$) the different behaviour is mainly due to the factor in the exponential decay. Thus, this is indicating that with \acrshort*{pbc} the exponential decay of the Casimir energy is not given by a mass value close to the one expected by the \acrshort*{nkp} but a much higher value that is closer to that of the lightest glueball.

In the logarithm scale plot \autoref{fig:Casimir_periodic_cond_log} we can see the same behaviour but accentuated, where the curve with $m=0.92$ is even farther from the simulation points as the distance between the boundaries grows. With the logarithm plot it can also be seen how the fit and the curve with $m=4.7$ starts to differ from each other, and how the results adjust better to the fitted curve. 

\begin{figure}[h]
	\centering
	\includegraphics[width=1\textwidth]{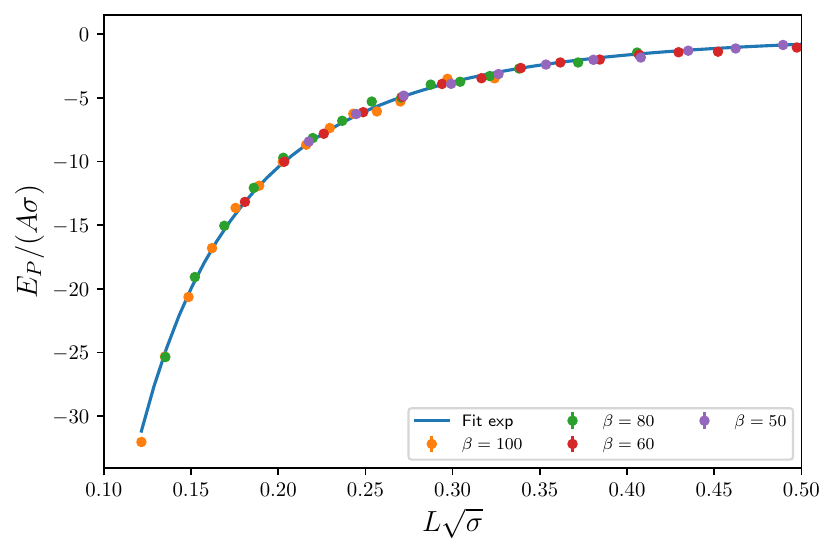}
	\caption{Dimensionless Casimir energy with \acrshort*{pbc} in a three dimensional lattice for different couplings and lattice sizes with parameters $N_A=96$, $N_{L_P}=96$, $N_{L}=8-24$ and $\beta=20-100$. The fit is given by the formula \eqref{eq:expp_fit_21_periodic}.}
	\label{fig:Casimir_periodic_cond_exp_lin}
\end{figure}

\begin{figure}[H]
	\centering
	\includegraphics[width=1\textwidth]{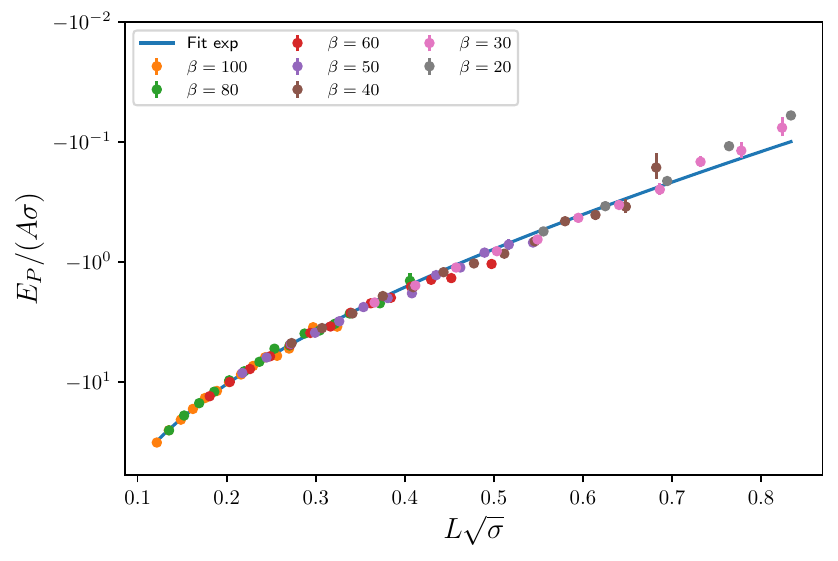}
	\caption{Dimensionless Casimir energy with \acrshort*{pbc} in a three dimensional lattice for different couplings and lattice sizes with parameters $N_A=96$, $N_{L_P}=96$, $N_{L}=8-24$ and $\beta=20-100$. The fit is given by the formula \eqref{eq:expp_fit_21_periodic}.}
	\label{fig:Casimir_periodic_cond_exp}
\end{figure}

\pagebreak
It is also worth to fit the results to an exponential of the form
\begin{equation}\label{eq:expp_fit_21_periodic}
	\frac{E_P(L\sqrt{\sigma})}{A\sigma}=-\frac{C}{(L\sqrt{\sigma})^\nu}e^{-\mu L\sqrt{\sigma}}
\end{equation}
following the same procedure. The best fit result gives the values of the parameters $$C=1.64\pm 0.09, \mu=3.72\pm 0.08 \ \mathrm{and}\  \nu=1.61\pm 0.02,$$  obtaining $\chi^2=4.2$. The value of the mass $\mu$ is close to the one we obtained in the previous fit $m=3.97$ and very far from the mass in the \acrshort*{nkp} $m_{\mathrm{K}}=0.92\sqrt{\sigma}$. Like in the \acrshort*{pccbc} case we get a value for the exponent $\nu$ 	which is an intermediate value between the asymptotic limit with $mL\sqrt{\sigma}$ of the massive scalar formula \eqref{eq:Casimir_asymptotic_periodic} and the massless case \eqref{eq:Casimir_massless_per_2d}. In the Figures
\ref{fig:Casimir_periodic_cond_exp_lin} (linear scale) and \ref{fig:Casimir_periodic_cond_exp} (logarithmic scale) we can see how this exponential function fits well the simulation results in a wide range of distances and replicates well the decay of this Casimir energy.

In summary, we observe how the Casimir energy of an $SU(2)$ pure gauge theory  with \acrshort*{pbc} exhibits an exponential decay characteristic of massive theories. This Casimir energy in $SU(2)$ seems to fit the Casimir energy of three massive scalar fields with a mass around $m_p=3.97\sqrt{\sigma}$ that is not compatible with the expected value in the \acrshort*{nkp}  $m_{\mathrm{K}}=0.92 \sqrt{\sigma}$ and is a lot closer to the lightest glueball $m_g\simeq 4.7\sqrt{\sigma}$, although still below this value.
\subsection{Continuum limit}
\begin{figure}[H]
	\centering
	\includegraphics[width=1\textwidth]{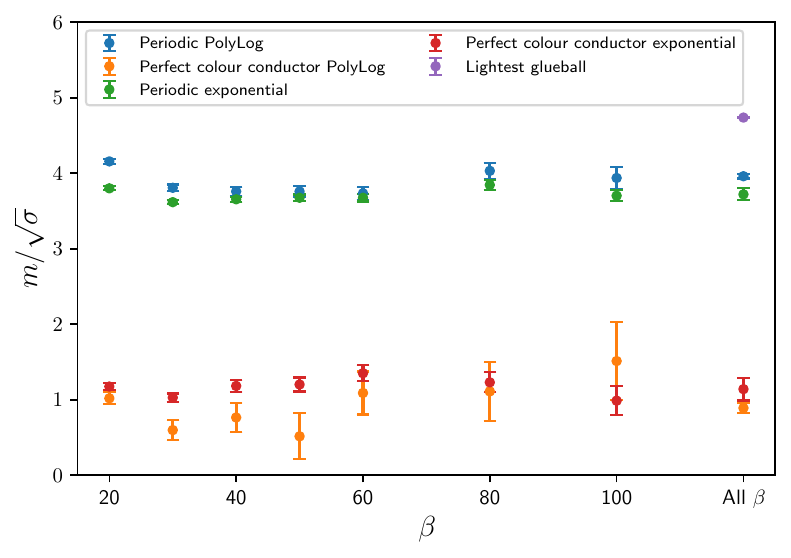}
	\caption{Mass in units of the string tension that drives the decay of the Casimir energy for the different $\beta$ and boundary conditions using the massive scalar field formula \eqref{eq:su2_neumann_fit} and \eqref{eq:exp_fit_21_periodic}, and the exponential fits \eqref{eq:exp_fit_21_perf} and \eqref{eq:expp_fit_21_periodic}. The continuum limit of the lightest glueball from \cite{athenodorou2017n} is also shown.}
	\label{fig:Casimir_betas_2d}
\end{figure}

\textcolor{black}{Finally we analyze the continuum limit of the masses obtained in this section, by computing them for each $\beta$ independently and comparing with the result we obtained fitting all the values of $\beta$ together. In \autoref{fig:Casimir_betas_2d} we plot the results for the masses that drive the exponential decay of the Casimir energy for each $\beta$ independently and, for different boundary conditions and fitting formulas. The results using formulas \eqref{eq:su2_neumann_fit} and \eqref{eq:exp_fit_21_periodic} are labeled as ``PolyLog'', and the ones obtained with the exponential formulas \eqref{eq:exp_fit_21_perf} and \eqref{eq:expp_fit_21_periodic} are labeled as `exponential'' in the plots. We can observe in the plot how for all values of $\beta$ the two boundary conditions have clearly different mass values and are consistent for the different values of $\beta$ and fitting formulas. Thus, we see how close we are to the continuum limit in the estimation of the mass that drives the exponential decay of the Casimir, and our precision allows us to clearly differentiate the values for each boundary condition and how both are far from the lightest glueball.  }

\textcolor{black}{Notice that this comparison raises some problems since the data for each $\beta$ have a different range of physical distances between the walls, and one would expect that in the points with a larger physical distance the mass value is more relevant to the fitting due to the enhanced effect of the exponential in the curve. Actually, this is the behaviour that the plot shows since the masses with the higher $\beta$ (smaller physical distances) have a larger error. Ideally one would have all the points with the different values of $\beta$ at the same physical distances (we compare those in plots \autoref{fig:Casimir_cond_sizey} and \autoref{fig:Casimir_periodic_sizey}) and the masses would approach a continuum limit as $\beta$ grows. Unfortunately, due to the exponential decay present in the Casimir energy it is not computationally feasible to obtain statistical errors small enough to compute the Casimir energy at the required lattice sizes for the larger values of $\beta$.}

\section{Finite volume errors}\label{sec:su21_lattice_effects}
Let us now analyze the errors associated to the finite volume of the lattice by comparing the results in \autoref{sec:Casimir_21} with the ones obtained by using smaller lattices. First, we analyze the behaviour of the mean energy when the timelike and longitudinal spatial dimensions $N_A$ as well as the extra transverse dimension $N_{L_0}$ we use to cancel the bulk contributions are smaller than the ones used in the previous section, in particular we set $N_A=64$ and $N_{L_0}=64$.

\begin{figure}[h]
	\centering
	\includegraphics[width=1\textwidth]{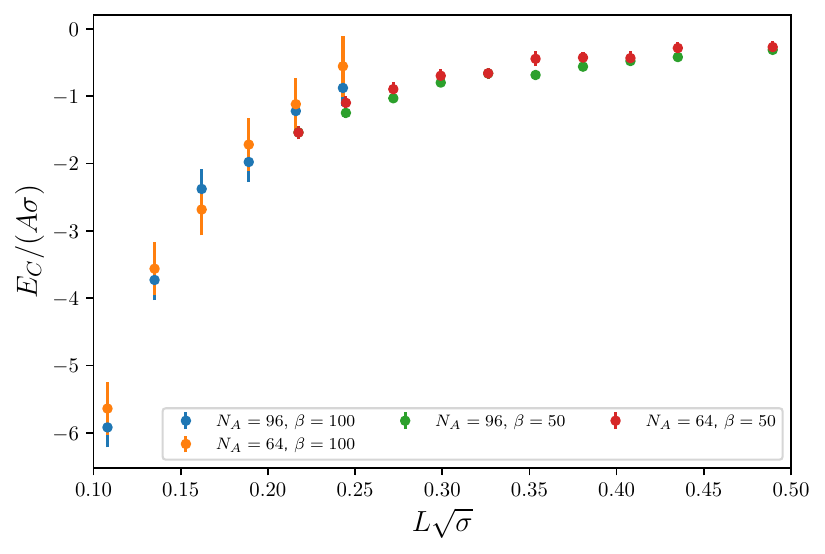}
	\caption{Dimensionless Casimir energy with \acrshort*{pccbc} in a three dimensional lattice for different couplings and sizes with parameters $N_A=\{64,96\}$, $N_{L_P}=\{64,96\}$, $N_{L_C}=48$, $N_{L}=8-20$ and $\beta=20-100$.}
	\label{fig:Casimir_perf_size_cond}
\end{figure}

\begin{figure}[h]
	\centering
	\includegraphics[width=1\textwidth]{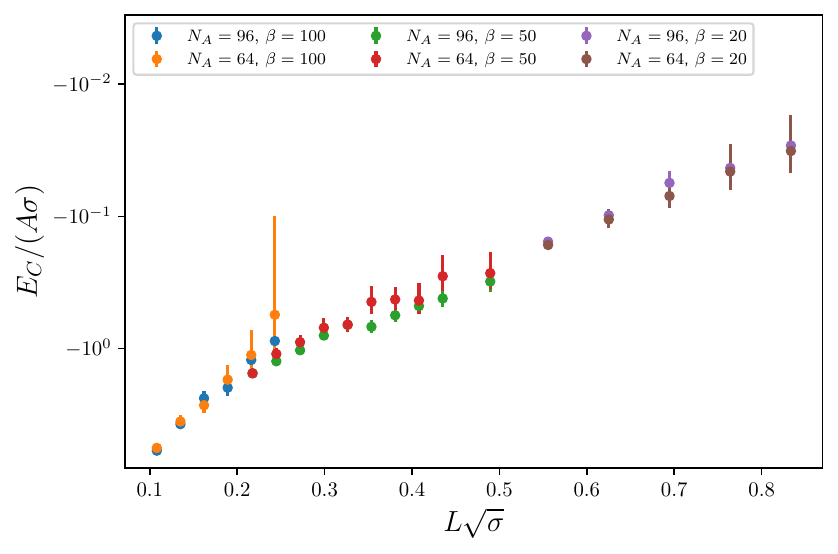}
	\caption{Dimensionless Casimir energy in logarithmic scale with \acrshort*{pccbc} in a three dimensional lattice for different couplings and sizes with parameters $N_A=\{64,96\}$, $N_{L_P}=\{64,96\}$, $N_{L_C}=48$, $N_{L}=8-20$ and $\beta=20-100$.}
	\label{fig:Casimir_perfc_size_cond_log}
\end{figure}

In \autoref{fig:Casimir_perf_size_cond} and \autoref{fig:Casimir_perfc_size_cond_log} we plot the comparison with different sizes of $N_A$ and $N_{L_0}$ for the largest, smallest and a medium value of $\beta$ we used in \autoref{sec:Casimir_21} for \acrshort*{pccbc}. The data used for computing these new values of the Casimir energy is shown in \autoref{tab:Energy_lattice_size_periodic}. In the linear plot \autoref{fig:Casimir_perf_size_cond} we can see how all the points with $\beta=100,50$ are compatible for the different lattice sizes. In the  logarithmic scale \autoref{fig:Casimir_perfc_size_cond_log} we can appreciate how also the points with $\beta=20$ match for the different lattices sizes. Also, by computing the difference between the Casimir energies with $N_A=64$ and $N_A=96$ we obtain that every point is within 1.5 sigmas from each other, thus these finite volume errors are smaller than the statistical error we reach.

\pagebreak

\begin{figure}[h]
	\centering
	\includegraphics[width=1\textwidth]{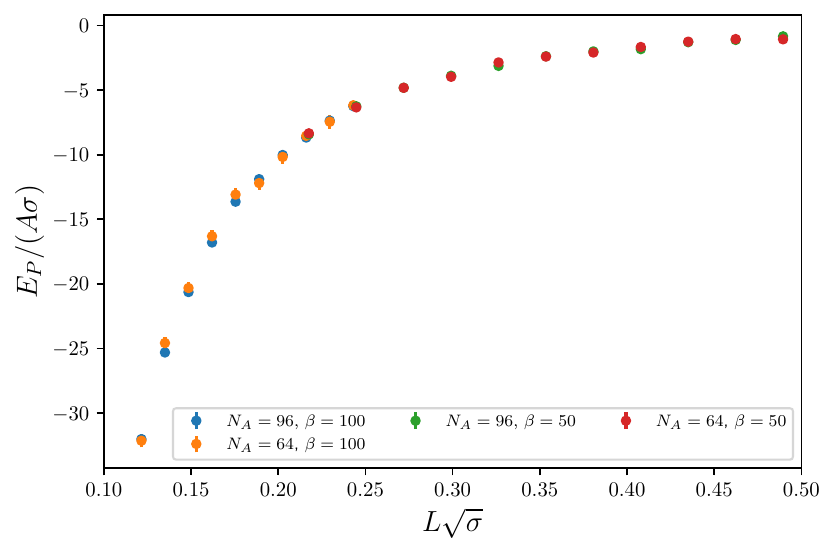}
	\caption{Dimensionless Casimir energy with \acrshort*{pbc} in a three dimensional lattice for different couplings and sizes with parameters $N_A=\{64,96\}$, $N_{L_P}=\{64,96\}$, $N_{L}=8-24$ and $\beta=20-100$.}
	\label{fig:Casimir_periodic_size_cond}
\end{figure}

\begin{figure}[H]
	\centering
	\includegraphics[width=1\textwidth]{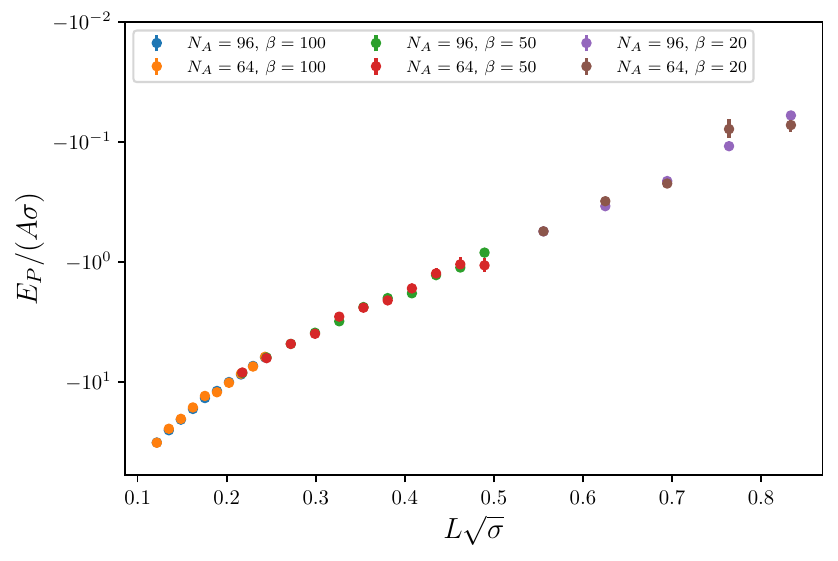}
	\caption{Dimensionless Casimir energy in logarithmic scale with \acrshort*{pbc} in a three dimensional lattice for different couplings and sizes with parameters $N_A=\{64,96\}$, $N_{L_P}=\{64,96\}$, $N_{L}=8-24$ and $\beta=20-100$.}
	\label{fig:Casimir_periodic_size_cond_log}
\end{figure}

In \autoref{fig:Casimir_periodic_size_cond} and \autoref{fig:Casimir_periodic_size_cond_log} we display the same plot but for \acrshort*{pbc}. The data used for computing these new values of the Casimir energy are presented in \autoref{tab:Energy_lattice_size_perf_cond}. Again, we see how the points from the different lattice sizes are compatible with each other for the different coupling values and distances. In particular, the difference between the points with the different lattices is always below two sigmas. Thus, as in the \acrshort*{pccbc} case, the errors associated to the finite size of $N_A$ and $N_{L_0}$ are smaller than our statistical error.
\vspace{5cm}


Thus, we can conclude that by using $N_A=96$ we have a large enough lattice such that the finite volume errors are negligible in comparison with the statistical error of the Casimir energy for both boundary conditions and for 
all the range of $\beta$ we analyze.

\begin{figure}[H]
	\centering
	\includegraphics[width=1\textwidth]{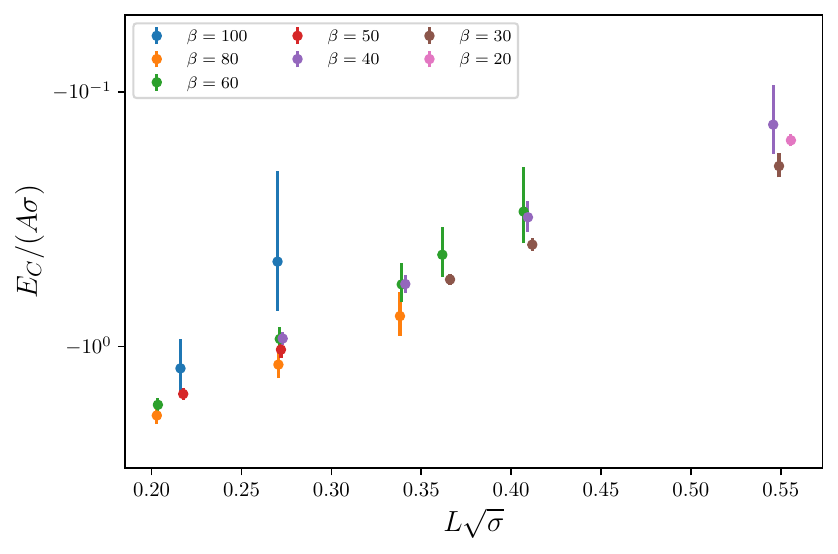}
	\caption{Dimensionless Casimir energy in logarithmic scale with \acrshort*{pccbc} in a three dimensional lattice for different couplings and sizes with parameters $N_A=96$, $N_{L_P}=96$, $N_{L_C}=48$, $N_{L}=8-24$ and $\beta=20-100$. Only values which have another point close in physical distance $L\sqrt{\sigma}$ but different $N_L$ are plotted.}
	\label{fig:Casimir_cond_sizey}
\end{figure}

Now we analyze the error due to the finite size $N_L$ of the transverse direction. From the results in \autoref{sec:Casimir_21} we can already conclude that the range of values that we use (from $N_L=8$ to 24) does not have large errors since all the points with these different values of $N_L$ follow the same behaviour. We can look at this matter in more detail by using the dependence of the string tension on $\beta$ \eqref{eq:string_tension}. When we consider large values of $\beta$ this dependence is proportional to the inverse of $\beta$ i.e. $a\sqrt{\sigma}\sim \beta^{-1}$. Thus, we can find different values of $N_L$ that correspond approximately to the same physical distance. For example, if we take $N_L=20$ with $\beta=100$ we get $L\sqrt{\sigma}=0.270$, whereas if we choose $N_L=10$ and $\beta=50$ we have $L\sqrt{\sigma}=0.272$. We can see how we get very close values for the physical distance which allows us to compare the behaviour with different lattice sizes. In \autoref{fig:Casimir_cond_sizey} we display only some values of the Casimir energy where there are two or more points with very close physical distance $\sqrt{\sigma}L$ and different $\beta$ for \acrshort*{pccbc}. These points are compatible within the statistical error, for all the range of $N_L$ (8-20). 
\pagebreak

\begin{figure}[H]
	\centering
	\includegraphics[width=1\textwidth]{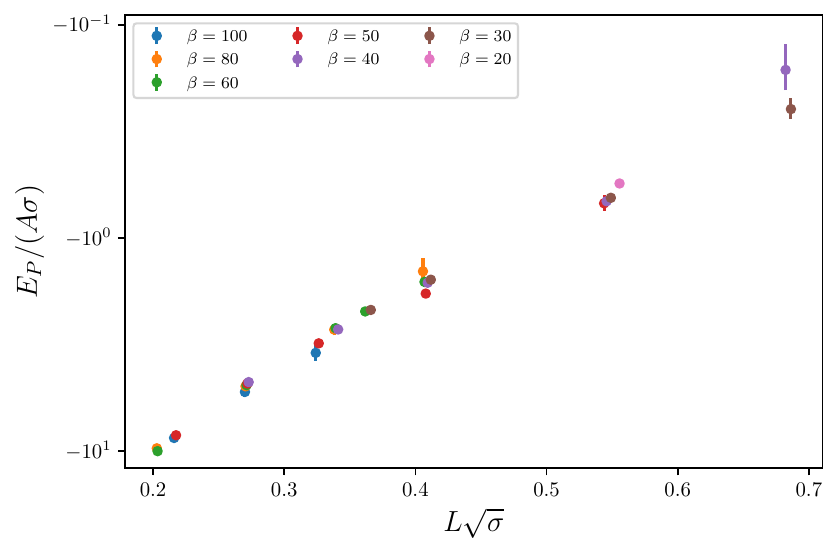}
	\caption{Dimensionless Casimir energy in logarithmic scale with \acrshort*{pbc} in a three dimensional lattice for different couplings and sizes with parameters $N_A=96$, $N_{L_P}=96$, $N_{L}=8-24$ and $\beta=20-100$. Only values which have another point close in physical distance $L\sqrt{\sigma}$ but different $N_L$ are displayed.}
	\label{fig:Casimir_periodic_sizey}
\end{figure}

In \autoref{fig:Casimir_periodic_sizey} we display the same plot but with \acrshort*{pbc}. We find the same behaviour where all the points at close physical distance have Casimir energies which are compatible with each other. Thus, we get a very good indicator that the range that we use for the values of $N_L$ and $\beta$ does not produce errors due to the size of $N_L$ that are significant in comparison with the statistical error.

     \chapter{Casimir energy of gauge fields in 3+1 dimensions}\label{chp:su2_31}
     In this chapter we address the non-perturbative calculation of the Casimir energy for a pure gauge $SU(2)$ theory in 3+1 dimensions by \acrshort*{mc} lattice simulations. We also use two different boundary conditions that belong to each of the families in the classification according to the exponential decay \eqref{eq:rate} we found in \autoref{chp:scalar_cont}, i.e. \acrshort*{pbc} and \acrshort*{pccbc}. We fit this Casimir energy to an exponential decay to find which mass drives its behaviour for large distances between the Casimir plates 
and compare it with the lightest glueball of the theory.

In \autoref{sec:energy_31} we analyze which observable should be used to compute the vacuum energy on the lattice. In \autoref{sec:therm_31} we analyze the thermalization and autocorrelation of such an energy observable in our \acrshort*{mc} simulations. In \autoref{sec:reno_31}  we test the renormalization procedure that will be used to compute the Casimir energy. In  \autoref{sec:Casimir_31} we finally obtain the Casimir energy and analyze its dependence on  the distance between the two boundary walls. Finally, in \autoref{sec:su31_lattice_effects} we analyze the finite volume errors of our calculation of the Casimir energy.

\section{The internal energy}\label{sec:energy_31}
Like in the 2+1 dimensional case the best observable to obtain the internal energy on the lattice is the first component of the energy-momentum tensor. In the 3+1 dimensional gauge theories it is given by
\begin{equation}
	T_{00}=\frac{1}{g^2} \text{tr}\left(F^2_{12}(x)+F^2_{23}( x)+F^2_{13}( x)-F^2_{01}( x)-F^2_{02}( x)-F^2_{03}( x)\right)),
\end{equation}
from which the energy is computed by integrating in the space dimensions
\begin{equation}
	\mathcal E=\frac{1}{g^2}\int d^3\mathbf x\ \text{tr}\left(F^2_{12}( x)+F^2_{23}(  x)+F^2_{13}( x)-F^2_{01}( x)-F^2_{02}( x)-F^2_{03}( x)\right)).
\end{equation}
Again, we can replace the integral by the sum over the lattice \eqref{eq:integral_sum} and use the relationship between the plaquettes and the field strength tensor \eqref{eq:plaquette_F}, to write the energy on the lattice as
\begin{equation}\label{eq:energy_tensor_31_gauge}
	\mathcal E=-\frac{\beta}{aN_0}\sum_{ n \in \Lambda}\left(\frac{1}{2}\text{tr}\left(P_{12}( n)+P_{23}( n)+P_{13}( n)-P_{01}( n)-P_{02}( n)-P_{03}( n)\right)\right),
\end{equation} 
where we are averaging over the temporal side of the lattice.  In a similar way as in the 2+1 case, since we are interested in computing the Casimir energy the particular sizes of the temporal side and the two spatial dimensions which are parallel to the boundary walls are not relevant as long as that they are large enough in comparison with the transverse spatial direction. This allows us to fix these sizes to the same value $N_0=N_1=N_2\equiv N_A$, which makes the following mean values of the plaquettes equivalent: $\braket{P_{12}}=\braket{P_{01}}=\braket{P_{02}}$ and $\braket{P_{13}}=\braket{P_{23}}=\braket{P_{03}}$. Thus, the difference of mean values of plaquettes simplifies to just a difference of two, one that involves the transverse spatial direction and another that does not. Instead of just computing this difference with two different plaquettes, we can compute each type of plaquette as the average value of the three equivalent ones, i.e
\begin{equation}\label{eq:plaq_dif}
	\braket{P_{12}+P_{23}+P_{13}-P_{01}-P_{02}-P_{03}}=\braket{\frac{P_{03}+P_{23}+P_{13}}{3}-\frac{P_{01}+P_{02}+P_{12}}{3}}\equiv \braket{P_d}
\end{equation}
where we have omitted the dependence with the lattice position $n$, and defined the difference of the average plaquettes that involve the transverse spatial dimension and the average of those that do not as $P_d( n)$. Obtaining the difference of plaquettes as this combination instead of simply computing the difference will reduce the statistical error without adding any extra computation apart from computing all the plaquettes in each measurement, which is much less demanding than the \acrshort*{mc} updates. Therefore, the internal energy is given by
\begin{equation}\label{eq:Energy_31}
	\braket{\mathcal E(N_A,N_L,\beta)}=-\frac{u^4\beta}{2aN_0}\braket{\sum_{ n \in \Lambda}\text{tr}\left(P_d(n)\right)}.
\end{equation}
where we have also applied the same smoothing procedure we used in 2+1 dimensions with the average plaquette \eqref{eq:improved_coupling}. Notice how if we set all the directions the same size and the same boundary conditions the mean value of the energy is zero, this will be relevant when analyzing the renormalization procedure.
\vspace{2cm}
\pagebreak

\section{Thermalization and autocorrelation }\label{sec:therm_31}
\begin{figure}[H]
	\centering
	\includegraphics[width=1\textwidth]{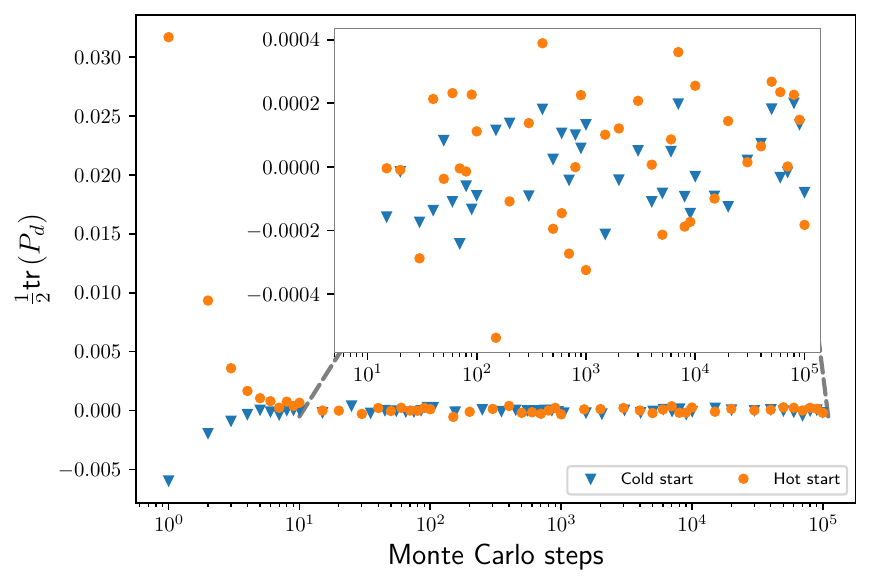}
	\caption{Evolution of the average over the lattice of the plaquette difference $P_d$ \eqref{eq:plaq_dif} with the \acrshort*{mc} steps in logarithmic scale from hot and cold initial configurations for a lattice with $N_A=48$, $N_{L}=16$, $\beta=2.7$ and \acrshort*{pbc}.}
	\label{fig:Energy_therm_both_per_31}
\end{figure}

Let us now analyze the thermalization and autocorrelation on this 3+1 dimensional case. First of all, we check how the difference of plaquettes  $P_d$ that defines the energy observable \eqref{eq:Energy_31} evolves from two different initial configurations (hot and cold starting scenarios). In \autoref{fig:Energy_therm_both_per_31} we plot this evolution in a lattice with \acrshort*{pbc} in every direction. It can be seen how when reaching around 200 \acrshort*{mc} steps both values converge and start fluctuating around the expected average. This behaviour can be observed more clearly in the zoomed part of the plot.

\begin{figure}[H]
	\centering
	\includegraphics[width=1\textwidth]{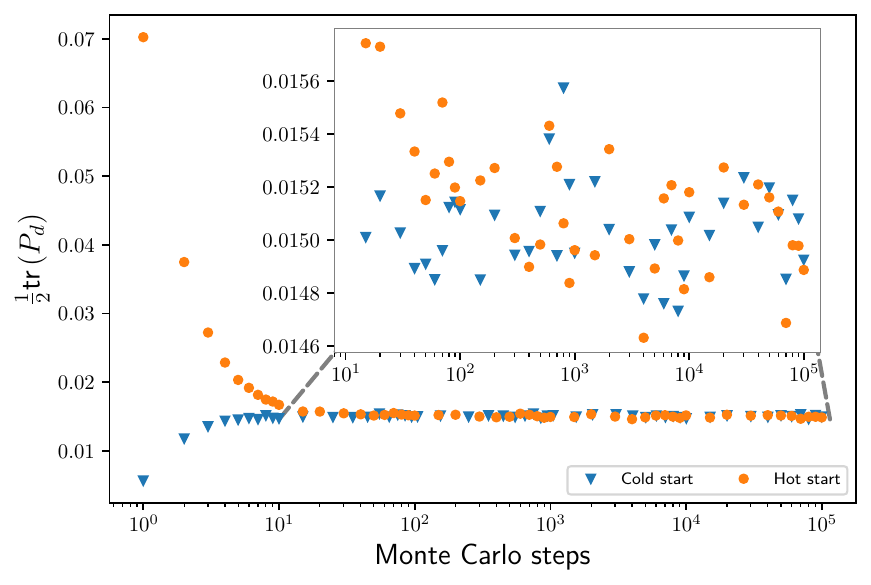}
	\caption{Evolution of the average over the lattice of the plaquette difference $P_d$ \eqref{eq:plaq_dif} with the \acrshort*{mc} steps in logarithmic scale from hot and cold initial configurations for a lattice with $N_A=48$, $N_{L}=16$, $\beta=2.7$ and \acrshort*{pccbc} on the transverse spatial dimension.}
	\label{fig:Energy_therm_both_dir_31}
\end{figure}

In the case that we have \acrshort*{pccbc} on the transverse spatial dimension the same behaviour can be observed (\autoref{fig:Energy_therm_both_dir_31}), where after 200 \acrshort*{mc} steps both initial configurations converge to the same mean value. Therefore, we can conclude from these data that a value around 10000 steps for the thermalization is large enough for both boundary conditions.

\begin{figure}[H]
	\centering
	\includegraphics[width=0.9\textwidth]{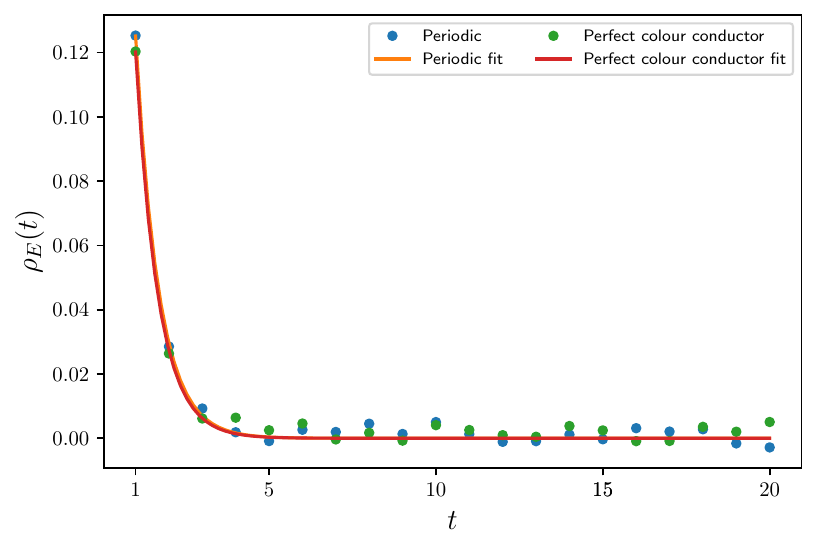}
	\caption{Normalized autocorrelation of the mean value $\braket{P_d}$ for a lattice where $N_A=48$, $N_{L}=10$ and $\beta=2.7$ with \acrshort*{pbc} or \acrshort*{pccbc} in the transverse spatial dimension.}
	\label{fig:Energy_auto_31}
\end{figure}

Let us now focus on the autocorrelation \eqref{eq:autocorrelation} of the observable $P_d$ after thermalization. In \autoref{fig:Energy_auto_31} we display the normalized autocorrelation for both \acrshort*{pbc} and \acrshort*{pccbc}. It can be shown how for consecutive measures the correlation is $\rho_E(1)\simeq 0.12$, and for reaching  a value  with $\rho_E(t)<0.05$ we need a distance of three \acrshort*{mc} steps $t=3$\footnote{Notice how this autocorrelation is larger than in the 2+1 dimensional case that was $\rho_E(1)\simeq 0.05$. This difference is due to simulation procedure used, in the 2+1 case the autocorrelation data was obtained by simulations using \acrshort*{gpu}s and, as it was explained in \autoref{sec:run_sim}, in this method we update simultaneously half of the lattice. This requires to generate $N/2$ independent sequences of random numbers, whereas in this case we have used simulations on \acrshort*{cpu}s to compute the autocorrelation. In these \acrshort*{cpu} simulations only $N_A/2$ plaquettes are updated simultaneously on the lattice, and thereby, just $N_A/2$ independent sequences of random numbers are needed.}. If we compute the integrated correlation time \eqref{eq:integrated_auto}, we get $\tau_{E,\text{int}}\simeq 0.68$ and $\tau_{E,\text{int}}\simeq 0.69$ for \acrshort*{pbc} and \acrshort*{pccbc} respectively. These are relevant values that would indicate the we have some (although small) correlation in the measurements. Thus, we will apply some \acrshort*{mc} steps (3) between measures to reduce this correlation. If we compute again the integrated correlation time but introducing these three steps between \acrshort*{mc} measurements, we obtain $\tau_{E,\text{int}}\simeq 0.51$ for \acrshort*{pbc} and $\tau_{E,\text{int}}\simeq 0.52$ for \acrshort*{pccbc}, which show that there is barely any autocorrelation left ($\tau_{E,\text{int}}= 0.5$ is the value for fully independent measures).

Finally, we can compute the exponential correlation time \eqref{eq:exp_correlated} by fitting the normalized correlation in \autoref{fig:Energy_auto_31} to an exponential. We get $\tau_{E,\text{exp}}\simeq 0.68$ with \acrshort*{pbc} and $\tau_{E,\text{exp}}\simeq 0.69$ for \acrshort*{pccbc}. These results are compatible with our previous analysis of the thermalization, where $1000 \tau_{E,\text{exp}}\simeq 1000$ steps are a good baseline, and at least $10000 \tau_{E,\text{exp}}\simeq 10000$ steps will be used.

All these results were obtained using the heat-bath algorithm, similarly to the 2+1 dimensional case, this is more efficient than using the Metropolis algorithm for the \acrshort*{mc} updates since the Metropolis algorithm produces a much larger correlation between the measures and demands adding a lot of \acrshort*{mc} steps between them. Also, as we mentioned on this subsection, when using the \acrshort*{cpu}s algorithm we have a higher correlation than in the \acrshort*{gpu}s simulations, thus requiring us to introduce some \acrshort*{mc} steps between measures to reduce the autocorrelation between the consecutive measures we consider for obtaining the results.

\newpage

\section{Renormalization}\label{sec:reno_31}
Let us analyze the internal energy behaviour with the size of the transverse spatial dimension $N_L$, with the goal of subtracting the divergences (finite on the lattice) of the energy and compute the Casimir energy in the continuum limit. In 2+1 dimensions we found that energy has the same vacuum structure as the scalar field \eqref{eq:energy_su2_21}, where there is a bulk contribution $C_0(\beta)$ that is common for both boundary conditions and the boundary contribution $C_1(\beta)$ that is zero in the \acrshort*{pbc} case. Also, this energy density observable on 2+1 dimensions was given by just the average of one plaquette \eqref{eq:Energy_su2_21}, therefore, this observed behaviour of the energy in 2+1 dimensions is that of one plaquette. In the 3+1 dimensional case, the energy is defined by the difference of two plaquettes \eqref{eq:Energy_31}. Thus, with the same expected behaviour for the plaquette, the difference of them should cancel the bulk contribution since it is independent of the boundary condition or the lattice size, and just the boundary term $C_1(\beta)$ and the Casimir contribution $E_U(N_L,\beta)$ should remain, i.e.
\begin{equation}\label{eq:energy_su2_31}
\braket{\mathcal E_U(N_L,\beta)}=C_1(\beta) N_A+E_U(N_L,\beta)+\ldots.
\end{equation}

\subsection{Periodic boundary conditions}
\begin{figure}[H]
	\centering
	\includegraphics[width=1\textwidth]{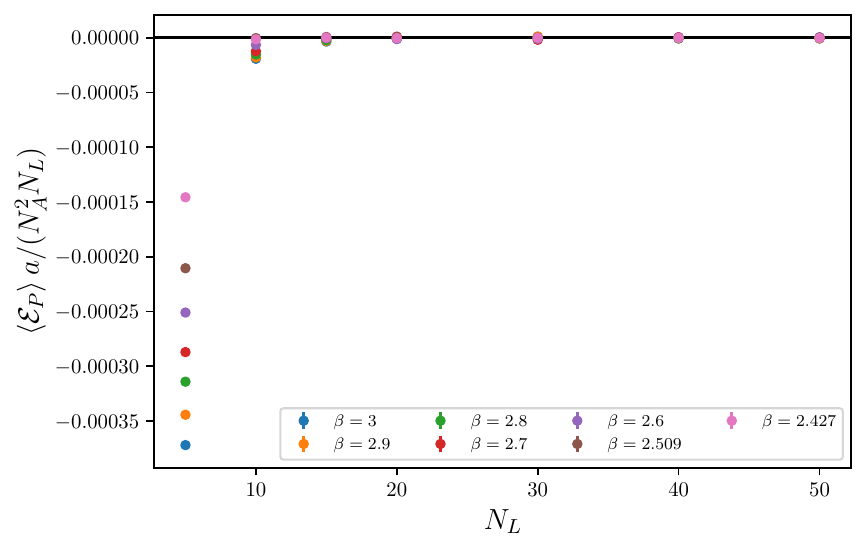}
	\caption{Energy density in a pure $SU(2)$ gauge theory in a four dimensional lattice with \acrshort*{pbc} for different lattices and coupling values. The chosen parameters of the lattice are $N_A=48$, $N_{L}=5-50$ and $\beta=2.427-3$.}
	\label{fig:Energy_periodic_su31}
\end{figure}

First, let us analyze this behaviour for \acrshort*{pbc}. As we just explained, we expect the bulk term to cancel, and since we are in \acrshort*{pbc}, that the boundary term also does. Thus, the energy should be the Casimir energy without the need of subtracting any other quantity. In \autoref{fig:Energy_periodic_su31} we plot the energy density for different sizes $N_L$ and coupling constants $\beta$ (the data used is shown in \autoref{tab:Energy_renom_periodic_1_31} and \autoref{tab:Energy_renom_periodic_2_31}). We can see the expected behaviour in the figure, where for the larger sizes the energy goes to zero whereas on the smaller sizes the Casimir energy can be observed. This means that the internal energy on the lattices is zero in the limit where all  lattice sizes goes to infinity, and is also consistent with the fact that when setting all directions the same size the expected value of the energy is also zero since we are subtracting two quantities which are totally equivalent. Thus, for computing the Casimir energy with \acrshort*{pbc} we do not need any renormalization procedure and we can directly obtain the Casimir energy from \eqref{eq:Energy_31}. A similar behaviour where the bulk contribution in the first component of the energy momentum tensor in 3+1 dimensions cancels was observed in Ref. \cite{chernodub2023boundary} for $SU(3)$. 

\subsection{Perfect colour conductor boundary conditions}
\begin{figure}[H]
	\centering
	\includegraphics[width=1\textwidth]{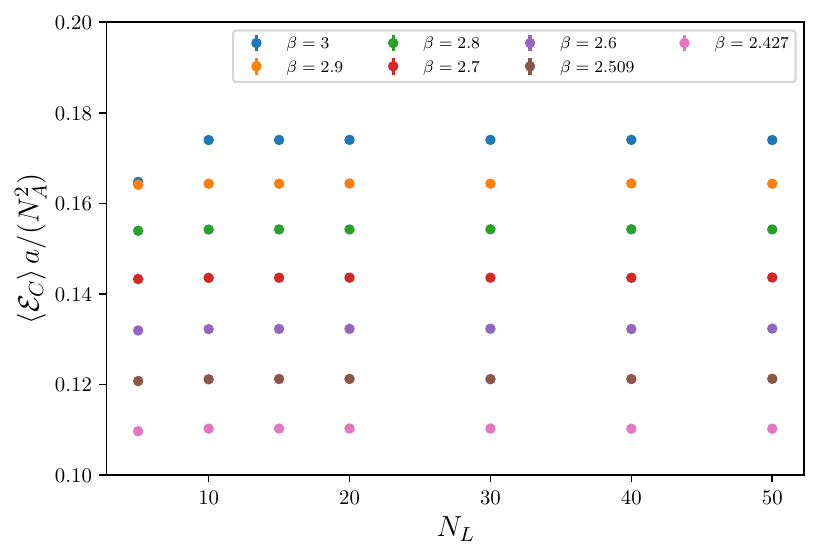}
	\caption{Energy in a pure $SU(2)$ gauge theory in a four dimensional lattice with \acrshort*{pccbc} on the transverse spatial dimension for different lattice sizes and coupling values. The chosen parameters of the lattice are $N_A=48$, $N_{L}=5-50$, and $\beta=2.427-3$.}
	\label{fig:Energy_conductor_su31}
\end{figure}

Now, let us  analyze the case with \acrshort*{pccbc} on the transverse spatial dimension. In this framework, we expect the cancellation of the bulk contribution as in the \acrshort*{pbc} case except for a boundary contribution which is dominated by the self-energy of the boundary walls. In \autoref{fig:Energy_conductor_su31} we display the vacuum energy  for different transverse lattice sizes $N_L$ and $\beta$ (see \autoref{tab:Energy_renom_conductor_1_31} and \autoref{tab:Energy_renom_conductor_2_31} to find the results used in the plots). We observe how the values for each $\beta$ seem to be independent of $N_L$ in the larger lattices, whereas when $N_L$ is smaller we can see fluctuations due to the Casimir energy. This is precisely the expected behaviour when there is a boundary term $C_1(\beta)$ that dominates the energy behaviour. Also, this means that this boundary term can be calculated as
\begin{equation}
	C_1(\beta)=\lim_{N_{L_0}\rightarrow\infty}\braket{\mathcal E_C(N_{L_0},\beta)}/N_A^2
\end{equation}
where we are also considering $N_A$ as a large value.
\vspace{1cm}
\section{Casimir energy}\label{sec:Casimir_31}
As in \autoref{chp:su2_21}, we use the string tension $\sigma$ as the physical scale to express the Casimir energy as a dimensionless quantity and also the physical distance between the boundary walls in the same terms. In the $3+1$ dimensional case, there is no equivalent phenomenological fit for the dependence of the string tension with the coupling constant $\beta$. Instead, we use the values obtained in Ref. \cite{athenodorou2021n} for some particular values of the coupling constant $\beta=\{2.427, 2.509,2.6,2.7,2.8\}$. Also, we compute it for two more values $\beta=\{2.9, 3\}$  by using the correlation of Polyakov loops and a multilevel simulation algorithm \cite{parisi1983measurement,luscher2001locality,meyer2003locality}.

\subsection{Periodic boundary conditions}
\begin{figure}[H]
	\centering
	\includegraphics[width=1\textwidth]{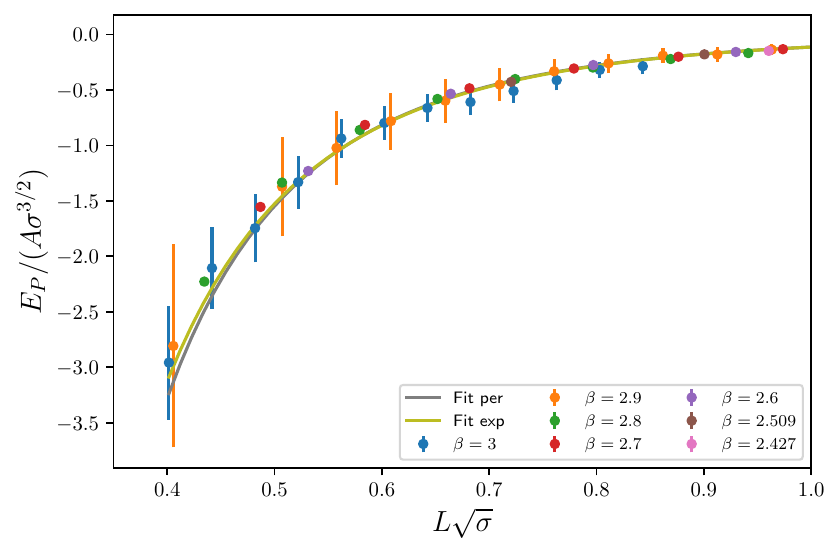}
	\caption{Dimensionless Casimir energy for \acrshort*{pbc} in a pure $SU(2)$ gauge theory and a four dimensional lattice for different couplings and lattice sizes with parameters $N_A=48$,  $N_{L}=4-21$ and $\beta=2.427-3$. The fits correspond to the formulas \eqref{eq:Casimir_periodic_31_scalar_fit} and \eqref{eq:Casimir_periodic_31_exp_fit}.}
	\label{fig:Casimir_periodic_31}
\end{figure}

\begin{figure}[H]
	\centering
	\includegraphics[width=1\textwidth]{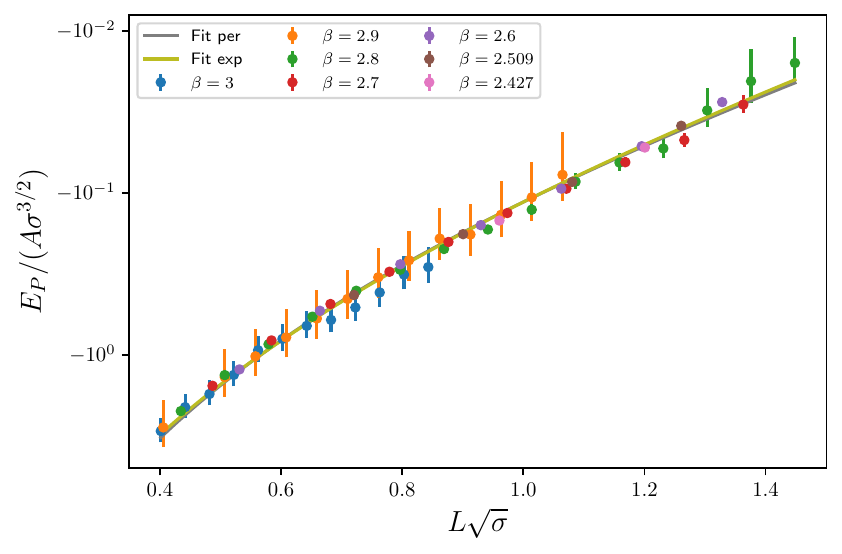}
	\caption{Dimensionless Casimir energy in logarithmic scale for \acrshort*{pbc} in a pure $SU(2)$ gauge theory and a four dimensional lattice for different couplings and lattice sizes with parameters $N_A=48$,  $N_{L}=4-21$ and $\beta=2.427-3$. The different fits correspond to the formulas \eqref{eq:Casimir_periodic_31_scalar_fit} and \eqref{eq:Casimir_periodic_31_exp_fit}.}
	\label{fig:Casimir_periodic_31_log}
\end{figure}

First, let us start with the case of \acrshort*{pbc} also in the transverse direction. As was shown in \autoref{sec:reno_31}, with \acrshort*{pbc} the energy observable defined as \eqref{eq:Energy_31} gives directly the Casimir energy. Thus, it can be directly computed as
\begin{equation}
	E_P(N_L,\beta)=\braket{\mathcal E_P(N_L,\beta)},
\end{equation}
where we take $N_A$ large in comparison with $N_L$. In \autoref{fig:Casimir_periodic_31} (linear scale) and \autoref{fig:Casimir_periodic_31_log} (logarithmic scale) we plot the Casimir energy for different values of $\beta$  and $N_L$ with the data given in \autoref{tab:Energy_periodic_1_31}, \autoref{tab:Energy_periodic_2_31} and \autoref{tab:Energy_periodic_3_31}. It can clearly be seen how the Casimir energy shows an exponential decay with the distance between the boundary walls. All the points with the different $\beta$ and $N_L$ follow the same behaviour, which indicates that the discretization effects in the transverse direction are under control and that we are approaching the continuum limit. We do two different fits to the simulation results. First, we fit to the formula of a massive scalar field in 3+1 dimensions with \acrshort*{pbc} \eqref{eq:Casimir_periodic_31_scalar}
\begin{equation}\label{eq:Casimir_periodic_31_scalar_fit}
	\frac{E_P(L\sqrt{\sigma})}{A\sigma^{3/2}}=-G\frac{(m/\sqrt \sigma)^{2}}{2\pi^{2}L\sqrt\sigma}\sum_{j=1}^{\infty}\frac{K_{2}(jmL\sqrt{\sigma})}{j^{2}},
\end{equation}
where the free parameters are the multiplicative constant $G$ and the mass $m$, and $A=a^2 N_A^2$ is the area of the boundary walls. This curve is labeled as ``Fit per'' in the plot. To obtain the best fit parameters we use a non-linear least square method that minimizes the same $\chi^2$ as in the $2+1$ dimensional case, i.e. 
\begin{equation}\label{eq:xi_square2}
	\chi ^2=\sum_{i,j=1}^n\text{cov}(E)^{-1}_{ij}(E_i-f_i)(E_j-f_j)/(n-p),
\end{equation} 
where $\text{cov}(E)_{ij}$ is the covariance matrix between the $n$ Casimir energy results $E_i$, $f_i$ the output of the fitting function for each point and $p$ the number of free parameters. The best fit parameters are $$G=2.290\pm0.008 \ \mathrm{and}\  m=2.086\pm0.007,$$ that reaches   $\chi^2=145$. We can see in the plots \autoref{fig:Casimir_periodic_31} and \autoref{fig:Casimir_periodic_31_log} how the fit follows the simulation results in all the range of distances $L\sqrt{\sigma}$ and reproduces the expected exponential decay. If we compare the value obtained for the mass, we can see that is much lower than the lightest glueball which is $m_g=3.78 \sqrt{\sigma}\pm 0.02$ \cite{athenodorou2021n,lucini2001n} in this  case.

We also fit the simulation results to an exponential that we label as ``Fit exp'' in the plot, and given by  the following formula
\begin{equation}\label{eq:Casimir_periodic_31_exp_fit}
	\frac{E_P(L\sqrt{\sigma})}{A\sigma^{3/2}}=-\frac{C}{(\sqrt{\sigma}L)^\nu}e^{-\mu L\sqrt{\sigma}}
\end{equation}
where the parameters of the fit are: the multiplicative constant $C$, the mass of the exponential decay $\mu$ and exponent $\nu$  of the dimensionless distance $\sqrt{\sigma}L$. The best fit is found for the following parameters: 
$$C=0.79\pm 0.02, \mu=1.94\pm0.02\ {\mathrm{and}}\ \nu=2.34\pm 0.02.$$ 
Using these parameters we reach $\chi^2=161$. We obtain a similar value for the mass than the previous fit with the scalar field formula $m=2.086$, and both of them are smaller than the lightest glueball $m_g\simeq3.78\sqrt{\sigma}$. Moreover, the value obtained for the exponent $\nu=2.34$ is an intermediate value between the asymptotic limit $mL\sqrt\sigma\rightarrow\infty$ on the massive scalar formula \eqref{eq:Casimir_asymptotic_periodic} $\nu=3/2$, and the massless limit \eqref{eq:massless_3d_per} $\nu=3$, which is consistent with the fact that we are in an intermediate range for the value of the exponent $\mu L\sqrt\sigma$.

\subsection{Perfect colour conductor boundary conditions}
When the boundaries have \acrshort*{pccbc} on the transverse direction we only have to cancel the boundary term $C_1(\beta)$ since the energy observable does not have the bulk contribution as was shown in \autoref{sec:reno_31}. Following the example of the 2+1 dimensional case, we can eliminate the contribution of this term by subtracting the energy of another lattice with the same parameters but a much larger transverse direction. Therefore, we can define the Casimir energy with \acrshort*{pccbc} as
\begin{equation}\label{eq:Casimir_cond_31}
	E_C(N_L,\beta)=\lim_{N_{L_0}\rightarrow \infty}\braket{\mathcal E_C(N_L,\beta)}-\braket{\mathcal E_C(N_{L_0},\beta)},
\end{equation}
where we  also consider the other directions of the lattice $N_A$ larger than the transverse one $N_L$. In the simulations we  set $N_{L_0}<N_A$ for optimizing the computing time but large enough so the finite volume effects are small in comparison with the Casimir energy that we want to calculate. 
\begin{figure}[H]
	\centering
	\includegraphics[width=1\textwidth]{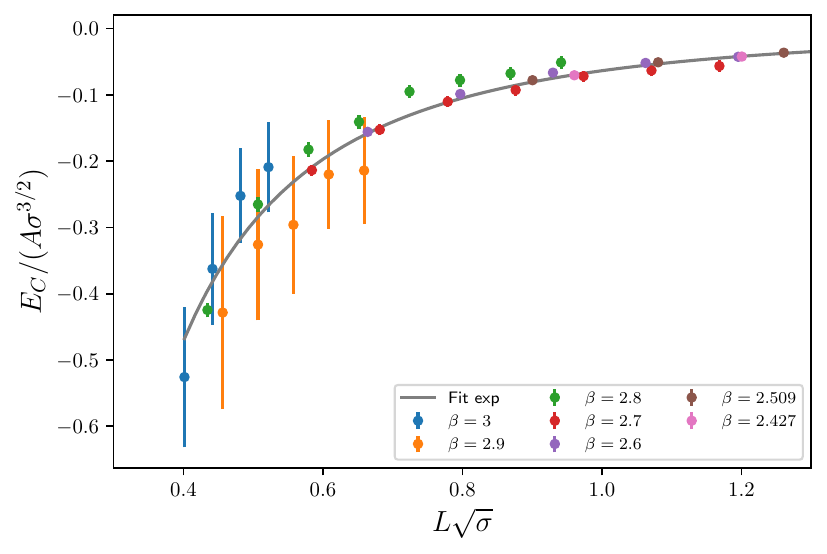}
	\caption{Dimensionless Casimir energy with \acrshort*{pccbc} on the transverse direction in a pure $SU(2)$ gauge theory and a four dimensional lattice for different couplings and lattice sizes with parameters $N_A=48$, $N_{L_0}=36$,  $N_{L}=4-16$ and $\beta=2.427-3$. The fit is given by the formula \eqref{eq:Casimir_cond_31_exp_fit}.}
	\label{fig:Casimir_cond_31}
\end{figure}

\begin{figure}[H]
	\centering
	\includegraphics[width=1\textwidth]{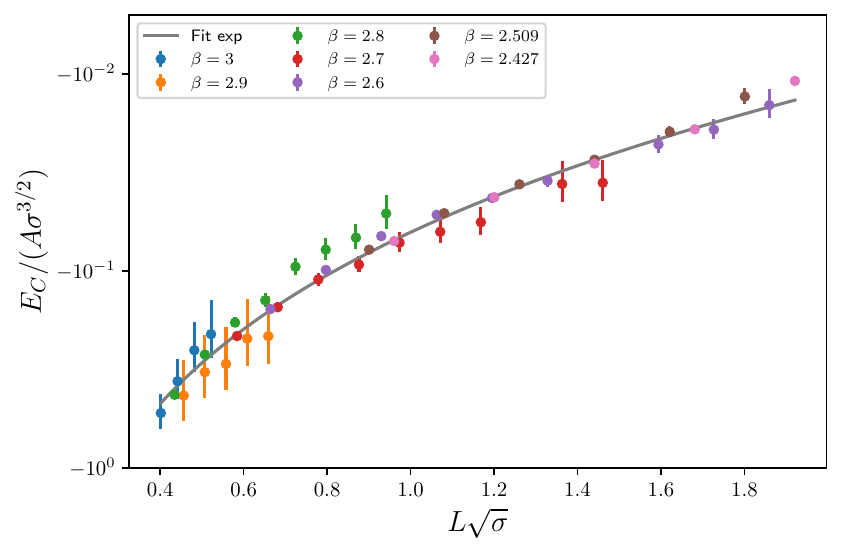}
	\caption{Dimensionless Casimir energy in logarithmic scale for \acrshort*{pccbc} on the transverse direction in a pure $SU(2)$ gauge theory and a four dimensional lattice for different and lattice sizes with parameters $N_A=48$, $N_{L_0}=36$,  $N_{L}=4-16$ and $\beta=2.427-3$. The fit is given by the formula \eqref{eq:Casimir_cond_31_exp_fit}.}
	\label{fig:Casimir_cond_31_log}
\end{figure}

In \autoref{fig:Casimir_cond_31} (linear scale) and \autoref{fig:Casimir_cond_31_log} (logarithmic scale) we plot the results of the Casimir energy using formula \eqref{eq:Casimir_cond_31} (the data used for the plots are listed in \autoref{tab:Energy_cond_1_31} and \autoref{tab:Energy_cond_2_31}). As in the previous cases, we can see how all the points for different coupling constants $\beta$ and distances between the boundary walls $N_L$ follow the same behaviour indicating that we are close to the continuum limit and there are not large discretization errors associated with the size of $N_L$. The decay of the Casimir energy seems to be much slower than in the \acrshort*{pbc} case. We can test this behaviour by fitting the results with the  exponential formula
\begin{equation}\label{eq:Casimir_cond_31_exp_fit}
	\frac{E_C(L\sqrt{\sigma})}{A\sigma^{3/2}}=-\frac{C}{(\sqrt{\sigma}L)^\nu}e^{-2\mu L\sqrt{\sigma}},
\end{equation}
where the parameters are the multiplicative constants $C$, the mass of the exponential decay $\mu$ and the exponent $\nu$. For computing the best fitting values we minimize the chi squared defined in \eqref{eq:xi_square2} and label the resulting curve as ``Fit exp'' in the plots. We obtain that the best fitting parameters are  
$$C= 0.081 \pm 0.003, \mu= 0.12 \pm 0.02 \ \mathrm{and}\  \nu= 2.03 \pm 0.04,$$ that reaches $\chi^2=112$. First of all, we highlight that this value for the mass is much smaller then the mass obtained with the \acrshort*{pbc} $m_p\simeq 2\sqrt{\sigma}$ and the glueball mass $m_g\simeq3.78\sqrt{\sigma}$. We remark the  very slow exponential decay for the Casimir energy for \acrshort*{pccbc}, and that at  short and medium distances, the decay is  dominated by the potential factor $\nu$. Notice, that the value obtained for the exponent $\nu$ is again an intermediate value between the asymptotic limit of the massive scalar field formula \eqref{eq:Casimir_asymptotic_Dirichlet} $\nu=3/2$ and the Casimir energy for the massless scalar field \eqref{eq:massless_3d_dir} $\nu=3$.

\subsection{Continuum limit}
\begin{figure}[H]
	\centering
	\includegraphics[width=1\textwidth]{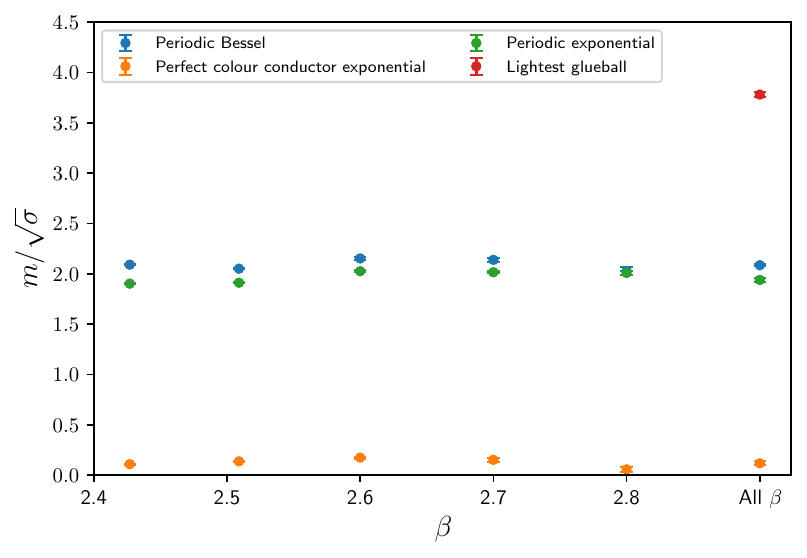}
	\caption{Mass in units of the string tension that drives the decay of the Casimir energy for the different $\beta$ and boundary conditions using the massive scalar field formula  \eqref{eq:Casimir_periodic_31_scalar_fit}, and the exponential fits \eqref{eq:Casimir_periodic_31_exp_fit} and \eqref{eq:Casimir_cond_31_exp_fit}. The continuum limit of the lightest glueball from \cite{athenodorou2021n} is also shown.}
	\label{fig:Casimir_betas_3d}
\end{figure}

\textcolor{black}{Now we analyze the continuum limit of the masses that drive the decay of the Casimir energy by computing them for each $\beta$ separately. In \autoref{fig:Casimir_betas_3d} we plot the masses we obtain for the different values of $\beta$ for both boundary conditions. The results using formulas \eqref{eq:Casimir_periodic_31_exp_fit} and \eqref{eq:Casimir_cond_31_exp_fit} are labeled as `exponential'', and the ones using \eqref{eq:Casimir_periodic_31_scalar_fit} as `Bessel''. We can observe in the plot how for all the values of $\beta$ the two boundary conditions have clearly different masses and are consistent within the different values of $\beta$ and fitting formulas. Therefore, we can argue that we are close to the continuum limit in the computation of the mass that drives the exponential decay of the Casimir energy with a precision that clearly allow us to differentiate the value for each boundary condition and with the lightest glueball of the theory.}
\textcolor{black}{As we mentioned in the two dimensional case, this continuum limit has the problem that when comparing data with different physical distances between the boundary walls, the results with higher values of $\beta$ have larger statistical errors.}

\pagebreak
\section{Finite volume errors}\label{sec:su31_lattice_effects}
In this section we shall show that the errors associated to the finite volume of the lattice are smaller than the statistical error of the Casimir energy. First, we analyze the errors associated to the finite size of the Euclidean time and longitudinal spatial directions $N_A$.

\begin{figure}[H]
	\centering
	\includegraphics[width=1\textwidth]{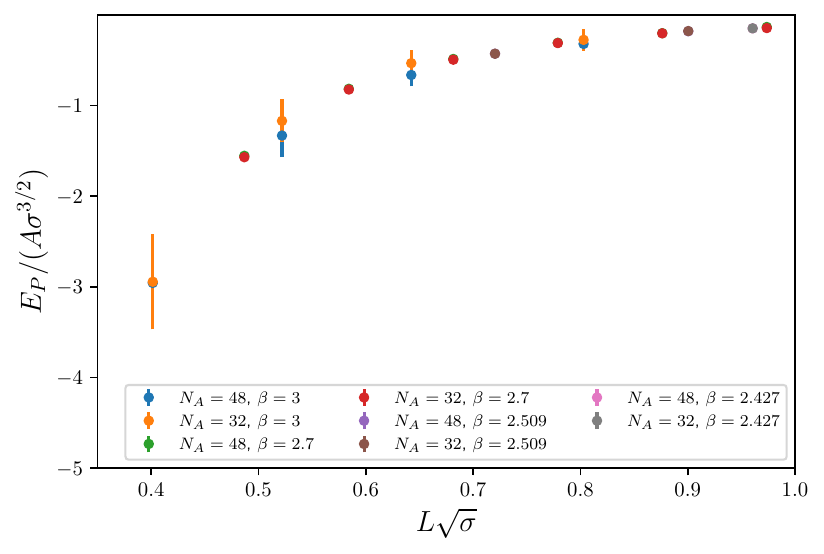}
	\caption{Dimensionless Casimir energy for \acrshort*{pbc} in a pure $SU(2)$ gauge theory and a four dimensional lattice for different couplings and lattice sizes with parameters $N_A=\{32,48\}$, $N_{L}=4-20$ and $\beta=2.427-3$.}
	\label{fig:Casimir_periodic_size_31}
\end{figure}

\begin{figure}[H]
	\centering
	\includegraphics[width=1\textwidth]{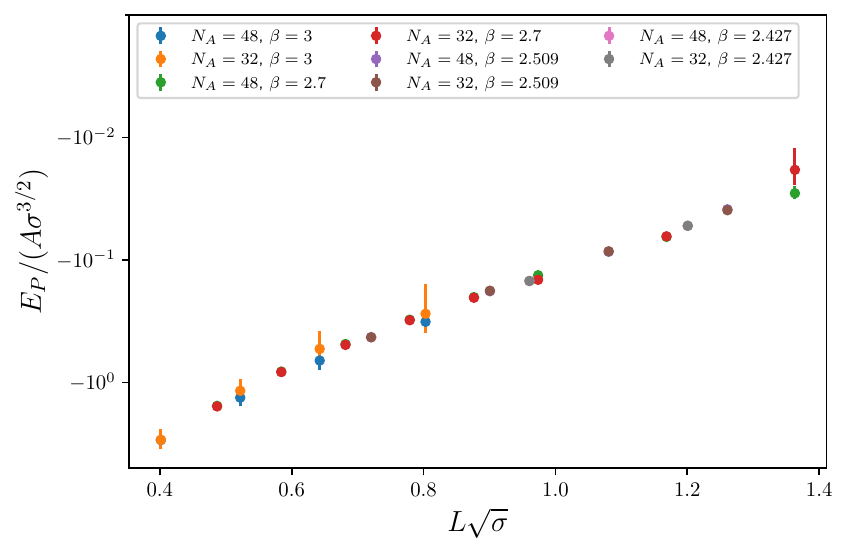}
	\caption{Dimensionless Casimir energy in logarithmic scale for \acrshort*{pbc} in a pure $SU(2)$ gauge theory and a four dimensional lattice for different couplings and lattice sizes with parameters $N_A=\{32,48\}$, $N_{L}=4-20$ and $\beta=2.427-3$.}
	\label{fig:Casimir_periodic_size_log_31}
\end{figure}

In \autoref{fig:Casimir_periodic_size_31} (linear scale) and \autoref{fig:Casimir_periodic_size_log_31} (logarithmic scale) we plot the Casimir energy for four different values of the coupling constant $\beta$ and two different values of $N_A=\{32,48\}$ with \acrshort*{pbc} (the corresponding data are listed in \autoref{tab:Energy_size_periodic_1_31}). We can see how all the points with different sizes $N_A$ are compatible within the statistical error, in particular, they are all within the $1.5\sigma$ domain. 
\vspace{3cm}
\vspace{5cm}

\begin{figure}[h]
	\centering
	\includegraphics[width=1\textwidth]{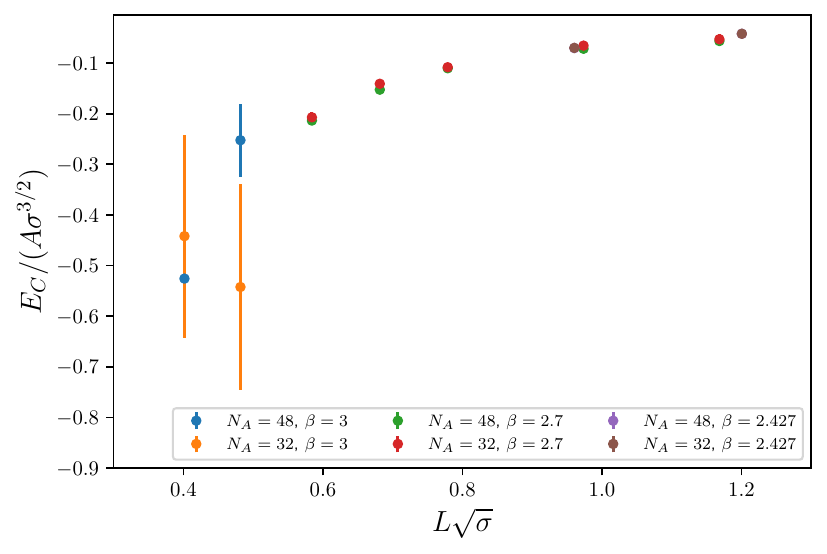}
	\caption{Dimensionless Casimir energy with \acrshort*{pccbc} on the transverse direction in a pure $SU(2)$ gauge theory in a four dimensional lattice for different couplings and lattice sizes with parameters $N_A=\{32,48\}$, $N_{L}=4-20$ and $\beta=2.427-3$.}
	\label{fig:Casimir_cond_size_31}
\end{figure}

\begin{figure}[H]
	\centering
	\includegraphics[width=1\textwidth]{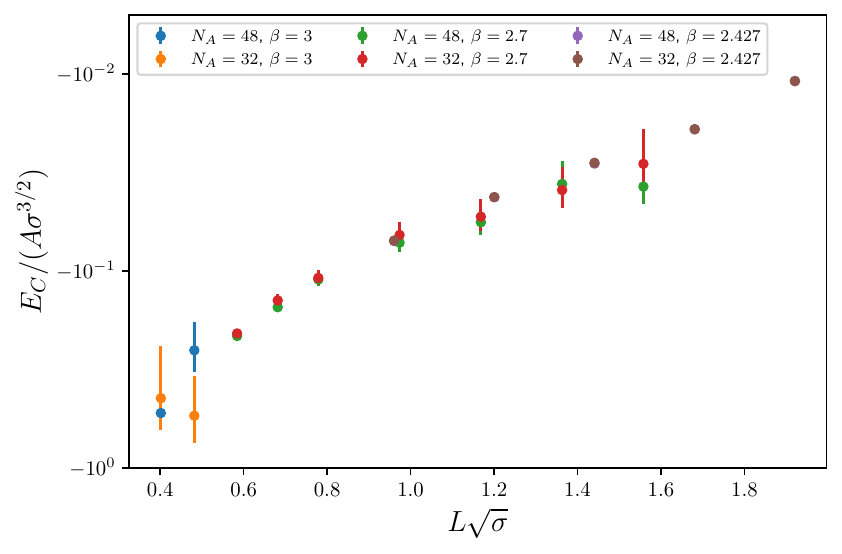}
	\caption{Dimensionless Casimir energy in logarithmic scale for \acrshort*{pccbc} on the transverse direction in a pure $SU(2)$ gauge theory and a four dimensional lattice for different couplings and lattice sizes with parameters $N_A=\{32,48\}$, $N_{L}=4-20$ and $\beta=2.427-3$.}
	\label{fig:Casimir_cond_size_log_31}
\end{figure}

We repeat the same analysis with \acrshort*{pccbc} in \autoref{fig:Casimir_cond_size_31} (linear scale) and \autoref{fig:Casimir_cond_size_log_31} (logarithmic scale) (the numeric values of the simulation results are listed in \autoref{tab:Energy_size_cond_1_31} and \autoref{tab:Energy_size_cond_2_31}). We can see how in this case all the points with the different lattices sizes $N_A=\{32,48\}$ are also compatible with each other considering the statistical error since their difference is in all cases lower than $1.5\sigma$. Thus, we can conclude that the errors associated with $N_A$ are under control in our simulations for both types of boundary conditions.
\vspace{5cm}

\begin{figure}[H]
	\centering
	\includegraphics[width=1\textwidth]{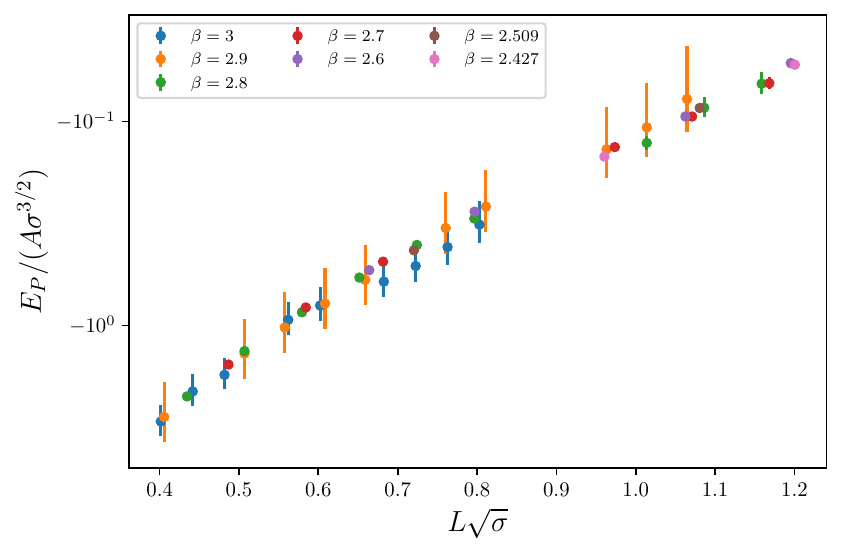}
	\caption{Dimensionless Casimir energy in logarithmic scale with \acrshort*{pbc}  in a pure $SU(2)$ gauge  theory and a four dimensional lattice for different couplings and lattice sizes with parameters $N_A=48$, $N_{L}=4-21$ and $\beta=2.427-3$. Only values which have a couple of  points within a close physical distance $L\sqrt{\sigma}$ but different $N_L$ are plotted. }
	\label{fig:Casimir_periodic_size_z_31}
\end{figure}

Finally, we analyze the behaviour of the results with the lattice size $N_L$. We follow the same strategy than in 2+1 dimensions, where we consider points that are close to each other in the scale of physical distances $L\sqrt{\sigma}$ but have different lattice sizes $N_L$, and compare the results they give for the Casimir energy.  This allows us to check if there are significant errors due to varying the transverse direction size $N_L$. In \autoref{fig:Casimir_periodic_size_z_31} we display the Casimir energy for \acrshort*{pbc} only when there are two or more points with a close enough physical distance. We can see how all the points seem compatible within error and there are no large errors even for the smallest case where $N_L=4$ for the lower values of $\beta$.

\begin{figure}[H]
	\centering
	\includegraphics[width=1\textwidth]{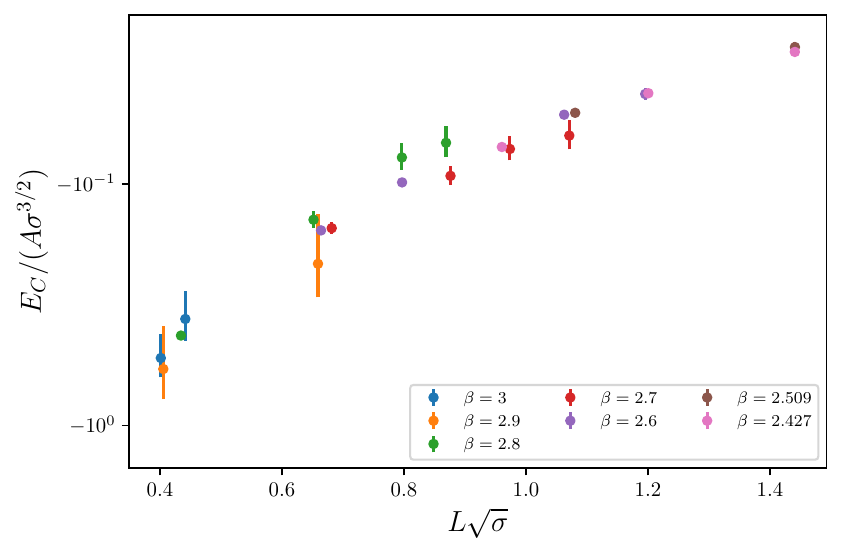}
	\caption{Dimensionless Casimir energy in logarithmic scale with \acrshort*{pccbc} on the transverse direction in a pure $SU(2)$ gauge theory and a four dimensional lattice for different couplings and lattice sizes with parameters $N_A=48$, $N_{L}=4-21$ and $\beta=2.427-3$. Only values which have a couple of  points within a close physical distance $L\sqrt{\sigma}$ but different $N_L$ are plotted. }
	\label{fig:Casimir_cond_size_z_31}
\end{figure}

\pagebreak
Now, we repeat the same calculation but with \acrshort*{pccbc} in \autoref{fig:Casimir_cond_size_z_31}. We observe how all the points with different $\beta$ and $N_L$ with similar physical distances are compatible with each other within the range of statistical errors. Thus, we have shown how for \acrshort*{pbc} and \acrshort*{pccbc} the different transverse lattices sizes $N_L$ that have been analyzed do not produce large errors in comparison with the statistical errors we achieve.

    \chapter{Conclusions}\label{chp:Ca}
    In this work we present non-perturbative calculations of the Casimir energy in various
pure non-abelian gauge theories using Monte Carlo simulations. By analyzing
the exponential decay of this  energy with the distance between the edges and
comparing it with the behaviour of the continuum limit, the mass scales
responsible for this behaviour have been obtained for different boundary conditions and compared with the masses of the glueballs in the theory.

We have studied the effective action of a massive scalar field confined between two
homogeneous and infinite walls in the low-temperature limit for  general boundary conditions. Using the zeta function renormalization method we have obtained the finite contributions to this effective action, including the Casimir energy for general boundary conditions characterized by the spectral function \eqref{eq:spectral_function}, generalizing that obtained in Ref. \cite{Boundary_general_2013}  for the massless case.

When analyzing the asymptotic behaviour of this Casimir energy with the effective distance
between the walls in the limit $mL \gg 1$, we have shown that it classifies the boundary conditions
into two distinct families depending on the rate of their exponential decay.
When the boundary condition involves independent  conditions on
each boundary  wall (e.g., Dirichlet and Neumann), the Casimir energy decays exponentially
with twice the mass $(\propto e^{-2mL})$. However, when the boundary condition
relates the values of the fields on the two walls with each other (e.g., periodic and anti-periodic), the exponential decay rate is given by the field mass $(\propto e^{-mL})$. In
summary, the first family of conditions has an exponential decay rate
of the Casimir energy that is twice that of the second family. These two families can be
characterized in the parametrization of boundary conditions by  $U(2)$  matrices \eqref{eq:parametrization}  
by their dependence on
$\sigma_1$. Those conditions that have rapid decay do not depend on 
$\sigma_1$ in this
parametrization, whereas those that have slower decay rates do depend on $\sigma_1$.
Furthermore, the same asymptotic behaviour is observed in the rest of the
terms of the free energy that depend on temperature. 

In \autoref{chp:scalar_lattice}, the Casimir energy was analyzed by
using the lattice formalism for massive scalar fields.
Five different boundary conditions were implemented on the lattice: periodic,
Dirichlet, anti-periodic, Neumann, and Zaremba; and their partition functions have been calculated analytically.
Analyzing the structure of the vacuum by studying the behaviour of the expectation value of the internal energy, we find the same type of behaviour as in the continuum limit  \eqref{eq:divergences_energy},
 where there is a contribution from the bulk that is proportional to the volume and independent
of the boundary conditions, in addition to a boundary term associated with the
existence of the walls that is proportional to the surface area of the walls, and the Casimir energy. Using the same prescription for the renormalization of the energy as in the
continuum limit \eqref{eq:zeta_re}, a Casimir energy is obtained that is compatible with the continuum formula
(\autoref{fig:Casimir_renormalization_lattice}) and with the same asymptotic behaviour that classifies the boundary conditions
into two distinct families.

However, we have explored the application of an alternative procedure for
renormalization, which is optimal for situations requiring numerical calculations.
In this procedure, we use the fact that the boundary term in periodic boundary conditions cancels out, which
allows us to calculate the contribution of the bulk using the energy of a large lattice
with periodic boundary conditions in all directions  \eqref{eq:bulk_periodic_scalar}. For boundary conditions
that do have a boundary term,  we have to subtract the energy of another
larger lattice with the same boundary conditions (see e.g. \eqref{eq:Casimir_Dirichlet_lattice}) where the Casimir effect is negligible.
The advantage of this method is that lattices with different sizes in the transverse direction
but the same mass have the same values for the  bulk  and boundary terms,
thus reducing the extra number of lattices that must be calculated to obtain the
Casimir energy. This alternative renormalization procedure is successful
for calculating the Casimir energy for any boundary condition, since we obtain  the same values  as in the continuum limit (Figures \ref{fig:Casimir_periodic_lattice}, \ref{fig:Casimir_Dirichlet_lattice}, \ref{fig:Casimir_antiperiodic_lattice}, \ref{fig:Casimir_Neumann_lattice} and \ref{fig:Casimir_Zaremba_lattice}). In particular, the property
of the continuum theory that the Casimir energy with Dirichlet and Neumann boundary conditions is the
same is also maintained in this calculation on the lattice. Furthermore, the correct
asymptotic behaviour is observed, where for Dirichlet, Neumann and Zaremba boundary conditions
the Casimir energy decays exponentially at twice the rate it does for periodic and anti-periodic boundary conditions.
The correct sign is also obtained, in which for periodic, Dirichlet, and Neumann boundary conditions  the Casimir energy is negative, whereas for
anti-periodic and Zaremba boundary conditions the Casimir energy is positive.
 Therefore, we can conclude that we have
successfully calculated the Casimir energy on the lattice for a massive scalar field
using the energy-momentum tensor and an optimal renormalization method for numerical calculations of this theory.

In \autoref{chp:su2_21}, we analyze the Casimir energy results for the Yang-Mills SU(2) theory in
2+1 dimensions obtained by means of Monte Carlo simulations.
We use two different boundary conditions in the transverse direction: periodic 
and perfect colour conductor boundary conditions. The same vacuum structure is observed
in the energy as in the scalar case, where there is a contribution from the bulk that is common  for both boundary conditions and a boundary term that cancels out for
 periodic boundary conditions (Figures \ref{fig:Energy_periodic_su21} and \ref{fig:Energy_conductor_su21}). This allows us to apply the
same renormalization strategy as in the scalar case to obtain the Casimir energy
on the lattice for the gauge fields.

For both boundary conditions, an exponential decay of the Casimir energy with the distance between the walls, characteristic of massive theories is observed. Adjusting the results obtained with the perfect colour conductor boundary conditions
(Figures \ref{fig:Casimir_perfect_cond} and \ref{fig:Casimir_perfect_cond_log}) to the Casimir energy formula of the massive scalar field with
Neumann boundary conditions in the continuum \eqref{eq:Cas_dir_2d}, we obtain that the best
parameters are $m=(0.89\pm 0.07) \sqrt{\sigma}$ and $G=3.26\pm 0.06$. This result is consistent
with the claim that the Casimir energy for perfect colour conductor boundary conditions is     given
by the Casimir energy of three scalar fields with Neumann boundary conditions, which
is the behaviour that emerges from the Karabali-Nair parametrization \cite{karabali2018casimir} where three scalar fields with mass $m_{\mathrm{K}}=0.92 \sqrt{\sigma}$ parametrize the invariant degrees of freedom with $SU(2)$ and 2+1 dimensions. It is remarkable  how this mass value is much lower than the lightest glueball in the Yang-Mills theory ($m_g\simeq 4.7\sqrt{\sigma}$ \cite{teper1998n,lucini2002n,athenodorou2017n}), and how this exponential decay
exhibited by the Casimir energy behaviour cannot be due to
a mass as high as that of the glueball. The results also hold when
we perform the fit (Figures \ref{fig:Casimir_perfect_cond_exp_lin} y \ref{fig:Casimir_perfect_cond_exp})  directly to an exponential decay,
where the best fit gives a mass value $m=(1.14\pm 0.15)\sqrt{\sigma}$, which is still compatible with the
result of the previous fit and the expected mass in the Karabali-Nair parametrization. \textcolor{black}{This result with perfect colour conductor boundary conditions was already found in Ref. \cite{chernodub2018casimir}, where it was shown that a compatible mass with the Karabali-Nair parametrization drives the exponential decay of the Casimir energy \cite{karabali2018casimir}}.

However, for periodic boundary conditions the best adjustment to the massive scalar field formula \eqref{eq:Cas_per_2d} (Figures \ref{fig:Casimir_periodic_cond} and \ref{fig:Casimir_periodic_cond_log}) gives
$G=2.90\pm 0.02$ and $m=(3.97\pm 0.03) \sqrt{\sigma}$, whereas if the fit is made with
an exponential (Figures \ref{fig:Casimir_periodic_cond_exp_lin}  and  \ref{fig:Casimir_periodic_cond_exp}), a similar result is obtained for the mass $m=(3.72\pm 0.08)\sqrt{\sigma}$. In both cases, the result is compatible with the fact that this
Casimir energy may be due to the contributions of three scalar fields with equal mass,
but which is much greater than expected in the Karabali-Nair parametrization. Even so,
the mass obtained is still less than that of the lightest glueball ($4.7\sqrt{\sigma}$).

From the results in both boundary conditions it can be concluded that it excludes the description
by an effective free scalar field, since this establishes a very different ratio between the two energies than the one observed. The masses obtained for the perfect colour conductor are not twice than those found with periodic boundary conditions, but actually less than half of their value. This fact also allows to exclude the validity of the conjecture formulated by D. Karabali and V. P. Nair. The discrepancy between the mass obtained from this Casimir energy and that expected in
the Karabali-Nair parametrization in periodic boundary conditions suggests that the validity of the approximations used to obtain this parametrization depends on the boundary conditions.

In \autoref{chp:su2_31}, we analyze the results of the Casimir energy in the case of Yang-Mills $SU(2)$ theories
in 3+1 dimensions. A key difference between this case and that of 2+1 dimensions
is the form of the first component of the energy-momentum tensor. In the
3+1 dimensional lattice, this is given by the difference between the three plaquettes
involving the time direction and the three that do not. This difference causes
the energy to tend to zero in the limit of large lattices and periodic boundary conditions
(\autoref{fig:Energy_periodic_su31}). Furthermore, this functional form of the energy-momentum tensor cancels
the contribution of the bulk to the energy and simplifies the renormalization process. \textcolor{black}{This cancellation of the bulk contribution in 3+1 dimensions was previously noticed in Ref. \cite{chernodub2023boundary} for $SU(3)$ with perfect colour conductor boundary conditions.} Due to this cancellation, when using periodic boundary conditions, the Casimir energy
is obtained directly from the expectation value of the energy \eqref{eq:Energy_31}.

The Casimir energy with periodic boundary conditions in 3+1 dimensions also exhibits an exponential decay with the distance between the walls (Figures \ref{fig:Casimir_periodic_31} and \ref{fig:Casimir_periodic_31_log}) with
a rate given by  $G=2.290\pm0.008$ and $m=(2.086\pm0.007)\sqrt{\sigma}$ by means of an
asymptotic fit to the behaviour of a scalar field, suggesting that the fit is
compatible with the description using two scalar fields of the same mass instead of
three as in 2+1 dimensions. If the fit is performed (Figures \ref{fig:Casimir_periodic_31} and \ref{fig:Casimir_periodic_31_log}) using
a pure exponential the result is $m=(1.94\pm0.02)\sqrt{\sigma}$, a value close to that obtained with the previous fit. In both cases, the mass value is much smaller than the lightest glueball mass
$m_g\simeq3.78 \sqrt{\sigma}$ \cite{athenodorou2021n,lucini2001n}.

Finally, we calculate the Casimir energy for perfect colour conductor boundary conditions
in 3+1 dimensions. In this case, the exponential decay of the Casimir energy
with the distance between the walls is much slower than in the previous cases
(Figures \ref{fig:Casimir_cond_31} and  \ref{fig:Casimir_cond_31_log}). This behaviour can be seen by adjusting this Casimir energy
to an exponential decay for which the best fit is obtained with a
mass $m= (0.12 \pm 0.02)\sqrt{\sigma}$, that is much smaller than that obtained for periodic boundary conditions and the lightest glueball mass.

The results in 2+1 and 3+1 dimensions show similar behaviour. The
mass that generates the exponential decay for Casimir energy is not the glueball mass. Moreover it  is always lower. The same pattern is found
in the Casimir energy, with perfect colour conductor boundary conditions it decays slower than for periodic boundary conditions in both 2+1 and 3+1
dimensional cases. This is the opposite of what would be expected from the description
of the behaviour of low-energy theories by means of a scalar field, since
periodic boundary conditions belong to the family with slower exponential decay,
 while perfect colour conductor boundary conditions (which are analogous to Dirichlet
in the scalar field) should belong to the family with the fastest exponential decay. This behaviour shows that the mass dependence that
generates the exponential decay of the Casimir energy in non-abelian gauge theories for the different boundary conditions is opposite to the behaviour generated by scalar field theories. Thus, we can exclude any possibility of description of the low energy regime of gauge theories in terms of free scalar fields  in spite of the fact the lightest glueball is scalar.

Extending these simulations to more types of boundary conditions could help
to better understand this problem and confirm the dependence of this mass driving the
exponential decay of the Casimir energy on the boundary conditions. Furthermore,
performing a similar analysis for $SU(3)$ could also clarify this situation. \textcolor{black}{In Ref. \cite{chernodub2023boundary} 
the case of a 3+1 dimensional gauge theory  with $SU(3)$  and perfect colour conductor boundary conditions has already been calculated,
and it was found that the exponential decay of the Casimir energy is driven by a mass $m=(1.0\pm 0.1)\sqrt{\sigma}$ which is much
lower than that of the lightest glueball $m_g\simeq 3.405\sqrt{\sigma}$ \cite{athenodorou2021n}, suggesting a behaviour similar than that of the $SU(2)$ case presented in this work.}


    \begin{appendices}
    	\chapter{Modified Bessel functions of the second kind}\label{sec:Bessel}
In this Appendix we are going to include some of the calculations involving the modified Bessel functions of the second kind $K_q(z)$ that are used in \autoref{chp:scalar_cont}. These functions are defined by being solutions of the modified Bessel's equation \cite{NIST:DLMF}
\begin{equation}
z^2\frac{d^2 w}{dz^2}+z\frac{dw}{dz}-(z^2+q^2)w=0
\end{equation}
and are characterized by its asymptotic behaviour 
\begin{equation}\label{eq:Bessel_asymptotic}
	K_q(z)= \sqrt{\frac{\pi}{2z}}e^{-z}\left(1+O\left(\frac{1}{z}\right)\right),
\end{equation}
as $z\rightarrow\infty$. Also, it will be of interest the behaviour when the argument goes to zero $z\rightarrow 0$ for some calculations in the massless limit
\begin{equation}\label{eq:Bessel_asymptotic_zero}
	K_q(z)=\frac{2^{q-1}\Gamma(q)}{z^q}\left(1+O(z^2)+O(z^{2q})\right).
\end{equation}
\section{Integral representation}\label{sec:Bessel_integral}
In this section we explore how we can use the integral representation of these Bessel functions
\begin{equation}\label{eq:Bessel_Integral}
	K_q(z)=	\frac{\sqrt{\pi}}{(q-1/2)!}\left(\frac{1}{2}z\right)^q\int_{1}^\infty e^{-zt}\left(t^2-1\right)^{q-1/2}dt,
\end{equation}
which is valid when $Re(q)>\frac{1}{2}$ and the absolute value of the phase of $z$ is smaller than $\pi/2$, for simplifying the integrals we obtain as a result of the Casimir energy for the particular boundary conditions analyzed in \autoref{sec:Particular_conditions}.

For \acrshort*{dbc} the integrals in \eqref{eq:Casimir_Dirichlet_integral} for the Casimir energy have the form
\begin{equation}\label{eq:Dirichlet_integral}
	I_D=\int_{m}^{\infty}\left(k^2-m^2\right)^{n}(1-\coth(kL))\ dk,
\end{equation}
we can expand the spectral term (second one) in a power series as
\begin{equation}
	1-\coth(kL)=-2\sum_{j=1}^\infty e^{-2jkL}
\end{equation}
when $kL\not =0$, i.e. the massive case. Inserting this power series into the integral \eqref{eq:Dirichlet_integral} we arrive at
\begin{equation}
	I_D=-2\sum_{j=1}^\infty\int_{m}^{\infty}\left(k^2-m^2\right)^{n}e^{-2jkL}\ dk,
\end{equation}
which we can trivially identify with the integral representation of the Bessel function \eqref{eq:Bessel_Integral} as
\begin{equation}\label{eq:Bessel_Dirichlet_result}
	I_D=-2\left(\frac{m}{L}\right)^{n+\frac{1}{2}}\frac{n!}{\sqrt \pi}\sum_{j=1}^\infty\frac{1}{j^{n+\frac{1}{2}}}K_{n+\frac{1}{2}}(2jmL),
\end{equation}
where we have used the change of variables; $t=k/m$, $z=2jmL$ and $q=n+1/2$.

For the \acrshort*{pbc}, the integral of the Casimir energy has the form
\begin{equation}
	I_P=\int_{m}^{\infty}\left(k^2-m^2\right)^{n}(1-\coth(kL/2))\ dk
\end{equation}
which is the same as the \acrshort*{dbc} case but with half the distance between the walls $L$. Therefore, the result in terms of a sum of Bessel functions is just the same as the \acrshort*{dbc} case taking into account the contribution of this factor two in $L$ when doing the change of variables, obtaining
\begin{equation}\label{eq:Bessel_periodic_result}
	I_P=-2^{n+\frac{3}{2}}\left(\frac{m}{L}\right)^{n+\frac{1}{2}}\frac{n!}{\sqrt \pi}\sum_{j=1}^\infty\frac{1}{j^{n+\frac{1}{2}}}K_{n+\frac{1}{2}}(jmL).
\end{equation}

For the \acrshort*{abc} case, the integral we want to simplify is 
\begin{equation}
	I_A=\int_{m}^{\infty}\left(k^2-m^2\right)^{n}(1-\tanh(kL/2))\ dk,
\end{equation}
we can take the same approach as before of using the power series of the spectral term taking advantage of working in the $m\not =0$ case
\begin{equation}
	1-\tanh(kL/2)=2\sum_{j=1}^\infty (-1)^{j+1} e^{-jkL}.
\end{equation}
Therefore, the integral of the Casimir energy can be expressed as
\begin{equation}
I_A=2\sum_{j=1}^{\infty}(-1)^{j+1}\int_{m}^{\infty}(k^2-m^2)^ne^{-jkL}\ dk,
\end{equation}
in which we can use a similar change of variables as in the previous cases to arrive at
\begin{equation}\label{eq:Bessel_anti_result}
	I_A=2^{n+\frac{3}{2}}\left(\frac{m}{L}\right)^{n+\frac{1}{2}}\frac{n!}{\sqrt \pi}\sum_{j=1}^\infty\frac{(-1)^{j+1}}{j^{n+\frac{1}{2}}}K_{n+\frac{1}{2}}(jmL).
\end{equation}
Finally, in \acrshort*{zbc}, the integral we have is
\begin{equation}
	I_Z=\int_{m}^{\infty}\left(k^2-m^2\right)^{n}(1-\tanh(kL))\ dk,
\end{equation}
that is the same expression than in \acrshort*{abc} but with double distance between the walls $L$. Taking this into account, the integral in terms of Bessel functions is given by
\begin{equation}\label{eq:Bessel_zarembat_result}
	I_Z=2\left(\frac{m}{L}\right)^{n+\frac{1}{2}}\frac{n!}{\sqrt \pi}\sum_{j=1}^\infty\frac{(-1)^{j+1}}{j^{n+\frac{1}{2}}}K_{n+\frac{1}{2}}(2jmL).
\end{equation}
\section{Relation with Polylogarithms}
In this section we shall use the polylogarithm function to simplify the sum of Bessel functions obtained in the 2+1 dimensions Casimir energy and 3+1 dimensions free energy. The Polylogarithm function is defined by the power series
\begin{equation}\label{eq:Polylog}
	\text{Li}_s(z)=\sum_{k=1}^\infty\frac{z^k}{k^s}
\end{equation}
and is valid for any complex $s$ and $|z|<1$, but it can be extended to $|z|\geq1$ by analytic continuation.
We write explicitly the Bessel function we obtain in the two dimensional formulas for the Casimir energy \eqref{eq:Dirichlet_2d} and use for the particular case of $q=3/2$ the expression \eqref{eq:bessel_semi_integer}
\begin{equation}
	K_{3/2}(x)=\sqrt{\frac{\pi}{2x}}e^{-x}\left(1+\frac{1}{x}\right).
\end{equation}
Now, we can insert this expression in the sum of Bessel functions that gives the Casimir energy \eqref{eq:Dirichlet_2d}
\begin{align}
	\sum_{j=1}^{\infty}\frac{K_{3/2}(jmL)}{j^{3/2}}=&\frac{\sqrt{\pi}}{\sqrt 2(mL)^{3/2}}\left(mL\sum_{j=1}^\infty\frac{e^{-jmL}}{j^{2}}+\sum_{j=1}^\infty\frac{e^{-jmL}}{j^{3}}\right)\nonumber\\ \label{eq:Bessel_poly_dir}
	=&\frac{\sqrt{\pi}}{\sqrt 2(mL)^{3/2}}\left(mL\ \text{Li}_2\left(e^{-mL}\right)+\text{Li}_3\left(e^{-mL}\right)\right),
\end{align}
this formula can be used for simplifying the 2+1 dimensional Casimir energy in \acrshort*{dbc}/\acrshort*{nbc} (with an extra factor two in the argument of the Bessel function) and \acrshort*{pbc}. In the case of \acrshort*{abc} and \acrshort*{zbc} the sum of Bessel functions has the following form instead
\begin{align}
	\sum_{j=1}^{\infty}\frac{(-1)^{j+1}}{j^{3/2}}K_{3/2}(jmL)=&-\frac{\sqrt{\pi}}{\sqrt 2(mL)^{3/2}}\left(mL\sum_{j=1}^\infty\frac{(-1)^{j}e^{-jmL}}{j^{2}}+\sum_{j=1}^\infty\frac{(-1)^{j}e^{-jmL}}{j^{3}}\right)\nonumber\\ \label{eq:Bessel_poly_anti}
	&-\frac{\sqrt{\pi}}{\sqrt 2(mL)^{3/2}}\left(mL\ \text{Li}_2\left(-e^{-mL}\right)+\text{Li}_3\left(-e^{-mL}\right)\right),
\end{align}
which again we can use to simplify the expressions of the 2+1 dimensional Casimir energy Casimir energy for \acrshort*{abc} and \acrshort*{zbc} (with an extra factor two in the argument of the Bessel function).

\chapter{Results of $SU(2)$ simulations in 2+1 dimensions}\label{ch:21_values}
In this Appendix we show explicitly the numerical values we used for the results presented in \autoref{chp:su2_21}. The statistical error is shown in parenthesis and corresponds to the last decimal places, i.e $0.467\pm 0.019\equiv 0.467(19)$.
\section{Renormalization}
First, we have the simulations we used for testing the energy observable behaviour in \autoref{sec:reno_21}. For this, we do not need much statistical precision as when we actually compute the Casimir energy. In the smaller lattices higher statistic is used since they will also be used for computing the Casimir energy.
\begin{table}[H]
	\centering
	\begin{tabular}{|c|c|c|c|c|c|}
		\hline
		\rule{0pt}{12pt}$\beta$& $N_0\times N_1\times N_2$ & $N_{\text{therm}}$& $N_{\text{measure}}$ &$\frac{1}{2}\braket{\text{tr}\left(P_{01}\right)}-1$ & $\frac{1}{2}\braket{\text{tr }(P_{\mu\nu})}$\\
		\hline
		100&$96\times96\times10$ &$5\cdot 10^4$ & $5\cdot 10^5$& -0.01002840(4) & 0.98997676(3)\\\hline
		100&$96\times96\times20$ &$5\cdot 10^4$ & $7\cdot 10^5$& -0.01002409(2) & 0.989976459(16)\\\hline
		100&$96\times96\times30$ &$10^4$ & $10^4$& -0.01002353(16) & 0.98997659(11)\\\hline
		100&$96\times96\times40$ &$10^4$ & $10^4$& -0.01002367(14) & 0.98997630(9)\\\hline
		100&$96\times96\times50$ &$10^4$ & $10^4$& -0.01002357(13) & 0.98997638(9)\\\hline
		100&$96\times96\times60$ &$10^4$ & $10^4$& -0.01002350(12) & 0.98997632(8)\\\hline
		100&$96\times96\times70$ &$10^4$ & $10^4$& -0.01002357(11) & 0.98997642(7)\\\hline
		100&$96\times96\times80$ &$10^4$ & $10^4$& -0.01002332(10) & 0.98997657(7)\\\hline
		100&$96\times96\times90$ &$10^4$ & $10^4$& -0.01002347(10) & 0.98997646(6)\\\hline
		100&$96\times96\times100$ &$10^4$ & $10^4$& -0.01002359(9) & 0.98997644(6)\\\hline
	\end{tabular}
	\caption{Mean value of the plaquette $P_{01}$ and the average of all the plaquettes $P_{\mu\nu}$ with coupling constant $\beta=100$ and lattice sizes $N_0N_1N_2$. \acrshort*{pbc} are used in every direction. $N_{\text{therm}}$ and $N_{\text{measure}}$ are the \acrshort*{mc} steps used for thermalization and measuring respectively.}
	\label{tab:Energy_renom_periodic_1}
\end{table}
\begin{table}[H]
	\centering
	\begin{tabular}{|c|c|c|c|c|c|}
		\hline
		\rule{0pt}{12pt}$\beta$& $N_0\times N_1\times N_2$ & $N_{\text{therm}}$& $N_{\text{measure}}$ &$\frac{1}{2}\braket{\text{tr}\left(P_{01}\right)}-1$ & $\frac{1}{2}\braket{\text{tr }(P_{\mu\nu})}$\\
		\hline
		80&$96\times96\times10$ &$5\cdot 10^4$ & $5\cdot 10^5$& -0.01254270(5) & 0.98746351(3)\\\hline
		80&$96\times96\times20$ &$5\cdot 10^4$ & $7\cdot 10^5$& -0.01253755(3) & 0.98746306(2)\\\hline
		80&$96\times96\times30$ &$10^4$ & $10^4$& -0.0125374(2) & 0.98746259(14)\\\hline
		80&$96\times96\times40$ &$10^4$ & $10^4$& -0.01253711(18) & 0.98746293(12)\\\hline
		80&$96\times96\times50$ &$10^4$ & $10^4$& -0.01253690(16) & 0.98746294(11)\\\hline
		80&$96\times96\times60$ &$10^4$ & $10^4$& -0.01253704(15) & 0.98746297(10)\\\hline
		80&$96\times96\times70$ &$10^4$ & $10^4$& -0.01253719(13) & 0.98746297(9)\\\hline
		80&$96\times96\times80$ &$10^4$ & $10^4$& -0.01253693(12) & 0.98746297(8)\\\hline
		80&$96\times96\times90$ &$10^4$ & $10^4$& -0.01253688(12) & 0.98746304(8)\\\hline
		80&$96\times96\times100$ &$10^4$ & $10^4$& -0.01253702(11) & 0.98746297(8)\\\hline
		60&$96\times96\times10$ &$5\cdot 10^4$ & $5\cdot 10^5$& -0.01674001(7) & 0.98326789(4)\\\hline
		60&$96\times96\times20$ &$5\cdot 10^4$ & $7\cdot 10^5$& -0.01673352(4) & 0.98326731(3)\\\hline
		60&$96\times96\times30$ &$10^4$ & $10^4$& -0.0167328(3) & 0.98326721(19)\\\hline
		60&$96\times96\times40$ &$10^4$ & $10^4$& -0.0167329(2) & 0.98326730(16)\\\hline
		60&$96\times96\times50$ &$10^4$ & $10^4$& -0.0167327(2) & 0.98326721(14)\\\hline
		60&$96\times96\times60$ &$10^4$ & $10^4$& -0.01673309(19) & 0.98326717(13)\\\hline
		60&$96\times96\times70$ &$10^4$ & $10^4$& -0.01673296(18) & 0.98326683(12)\\\hline
		60&$96\times96\times80$ &$10^4$ & $10^4$& -0.01673263(17) & 0.98326709(11)\\\hline
		60&$96\times96\times90$ &$10^4$ & $10^4$& -0.01673264(16) & 0.98326732(11)\\\hline
		60&$96\times96\times100$ &$10^4$ & $10^4$& -0.01673285(15) & 0.98326703(10)\\\hline
		50&$96\times96\times10$ &$5\cdot 10^4$ & $2.5\cdot 10^5$& -0.02010364(11) & 0.97990539(8)\\\hline
		50&$96\times96\times20$ &$5\cdot 10^4$ & $9\cdot 10^5$& -0.02009647(4) & 0.97990435(3)\\\hline
		50&$96\times96\times30$ &$10^4$ & $10^4$& -0.0200953(3) & 0.9799044(2)\\\hline
		50&$96\times96\times40$ &$10^4$ & $10^4$& -0.0200961(3) & 0.97990404(19)\\\hline
		50&$96\times96\times50$ &$10^4$ & $10^4$& -0.0200956(3) & 0.97990405(17)\\\hline
		50&$96\times96\times60$ &$10^4$ & $10^4$& -0.0200959(2) & 0.97990397(16)\\\hline
		50&$96\times96\times70$ &$10^4$ & $10^4$& -0.0200959(2) & 0.97990405(14)\\\hline
		50&$96\times96\times80$ &$10^4$ & $10^4$& -0.0200956(2) & 0.97990428(14)\\\hline
		50&$96\times96\times90$ &$10^4$ & $10^4$& -0.02009596(19) & 0.97990402(13)\\\hline
		50&$96\times96\times100$ &$10^4$ & $10^4$& -0.02009571(18) & 0.97990443(12)\\\hline
		40&$96\times96\times10$ &$5\cdot 10^4$ &$5\cdot 10^5$& -0.02515960(10) & 0.97485086(7)\\\hline
		40&$96\times96\times20$ &$5\cdot 10^4$ & $7\cdot 10^5$& -0.02515121(6) & 0.97484952(4)\\\hline
		40&$96\times96\times30$ &$10^4$ & $10^4$& -0.0251516(4) & 0.9748486(3)\\\hline
		40&$96\times96\times40$ &$10^4$ & $10^4$& -0.0251507(4) & 0.9748494(2)\\\hline
		40&$96\times96\times50$ &$10^4$ & $10^4$& -0.0251509(3) & 0.9748490(2)\\\hline
		40&$96\times96\times60$ &$10^4$ & $10^4$& -0.0251508(3) & 0.9748492(2)\\\hline
		40&$96\times96\times70$ &$10^4$ & $10^4$& -0.0251511(3) & 0.97484848(18)\\\hline
		40&$96\times96\times80$ &$10^4$ & $10^4$& -0.0251514(3) & 0.97484874(17)\\\hline
		40&$96\times96\times90$ &$10^4$ & $10^4$& -0.0251511(2) & 0.97484917(16)\\\hline
		40&$96\times96\times100$ &$10^4$ & $10^4$& -0.0251510(2) & 0.97484892(15)\\\hline
	\end{tabular}
	\caption{Mean value of the plaquette $P_{01}$ and the average of all the plaquettes $P_{\mu\nu}$ with couplings constants $\beta=\{80,60,50,40\}$, lattice sizes $N_0N_1N_2$ and \acrshort*{pbc}. $N_{\text{therm}}$ and $N_{\text{measure}}$ are the \acrshort*{mc} steps used for thermalization and measuring respectively.}
	\label{tab:Energy_renom_periodic_2}
\end{table}
\begin{table}[H]
	\centering
	\begin{tabular}{|c|c|c|c|c|c|}
		\hline
		\rule{0pt}{12pt}$\beta$& $N_0\times N_1\times N_2$ & $N_{\text{therm}}$& $N_{\text{measure}}$ &$\frac{1}{2}\braket{\text{tr}\left(P_{01}\right)}-1$ & $\frac{1}{2}\braket{\text{tr }(P_{\mu\nu})}$\\
		\hline
		30&$96\times96\times10$ &$5\cdot 10^4$ & $5\cdot 10^5$& -0.03361423(13) & 0.96639808(9)\\\hline
		30&$96\times96\times20$ &$10^4$ & $10^4$& -0.0336052(7) & 0.9663952(5)\\\hline
		30&$96\times96\times30$ &$10^4$ & $10^4$& -0.0336052(5) & 0.9663948(4)\\\hline
		30&$96\times96\times40$ &$10^4$ & $10^4$& -0.0336052(5) & 0.9663948(3)\\\hline
		30&$96\times96\times50$ &$10^4$ & $10^4$& -0.0336052(4) & 0.9663946(3)\\\hline
		30&$96\times96\times60$ &$10^4$ & $10^4$& -0.0336053(4) & 0.9663949(3)\\\hline
		30&$96\times96\times70$ &$10^4$ & $10^4$& -0.0336053(4) & 0.9663944(2)\\\hline
		30&$96\times96\times80$ &$10^4$ & $10^4$& -0.0336058(3) & 0.9663946(2)\\\hline
		30&$96\times96\times90$ &$10^4$ & $10^4$& -0.0336055(3) & 0.9663944(2)\\\hline
		30&$96\times96\times100$ &$10^4$ & $10^4$& -0.0336050(3) & 0.9663949(2)\\\hline
		20&$96\times96\times10$ &$5\cdot 10^4$ & $5\cdot 10^5$& -0.0506346(2) & 0.94937833(14)\\\hline
		20&$96\times96\times20$ &$5\cdot 10^4$ & $5\cdot 10^5$& -0.05062833(17) & 0.94937162(10)\\\hline
		20&$96\times96\times30$ &$10^4$ & $10^4$& -0.0506276(8) & 0.9493718(6)\\\hline
		20&$96\times96\times40$ &$10^4$ & $10^4$& -0.0506289(7) & 0.9493721(5)\\\hline
		20&$96\times96\times50$ &$10^4$ & $10^4$& -0.0506286(6) & 0.9493718(4)\\\hline
		20&$96\times96\times60$ &$10^4$ & $10^4$& -0.0506266(6) & 0.9493727(4)\\\hline
		20&$96\times96\times70$ &$10^4$ & $10^4$& -0.0506279(5) & 0.9493715(4)\\\hline
		20&$96\times96\times80$ &$10^4$ & $10^4$& -0.0506287(5) & 0.9493713(3)\\\hline
		20&$96\times96\times90$ &$10^4$ & $10^4$& -0.0506282(5) & 0.9493712(3)\\\hline
		20&$96\times96\times100$ &$10^4$ & $10^4$& -0.0506300(5) & 0.9493707(3)\\\hline
	\end{tabular}
	\caption{Mean value of the plaquette $P_{01}$ and the average of all the plaquettes $P_{\mu\nu}$ with couplings constants $\beta=\{30,20\}$, lattice sizes $N_0N_1N_2$ and \acrshort*{pbc}. $N_{\text{therm}}$ and $N_{\text{measure}}$ are the \acrshort*{mc} steps used for thermalization and measuring respectively.}
	\label{tab:Energy_renom_periodic_3}
\end{table}

\begin{table}[H]
	\centering
	\begin{tabular}{|c|c|c|c|c|c|}
		\hline
		\rule{0pt}{12pt}$\beta$& $N_0\times N_1\times N_2$ &$N_{\text{therm}}$& $N_{\text{measure}}$&$\frac{1}{2}\braket{\text{tr}\left(P_{01}\right)}-1$ & $\frac{1}{2}\braket{\text{tr }P_{\mu\nu}}$\\
		\hline
		100&$96\times96\times10$ &$5\cdot 10^4$ & $7\cdot 10^5$& -0.00892281(3) & 0.99047860(2)\\\hline
		100&$96\times96\times20$ &$10^4$ & $1.8\cdot 10^6$& -0.009472891(15) & 0.990227535(10)\\\hline
		100&$96\times96\times30$ &$10^4$ & $10^4$& -0.00965633(16) & 0.99014395(11)\\\hline
		100&$96\times96\times40$ &$10^4$ & $10^4$& -0.00974836(14) & 0.99010197(9)\\\hline
		100&$96\times96\times50$ &$10^4$ & $10^4$& -0.00980318(13) & 0.99007676(8)\\\hline
		100&$96\times96\times60$ &$10^4$ & $10^4$& -0.00983990(11) & 0.99006017(8)\\\hline
		100&$96\times96\times70$ &$10^4$ & $10^4$& -0.00986622(11) & 0.99004814(7)\\\hline
		100&$96\times96\times80$ &$10^4$ & $10^4$& -0.00988590(10) & 0.99003921(7)\\\hline
		100&$96\times96\times90$ &$10^4$ & $10^4$& -0.00990125(9) & 0.99003218(6)\\\hline
		100&$96\times96\times100$ &$10^4$ & $10^4$& -0.00991360(9) & 0.99002655(6)\\\hline
	\end{tabular}
	\caption{Mean value of the plaquette $P_{01}$ and the average of all the plaquettes $P_{\mu\nu}$ with couplings constant $\beta=100$ and lattice sizes $N_0N_1N_2$. The \acrshort*{pccbc} are imposed in the second spatial direction $N_2$. $N_{\text{therm}}$ and $N_{\text{measure}}$ are the \acrshort*{mc} steps used for thermalization and measuring respectively.}
	\label{tab:Energy_renom_cond_1}
\end{table}
\begin{table}[H]
	\centering
	\begin{tabular}{|c|c|c|c|c|c|}
		\hline
		\rule{0pt}{12pt}$\beta$& $N_0\times N_1\times N_2$ & $N_{\text{therm}}$& $N_{\text{measure}}$&$\frac{1}{2}\braket{\text{tr}\left(P_{01}\right)}-1$ & $\frac{1}{2}\braket{\text{tr }P_{\mu\nu}}$\\
		\hline
		80&$96\times96\times10$ &$5\cdot10^4$ & $7\cdot10^5$& -0.01115970(4) & 0.98809144(3)\\\hline
		80&$96\times96\times20$ &$5\cdot10^4$ & $2.3\cdot10^6$& -0.011848029(16) & 0.987777212(11)\\\hline
		80&$96\times96\times30$ &$10^4$ & $10^4$& -0.0120774(2) & 0.98767259(14)\\\hline
		80&$96\times96\times40$ &$10^4$ & $10^4$& -0.01219265(17) & 0.98762014(12)\\\hline
		80&$96\times96\times50$ &$10^4$ & $10^4$& -0.01226157(16) & 0.98758842(11)\\\hline
		80&$96\times96\times60$ &$10^4$ & $10^4$& -0.01230751(14) & 0.98756753(10)\\\hline
		80&$96\times96\times70$ &$10^4$ & $10^4$& -0.01234014(13) & 0.98755286(9)\\\hline
		80&$96\times96\times80$ &$10^4$ & $10^4$& -0.01236488(12) & 0.98754149(8)\\\hline
		80&$96\times96\times90$ &$10^4$ & $10^4$& -0.01238401(12) & 0.98753269(8)\\\hline
		80&$96\times96\times100$ &$10^4$ & $10^4$& -0.01239937(11) & 0.98752580(8)\\\hline
	    60&$96\times96\times10$ &$5\cdot10^4$ & $7\cdot10^5$& -0.01489365(5) & 0.98410665(4)\\\hline
	    60&$96\times96\times20$ &$10^4$ & $1.5\cdot10^6$& -0.01581274(3) & 0.983686907(18)\\\hline
	    60&$96\times96\times30$ &$10^4$ & $10^4$& -0.01611947(3) & 0.98354687(2)\\\hline
	    60&$96\times96\times40$ &$10^4$ & $10^4$& -0.0162727(2) & 0.98347694(16)\\\hline
	    60&$96\times96\times50$ &$10^4$ & $10^4$& -0.0163644(2) & 0.98343516(14)\\\hline
	    60&$96\times96\times60$ &$10^4$ & $10^4$& -0.01642598(19) & 0.98340709(13)\\\hline
	    60&$96\times96\times70$ &$10^4$ & $10^4$& -0.01646991(18) & 0.98338711(12)\\\hline
	    60&$96\times96\times80$ &$10^4$ & $10^4$& -0.01650288(16) & 0.98337201(11)\\\hline
	    60&$96\times96\times90$ &$10^4$ & $10^4$& -0.01652838(16) & 0.98336034(10)\\\hline
	    60&$96\times96\times100$ &$10^4$ & $10^4$& -0.01654888(15) & 0.98335104(10)\\\hline
		50&$96\times96\times10$ &$10^4$ & $4\cdot10^5$& -0.01788588(8) & 0.98091316(6)\\\hline
		50&$96\times96\times20$ &$10^4$ & $2.1\cdot10^6$& -0.01899035(3) & 0.980408593(18)\\\hline
		50&$96\times96\times30$ &$10^4$ & $10^4$& -0.01935875(3) & 0.98024048(2)\\\hline
		50&$96\times96\times40$ &$10^4$ & $10^4$& -0.0195428(3) & 0.98015652(19)\\\hline
		50&$96\times96\times50$ &$10^4$ & $10^4$& -0.0196533(3) & 0.98010585(17)\\\hline
		50&$96\times96\times60$ &$10^4$ & $10^4$& -0.0197271(2) & 0.98007228(16)\\\hline
		50&$96\times96\times70$ &$10^4$ & $10^4$& -0.0197799(2) & 0.98004823(14)\\\hline
		50&$96\times96\times80$ &$10^4$ & $10^4$& -0.0198197(2) & 0.98003015(14)\\\hline
		50&$96\times96\times90$ &$10^4$ & $10^4$& -0.01985004(19) & 0.98001615(13)\\\hline
		50&$96\times96\times100$ &$10^4$ & $10^4$& -0.01987490(18) & 0.98000492(12)\\\hline
		40&$96\times96\times10$ &$10^4$ & $5\cdot10^5$& -0.02238323(9) & 0.97611314(7)\\\hline
		40&$96\times96\times20$ &$10^4$ & $10^4$& -0.0237661(5) & 0.9754811(3)\\\hline
		40&$96\times96\times30$ &$10^4$ & $10^4$& -0.0242273(4) & 0.9752707(3)\\\hline
		40&$96\times96\times40$ &$10^4$ & $10^4$& -0.0244579(4) & 0.9751652(2)\\\hline
		40&$96\times96\times50$ &$10^4$ & $10^4$& -0.0245968(3) & 0.9751018(2)\\\hline
		40&$96\times96\times60$ &$10^4$ & $10^4$& -0.0246896(3) & 0.97505963(19)\\\hline
		40&$96\times96\times70$ &$10^4$ & $10^4$& -0.0247551(3) & 0.97502969(18)\\\hline
		40&$96\times96\times80$ &$10^4$ & $10^4$& -0.0248051(2) & 0.97500661(17)\\\hline
		40&$96\times96\times90$ &$10^4$ & $10^4$& -0.0248430(2) & 0.97498941(16)\\\hline
		40&$96\times96\times100$ &$10^4$ & $10^4$& -0.0248743(2) & 0.97497514(15)\\\hline
	\end{tabular}
	\caption{Mean value of the plaquette $P_{01}$ and the average of all the plaquettes $P_{\mu\nu}$ with different coupling constants $\beta=\{80,60,50,40\}$ and lattice sizes $N_0N_1N_2$. The \acrshort*{pccbc} are imposed in the second spatial direction $N_2$. $N_{\text{therm}}$ and $N_{\text{measure}}$ are the \acrshort*{mc} steps used for thermalization and measuring respectively.}
	\label{tab:Energy_renom_cond_2}
\end{table}
\begin{table}[H]
	\centering
	\begin{tabular}{|c|c|c|c|c|c|}
		\hline
		\rule{0pt}{12pt}$\beta$& $N_0\times N_1\times N_2$ & $N_{\text{therm}}$& $N_{\text{measure}}$ &$\frac{1}{2}\braket{\text{tr}\left(P_{01}\right)}-1$ & $\frac{1}{2}\braket{\text{tr }P_{\mu\nu}}$\\
		\hline
		30&$96\times96\times10$ &$10^4$ & $6\cdot10^5$& -0.02990253(12) & 0.96808708(8)\\\hline
		30&$96\times96\times20$ &$10^4$ & $10^4$& -0.0317519(7) & 0.9672408(4)\\\hline
		30&$96\times96\times30$ &$10^4$ & $10^4$& -0.0323699(5) & 0.9669592(4)\\\hline
		30&$96\times96\times40$ &$10^4$ & $10^4$& -0.0326791(5) & 0.9668183(3)\\\hline
		30&$96\times96\times50$ &$10^4$ & $10^4$& -0.0328646(4) & 0.9667332(3)\\\hline
		30&$96\times96\times60$ &$10^4$ & $10^4$& -0.0329882(4) & 0.9666766(3)\\\hline
		30&$96\times96\times70$ &$10^4$ & $10^4$& -0.0330765(4) & 0.9666361(2)\\\hline
		30&$96\times96\times80$ &$10^4$ & $10^4$& -0.0331418(3) & 0.9666063(2)\\\hline
		30&$96\times96\times90$ &$10^4$ & $10^4$& -0.0331937(3) & 0.9665824(2)\\\hline
		30&$96\times96\times100$ &$10^4$ & $10^4$& -0.0332347(3) & 0.9665637(2)\\\hline
		20&$96\times96\times10$ &$10^4$ & $3\cdot10^5$& -0.0450359(2) & 0.95193126(17)\\\hline
		20&$96\times96\times20$ &$10^4$ & $10^4$& -0.0478311(10) & 0.9506509(7)\\\hline
		20&$96\times96\times30$ &$10^4$ & $10^4$& -0.0487633(8) & 0.9502246(6)\\\hline
		20&$96\times96\times40$ &$10^4$ & $10^4$& -0.0492300(7) & 0.9500110(5)\\\hline
		20&$96\times96\times50$ &$10^4$ & $10^4$& -0.0495099(6) & 0.9498824(4)\\\hline
		20&$96\times96\times60$ &$10^4$ & $10^4$& -0.0496959(6) & 0.9497979(4)\\\hline
		20&$96\times96\times70$ &$10^4$ & $10^4$& -0.0498284(5) & 0.9497366(4)\\\hline
		20&$96\times96\times80$ &$10^4$ & $10^4$& -0.0499294(5) & 0.9496912(3)\\\hline
		20&$96\times96\times90$ &$10^4$ & $10^4$& -0.0500065(5) & 0.9496559(3)\\\hline
		20&$96\times96\times100$ &$10^4$ & $10^4$& -0.0500686(5) & 0.9496277(3)\\\hline
	\end{tabular}
	\caption{Mean value of the plaquette $P_{01}$ and the average of all the plaquettes $P_{\mu\nu}$ with different coupling constants $\beta=\{30,20\}$ and lattice sizes $N_0N_1N_2$. The \acrshort*{pccbc} are imposed in the second spatial direction $N_2$. $N_{\text{therm}}$ and $N_{\text{measure}}$ are the \acrshort*{mc} steps used for thermalization and measuring respectively.}
	\label{tab:Energy_renom_cond_3}
\end{table}

\section{Casimir energy}\label{sec:Table_Cas_21}
In this section we show the simulations results that are used for computing the Casimir energy in \autoref{sec:Casimir_21}. In this, a lot more statistic is required than in the previous section since we need much smaller errors. As was mentioned in the main text in \autoref{sec:Casimir_21}, for the cases where we have a cubic lattice (all the directions have equal size) and \acrshort*{pbc} we use the average value of the plaquette in all directions to compute $\frac{1}{2}\braket{\text{tr}\left(P_{01}\right)}-1$ since the three directions are equivalent instead of just computing the average value with $P_{01}$, thus we use formula \eqref{eq:periodic_square_lattice_21} to compute the energy in the cubic lattices.  Notice how the lattice where the most statistics is used are the lattices with size $96\times96\times 48$ with \acrshort*{pccbc}, these are the ones used to cancel the boundary term and the largest source of error in the computation of the Casimir energy with these boundary conditions.
\begin{table}[H]
	\centering
	\begin{tabular}{|c|c|c|c|c|c|}
		\hline
		\rule{0pt}{12pt}$\beta$& $N_0\times N_1\times N_2$ & $N_{\text{therm}}$& $N_{\text{measure}}$ &$\frac{1}{2}\braket{\text{tr}\left(P_{01}\right)}-1$ & $\frac{1}{2}\braket{\text{tr }P_{\mu\nu}}$\\
		\hline
		100&$96\times96\times8$ &$5\cdot10^4$ & $7\cdot10^5$& -0.00864813(4) & 0.99060417(2)\\\hline
		100&$96\times96\times9$ &$5\cdot10^4$ & $5\cdot 10^5$& -0.00880070(4) & 0.99053444(3)\\\hline
		100&$96\times96\times10$ &$5\cdot10^4$ & $7\cdot10^5$& -0.00892281(3) & 0.99047860(2)\\\hline
		100&$96\times96\times11$ &$5\cdot10^4$ & $5\cdot10^5$& -0.00902267(4) & 0.99043301(3)\\\hline
		100&$96\times96\times12$ &$5\cdot10^4$ & $8\cdot10^5$& -0.00910606(3) & 0.990394929(19)\\\hline
		100&$96\times96\times13$ &$5\cdot10^4$ & $10^6$& -0.00917660(2) & 0.990362727(16)\\\hline
		100&$96\times96\times14$ &$5\cdot10^4$ & $8\cdot10^5$& -0.00923708(3) & 0.990335084(18)\\\hline
		100&$96\times96\times15$ &$5\cdot10^4$ & $10^6$& -0.00928944(2) & 0.990311212(15)\\\hline
		100&$96\times96\times16$ &$5\cdot10^4$ & $1.1\cdot 10^6$& -0.00933530(2) & 0.990290292(14)\\\hline
		100&$96\times96\times17$ &$5\cdot10^4$ & $1.5\cdot10^6$& -0.009375736(17) & 0.990271836(12)\\\hline
		100&$96\times96\times18$ &$5\cdot10^4$ & $1.7\cdot10^6$& -0.009411745(16) & 0.990255429(11)\\\hline
		100&$96\times96\times19$ &$5\cdot10^4$ & $1.5\cdot10^6$& -0.009443894(16) & 0.990240749(11)\\\hline
		100&$96\times96\times20$ &$5\cdot10^4$ & $1.8\cdot10^6$& -0.009472891(15) & 0.990227535(10)\\\hline
		100&$96\times96\times48$ &$5\cdot10^4$ & $4.2\cdot10^6$& -0.009794115(6) & 0.990081024(4)\\\hline
		80&$96\times96\times8$ &$5\cdot10^4$ & $7\cdot10^5$& -0.01081605(4) & 0.98824856(3)\\\hline
		80&$96\times96\times9$ &$5\cdot10^4$ & $5\cdot10^5$& -0.01100699(5) & 0.98816129(3)\\\hline
		80&$96\times96\times10$ &$5\cdot10^4$ & $7\cdot10^5$& -0.01115970(4) & 0.98809144(3)\\\hline
		80&$96\times96\times11$ &$5\cdot10^4$ & $10^6$& -0.01128476(3) & 0.98803433(2)\\\hline
		80&$96\times96\times12$ &$5\cdot10^4$ & $2\cdot10^6$& -0.01138903(2) & 0.987986682(15)\\\hline
		80&$96\times96\times13$ &$5\cdot10^4$ & $10^6$& -0.01147723(3) & 0.98794644(2)\\\hline
		80&$96\times96\times14$ &$5\cdot10^4$ & $1.8\cdot10^6$& -0.01155292(2) & 0.987911866(15)\\\hline
		80&$96\times96\times15$ &$5\cdot10^4$ & $10^6$& -0.01161851(3) & 0.987881931(19)\\\hline
		80&$96\times96\times16$ &$5\cdot10^4$ & $1.7\cdot10^6$& -0.01167587(2) & 0.987855766(14)\\\hline
		80&$96\times96\times17$ &$5\cdot10^4$ & $1.5\cdot10^6$& -0.01172648(2) & 0.987832657(15)\\\hline
		80&$96\times96\times18$ &$5\cdot10^4$ & $2.3\cdot10^6$& -0.011771485(17) & 0.987812144(12)\\\hline
		80&$96\times96\times19$ &$5\cdot10^4$ & $1.5\cdot10^6$& -0.01181177(2) & 0.987793768(14)\\\hline
		80&$96\times96\times20$ &$5\cdot10^4$ & $2.3\cdot10^6$& -0.011848029(16) & 0.987777212(11)\\\hline
		80&$96\times96\times48$ &$5\cdot10^4$ & $5\cdot10^6$& -0.012249893(7) & 0.987593914(5)\\\hline
		60&$96\times96\times8$ &$5\cdot10^4$ & $7\cdot10^5$& -0.01443463(6) & 0.98431655(4)\\\hline
		60&$96\times96\times9$ &$5\cdot10^4$ & $5\cdot10^5$& -0.01468956(7) & 0.98419991(5)\\\hline
		60&$96\times96\times10$ &$5\cdot10^4$ & $7\cdot10^5$& -0.01489365(5) & 0.98410665(4)\\\hline
		60&$96\times96\times11$ &$5\cdot10^4$ & $10^6$& -0.01506058(4) & 0.98403035(3)\\\hline
		60&$96\times96\times12$ &$5\cdot10^4$ & $10^6$& -0.01519982(4) & 0.98396674(3)\\\hline
		60&$96\times96\times13$ &$5\cdot10^4$ & $10^6$& -0.01531770(4) & 0.98391285(3)\\\hline
		60&$96\times96\times14$ &$5\cdot10^4$ & $8\cdot10^5$& -0.01541869(4) & 0.98386677(3)\\\hline
		60&$96\times96\times15$ &$5\cdot10^4$ & $10^6$& -0.01550621(4) & 0.98382680(3)\\\hline
		60&$96\times96\times16$ &$5\cdot10^4$ & $1.2\cdot10^6$& -0.01558280(3) & 0.98379185(2)\\\hline
		60&$96\times96\times17$ &$5\cdot10^4$ & $1.5\cdot10^6$& -0.01565049(3) & 0.98376099(2)\\\hline
		60&$96\times96\times18$ &$5\cdot10^4$ & $1.2\cdot10^6$& -0.01571052(3) & 0.98373354(2)\\\hline
		60&$96\times96\times48$ &$5\cdot10^4$ & $3.8\cdot10^6$& -0.016349447(11) & 0.983441984(7)\\\hline
	\end{tabular}
	\caption{Mean value of the plaquette $P_{01}$ and the average of all the plaquettes $P_{\mu\nu}$ with different coupling constants $\beta=\{100,80,60\}$ and lattice sizes $N_0N_1N_2$. The \acrshort*{pccbc} are imposed in the second spatial direction $N_2$. $N_{\text{therm}}$ and $N_{\text{measure}}$ are the \acrshort*{mc} steps used for thermalization and measuring respectively.}
	\label{tab:Casimir_21_cond_1}
\end{table}
\begin{table}[H]
	\centering
	\begin{tabular}{|c|c|c|c|c|c|}
		\hline
		\rule{0pt}{12pt}$\beta$& $N_0\times N_1\times N_2$ & $N_{\text{therm}}$& $N_{\text{measure}}$ &$\frac{1}{2}\braket{\text{tr}\left(P_{01}\right)}-1$ & $\frac{1}{2}\braket{\text{tr }P_{\mu\nu}}$\\
		\hline
		50&$96\times96\times8$ &$10^4$ & $2.5\cdot10^5$& -0.01733440(12) & 0.98116550(8)\\\hline
		50&$96\times96\times9$ &$5\cdot10^4$ & $5\cdot10^5$& -0.01764071(8) & 0.98102528(6)\\\hline
		50&$96\times96\times10$ &$10^4$ & $4\cdot10^5$& -0.01788588(8) & 0.98091316(6)\\\hline
		50&$96\times96\times11$ &$5\cdot10^4$ & $10^6$& -0.01808646(5) & 0.98082148(4)\\\hline
		50&$96\times96\times12$ &$5\cdot10^4$ & $9\cdot10^5$& -0.01825373(5) & 0.98074502(4)\\\hline
		50&$96\times96\times13$ &$5\cdot10^4$ & $10^6$& -0.01839552(5) & 0.98068022(3)\\\hline
		50&$96\times96\times14$ &$5\cdot10^4$ & $10^6$& -0.01851678(4) & 0.98062483(3)\\\hline
		50&$96\times96\times15$ &$5\cdot10^4$ & $10^6$& -0.01862197(4) & 0.98057681(3)\\\hline
		50&$96\times96\times16$ &$5\cdot10^4$ & $1.2\cdot10^6$& -0.01871403(4) & 0.98053474(3)\\\hline
		50&$96\times96\times17$ &$5\cdot10^4$ & $1.5\cdot10^6$& -0.01879516(3) & 0.98049775(2)\\\hline
		50&$96\times96\times18$ &$5\cdot10^4$ & $1.6\cdot10^6$& -0.01886748(3) & 0.98046474(2)\\\hline
		50&$96\times96\times48$ &$5\cdot10^4$ & $4.4\cdot10^6$& -0.019635147(12) & 0.980114326(8)\\\hline
		40&$96\times96\times8$ &$10^4$ & $2\cdot10^5$& -0.02169276(16) & 0.97642915(12)\\\hline
		40&$96\times96\times9$ &$5\cdot10^4$ & $6\cdot10^5$& -0.02207617(9) & 0.97625367(6)\\\hline
		40&$96\times96\times10$ &$10^4$ & $5\cdot10^5$& -0.02238323(9) & 0.97611314(7)\\\hline
		40&$96\times96\times11$ &$5\cdot10^4$ & $1.1\cdot10^6$& -0.02263437(6) & 0.97599834(4)\\\hline
		40&$96\times96\times12$ &$5\cdot10^4$ & $8\cdot10^5$& -0.02284383(7) & 0.97590270(5)\\\hline
		40&$96\times96\times13$ &$5\cdot10^4$ & $1.1\cdot10^6$& -0.02302126(6) & 0.97582155(4)\\\hline
		40&$96\times96\times14$ &$5\cdot10^4$ & $9\cdot10^5$& -0.02317313(6) & 0.97575208(4)\\\hline
		40&$96\times96\times15$ &$5\cdot10^4$ & $1.1\cdot10^6$& -0.02330499(5) & 0.97569186(4)\\\hline
		40&$96\times96\times16$ &$5\cdot10^4$ & $1.2\cdot10^6$& -0.02342025(5) & 0.97563920(3)\\\hline
		40&$96\times96\times17$ &$5\cdot10^4$ & $10^6$& -0.02352202(5) & 0.97559268(4)\\\hline
		40&$96\times96\times18$ &$5\cdot10^4$ & $9\cdot10^5$& -0.02361245(5) & 0.97555147(4)\\\hline
		40&$96\times96\times48$ &$5\cdot10^4$ & $4.1\cdot10^6$& -0.024573955(16) & 0.975112335(11)\\\hline
		30&$96\times96\times8$ &$10^4$ & $3\cdot10^5$& -0.02897953(18) & 0.96850998(13)\\\hline
		30&$96\times96\times9$ &$5\cdot10^4$ & $5\cdot10^5$& -0.02949220(13) & 0.96827500(9)\\\hline
		30&$96\times96\times10$ &$10^4$ & $6\cdot10^5$& -0.02990253(12) & 0.96808708(8)\\\hline
		30&$96\times96\times11$ &$5\cdot10^4$ & $10^6$& -0.03023860(9) & 0.96793334(6)\\\hline
		30&$96\times96\times12$ &$5\cdot10^4$ & $9\cdot10^5$& -0.03051913(9) & 0.96780493(6)\\\hline
		30&$96\times96\times13$ &$5\cdot10^4$ & $10^6$& -0.03075623(8) & 0.96769656(6)\\\hline
		30&$96\times96\times14$ &$5\cdot10^4$ & $9\cdot10^5$& -0.03095956(8) & 0.96760353(6)\\\hline
		30&$96\times96\times15$ &$5\cdot10^4$ & $10^6$& -0.03113570(7) & 0.96752294(5)\\\hline
		30&$96\times96\times16$ &$5\cdot10^4$ & $3\cdot10^5$& -0.03129012(13) & 0.96745218(9)\\\hline
		30&$96\times96\times48$ &$5\cdot10^4$ & $4.5\cdot10^6$& -0.03283346(2) & 0.966747160(14)\\\hline
		20&$96\times96\times8$ &$ 10^4$ & $7\cdot 10^5$& -0.04364151(18) & 0.95257090(12)\\\hline
		20&$96\times96\times9$ &$5\cdot 10^4$ & $5\cdot 10^5$& -0.0444159(2) & 0.95221562(14)\\\hline
		20&$96\times96\times10$ &$ 10^4$ & $3\cdot 10^5$& -0.0450359(2) & 0.95193126(17)\\\hline
		20&$96\times96\times11$ &$5\cdot10^4$ & $10^6$& -0.04554397(13) & 0.95169847(9)\\\hline
		20&$96\times96\times12$ &$5\cdot 10^4$ & $5\cdot 10^5$& -0.04596732(15) & 0.95150418(10)\\\hline
		20&$96\times96\times48$ &$5\cdot10^4$ & $5.1 \cdot 10^6$& -0.04946288(3) & 0.949904463(19)\\\hline
	\end{tabular}
	\caption{Mean value of the plaquette $P_{01}$ and the average of all the plaquettes $P_{\mu\nu}$ with different coupling constants $\beta=\{50,40,30,20\}$ and lattice sizes $N_0N_1N_2$. The \acrshort*{pccbc} are imposed in the second spatial direction $N_2$. $N_{\text{therm}}$ and $N_{\text{measure}}$ are the \acrshort*{mc} steps used for thermalization and measuring respectively.}
	\label{tab:Casimir_21_cond_2}
\end{table}

\begin{table}[H]
	\centering
	\begin{tabular}{|c|c|c|c|c|c|}
		\hline
		\rule{0pt}{12pt}$\beta$& $N_0\times N_1\times N_2$ &  $N_{\text{therm}}$& $N_{\text{measure}}$ &$\frac{1}{2}\braket{\text{tr}\left(P_{01}\right)}-1$ & $\frac{1}{2}\braket{\text{tr }P_{\mu\nu}}$\\
		\hline
		100&$96\times96\times9$ &$5\cdot10^4$ & $3\cdot10^5$& -0.01003035(6) & 0.98997683(4)\\\hline
		100&$96\times96\times10$ &$5\cdot10^4$ & $5\cdot10^5$& -0.01002840(4) & 0.98997676(3)\\\hline
		100&$96\times96\times11$ &$5\cdot10^4$ & $3\cdot10^5$& -0.01002716(5) & 0.98997657(3)\\\hline
		100&$96\times96\times12$ &$5\cdot10^4$ & $5\cdot10^5$& -0.01002625(4) & 0.98997659(2)\\\hline
		100&$96\times96\times13$ &$5\cdot10^4$ & $3\cdot10^5$& -0.01002559(5) & 0.98997658(3)\\\hline
		100&$96\times96\times14$ &$5\cdot10^4$ & $7\cdot10^5$& -0.01002521(3) & 0.989976549(19)\\\hline
		100&$96\times96\times15$ &$5\cdot10^4$ & $3\cdot10^5$& -0.01002486(4) & 0.98997654(3)\\\hline
		100&$96\times96\times16$ &$5\cdot10^4$ & $7\cdot10^5$& -0.01002462(3) & 0.989976551(18)\\\hline
		100&$96\times96\times17$ &$5\cdot10^4$ & $3\cdot10^5$& -0.01002442(4) & 0.98997656(3)\\\hline
		100&$96\times96\times18$ &$5\cdot10^4$ & $7\cdot10^5$& -0.01002425(3) & 0.989976510(17)\\\hline
		100&$96\times96\times19$ &$5\cdot10^4$ & $10^6$& -0.01002420(2) & 0.989976469(14)\\\hline
		100&$96\times96\times20$ &$5\cdot10^4$ & $7\cdot10^5$& -0.01002409(2) & 0.989976459(16)\\\hline
		100&$96\times96\times22$ &$5\cdot10^4$ & $7\cdot10^5$& -0.01002390(2) & 0.989976485(15)\\\hline
		100&$96\times96\times24$ &$5\cdot10^4$ & $7\cdot10^4$& -0.01002387(2) & 0.989976430(15)\\\hline
		100&$96\times96\times96$ &$5\cdot10^4$ & $4.5\cdot10^5$& -0.010023593(9) & 0.989976407(9)\\\hline
		80&$96\times96\times8$ &$5\cdot10^4$ & $5\cdot10^5$& -0.01254895(6) & 0.98746371(4)\\\hline
		80&$96\times96\times9$ &$5\cdot10^4$ & $3\cdot10^5$& -0.01254500(7) & 0.98746360(5)\\\hline
		80&$96\times96\times10$ &$5\cdot10^4$ & $5\cdot10^5$& -0.01254270(5) & 0.98746351(3)\\\hline
		80&$96\times96\times11$ &$5\cdot10^4$ & $3\cdot10^5$& -0.01254116(6) & 0.98746339(4)\\\hline
		80&$96\times96\times12$ &$5\cdot10^4$ & $5\cdot10^5$& -0.01254008(5) & 0.98746329(3)\\\hline
		80&$96\times96\times13$ &$5\cdot10^4$ & $3\cdot10^5$& -0.01253940(6) & 0.98746317(4)\\\hline
		80&$96\times96\times14$ &$5\cdot10^4$ & $7\cdot10^5$& -0.01253886(4) & 0.98746325(2)\\\hline
		80&$96\times96\times15$ &$5\cdot10^4$ & $3\cdot10^5$& -0.01253836(5) & 0.98746320(4)\\\hline
		80&$96\times96\times16$ &$5\cdot10^4$ & $7\cdot10^5$& -0.01253821(3) & 0.98746315(2)\\\hline
		80&$96\times96\times17$ &$5\cdot10^4$ & $3\cdot10^5$& -0.01253791(5) & 0.98746316(3)\\\hline
		80&$96\times96\times18$ &$5\cdot10^4$ & $7\cdot10^5$& -0.01253782(3) & 0.98746311(2)\\\hline
		80&$96\times96\times19$ &$5\cdot10^4$ & $5\cdot10^45$& -0.01253769(4) & 0.98746314(2)\\\hline
		80&$96\times96\times20$ &$5\cdot10^4$ & $7\cdot10^5$& -0.01253755(3) & 0.98746306(2)\\\hline
		80&$96\times96\times22$ &$5\cdot10^4$ & $7\cdot10^5$& -0.01253742(3) & 0.987463048(19)\\\hline
		80&$96\times96\times24$ &$5\cdot10^4$ & $7\cdot10^5$& -0.01253726(3) & 0.987463056(18)\\\hline
		80&$96\times96\times96$ &$5\cdot10^4$ & $1.5\cdot10^5$& -0.012537040(10) & 0.987462960(10)\\\hline
	\end{tabular}
	\caption{Mean value of the plaquette $P_{01}$ and the average of all the plaquettes $P_{\mu\nu}$ with different coupling constants $\beta=\{100,80\}$ and lattice sizes $N_0N_1N_2$. \acrshort*{pbc} are used in every direction. $N_{\text{therm}}$ and $N_{\text{measure}}$ are the \acrshort*{mc} steps used for thermalization and measuring respectively.}
	\label{tab:Casimir_21_periodic_1}
\end{table}
\begin{table}[H]
	\centering
	\begin{tabular}{|c|c|c|c|c|c|}
		\hline
		\rule{0pt}{12pt}$\beta$& $N_0\times N_1\times N_2$ &  $N_{\text{therm}}$& $N_{\text{measure}}$ &$\frac{1}{2}\braket{\text{tr}\left(P_{01}\right)}-1$ & $\frac{1}{2}\braket{\text{tr }P_{\mu\nu}}$\\
		\hline
		60&$96\times96\times8$ &$5\cdot10^4$ & $5\cdot10^5$& -0.01674789(8) & 0.98326836(5)\\\hline
		60&$96\times96\times9$ &$5\cdot10^4$ & $3\cdot10^5$& -0.01674303(9) & 0.98326807(6)\\\hline
		60&$96\times96\times10$ &$5\cdot10^4$ & $5\cdot10^5$& -0.01674001(7) & 0.98326789(4)\\\hline
		60&$96\times96\times11$ &$5\cdot10^4$ & $3\cdot10^5$& -0.01673796(8) & 0.98326776(6)\\\hline
		60&$96\times96\times12$ &$5\cdot10^4$ & $5\cdot10^5$& -0.01673662(6) & 0.98326769(4)\\\hline
		60&$96\times96\times13$ &$5\cdot10^4$ & $3\cdot10^4$& -0.01673563(8) & 0.98326765(5)\\\hline
		60&$96\times96\times14$ &$5\cdot10^4$ & $5.5\cdot10^5$& -0.01673514(5) & 0.98326751(4)\\\hline
		60&$96\times96\times15$ &$5\cdot10^4$ & $3\cdot10^5$& -0.01673450(7) & 0.98326748(5)\\\hline
		60&$96\times96\times16$ &$5\cdot10^4$ & $7\cdot10^5$& -0.01673415(4) & 0.98326747(3)\\\hline
		60&$96\times96\times17$ &$5\cdot10^4$ & $3\cdot10^5$& -0.01673396(7) & 0.98326737(4)\\\hline
		60&$96\times96\times18$ &$5\cdot10^4$ & $7\cdot10^5$& -0.01673371(4) & 0.98326729(3)\\\hline
		60&$96\times96\times19$ &$5\cdot10^4$ & $5\cdot10^5$& -0.01673357(5) & 0.98326730(3)\\\hline
		60&$96\times96\times20$ &$5\cdot10^4$ & $6\cdot10^5$& -0.01673352(4) & 0.98326731(3)\\\hline
		60&$96\times96\times22$ &$5\cdot10^4$ & $7\cdot10^5$& -0.01673332(4) & 0.98326720(3)\\\hline
		60&$96\times96\times96$ &$5\cdot10^4$ & $5\cdot10^5$& -0.016732894(15) & 0.983267106(15)\\\hline
		50&$96\times96\times8$ &$5\cdot10^4$ & $2.5\cdot10^5$& -0.02011281(13) & 0.97990592(9)\\\hline
		50&$96\times96\times9$ &$5\cdot10^4$ & $3\cdot10^5$& -0.02010707(11) & 0.97990572(7)\\\hline
		50&$96\times96\times10$ &$5\cdot10^4$ & $2.5\cdot10^5$& -0.02010364(11) & 0.97990539(8)\\\hline
		50&$96\times96\times11$ &$5\cdot10^4$ & $3\cdot10^5$& -0.02010158(10) & 0.97990498(7)\\\hline
		50&$96\times96\times12$ &$5\cdot10^4$ & $2.5\cdot10^5$& -0.02010009(10) & 0.97990481(7)\\\hline
		50&$96\times96\times13$ &$5\cdot10^4$ & $3\cdot10^5$& -0.02009885(9) & 0.97990485(6)\\\hline
		50&$96\times96\times14$ &$5\cdot10^4$ & $2.5\cdot10^5$& -0.02009821(10) & 0.97990469(6)\\\hline
		50&$96\times96\times15$ &$5\cdot10^4$ & $3\cdot10^5$& -0.02009786(9) & 0.97990450(6)\\\hline
		50&$96\times96\times16$ &$5\cdot10^4$ & $7.5\cdot10^5$& -0.02009720(5) & 0.97990458(3)\\\hline
		50&$96\times96\times17$ &$5\cdot10^4$ & $6\cdot10^5$& -0.02009696(6) & 0.97990452(4)\\\hline
		50&$96\times96\times18$ &$5\cdot10^4$ & $7.5\cdot10^5$& -0.02009666(5) & 0.97990454(3)\\\hline
		50&$96\times96\times19$ &$5\cdot10^4$ & $5\cdot10^5$& -0.02009652(6) & 0.97990447(4)\\\hline
		50&$96\times96\times20$ &$5\cdot10^4$ & $9\cdot10^5$& -0.02009647(4) & 0.97990435(3)\\\hline
		50&$96\times96\times96$ &$5\cdot10^4$ & $5.5\cdot10^5$& -0.020095913(17) & 0.979904087(17)\\\hline
		40&$96\times96\times8$ &$5\cdot10^4$ & $5\cdot10^5$& -0.02517005(11) & 0.97485192(8)\\\hline
		40&$96\times96\times9$ &$5\cdot10^4$ & $3\cdot10^5$& -0.02516368(14) & 0.97485124(9)\\\hline
		40&$96\times96\times10$ &$5\cdot10^4$ & $5\cdot10^5$& -0.02515960(10) & 0.97485086(7)\\\hline
		40&$96\times96\times11$ &$5\cdot10^4$ & $3\cdot10^5$& -0.02515658(12) & 0.97485082(8)\\\hline
		40&$96\times96\times12$ &$5\cdot10^4$ & $5\cdot10^5$& -0.02515530(9) & 0.97485026(6)\\\hline
		40&$96\times96\times13$ &$5\cdot10^4$ & $3\cdot10^5$& -0.02515396(11) & 0.97485022(8)\\\hline
		40&$96\times96\times14$ &$5\cdot10^4$ & $5.5\cdot10^5$& -0.02515331(8) & 0.97484993(5)\\\hline
		40&$96\times96\times15$ &$5\cdot10^4$ & $3\cdot10^5$& -0.02515277(11) & 0.97484979(7)\\\hline
		40&$96\times96\times16$ &$5\cdot10^4$ & $9\cdot10^5$& -0.02515230(6) & 0.97484969(4)\\\hline
		40&$96\times96\times17$ &$5\cdot10^4$ & $6\cdot10^5$& -0.02515181(7) & 0.97484964(5)\\\hline
		40&$96\times96\times18$ &$5\cdot10^4$ & $7\cdot10^5$& -0.02515167(6) & 0.97484955(4)\\\hline
		40&$96\times96\times19$ &$5\cdot10^4$ & $5\cdot10^5$& -0.02515153(7) & 0.97484946(5)\\\hline
		40&$96\times96\times20$ &$5\cdot10^4$ & $7\cdot10^4$& -0.02515121(6) & 0.97484952(4)\\\hline
		40&$96\times96\times96$ &$5\cdot10^4$ & $5\cdot10^5$& -0.02515094(2) & 0.97484906(2)\\\hline
	\end{tabular}
	\caption{Mean value of the plaquette $P_{01}$ and the average of all the plaquettes $P_{\mu\nu}$ with different coupling constants $\beta=\{60,50,40\}$ and lattice sizes $N_0N_1N_2$. \acrshort*{pbc} are used in every direction. $N_{\text{therm}}$ and $N_{\text{measure}}$ are the \acrshort*{mc} steps used for thermalization and measuring respectively.}
	\label{tab:Casimir_21_periodic_2}
\end{table}
\begin{table}[H]
	\centering
	\begin{tabular}{|c|c|c|c|c|c|}
		\hline
		\rule{0pt}{12pt}$\beta$& $N_0\times N_1\times N_2$ &  $N_{\text{therm}}$& $N_{\text{measure}}$ &$\frac{1}{2}\braket{\text{tr}\left(P_{01}\right)}-1$ & $\frac{1}{2}\braket{\text{tr }P_{\mu\nu}}$\\
		\hline
		30&$96\times96\times8$ &$5\cdot10^4$ & $5\cdot10^5$& -0.03362710(15) & 0.96639984(10)\\\hline
		30&$96\times96\times9$ &$5\cdot10^4$ & $3\cdot10^5$& -0.03361929(18) & 0.96639871(12)\\\hline
		30&$96\times96\times10$ &$5\cdot10^4$ & $5\cdot10^5$& -0.03361423(13) & 0.96639808(9)\\\hline
		30&$96\times96\times11$ &$5\cdot10^4$ & $3\cdot10^5$& -0.03361124(17) & 0.96639738(11)\\\hline
		30&$96\times96\times12$ &$5\cdot10^4$ & $5\cdot10^5$& -0.03360966(12) & 0.96639671(8)\\\hline
		30&$96\times96\times13$ &$5\cdot10^4$ & $5\cdot10^5$& -0.03360798(15) & 0.96639674(10)\\\hline
		30&$96\times96\times14$ &$5\cdot10^4$ & $1.2\cdot10^6$& -0.03360725(7) & 0.96639617(5)\\\hline
		30&$96\times96\times15$ &$5\cdot10^4$ & $3\cdot10^5$& -0.03360667(14) & 0.96639589(10)\\\hline
		30&$96\times96\times16$ &$5\cdot10^4$ & $10^6$& -0.03360607(8) & 0.96639581(5)\\\hline
		30&$96\times96\times17$ &$5\cdot10^4$ & $8\cdot10^5$& -0.03360589(8) & 0.96639561(5)\\\hline
		30&$96\times96\times18$ &$5\cdot10^4$ & $1.7\cdot10^6$& -0.03360568(5) & 0.96639533(4)\\\hline
		30&$96\times96\times96$ &$5\cdot10^4$ & $6.5\cdot10^5$& -0.03360534(3) & 0.96639466(3)\\\hline
		20&$96\times96\times8$ &$5\cdot10^4$ & $5\cdot10^5$& -0.0506489(2) & 0.94938325(15)\\\hline
		20&$96\times96\times9$ &$5\cdot10^4$ & $3\cdot10^5$& -0.0506396(3) & 0.94938047(19)\\\hline
		20&$96\times96\times10$ &$5\cdot10^4$ & $5\cdot10^4$& -0.0506346(2) & 0.94937833(14)\\\hline
		20&$96\times96\times11$ &$5\cdot10^4$ & $8\cdot10^5$& -0.05063122(15) & 0.94937680(10)\\\hline
		20&$96\times96\times12$ &$5\cdot10^4$ & $1.8\cdot10^6$& -0.05062979(10) & 0.94937508(7)\\\hline
		20&$96\times96\times96$ &$5\cdot10^4$ & $7.5\cdot10^5$& -0.05062831(4) & 0.94937169(4)\\\hline
	\end{tabular}
	\caption{Mean value of the plaquette $P_{01}$ and the average of all the plaquettes $P_{\mu\nu}$ for different couplings constants $\beta=\{30,20\}$ and lattice sizes $N_0N_1N_2$. \acrshort*{pbc} are used in every direction. $N_{\text{therm}}$ and $N_{\text{measure}}$ are the \acrshort*{mc} steps used for thermalization and measuring respectively.}
	\label{tab:Casimir_21_periodic_3}
\end{table}
\pagebreak
\section{Finite volume errors}

In this section we show the simulation results with the smaller lattices used in \autoref{sec:su21_lattice_effects} for checking the errors associated with the finite size for \acrshort*{pccbc} and \acrshort*{pbc}. Like we did in \autoref{sec:Table_Cas_21} when we have a \acrshort*{pbc} cubic lattice we compute the energy with the average of the plaquettes in all directions, i.e. \eqref{eq:periodic_square_lattice_21}.

\begin{table}[H]
	\centering
	\begin{tabular}{|c|c|c|c|c|c|}
		\hline
		\rule{0pt}{12pt}$\beta$& $N_0\times N_1\times N_2$ & $N_{\text{therm}}$& $N_{\text{measure}}$ &$\frac{1}{2}\braket{\text{tr}\left(P_{01}\right)}-1$ & $\frac{1}{2}\braket{\text{tr }P_{\mu\nu}}$\\
		\hline
		100&$64\times64\times9$ &$5\cdot 10^4$ & $3\cdot 10^5$& -0.01003038(8) & 0.98997681(6)\\\hline
		100&$64\times64\times10$ &$5\cdot 10^4$ & $3\cdot 10^5$& -0.01002827(8) & 0.98997684(5)\\\hline
		100&$64\times64\times11$ &$5\cdot 10^4$ & $3\cdot 10^5$& -0.01002711(7) & 0.98997664(5)\\\hline
		100&$64\times64\times12$ &$5\cdot 10^4$ & $3\cdot 10^5$& -0.01002618(7) & 0.98997658(5)\\\hline
		100&$64\times64\times13$ &$5\cdot 10^4$ & $3\cdot 10^5$& -0.01002551(7) & 0.98997653(5)\\\hline
		100&$64\times64\times14$ &$5\cdot 10^4$ & $3\cdot 10^5$& -0.01002525(7) & 0.98997648(4)\\\hline
		100&$64\times64\times15$ &$5\cdot 10^4$ & $3\cdot 10^5$& -0.01002489(6) & 0.98997658(4)\\\hline
		100&$64\times64\times16$ &$5\cdot 10^4$ & $8\cdot 10^5$& -0.01002461(4) & 0.98997654(3)\\\hline
		100&$64\times64\times17$ &$5\cdot 10^4$ & $3\cdot 10^5$& -0.01002443(6) & 0.98997656(4)\\\hline
		100&$64\times64\times18$ &$5\cdot 10^4$ & $8\cdot 10^5$& -0.01002425(4) & 0.98997651(2)\\\hline
		100&$64\times64\times64$ &$5\cdot 10^4$ & $10^6$& -0.010023601(11) & 0.989976399(11)\\\hline
		50&$64\times64\times8$ &$5\cdot 10^4$ & $3\cdot 10^5$& -0.02011260(17) & 0.97990599(9)\\\hline
		50&$64\times64\times9$ &$5\cdot 10^4$ & $3\cdot 10^5$& -0.02010715(16) & 0.97990557(11)\\\hline
		50&$64\times64\times10$ &$5\cdot 10^4$ & $3\cdot 10^5$& -0.02010359(16) & 0.97990548(10)\\\hline
		50&$64\times64\times11$ &$5\cdot 10^4$ & $3\cdot 10^5$& -0.02010164(15) & 0.97990502(10)\\\hline
		50&$64\times64\times12$ &$5\cdot 10^4$ & $3\cdot 10^5$& -0.02009967(14) & 0.97990500(10)\\\hline
		50&$64\times64\times13$ &$5\cdot 10^4$ & $3\cdot 10^5$& -0.02009882(14) & 0.97990497(9)\\\hline
		50&$64\times64\times14$ &$5\cdot 10^4$ & $3\cdot 10^5$& -0.02009825(13) & 0.97990470(9)\\\hline
		50&$64\times64\times15$ &$5\cdot 10^4$ & $3\cdot 10^5$& -0.02009763(13) & 0.97990463(9)\\\hline
		50&$64\times64\times16$ &$5\cdot 10^4$ & $3\cdot 10^5$& -0.02009710(12) & 0.97990466(8)\\\hline
		50&$64\times64\times17$ &$5\cdot 10^4$ & $3\cdot 10^5$& -0.02009684(12) & 0.97990453(8)\\\hline
		50&$64\times64\times18$ &$5\cdot 10^4$ & $3\cdot 10^5$& -0.02009680(12) & 0.97990448(8)\\\hline
		50&$64\times64\times64$ &$5\cdot 10^4$ & $10^6$& -0.02009585(2) & 0.97990415(2)\\\hline
		20&$64\times64\times8$ &$5\cdot 10^4$ & $3\cdot 10^5$& -0.0506491(4) & 0.9493831(3)\\\hline
		20&$64\times64\times9$ &$5\cdot 10^4$ & $3\cdot 10^5$& -0.0506387(4) & 0.9493810(3)\\\hline
		20&$64\times64\times10$ &$5\cdot 10^4$ & $3\cdot 10^5$& -0.0506350(4) & 0.9493779(3)\\\hline
		20&$64\times64\times11$ &$5\cdot 10^4$ & $3\cdot 10^5$& -0.0506306(4) & 0.9493772(3)\\\hline
		20&$64\times64\times12$ &$5\cdot 10^4$ & $8\cdot 10^5$& -0.0506302(2) & 0.94937526(15)\\\hline
		20&$64\times64\times64$ &$5\cdot 10^4$ & $5\cdot 10^5$& -0.05062846(8) & 0.94937154(8)\\\hline
	\end{tabular}
	\caption{Mean value of the plaquette $P_{01}$ and the average of all the plaquettes $P_{\mu\nu}$ with couplings constants $\beta=\{100,50,20\}$ and lattice sizes $N_0N_1N_2$. \acrshort*{pbc} are used in every direction. $N_{\text{therm}}$ and $N_{\text{measure}}$ are the \acrshort*{mc} steps used for thermalization and measuring respectively.}
	\label{tab:Energy_lattice_size_periodic}
\end{table}

\begin{table}[H]
	\centering
	\begin{tabular}{|c|c|c|c|c|c|}
		\hline
		\rule{0pt}{12pt}$\beta$& $N_0\times N_1\times N_2$ & $N_{\text{therm}}$& $N_{\text{measure}}$ &$\frac{1}{2}\braket{\text{tr}\left(P_{01}\right)}-1$ & $\frac{1}{2}\braket{\text{tr }P_{\mu\nu}}$\\
		\hline
		100&$64\times64\times8$ &$5\cdot 10^4$ & $10^6$& -0.00864809(4) & 0.99060424(3)\\\hline
		100&$64\times64\times10$ &$5\cdot 10^4$ & $10^6$& -0.00892280(4) & 0.99047869(3)\\\hline
		100&$64\times64\times12$ &$5\cdot 10^4$ & $2\cdot 10^6$& -0.00910613(3) & 0.99039492(2)\\\hline
		100&$64\times64\times14$ &$5\cdot 10^4$ & $2\cdot 10^6$& -0.00923706(3) & 0.99033516(2)\\\hline
		100&$64\times64\times16$ &$5\cdot 10^4$ & $2\cdot 10^6$& -0.00933531(3) & 0.99029032(2)\\\hline
		100&$64\times64\times18$ &$5\cdot 10^4$ & $1.5\cdot 10^6$& -0.00941173(4) & 0.99025546(3)\\\hline
		100&$64\times64\times48$ &$10^5$ & $4\cdot 10^6$& -0.009794126(10) & 0.990081035(7)\\\hline
		50&$64\times64\times8$ &$5\cdot 10^4$ & $5\cdot 10^5$& -0.01733471(12) & 0.98116553(9)\\\hline
		50&$64\times64\times9$ &$5\cdot 10^4$ & $5\cdot 10^5$& -0.01764072(12) & 0.98102542(8)\\\hline
		50&$64\times64\times10$ &$5\cdot 10^4$ & $5\cdot 10^5$& -0.01788591(11) & 0.98091320(8)\\\hline
		50&$64\times64\times11$ &$5\cdot 10^4$ & $5\cdot 10^5$& -0.01808652(11) & 0.98082162(8)\\\hline
		50&$64\times64\times12$ &$5\cdot 10^4$ & $2\cdot 10^6$& -0.01825392(6) & 0.98074506(4)\\\hline
		50&$64\times64\times13$ &$5\cdot 10^4$ & $5\cdot 10^5$& -0.01839534(10) & 0.98068065(7)\\\hline
		50&$64\times64\times14$ &$5\cdot 10^4$ & $2\cdot 10^6$& -0.01851678(5) & 0.98062493(4)\\\hline
		50&$64\times64\times15$ &$5\cdot 10^4$ & $5\cdot 10^5$& -0.01862206(9) & 0.98057695(7)\\\hline
		50&$64\times64\times16$ &$5\cdot 10^4$ & $2\cdot 10^6$& -0.01871403(5) & 0.98053489(4)\\\hline
		50&$64\times64\times18$ &$5\cdot 10^4$ & $2.5\cdot 10^6$& -0.01886755(4) & 0.98046477(3)\\\hline
		50&$64\times64\times48$ &$10^5$ & $5\cdot 10^6$& -0.019635150(17) & 0.980114360(12)\\\hline
		20&$64\times64\times8$ &$5\cdot 10^4$ & $10^6$& -0.0436416(2) & 0.9525718(2)\\\hline
		20&$64\times64\times9$ &$5\cdot 10^4$ & $5\cdot 10^5$& -0.0444160(3) & 0.9522161(2)\\\hline
		20&$64\times64\times10$ &$5\cdot 10^4$ & $10^6$& -0.0450362(2) & 0.95193193(14)\\\hline
		20&$64\times64\times11$ &$5\cdot 10^4$ & $5\cdot 10^5$& -0.0455439(3) & 0.95169923(19)\\\hline
		20&$64\times64\times12$ &$5\cdot 10^4$ & $2\cdot 10^6$& -0.04596731(13) & 0.95150481(10)\\\hline
		20&$64\times64\times48$ &$10^5$ & $2.5\cdot 10^6$& -0.04946297(6) & 0.94990444(4)\\\hline
	\end{tabular}
	\caption{Mean value of the plaquette $P_{01}$ and the average of all the plaquettes $P_{\mu\nu}$ with couplings constants $\beta=\{100,50,20\}$ and lattice sizes $N_0N_1N_2$. The \acrshort*{pccbc} are imposed in the second spatial direction $N_2$. $N_{\text{therm}}$ and $N_{\text{measure}}$ are the \acrshort*{mc} steps used for thermalization and measuring respectively.}
	\label{tab:Energy_lattice_size_perf_cond}
\end{table}

\chapter{Results of $SU(2)$ simulations in 3+1 dimensions}\label{ch:31_values}
In this Appendix we show explicitly the numerical values of the \acrshort*{mc} simulations in 3+1 dimensions that were used for obtaining the results presented in \autoref{chp:su2_31}. As we did in the previous Appendix, the statistical error is shown in parenthesis and corresponds to the last decimal places, i.e $0.967\pm 0.009\equiv 0.967(9)$.
\section{Renormalization}\label{sec:Table_reno_31}
In this first section, we present the data that was used for for analyzing the renormalization of the energy observable in \autoref{sec:reno_31}. Since we do not need much precision for this matter, a lot less statistics is used in comparison to the data that will be used for computing the Casimir energy. 
\begin{table}[H]
	\centering
	\begin{tabular}{|c|c|c|c|c|c|}
		\hline
		\rule{0pt}{12pt}$\beta$& $N_0\times N_1\times N_2\times N_3$ & $N_{\text{therm}}$& $N_{\text{measure}}$ &$\frac{1}{2}\braket{\text{tr }P_d}$ & $\frac{1}{2}\braket{\text{tr }P_{\mu\nu}}$\\
		\hline
		3&$48\times48\times48\times5$ &$10^4$ & $10^4$& -0.000453(3) & 0.7231975(18)\\\hline
		3&$48\times48\times48\times10$ &$8\cdot 10^4$ & $10^5$& -0.0000233(6) & 0.7231602(4)\\\hline
		3&$48\times48\times48\times15$ &$1.5\cdot10^4$ & $2\cdot10^5$& -0.0000042(3) & 0.7231589(2)\\\hline
		3&$48\times48\times48\times20$ &$1.5\cdot10^4$ & $4\cdot10^5$& -0.0000013(2) & 0.72315735(14)\\\hline
		3&$48\times48\times48\times30$ &$10^4$ & $10^4$& 0.0000008(10) & 0.7231560(7)\\\hline
		3&$48\times48\times48\times40$ &$10^4$ & $10^4$& -0.0000002(9) & 0.7231560(6)\\\hline
		3&$48\times48\times48\times50$ &$10^4$ & $10^4$& 0.0000005(8) & 0.7231557(6)\\\hline
	\end{tabular}
	\caption{Mean value of the difference plaquette $P_{d}$ \eqref{eq:plaq_dif} and average of all the plaquettes $P_{\mu\nu}$ with coupling constant $\beta=3$ and lattice sizes $N_0N_1N_2N_3$. \acrshort*{pbc} are used in every direction. $N_{\text{therm}}$ and $N_{\text{measure}}$ are the \acrshort*{mc} steps used for thermalization and measuring respectively.}
	\label{tab:Energy_renom_periodic_1_31}
\end{table}
\begin{table}[H]
	\centering
	\begin{tabular}{|c|c|c|c|c|c|}
		\hline
		\rule{0pt}{12pt}$\beta$& $N_0\times N_1\times N_2\times N_3$ & $N_{\text{therm}}$& $N_{\text{measure}}$ &$\frac{1}{2}\braket{\text{tr }P_d}$ & $\frac{1}{2}\braket{\text{tr }P_{\mu\nu}}$\\
		\hline
		2.9&$48\times48\times48\times5$ &$10^4$ & $10^4$& -0.000462(3) & 0.7118790(19)\\\hline
		2.9&$48\times48\times48\times10$ &$8\cdot 10^4$ & $10^5$& -0.0000240(6) & 0.7118381(4)\\\hline
		2.9&$48\times48\times48\times15$ &$1.5\cdot 10^5$ & $2\cdot 10^5$& -0.0000039(3) & 0.7118332(2)\\\hline
		2.9&$48\times48\times48\times20$ &$1.5\cdot 10^5$ & $4\cdot 10^5$& -0.0000009(2) & 0.71183094(15)\\\hline
		2.9&$48\times48\times48\times30$ &$10^4$ & $10^4$& 0.0000016(11) & 0.7118289(8)\\\hline
		2.9&$48\times48\times48\times40$ &$10^4$ & $10^4$& 0.0000005(9) & 0.7118296(7)\\\hline
		2.9&$48\times48\times48\times50$ &$10^4$ & $10^4$& -0.0000008(8) & 0.7118292(6)\\\hline
		2.8&$48\times48\times48\times5$ &$10^4$ & $10^4$& -0.000469(3) & 0.699463(2)\\\hline
		2.8&$48\times48\times48\times10$ &$8\cdot 10^4$ & $10^5$& -0.0000228(6) & 0.6993991(4)\\\hline
		2.8&$48\times48\times48\times15$ &$1.5\cdot 10^5$ & $2\cdot 10^5$& -0.0000032(4) & 0.6993905(3)\\\hline
		2.8&$48\times48\times48\times20$ &$1.5\cdot 10^5$ & $10^6$& -0.00000045(14) & 0.69938477(10)\\\hline
		2.8&$48\times48\times48\times30$ &$10^4$ & $10^4$& -0.0000004(11) & 0.6993834(8)\\\hline
		2.8&$48\times48\times48\times40$ &$10^4$ & $10^4$& -0.0000010(10) & 0.6993838(7)\\\hline
		2.8&$48\times48\times48\times50$ &$10^4$ & $10^4$& -0.0000008(9) & 0.6993836(6)\\\hline
		2.7&$48\times48\times48\times5$ &$8\cdot 10^4$ & $8\cdot 10^4$& -0.0004813(10) & 0.6857049(7)\\\hline
		2.7&$48\times48\times48\times10$ &$8\cdot 10^4$ & $10^5$& -0.0000206(6) & 0.6856067(5)\\\hline
		2.7&$48\times48\times48\times15$ &$10^5$ & $5\cdot 10^5$& -0.0000007(2) & 0.68557787(17)\\\hline
		2.7&$48\times48\times48\times20$ &$10^4$ & $10^4$& 0.0000017(14) & 0.6855752(10)\\\hline
		2.7&$48\times48\times48\times30$ &$10^4$ & $10^4$& -0.0000030(12) & 0.6855732(9)\\\hline
		2.7&$48\times48\times48\times40$ &$10^4$ & $10^4$& 0.0000009(10) & 0.6855716(7)\\\hline
		2.7&$48\times48\times48\times50$ &$10^4$ & $10^4$& -0.0000004(9) & 0.6855755(7)\\\hline
		2.6&$48\times48\times48\times5$ &$8\cdot 10^4$ & $8\cdot 10^4$& -0.0004790(11) & 0.6702687(8)\\\hline
		2.6&$48\times48\times48\times10$ &$7\cdot 10^4$ & $8\cdot 10^4$& -0.0000123(8) & 0.6700529(6)\\\hline
		2.6&$48\times48\times48\times15$ &$10^4$ & $10^4$& -0.0000008(17) & 0.6700138(13)\\\hline
		2.6&$48\times48\times48\times20$ &$10^4$ & $10^4$& -0.0000021(15) & 0.6700100(11)\\\hline
		2.6&$48\times48\times48\times30$ &$10^4$ & $10^4$& -0.0000017(12) & 0.6700123(9)\\\hline
		2.6&$48\times48\times48\times40$ &$10^4$ & $10^4$& 0.0000005(11) & 0.6700112(8)\\\hline
		2.6&$48\times48\times48\times50$ &$10^4$ & $10^4$& 0.0000005(9) & 0.6700118(7)\\\hline
		2.509&$48\times48\times48\times5$ &$10^5$ & $10^5$& -0.0004593(10) & 0.6542483(8)\\\hline
		2.509&$48\times48\times48\times10$ &$1.6\cdot 10^5$ & $4\cdot 10^5$& -0.0000007(3) & 0.6537249(3)\\\hline
		2.509&$48\times48\times48\times15$ &$10^4$ & $10^4$& 0.0000017(18) & 0.6537244(14)\\\hline
		2.509&$48\times48\times48\times20$ &$10^4$ & $10^4$& 0.0000016(16) & 0.6537267(12)\\\hline
		2.509&$48\times48\times48\times30$ &$10^4$ & $10^4$& 0.0000003(13) & 0.6537203(10)\\\hline
		2.509&$48\times48\times48\times40$ &$10^4$ & $10^4$& -0.0000008(11) & 0.6537247(9)\\\hline
		2.509&$48\times48\times48\times50$ &$10^4$ & $10^4$& 0.0000004(10) & 0.6537243(8)\\\hline
		2.427&$48\times48\times48\times5$ &$8\cdot 10^4$ & $10^5$& -0.0003659(10) & 0.6372941(8)\\\hline
		2.427&$48\times48\times48\times10$ &$10^4$ & $10^4$& -0.000004(2) & 0.6364323(19)\\\hline
		2.427&$48\times48\times48\times15$ &$10^4$ & $10^4$& 0.0000004(19) & 0.6364328(15)\\\hline
		2.427&$48\times48\times48\times20$ &$10^4$ & $10^4$& -0.0000006(17) & 0.6364379(13)\\\hline
		2.427&$48\times48\times48\times30$ &$10^4$ & $10^4$& -0.0000001(13) & 0.6364345(11)\\\hline
		2.427&$48\times48\times48\times40$ &$10^4$ & $10^4$& -0.0000004(12) & 0.6364343(10)\\\hline
		2.427&$48\times48\times48\times50$ &$10^4$ & $10^4$& -0.0000006(10) & 0.6364327(8)\\\hline
	\end{tabular}
	\caption{Mean value of the difference plaquette $P_{d}$ \eqref{eq:plaq_dif} and average of all the plaquettes $P_{\mu
			\nu}$ with coupling constants $\beta=\{2.9,2.8,2.7,2.6,2.509,2.427\}$ and lattice sizes $N_0N_1N_2N_3$. \acrshort*{pbc} are used in every direction. $N_{\text{therm}}$ and $N_{\text{measure}}$ are the \acrshort*{mc} steps used for thermalization and measuring respectively.}
	\label{tab:Energy_renom_periodic_2_31}
\end{table}
\begin{table}[H]
	\centering
	\begin{tabular}{|c|c|c|c|c|c|}
		\hline
		\rule{0pt}{12pt}$\beta$& $N_0\times N_1\times N_2\times N_3$ & $N_{\text{therm}}$& $N_{\text{measure}}$ &$\frac{1}{2}\braket{\text{tr }P_d}$ & $\frac{1}{2}\braket{\text{tr }P_{\mu\nu}}$\\
		\hline
		3&$48\times48\times48\times5$ &$10^4$ & $10^4$& 0.040160(2) & 0.6144497(14)\\\hline
		3&$48\times48\times48\times10$ &$10^5$ & $6\cdot 10^5$& 0.0212064(2) & 0.74294057(16)\\\hline
		3&$48\times48\times48\times15$ &$10^4$ & $10^4$& 0.0141397(14) & 0.7363472(10)\\\hline
		3&$48\times48\times48\times20$ &$10^4$ & $10^4$& 0.0106058(12) & 0.7330501(9)\\\hline
		3&$48\times48\times48\times30$ &$10^4$ & $10^4$& 0.0070705(10) & 0.7297517(7)\\\hline
		3&$48\times48\times48\times40$ &$10^4$ & $10^4$& 0.0053031(9) & 0.7281025(6)\\\hline
		3&$48\times48\times48\times50$ &$10^4$ & $10^4$& 0.0042412(8) & 0.7271150(6)\\\hline
		2.9&$48\times48\times48\times5$ &$10^4$ & $10^4$& 0.044085(2) & 0.7532614(17)\\\hline
		2.9&$48\times48\times48\times10$ &$10^5$ & $2\cdot10^5$& 0.0220734(4) & 0.7325600(3)\\\hline
		2.9&$48\times48\times48\times15$ &$10^4$ & $10^4$& 0.0147166(15) & 0.7256528(11)\\\hline
		2.9&$48\times48\times48\times20$ &$10^4$ & $10^4$& 0.0110394(13) & 0.7221956(9)\\\hline
		2.9&$48\times48\times48\times30$ &$10^4$ & $10^4$& 0.0073582(11) & 0.7187405(8)\\\hline
		2.9&$48\times48\times48\times40$ &$10^4$ & $10^4$& 0.0055202(9) & 0.7170121(7)\\\hline
		2.9&$48\times48\times48\times50$ &$10^4$ & $10^4$& 0.0044149(8) & 0.7159770(6)\\\hline
		2.8&$48\times48\times48\times5$ &$10^4$ & $10^4$& 0.045966(3) & 0.7429635(18)\\\hline
		2.8&$48\times48\times48\times10$ &$10^5$ & $4.5\cdot10^5$& 0.0230224(3) & 0.72119379(20)\\\hline
		2.8&$48\times48\times48\times15$ &$10^4$ & $10^4$& 0.0153511(16) & 0.7139277(11)\\\hline
		2.8&$48\times48\times48\times20$ &$10^4$ & $10^4$& 0.0115131(13) & 0.7102936(10)\\\hline
		2.8&$48\times48\times48\times30$ &$10^4$ & $10^4$& 0.0076765(11) & 0.7066546(8)\\\hline
		2.8&$48\times48\times48\times40$ &$10^4$ & $10^4$& 0.0057575(9) & 0.7048390(7)\\\hline
		2.8&$48\times48\times48\times50$ &$10^4$ & $10^4$& 0.0046051(9) & 0.7037464(6)\\\hline
		2.7&$48\times48\times48\times5$ &$8\cdot10^4$ & $5\cdot10^4$& 0.0480541(12) & 0.7316438(8)\\\hline
		2.7&$48\times48\times48\times10$ &$1.6\cdot10^5$ & $2\cdot10^5$& 0.0240694(4) & 0.7086495(3)\\\hline
		2.7&$48\times48\times48\times15$ &$1.4\cdot10^5$ & $3\cdot10^5$& 0.0160500(3) & 0.7009656(2)\\\hline
		2.7&$48\times48\times48\times20$ &$10^4$ & $10^4$& 0.0120386(14) & 0.6971153(10)\\\hline
		2.7&$48\times48\times48\times30$ &$10^4$ & $10^4$& 0.0080255(12) & 0.6932687(8)\\\hline
		2.7&$48\times48\times48\times40$ &$10^4$ & $10^4$& 0.0060188(10) & 0.6913442(7)\\\hline
		2.7&$48\times48\times48\times50$ &$10^4$ & $10^4$& 0.0048169(9) & 0.6901919(7)\\\hline
		2.6&$48\times48\times48\times5$ &$8\cdot 10^4$ & $5\cdot10^4$& 0.0503664(12) & 0.7191176(9)\\\hline
		2.6&$48\times48\times48\times10$ &$10^5$ & $10^5$& 0.0252373(6) & 0.6946494(5)\\\hline
		2.6&$48\times48\times48\times15$ &$1.4\cdot10^5$ & $3\cdot10^5$& 0.0168326(3) & 0.6864398(2)\\\hline
		2.6&$48\times48\times48\times20$ &$10^4$ & $10^4$& 0.0126257(15) & 0.6823282(11)\\\hline
		2.6&$48\times48\times48\times30$ &$10^4$ & $10^4$& 0.0084187(12) & 0.6782233(9)\\\hline
		2.6&$48\times48\times48\times40$ &$10^4$ & $10^4$& 0.0063116(10) & 0.6761688(8)\\\hline
		2.6&$48\times48\times48\times50$ &$10^4$ & $10^4$& 0.0050525(9) & 0.6749382(7)\\\hline
		2.509&$48\times48\times48\times5$ &$8\cdot10^4$ & $5\cdot10^4$& 0.0527189(13) & 0.7064288(9)\\\hline
		2.509&$48\times48\times48\times10$ &$10^5$ & $10^5$& 0.0264421(7) & 0.6802340(5)\\\hline
		2.509&$48\times48\times48\times15$ &$10^4$ & $10^4$& 0.0176398(17) & 0.6713768(13)\\\hline
		2.509&$48\times48\times48\times20$ &$10^4$ & $10^4$& 0.0132302(15) & 0.6669598(12)\\\hline
		2.509&$48\times48\times48\times30$ &$10^4$ & $10^4$& 0.0088167(13) & 0.6625509(10)\\\hline
		2.509&$48\times48\times48\times40$ &$10^4$ & $10^4$& 0.0066134(11) & 0.6603428(8)\\\hline
		2.509&$48\times48\times48\times50$ &$10^4$ & $10^4$& 0.0052935(10) & 0.6590194(8)\\\hline
	\end{tabular}
	\caption{Mean value of the difference plaquette $P_{d}$ \eqref{eq:plaq_dif} and average of all the plaquettes $P_{\mu
			\nu}$ with coupling constants $\beta=\{3,2.9,2.8,2.7,2.6,2.509\}$ and lattice sizes $N_0N_1N_2N_3$. \acrshort*{pccbc} are imposed in the transverse direction $N_3$. $N_{\text{therm}}$ and $N_{\text{measure}}$ are the \acrshort*{mc} steps used for thermalization and measuring respectively.}
	\label{tab:Energy_renom_conductor_1_31}
\end{table}
\begin{table}[H]
	\centering
	\begin{tabular}{|c|c|c|c|c|c|}
		\hline
		\rule{0pt}{12pt}$\beta$& $N_0\times N_1\times N_2\times N_3$ & $N_{\text{therm}}$& $N_{\text{measure}}$ &$\frac{1}{2}\braket{\text{tr }P_d}$ & $\frac{1}{2}\braket{\text{tr }P_{\mu\nu}}$\\
		\hline
		2.427&$48\times48\times48\times5$ &$8\cdot10^4$ & $5\cdot10^4$& 0.0551082(13) & 0.6936542(10)\\\hline
		2.427&$48\times48\times48\times10$ &$1.4\cdot10^4$ & $3\cdot10^5$& 0.0276961(4) & 0.6651860(3)\\\hline
		2.427&$48\times48\times48\times15$ &$10^4$ & $10^4$& 0.0184678(18) & 0.6555926(15)\\\hline
		2.427&$48\times48\times48\times20$ &$10^4$ & $10^4$& 0.0138514(16) & 0.6507978(13)\\\hline
		2.427&$48\times48\times48\times30$ &$10^4$ & $10^4$& 0.0092337(13) & 0.6460119(10)\\\hline
		2.427&$48\times48\times48\times40$ &$10^4$ & $10^4$& 0.0069229(12) & 0.6436140(9)\\\hline
		2.427&$48\times48\times48\times50$ &$10^4$ & $10^4$& 0.0055388(10) & 0.6421802(8)\\\hline
	\end{tabular}
	\caption{Mean value of the difference plaquette $P_{d}$ \eqref{eq:plaq_dif} and average of all the plaquettes $P_{\mu
			\nu}$ with coupling constant $\beta=2.427$ and lattice sizes $N_0N_1N_2N_3$. \acrshort*{pccbc} are imposed in the transverse direction $N_3$. $N_{\text{therm}}$ and $N_{\text{measure}}$ are the \acrshort*{mc} steps used for thermalization and measuring respectively.}
	\label{tab:Energy_renom_conductor_2_31}
\end{table}
\section{Casimir energy}
In this section we show the simulations results that are used for computing the Casimir energy in \autoref{sec:Casimir_31}. In this, a lot more statistic is used than in \autoref{sec:Table_reno_31} since much smaller errors are required.

\begin{table}[H]
	\centering
	\begin{tabular}{|c|c|c|c|c|c|}
		\hline
		\rule{0pt}{12pt}$\beta$& $N_0\times N_1\times N_2\times N_3$ & $N_{\text{therm}}$& $N_{\text{measure}}$ &$\frac{1}{2}\braket{\text{tr }P_d}$ & $\frac{1}{2}\braket{\text{tr }P_{\mu\nu}}$\\
		\hline
		3&$48\times48\times48\times10$ &$8\cdot10^4$ & $10^5$& -0.0000233(6) & 0.7231602(4)\\\hline
		3&$48\times48\times48\times11$ &$10^5$ & $1.4\cdot10^5$& -0.0000151(5) & 0.7231606(3)\\\hline
		3&$48\times48\times48\times12$ &$10^5$ & $1.2\cdot10^5$& -0.0000115(5) & 0.7231605(3)\\\hline
		3&$48\times48\times48\times13$ &$1.2\cdot10^5$ & $1.6\cdot10^5$& -0.0000081(4) & 0.7231591(3)\\\hline
		3&$48\times48\times48\times14$ &$1.2\cdot10^5$ & $1.6\cdot10^5$& -0.0000053(4) & 0.7231583(3)\\\hline
		3&$48\times48\times48\times15$ &$1.5\cdot10^5$ & $2\cdot10^5$& -0.0000042(3) & 0.7231589(2)\\\hline
		3&$48\times48\times48\times16$ &$1.5\cdot10^5$ & $3\cdot10^5$& -0.0000033(3) & 0.72315815(18)\\\hline
		3&$48\times48\times48\times17$ &$1.5\cdot10^5$ & $3\cdot10^5$& -0.0000028(3) & 0.72315777(18)\\\hline
		3&$48\times48\times48\times18$ &$1.5\cdot10^5$ & $3\cdot10^5$& -0.0000022(2) & 0.72315786(17)\\\hline
		3&$48\times48\times48\times19$ &$1.5\cdot10^5$ & $3\cdot10^5$& -0.0000017(2) & 0.72315791(17)\\\hline
		3&$48\times48\times48\times20$ &$1.5\cdot10^5$ & $4\cdot10^5$& -0.0000013(2) & 0.72315735(14)\\\hline
		3&$48\times48\times48\times21$ &$1.5\cdot10^5$ & $4\cdot10^5$& -0.0000011(2) & 0.72315728(14)\\\hline
	\end{tabular}
	\caption{Mean value of the difference plaquette $P_{d}$ \eqref{eq:plaq_dif} and average of all the plaquettes $P_{\mu
		\nu}$ with coupling constant $\beta=3$ and lattice sizes $N_0N_1N_2N_3$. \acrshort*{pbc} are used in every direction. $N_{\text{therm}}$ and $N_{\text{measure}}$ are the \acrshort*{mc} steps used for thermalization and measuring respectively.}
	\label{tab:Energy_periodic_1_31}
\end{table}
\begin{table}[H]
	\centering
	\begin{tabular}{|c|c|c|c|c|c|}
		\hline
		\rule{0pt}{12pt}$\beta$& $N_0\times N_1\times N_2\times N_3$ & $N_{\text{therm}}$& $N_{\text{measure}}$ &$\frac{1}{2}\braket{\text{tr }P_d}$ & $\frac{1}{2}\braket{\text{tr }P_{\mu\nu}}$\\
		\hline
		2.9&$48\times48\times48\times8$ &$8\cdot 10^4$ & $10^5$& -0.0000614(7) & 0.7118438(5)\\\hline
		2.9&$48\times48\times48\times10$ &$8\cdot10^4$ & $10^5$& -0.0000240(6) & 0.7118381(4)\\\hline
		2.9&$48\times48\times48\times11$ &$10^5$ & $1.4\cdot10^5$& -0.0000163(5) & 0.7118364(3)\\\hline
		2.9&$48\times48\times48\times12$ &$10^5$ & $1.2\cdot10^5$& -0.0000114(5) & 0.7118355(4)\\\hline
		2.9&$48\times48\times48\times13$ &$1.2\cdot10^5$ & $1.6\cdot10^5$& -0.0000080(4) & 0.7118341(3)\\\hline
		2.9&$48\times48\times48\times14$ &$1.2\cdot10^5$ & $1.6\cdot10^5$& -0.0000057(4) & 0.7118334(3)\\\hline
		2.9&$48\times48\times48\times15$ &$1.5\cdot10^5$ & $2\cdot10^5$& -0.0000039(3) & 0.7118332(2)\\\hline
		2.9&$48\times48\times48\times16$ &$1.5\cdot10^5$ & $3\cdot10^5$& -0.0000029(3) & 0.71183245(19)\\\hline
		2.9&$48\times48\times48\times17$ &$1.5\cdot10^5$ & $3\cdot10^5$& -0.0000020(3) & 0.71183170(19)\\\hline
		2.9&$48\times48\times48\times18$ &$1.5\cdot10^5$ & $3\cdot10^5$& -0.0000018(3) & 0.71183209(18)\\\hline
		2.9&$48\times48\times48\times19$ &$1.5\cdot10^5$ & $3\cdot10^5$& -0.0000013(2) & 0.71183144(18)\\\hline
		2.9&$48\times48\times48\times20$ &$1.5\cdot10^5$ & $4\cdot10^5$& -0.0000009(2) & 0.71183094(15)\\\hline
		2.9&$48\times48\times48\times21$ &$1.5\cdot10^5$ & $4\cdot10^5$& -0.0000006(2) & 0.71183092(15)\\\hline
		2.8&$48\times48\times48\times6$ &$6\cdot10^4$ & $3.2\cdot10^5$& -0.0002105(4) & 0.6994310(3)\\\hline
		2.8&$48\times48\times48\times7$ &$6\cdot10^4$ & $3.2\cdot10^5$& -0.0001082(4) & 0.6994161(3)\\\hline
		2.8&$48\times48\times48\times8$ &$6\cdot10^4$ & $10^5$& -0.0000611(7) & 0.6994085(5)\\\hline
		2.8&$48\times48\times48\times9$ &$6\cdot10^4$ & $10^5$& -0.0000367(6) & 0.6994032(5)\\\hline
		2.8&$48\times48\times48\times10$ &$8\cdot10^4$ & $10^5$& -0.0000228(6) & 0.6993991(4)\\\hline
		2.8&$48\times48\times48\times11$ &$10^5$ & $1.4\cdot 10^5$& -0.0000154(5) & 0.6993967(4)\\\hline
		2.8&$48\times48\times48\times12$ &$10^5$ & $1.2\cdot 10^5$& -0.0000105(5) & 0.6993948(4)\\\hline
		2.8&$48\times48\times48\times13$ &$1.2\cdot10^5$ & $1.6\cdot 10^5$& -0.0000074(4) & 0.6993933(3)\\\hline
		2.8&$48\times48\times48\times14$ &$1.2\cdot10^5$ & $1.6\cdot 10^5$& -0.0000052(4) & 0.6993920(3)\\\hline
		2.8&$48\times48\times48\times15$ &$1.5\cdot10^5$ & $2\cdot 10^5$& -0.0000032(4) & 0.6993905(3)\\\hline
		2.8&$48\times48\times48\times16$ &$1.5\cdot10^5$ & $3\cdot 10^5$& -0.0000023(3) & 0.6993896(2)\\\hline
		2.8&$48\times48\times48\times17$ &$1.5\cdot10^5$ & $3\cdot 10^5$& -0.0000018(3) & 0.6993878(2)\\\hline
		2.8&$48\times48\times48\times18$ &$1.5\cdot10^5$ & $3\cdot 10^5$& -0.0000010(3) & 0.69938645(19)\\\hline
		2.8&$48\times48\times48\times19$ &$1.5\cdot10^5$ & $4\cdot 10^5$& -0.0000006(2) & 0.69938581(16)\\\hline
		2.8&$48\times48\times48\times20$ &$1.5\cdot10^5$ & $10^6$& -0.00000045(14) & 0.69938477(10)\\\hline
		2.7&$48\times48\times48\times5$ &$8\cdot10^4$ & $8\cdot 10^4$& -0.0004813(10) & 0.6857049(7)\\\hline
		2.7&$48\times48\times48\times6$ &$8\cdot10^4$ & $8\cdot 10^4$& -0.0002104(9) & 0.6856599(7)\\\hline
		2.7&$48\times48\times48\times7$ &$8\cdot10^4$ & $10^5$& -0.0001075(8) & 0.6856386(6)\\\hline
		2.7&$48\times48\times48\times8$ &$8\cdot10^4$ & $10^5$& -0.0000595(7) & 0.6856235(5)\\\hline
		2.7&$48\times48\times48\times9$ &$10^5$ & $10^5$& -0.0000346(7) & 0.6856131(5)\\\hline
		2.7&$48\times48\times48\times10$ &$8\cdot10^4$ & $10^5$& -0.0000206(6) & 0.6856067(5)\\\hline
		2.7&$48\times48\times48\times11$ &$8\cdot10^4$ & $10^5$& -0.0000133(6) & 0.6856002(4)\\\hline
		2.7&$48\times48\times48\times12$ &$8\cdot10^4$ & $10^5$& -0.0000084(6) & 0.6855948(4)\\\hline
		2.7&$48\times48\times48\times13$ &$8\cdot10^4$ & $10^5$& -0.0000056(6) & 0.6855882(4)\\\hline
		2.7&$48\times48\times48\times14$ &$1.2\cdot10^5$ & $2\cdot10^5$& -0.0000032(4) & 0.6855829(3)\\\hline
		2.6&$48\times48\times48\times4$ &$8\cdot10^4$ & $8\cdot10^4$& -0.0013770(12) & 0.6704354(9)\\\hline
		2.6&$48\times48\times48\times5$ &$8\cdot10^4$ & $8\cdot10^4$& -0.0004790(11) & 0.6702687(8)\\\hline
		2.6&$48\times48\times48\times6$ &$8\cdot10^4$ & $10^5$& -0.0002061(9) & 0.6701912(6)\\\hline
		2.6&$48\times48\times48\times7$ &$8\cdot10^4$ & $10^5$& -0.0001013(8) & 0.6701454(6)\\\hline
		2.6&$48\times48\times48\times8$ &$8\cdot10^4$ & $1.6\cdot10^5$& -0.0000527(6) & 0.6701116(4)\\\hline
		2.6&$48\times48\times48\times9$ &$8\cdot10^4$ & $10^5$& -0.0000257(7) & 0.6700828(5)\\\hline
		2.6&$48\times48\times48\times10$ &$8\cdot10^4$ & $8\cdot10^4$& -0.0000123(8) & 0.6700529(6)\\\hline
	\end{tabular}
		\caption{Mean value of the difference plaquette $P_{d}$ \eqref{eq:plaq_dif} and average of all the plaquettes $P_{\mu\nu}$ with coupling constants $\beta=\{2.9,2.8,2.7,2.6\}$ and lattice sizes $N_0N_1N_2N_3$. \acrshort*{pbc} are used in every direction. $N_{\text{therm}}$ and $N_{\text{measure}}$ are the \acrshort*{mc} steps used for thermalization and measuring respectively.}
	\label{tab:Energy_periodic_2_31}
\end{table}
\begin{table}[H]
	\centering
	\begin{tabular}{|c|c|c|c|c|c|}
		\hline
		\rule{0pt}{12pt}$\beta$& $N_0\times N_1\times N_2\times N_3$ & $N_{\text{therm}}$& $N_{\text{measure}}$ &$\frac{1}{2}\braket{\text{tr }P_d}$ & $\frac{1}{2}\braket{\text{tr }P_{\mu\nu}}$\\
		\hline
		2.509&$48\times48\times48\times4$ &$8\cdot10^4$ & $8\cdot10^4$& -0.0013640(12) & 0.6545488(9)\\\hline
		2.509&$48\times48\times48\times5$ &$8\cdot10^4$ & $10^5$& -0.0004593(10) & 0.6542483(8)\\\hline
		2.509&$48\times48\times48\times6$ &$8\cdot10^4$ & $10^5$& -0.0001821(9) & 0.6540760(7)\\\hline
		2.509&$48\times48\times48\times7$ &$8\cdot10^4$ & $10^5$& -0.0000704(8) & 0.6539322(6)\\\hline
		2.427&$48\times48\times48\times4$ &$8\cdot10^4$ & $8\cdot10^4$& -0.0012910(13) & 0.6379957(10)\\\hline
		2.427&$48\times48\times48\times5$ &$8\cdot10^4$ & $10^5$& -0.0003659(10) & 0.6372941(8)\\\hline
	\end{tabular}
\caption{Mean value of the difference plaquette $P_{d}$ \eqref{eq:plaq_dif} and average of all the plaquettes $P_{\mu\nu}$ with coupling constants $\beta=\{2.509,2.427\}$ and lattice sizes $N_0N_1N_2N_3$. \acrshort*{pbc} are used in every direction. $N_{\text{therm}}$ and $N_{\text{measure}}$ are the \acrshort*{mc} steps used for thermalization and measuring respectively.}
	\label{tab:Energy_periodic_3_31}
\end{table}

\begin{table}[H]
	\centering
	\begin{tabular}{|c|c|c|c|c|c|}
		\hline
		\rule{0pt}{12pt}$\beta$& $N_0\times N_1\times N_2\times N_3$ & $N_{\text{therm}}$& $N_{\text{measure}}$ &$\frac{1}{2}\braket{\text{tr }P_d}$ & $\frac{1}{2}\braket{\text{tr }P_{\mu\nu}}$\\
		\hline
		3&$48\times48\times48\times10$ &$10^5$ & $6\cdot10^5$& 0.0212064(2) & 0.74294057(16)\\\hline
		3&$48\times48\times48\times11$ &$10^5$ & $6\cdot10^5$& 0.0192797(2) & 0.74114248(15)\\\hline
		3&$48\times48\times48\times12$ &$10^5$ & $6\cdot10^5$& 0.0176738(2) & 0.73964398(15)\\\hline
		3&$48\times48\times48\times13$ &$10^5$ & $6\cdot10^5$& 0.0163145(2) & 0.73837618(14)\\\hline
		3&$48\times48\times48\times36$ &$1.5\cdot 10^5$ & $8\cdot10^5$& 0.00589181(11) & 0.72865307(7)\\\hline
		2.9&$48\times48\times48\times9$ &$10^5$ & $2\cdot10^5$& 0.0245240(4) & 0.7348620(3)\\\hline
		2.9&$48\times48\times48\times10$ &$10^5$ & $2\cdot10^5$& 0.0220734(4) & 0.7325600(3)\\\hline
		2.9&$48\times48\times48\times11$ &$10^5$ & $4.5\cdot10^5$& 0.0200672(3) & 0.73067655(18)\\\hline
		2.9&$48\times48\times48\times12$ &$10^5$ & $4.5\cdot10^5$& 0.0183961(2) & 0.72910661(18)\\\hline
		2.9&$48\times48\times48\times13$ &$10^5$ & $4.5\cdot10^5$& 0.0169811(2) & 0.72777753(17)\\\hline
		2.9&$48\times48\times48\times36$ &$1.5\cdot10^5$ & $3\cdot10^5$& 0.00613309(18) & 0.71758908(13)\\\hline
		2.8&$48\times48\times48\times6$ &$10^5$ & $2\cdot10^5$& 0.0383394(5) & 0.7357111(4)\\\hline
		2.8&$48\times48\times48\times7$ &$8\cdot10^4$ & $1.2\cdot10^5$& 0.0328753(6) & 0.7305297(5)\\\hline
		2.8&$48\times48\times48\times8$ &$10^5$ & $2\cdot10^5$& 0.0287717(5) & 0.7266405(3)\\\hline
		2.8&$48\times48\times48\times9$ &$10^5$ & $2\cdot10^5$& 0.0255775(4) & 0.7236146(3)\\\hline
		2.8&$48\times48\times48\times10$ &$10^5$ & $4.5\cdot10^5$& 0.0230224(3) & 0.7211938(2)\\\hline
		2.8&$48\times48\times48\times11$ &$10^5$ & $4.5\cdot10^5$& 0.0209303(3) & 0.71921232(19)\\\hline
		2.8&$48\times48\times48\times12$ &$10^5$ & $4.5\cdot10^5$& 0.0191866(3) & 0.71756130(18)\\\hline
		2.8&$48\times48\times48\times13$ &$10^5$ & $4.5\cdot10^5$& 0.0177114(2) & 0.71616351(18)\\\hline
		2.8&$48\times48\times48\times36$ &$1.5\cdot10^5$ & $6\cdot10^5$& 0.00639659(13) & 0.70544349(9)\\\hline
		
	\end{tabular}
	\caption{Mean value of the difference plaquette $P_{d}$ \eqref{eq:plaq_dif} and average of all the plaquettes $P_{\mu
			\nu}$ with coupling constants $\beta=\{3,2.9,2.8\}$ and lattice sizes $N_0N_1N_2N_3$. \acrshort*{pccbc} are imposed in the transverse direction $N_3$. $N_{\text{therm}}$ and $N_{\text{measure}}$ are the \acrshort*{mc} steps used for thermalization and measuring respectively.}
	\label{tab:Energy_cond_1_31}
\end{table}
\begin{table}[H]
	\centering
	\begin{tabular}{|c|c|c|c|c|c|}
		\hline
		\rule{0pt}{12pt}$\beta$& $N_0\times N_1\times N_2\times N_3$ & $N_{\text{therm}}$& $N_{\text{measure}}$ &$\frac{1}{2}\braket{\text{tr }P_d}$ & $\frac{1}{2}\braket{\text{tr }P_{\mu\nu}}$\\
		\hline
		2.7&$48\times48\times48\times6$ &$10^5$ & $10^5$& 0.0400791(8) & 0.7239914(6)\\\hline
		2.7&$48\times48\times48\times7$ &$10^5$ & $10^5$& 0.0343671(7) & 0.7185152(5)\\\hline
		2.7&$48\times48\times48\times8$ &$1.6\cdot10^5$ & $2\cdot10^5$& 0.0300793(5) & 0.7144070(3)\\\hline
		2.7&$48\times48\times48\times9$ &$10^5$ & $10^5$& 0.0267402(6) & 0.7112094(5)\\\hline
		2.7&$48\times48\times48\times10$ &$1.6\cdot10^5$ & $2\cdot10^5$& 0.0240694(4) & 0.7086495(3)\\\hline
		2.7&$48\times48\times48\times11$ &$10^5$ & $10^5$& 0.0218825(6) & 0.7065553(4)\\\hline
		2.7&$48\times48\times48\times12$ &$10^5$ & $10^5$& 0.0200598(6) & 0.7048090(4)\\\hline
		2.7&$48\times48\times48\times14$ &$10^5$ & $2\cdot10^5$& 0.0171964(4) & 0.7020641(3)\\\hline
		2.7&$48\times48\times48\times15$ &$1.4\cdot10^5$ & $3\cdot10^5$& 0.0160500(3) & 0.7009656(2)\\\hline
		2.7&$48\times48\times48\times36$ &$10^5$ & $10^5$& 0.0066890(3) & 0.6919859(2)\\\hline
		2.6&$48\times48\times48\times5$ &$8\cdot 10^4$ & $5\cdot10^4$& 0.0503664(12) & 0.7191176(9)\\\hline
		2.6&$48\times48\times48\times6$ &$10^5$ & $10^5$& 0.0420146(8) & 0.7109840(6)\\\hline
		2.6&$48\times48\times48\times7$ &$10^5$ & $10^5$& 0.0360331(8) & 0.7051587(6)\\\hline
		2.6&$48\times48\times48\times8$ &$10^5$ & $10^5$& 0.0315372(7) & 0.7007840(5)\\\hline
		2.6&$48\times48\times48\times9$ &$10^5$ & $10^5$& 0.0280376(7) & 0.6973776(5)\\\hline
		2.6&$48\times48\times48\times10$ &$10^5$ & $10^5$& 0.0252373(6) & 0.6946494(5)\\\hline
		2.6&$48\times48\times48\times12$ &$10^5$ & $1.5\cdot10^5$& 0.0210356(5) & 0.6905512(4)\\\hline
		2.6&$48\times48\times48\times13$ &$1.4\cdot10^5$ & $3\cdot10^5$& 0.0194187(3) & 0.6889721(2)\\\hline
		2.6&$48\times48\times48\times14$ &$1.4\cdot10^5$ & $1.5\cdot10^5$& 0.0180332(4) & 0.6876164(3)\\\hline
		2.6&$48\times48\times48\times36$ &$1.5\cdot10^5$ & $2\cdot10^5$& 0.0070147(2) & 0.67685452(18)\\\hline
		2.509&$48\times48\times48\times5$ &$8\cdot10^4$ & $5\cdot10^4$& 0.0527189(13) & 0.7064288(9)\\\hline
		2.509&$48\times48\times48\times6$ &$10^5$ & $10^5$& 0.0439898(8) & 0.6977500(6)\\\hline
		2.509&$48\times48\times48\times7$ &$8\cdot10^4$ & $10^5$& 0.0377321(8) & 0.6915236(6)\\\hline
		2.509&$48\times48\times48\times8$ &$10^5$ & $10^5$& 0.0330300(7) & 0.6868374(6)\\\hline
		2.509&$48\times48\times48\times9$ &$10^5$ & $10^5$& 0.0293708(7) & 0.6831770(5)\\\hline
		2.509&$48\times48\times48\times10$ &$10^5$ & $10^5$& 0.0264421(7) & 0.6802340(5)\\\hline
		2.509&$48\times48\times48\times36$ &$10^5$ & $10^5$& 0.0073496(4) & 0.6610770(3)\\\hline
		2.427&$48\times48\times48\times4$ &$8\cdot10^4$ & $5\cdot10^4$& 0.0686400(15) & 0.7074482(11)\\\hline
		2.427&$48\times48\times48\times5$ &$8\cdot10^4$ & $5\cdot10^4$& 0.0551082(13) & 0.6936542(10)\\\hline
		2.427&$48\times48\times48\times6$ &$8\cdot10^4$ & $8\cdot10^4$& 0.0460031(10) & 0.6843312(7)\\\hline
		2.427&$48\times48\times48\times7$ &$10^5$ & $1.5\cdot10^5$& 0.0394778(7) & 0.6775870(5)\\\hline
		2.427&$48\times48\times48\times8$ &$8\cdot10^4$ & $1.2\cdot10^5$& 0.0345789(7) & 0.6724497(5)\\\hline
		2.427&$48\times48\times48\times36$ &$1.5\cdot10^5$ & $2\cdot10^5$& 0.0076946(3) & 0.6444156(2)\\\hline
	\end{tabular}
		\caption{Mean value of the difference plaquette $P_{d}$ \eqref{eq:plaq_dif} and average of all the plaquettes $P_{\mu
			\nu}$ with coupling constants $\beta=\{2.7,2.6,2.509,2.427\}$ and lattice sizes $N_0N_1N_2N_3$. \acrshort*{pccbc} are imposed in the transverse direction $N_3$. $N_{\text{therm}}$ and $N_{\text{measure}}$ are the \acrshort*{mc} steps used for thermalization and measuring respectively.}
	\label{tab:Energy_cond_2_31}
\end{table}

\section{Finite volume errors}
In this section we show the simulation results with the smaller lattices used in \autoref{sec:su31_lattice_effects} for checking error associated to the finite volume of the lattice for \acrshort*{pccbc} and \acrshort*{pbc}.
\begin{table}[H]
	\centering
	\begin{tabular}{|c|c|c|c|c|c|}
		\hline
		\rule{0pt}{12pt}$\beta$& $N_0\times N_1\times N_2\times N_3$ & $N_{\text{therm}}$& $N_{\text{measure}}$ &$\frac{1}{2}\braket{\text{tr }P_d}$ & $\frac{1}{2}\braket{\text{tr }P_{\mu\nu}}$\\
		\hline
		3&$32\times32\times32\times10$ &$8\cdot10^4$ & $10^5$& -0.0000232(11) & 0.7231609(7)\\\hline
		3&$32\times32\times32\times13$ &$8\cdot10^4$ & $1.5\cdot10^5$& -0.0000071(8) & 0.7231588(5)\\\hline
		3&$32\times32\times32\times16$ &$8\cdot10^4$ & $2\cdot10^5$& -0.0000026(6) & 0.7231586(4)\\\hline
		3&$32\times32\times32\times20$ &$8\cdot10^4$ & $3\cdot10^5$& -0.0000011(4) & 0.7231582(3)\\\hline
		2.7&$32\times32\times32\times5$ &$8\cdot10^4$ & $10^5$& -0.0004861(17) & 0.6857041(12)\\\hline
		2.7&$32\times32\times32\times6$ &$8\cdot10^4$ & $8\cdot10^4$& -0.0002124(17) & 0.6856607(12)\\\hline
		2.7&$32\times32\times32\times7$ &$8\cdot10^4$ & $10^5$& -0.0001092(14) & 0.6856380(10)\\\hline
		2.7&$32\times32\times32\times8$ &$8\cdot10^4$ & $10^5$& -0.0000601(13) & 0.6856251(10)\\\hline
		2.7&$32\times32\times32\times9$ &$8\cdot10^4$ & $1.2\cdot10^5$& -0.0000350(11) & 0.6856133(8)\\\hline
		2.7&$32\times32\times32\times10$ &$8\cdot10^4$ & $10^5$& -0.0000225(12) & 0.6856063(9)\\\hline
		2.7&$32\times32\times32\times12$ &$8\cdot10^4$ & $2.5\cdot10^5$& -0.0000083(7) & 0.6855952(5)\\\hline
		2.7&$32\times32\times32\times14$ &$8\cdot10^4$ & $3\cdot10^5$& -0.0000020(7) & 0.6855845(5)\\\hline
		2.509&$32\times32\times32\times4$ &$8\cdot10^4$ & $10^5$& -0.001365(2) & 0.6545487(15)\\\hline
		2.509&$32\times32\times32\times5$ &$8\cdot10^4$ & $1.4\cdot10^5$& -0.0004549(15) & 0.6542499(12)\\\hline
		2.509&$32\times32\times32\times6$ &$8\cdot10^4$ & $10^5$& -0.0001807(17) & 0.6540776(13)\\\hline
		2.509&$32\times32\times32\times7$ &$8\cdot10^4$ & $2\cdot10^5$& -0.0000715(11) & 0.6539287(8)\\\hline
		2.427&$32\times32\times32\times4$ &$8\cdot10^4$ & $8\cdot10^4$& -0.001291(2) & 0.6379980(19)\\\hline
		2.427&$32\times32\times32\times5$ &$8\cdot10^4$ & $1.4\cdot10^5$& -0.0003662(16) & 0.6372945(13)\\\hline
	\end{tabular}
	\caption{Mean value of the difference plaquette $P_{d}$ \eqref{eq:plaq_dif} and average of all the plaquettes $P_{\mu\nu}$ with coupling constants $\beta=\{3,2.7,2.509,2.427\}$ and lattice sizes $N_0N_1N_2N_3$. \acrshort*{pbc} are used in every direction. $N_{\text{therm}}$ and $N_{\text{measure}}$ are the \acrshort*{mc} steps used for thermalization and measuring respectively.}
	\label{tab:Energy_size_periodic_1_31}
\end{table}
\begin{table}[H]
	\centering
	\begin{tabular}{|c|c|c|c|c|c|}
		\hline
		\rule{0pt}{12pt}$\beta$& $N_0\times N_1\times N_2\times N_3$ & $N_{\text{therm}}$& $N_{\text{measure}}$ &$\frac{1}{2}\braket{\text{tr }P_d}$ & $\frac{1}{2}\braket{\text{tr }P_{\mu\nu}}$\\
		\hline
		3&$32\times32\times32\times10$ &$8\cdot10^4$ & $10^5$& 0.0212079(10) & 0.7429408(7)\\\hline
		3&$32\times32\times32\times12$ &$8\cdot10^4$ & $1.4\cdot10^5$& 0.0176726(8) & 0.7396447(6)\\\hline
		3&$32\times32\times32\times32$ &$8\cdot10^4$ & $3\cdot10^5$& 0.0066286(3) & 0.7293399(2)\\\hline
	\end{tabular}
	\caption{Mean value of the difference plaquette $P_{d}$ \eqref{eq:plaq_dif} and average of all the plaquettes $P_{\mu
			\nu}$ with coupling constant $\beta=3$ and lattice sizes $N_0N_1N_2N_3$. \acrshort*{pccbc} are imposed in the transverse direction $N_3$. $N_{\text{therm}}$ and $N_{\text{measure}}$ are the \acrshort*{mc} steps used for thermalization and measuring respectively.}
	\label{tab:Energy_size_cond_1_31}
\end{table}
\begin{table}[H]
	\centering
	\begin{tabular}{|c|c|c|c|c|c|}
		\hline
		\rule{0pt}{12pt}$\beta$& $N_0\times N_1\times N_2\times N_3$ & $N_{\text{therm}}$& $N_{\text{measure}}$ &$\frac{1}{2}\braket{\text{tr }P_d}$ & $\frac{1}{2}\braket{\text{tr }P_{\mu\nu}}$\\
		\hline
		2.7&$32\times32\times32\times6$ &$8\cdot10^4$ & $8\cdot10^4$& 0.0400773(16) & 0.7239913(11)\\\hline
		2.7&$32\times32\times32\times7$ &$8\cdot10^4$ & $1.2\cdot10^5$& 0.0343667(12) & 0.7185177(9)\\\hline
		2.7&$32\times32\times32\times8$ &$8\cdot10^4$ & $1.6\cdot10^5$& 0.0300772(10) & 0.7144075(7)\\\hline
		2.7&$32\times32\times32\times10$ &$8\cdot10^4$ & $1.8\cdot10^5$& 0.0240683(8) & 0.7086506(6)\\\hline
		2.7&$32\times32\times32\times12$ &$8\cdot10^4$ & $2\cdot10^5$& 0.0200586(7) & 0.7048102(5)\\\hline
		2.7&$32\times32\times32\times14$ &$8\cdot10^4$ & $3\cdot10^5$& 0.0171946(6) & 0.7020660(4)\\\hline
		2.7&$32\times32\times32\times16$ &$10^5$ & $3\cdot10^5$& 0.0150463(5) & 0.7000059(4)\\\hline
		2.7&$32\times32\times32\times32$ &$10^5$ & $3\cdot10^5$& 0.0075245(4) & 0.6927877(3)\\\hline
		2.427&$32\times32\times32\times4$ &$8\cdot10^4$ & $8\cdot10^4$& 0.068639(2) & 0.7074499(16)\\\hline
		2.427&$32\times32\times32\times5$ &$8\cdot10^4$ & $10^5$& 0.0551054(17) & 0.6936575(13)\\\hline
		2.427&$32\times32\times32\times6$ &$8\cdot10^4$ & $10^5$& 0.0460017(16) & 0.6843311(12)\\\hline
		2.427&$32\times32\times32\times7$ &$8\cdot10^4$ & $1.4\cdot10^5$& 0.0394755(13) & 0.6775914(10)\\\hline
		2.427&$32\times32\times32\times8$ &$8\cdot10^4$ & $1.6\cdot10^5$& 0.0345766(11) & 0.6724577(9)\\\hline
		2.427&$32\times32\times32\times32$ &$10^5$ & $2\cdot10^5$& 0.0086559(5) & 0.6454136(4)\\\hline
	\end{tabular}
	\caption{Mean value of the difference plaquette $P_{d}$ \eqref{eq:plaq_dif} and average of all the plaquettes $P_{\mu
			\nu}$ with coupling constants $\beta=\{2.7,2.427\}$ and lattice sizes $N_0N_1N_2N_3$. \acrshort*{pccbc} are imposed in the transverse direction $N_3$. $N_{\text{therm}}$ and $N_{\text{measure}}$ are the \acrshort*{mc} steps used for thermalization and measuring respectively.}
	\label{tab:Energy_size_cond_2_31}
\end{table}
    \end{appendices}
\printbibliography[heading=bibintoc]

\cleardoublepage
\phantomsection
\addcontentsline{toc}{chapter}{\listfigurename}
{   \hypersetup{linkcolor=black}
	\listoffigures
}

\cleardoublepage
\phantomsection
\addcontentsline{toc}{chapter}{\listtablename}
{   \hypersetup{linkcolor=black}
	\listoftables
}

\end{document}